\documentclass[11pt,a4paper]{oxarticle}

\pdfoutput=1

\usepackage[letterpaper, body={6.5in, 9.25in}, top=1.0in, left=1.5in]{geometry}
\usepackage{amssymb}

\usepackage{titlesec}
\titlespacing*{\subsubsection}
{0pt}{5pt plus 1pt minus 1pt}{5pt plus 1pt}

\usepackage[usenames, dvipsnames]{xcolor}
\usepackage{graphicx}
\usepackage{float}
\usepackage{amscd}
\usepackage{amssymb}
\usepackage{longtable}
\usepackage[bbgreekl]{mathbbol}
\usepackage[utf8]{inputenc}
\usepackage[sort&compress, comma, square, numbers]{natbib}

\usepackage[bottom]{footmisc}
\usepackage{bm}
\usepackage{tikz}
\usepackage[pdfpagelabels=false]{hyperref}
\hypersetup{
	colorlinks,
	linkcolor={blue!50!black},
	citecolor={blue!50!black},
	urlcolor={blue!50!black}
}

\newcommand\nnfootnote[1]{%
  \begin{NoHyper}
  \renewcommand\thefootnote{}\footnote{#1}%
  \addtocounter{footnote}{-1}%
  \end{NoHyper}
}

\usepackage{tabularx}
\usepackage{caption}
\usepackage{cellspace}
\DeclareSymbolFontAlphabet{\mathbb}{AMSb}
\DeclareSymbolFontAlphabet{\mathbbl}{bbold}

\DeclareFontFamily{OT1}{pzc}{}
\DeclareFontShape{OT1}{pzc}{m}{it}{<-> s * [1.200] pzcmi7t}{}
\DeclareMathAlphabet{\mathpzc}{OT1}{pzc}{m}{it}

\newcommand{\cD}{\mathcal{D}}
\newcommand{\cE}{\mathcal{E}}

\newcommand{\cH}{\mathcal{H}}

\newcommand{\cL}{\mathcal{L}}
\newcommand{\cM}{\mathcal{M}}
\newcommand{\cN}{\mathcal{N}}
\newcommand{\cO}{\mathcal{O}}
\newcommand{\cP}{\mathcal{P}}

\newcommand{\cU}{\mathcal{U}}

\newcommand{\cW}{\mathcal{W}}

\DeclareFontFamily{U}{bbold}{}
\DeclareFontShape{U}{bbold}{m}{n}
{  <-5.5> s*[1.05] bbold5
	<5.5-6.5> s*[1.05] bbold6
	<6.5-7.5> s*[1.05] bbold7
	<7.5-8.5> s*[1.05] bbold8
	<8.5-9.5> s*[1.05] bbold9
	<9.5-11.5> s*[1.05] bbold10
	<11.5-16> s*[1.05] bbold12
	<16-> s*[1.05] bbold17
}{}

\newcommand{\IC}{\mathbb{C}}

\newcommand{\IE}{\mathbb{E}}
\newcommand{\IF}{\mathbb{F}}

\newcommand{\IH}{\mathbb{H}}
\newcommand{\II}{\mathbb{I}}

\newcommand{\IK}{\mathbb{K}}

\newcommand{\IP}{\mathbb{P}}
\newcommand{\IQ}{\mathbb{Q}}

\newcommand{\IS}{\mathbb{S}}

\newcommand{\IZ}{\mathbb{Z}}

\font\elevenrmfromseventeenrm = cmr17 at 11pt
\newcommand{\inbar}{\vrule height6.9pt depth-0.2pt width0.35pt}
\newcommand{\zero}{\hbox{{\elevenrmfromseventeenrm 0}\kern-3.5pt\inbar\kern1pt\inbar\kern2pt}}

\font\eightrmfromseventeenrm = cmr17 at 8pt
\newcommand{\ssinbar}{\vrule height5pt depth-0.1pt width0.3pt}
\newcommand{\sszero}{\hbox{{\eightrmfromseventeenrm 0}\kern-2.53pt\ssinbar\kern0.7pt\ssinbar\kern2pt}}

\newcommand{\mtB}{\text{B}}

\newcommand{\mtE}{\text{E}}
\newcommand{\mtF}{\text{F}}

\newcommand{\mtI}{\text{I}}

\newcommand{\mtM}{\text{M}}

\newcommand{\mtS}{\text{S}}
\newcommand{\mtT}{\text{T}}
\newcommand{\mtU}{\text{U}}
\newcommand{\mtV}{\text{V}}
\newcommand{\mtW}{\text{W}}

\newcommand{\1}{\mathbf{1}}

\font\eightrm=cmr8 at 8pt
\font\csc=cmcsc10

\newcommand{\beq}{\begin{equation}}
\newcommand{\eeq}{\end{equation}}
\newcommand{\beqnn}{\begin{equation*}}
\newcommand{\eeqnn}{\end{equation*}}
\newcommand{\bea}{\begin{eqnarray}}
\newcommand{\eea}{\end{eqnarray}}
\newcommand{\bean}{\begin{eqnarray*}}
	\newcommand{\eean}{\end{eqnarray*}}

\newcommand{\cicy}[2]{\begin{matrix} #1\end{matrix}\!\left[\begin{matrix}#2 \end{matrix}\right]}

\newcommand{\defineas}{:=}

\newcommand{\sla}[4]{{\left. #1 \right|_{#2}^{#3}\!{#4}}}

\newcommand{\place}[3]{\vbox to0pt{\kern-\parskip\kern-7pt
		\kern-#2truein\hbox{\kern#1truein #3}
		\vss}\nointerlineskip}
\newcommand{\capt}[3]{\parbox{#1}{\renewcommand{\baselinestretch}{1.0}
		\caption{\label{#2}\small\it #3}}}

\newcommand{\+}{\phantom{-}}
\renewcommand{\=}{\;=\;}

\newcommand{\fref}[1]{figure~\ref{#1}}
\newcommand{\tref}[1]{table~\ref{#1}}

\newcommand{\cref}[1]{Chapter \ref{#1}}

\DeclareFontFamily{U}{wncy}{}
\DeclareFontShape{U}{wncy}{m}{n}{<->wncyr10}{}
\DeclareSymbolFont{mcy}{U}{wncy}{m}{n}
\DeclareMathSymbol{\sha}{\mathord}{mcy}{"58}

\newcommand{\Tr}{\text{Tr}}
\newcommand{\SU}{\text{SU}}

\newcommand{\SL}{\text{SL}}
\newcommand{\GL}{\text{GL}}

\newcommand{\Teich}{\text{Teich}}
\newcommand{\Frob}{\text{Frob}}

\renewcommand{\Im}{\text{Im}}
\renewcommand{\Re}{\text{Re}}

\newcommand{\diag}{\text{diag}}

\newcommand{\HV}{{\text{HV}}}
\newcommand{\MHV}{{\text{H}\Lambda}}

\newcommand{\Fr}{\text{Fr}}

\newcommand{\me}{\text{e}}

\newcommand{\ii}{\text{i}}
\newcommand{\dd}{\text{d}}

\newcommand{\wt}[1]{\widetilde{#1}}

\makeatletter
\g@addto@macro\bfseries{\boldmath}

\def\blindfootnote{\xdef\@thefnmark{}\@footnotetext}
\makeatother

\numberwithin{equation}{section}

\newcommand{\dR}{\text{dR}}
\newcommand{\BB}{\text{B}}
\newcommand{\et}{\text{ét}}

\newcommand{\AK}{\text{AK3}}
\newcommand{\bmlog}{\text{\textbf{log}}\;}

\newcommand{\tablepreambleFourfold}[1]{
	\vspace{-0.2cm}
	\begin{center}
		\begin{longtable}{| >{\footnotesize$~} c <{~$} | >{~\footnotesize$~} c <{~$} |>{\centering\footnotesize $}p{3.7in}<{$}|}\hline
			p & \varphi & R_H^{(p)}(\HV_\varphi/\IZ_6,T) \tabularnewline[0.5pt] \hline\hline
			\endfirsthead
			\hline
			p & \varphi & R_H^{(p)}(\HV_\varphi/\IZ_6,T) \tabularnewline[0.5pt] \hline\hline
			\endhead
			\hline\hline 
			\multicolumn{3}{|r|}{{\footnotesize\sl Continued on the following page}}\tabularnewline[0.5pt] \hline
			\endfoot
			\hline
			\endlastfoot}
		
\newcommand{\tablepostamble}{\end{longtable}\end{center}}

\begin{document}
	\proofmodefalse
	
	\thispagestyle{empty}  
    \begin{flushright}MITP--23--071\end{flushright}
 
	\begin{center}
		\null\vskip0.2in
        {\Huge Modular Calabi--Yau Fourfolds \\[12pt]
        and Connections to M-Theory Fluxes}
\vskip1cm
{\csc Hans Jockers${}^1$, Sören Kotlewski${}^2$, and Pyry Kuusela${}^3$ \\[2cm]}

\nnfootnote{${}^1\,$jockers@uni-mainz.de \hfil
${}^2\,$soeren-kotlewski@t-online.de\hfil
${}^3\,$pyry.r.kuusela@gmail.com\hfil }

{\it PRISMA+ Cluster of Excellence \& Mainz Institute for Theoretical Physics\\ Johannes Gutenberg-Universität Mainz\\
55099 Mainz, Germany\\[1cm]}

\vfill
\textbf{Abstract}
\end{center}
\vskip-5pt
\begin{minipage}{\textwidth}
\baselineskip=15pt
\noindent 
In this work, we study the local zeta functions of Calabi--Yau fourfolds. This is done by developing arithmetic deformation techniques to compute the factor of the zeta function that is attributed to the horizontal four-form cohomology. This, in turn, is sensitive to the complex structure of the fourfold. Focusing mainly on examples of fourfolds with a single complex structure parameter, it is demonstrated that the proposed arithmetic techniques are both applicable and consistent. We present a Calabi--Yau fourfold for which a factor of the horizontal four-form cohomology further splits into two pieces of Hodge type $(4,0)+(2,2)+(0,4)$ and $(3,1)+(1,3)$. The latter factor corresponds to a weight-3 modular form, which allows expressing the value of the periods in terms of critical values of the $L$-function of this modular form, in accordance with Deligne's conjecture. The arithmetic considerations are related to M-theory Calabi--Yau fourfold compactifications with background four-form fluxes. We classify such background fluxes according to their Hodge type. For those fluxes associated to modular forms, we express their couplings in the low-energy effective action in terms of $L$-function values.

\vspace*{30pt}
\textit{December 2023}
\end{minipage}

\clearpage
\thispagestyle{empty}	

{\baselineskip=11pt
	\tableofcontents} 

\newpage
\thispagestyle{empty}
\section*{Notation}
\vskip-10pt
{
	\renewcommand{\arraystretch}{1.35}
	\centering
	\begin{tabularx}{0.99\textwidth}{|>{\hsize=.16\hsize\textwidth=\hsize}X|
			>{\hsize=0.76\hsize\textwidth=\hsize}X|>{\hsize=0.07\hsize\textwidth=\hsize}X|}
		\hline
		\textbf{Symbol} & \hfil\textbf{Definition/Description}\hfil & \hfil \textbf{Ref.}\\[3pt]
		\hline	\hline
		$\bm \varphi$
		& The coordinates $(\varphi_{1},\dots,\varphi_{h^{1,3}})$ on the complex structure space of a Calabi--Yau manifold $X_{\bm \varphi}$.
		& \eqref{eq:HVlocaleq} 	\\[4pt] \hline
		$\bm t$ & The K\"ahler parameters $(t_1,\ldots,t_{h^{1,1}})$ of a Calabi--Yau manifold $X$.& \eqref{eq:Kahler_class_HV} \\[4pt] \hline
		$\varpi$& The period vector of $X_{\bm \varphi}$ in the Frobenius basis.&\eqref{eq:frobenius_basis_definition}\\[4pt] \hline
		$\Pi$& The period vector of $X_{\bm \varphi}$ in the rational B-brane basis.&\eqref{eq:definiton_intergal_periods}\\[4pt] \hline
         $\theta$ & The logarithmic derivative $\varphi \partial_\varphi$.  & \eqref{eq:H_I_definition}\\[4pt] \hline
         $H^4_I(X,\IC)$ &The subspace $ H_I^4(X,\IC) \defineas \langle \partial \Omega \rangle_{\partial \in I}$ of the middle cohomology $H^4(X,\IC)$ of the Calabi--Yau fourfold $X$ which is generated by the action of the differential ideal $I$ on the holomorphic $(4,0)$-form $\Omega$. & \eqref{eq:H_I_definition} \\[4pt] \hline
	    $H^4_H(X,\IC)$ & The horizontal cohomology of the Calabi--Yau fourfold $X$. That is, the subspace of the horizontal cohomology defined by $ H_H^4(X,\IC) \defineas H^4_J(X,\IC)$ where $J = \langle \partial_{\varphi_1},\dots,\partial_{\varphi_{h^{1,3}}} \rangle$.
     & \eqref{eq:H_H_definition} \\[4pt] \hline
		$\IF_{p^n}$ & The finite field with $p^n$ elements. & \eqref{eq:zeta_definition}\\[4pt] \hline
       $\Fr_p|H^m(X)$ & The map $\Fr_{p^n}|H^m(X) : H^m(X) \to H^m(X)$ induced by the Frobenius map $\Frob_{p^n}: (x_1,\dots,x_k) \mapsto (x_1^{p^n},\dots,x_k^{p^n})$ on the ambient space. & \eqref{eq:definition_Fr} \\[4pt] \hline
        $\Teich_p(x)$ &The Teichm\"uller representative of the integer $x\in \IF_p$. & \eqref{eq:Teichmueller_property} \\[4pt] \hline
		$\zeta_p(X_{\bm \varphi},T)$ & The local zeta function of a Calabi--Yau manifold $X_{\bm \varphi}$.	& \eqref{eq:zeta_definition}
		\\[4pt] \hline
		$\mtU_p(\varphi)$ & The matrix representing the action of the inverse Frobenius map on the space $H_I^4(X_{\varphi}) \subset H^4(X_\varphi)$ of a Calabi--Yau manifold $X_\varphi$. &\eqref{eq:R_H_definition} \\[4pt] \hline
  		$R_I^{(p)}(X_\varphi,T)$ & The characteristic polynomial of $\mtU_p(\varphi)$ appearing in the the zeta function $\zeta_p(X_\varphi,T)$.& \eqref{eq:R_H_definition}	\\[4pt] \hline
        $\sigma$ & The matrix representation of the intersection product $\int_{X_\varphi} \theta^i\Omega \wedge \theta^j \Omega$ in the Frobenius basis. & \eqref{eq:wedge_product} \\[4pt] \hline     
        $\mtE(\varphi)$ & The matrix of periods of $X_\varphi$ and their derivatives in the Frobenius basis. & \eqref{eq:E_matrix_definition} \\[4pt] \hline
        $\HV_{\bm \varphi}$ & The Hulek--Verrill fourfold with complex structure moduli $\bm \varphi = (\varphi_1,\dots,\varphi_6)$. & \eqref{eq:HVlocaleq} \\[4pt] \hline
        $\MHV_{\bm t}$ & The mirror Hulek--Verrill fourfold with the Kähler structure specified by $\bm t = (t_1,\dots,t_6)$. & \eqref{eq:MirrorHV_iso}\\[4pt] \hline
        $\mathcal{E}_{(\varphi_0,\varphi_1,\varphi_2)}$ &The elliptic curve of Hulek--Verrill type which is characterised by the complex structure moduli $(\varphi_0,\varphi_1,\varphi_2)$.& \eqref{eq:definition_fibres} \\[4pt] \hline
		$\text{K3}_{\left(\varphi_0,\varphi_1,\varphi_2,\varphi_3\right)}$ &The polarised $K3$ surface of Hulek--Verrill type which is characterised by the complex structure moduli $(\varphi_0,\varphi_1,\varphi_2,\varphi_3)$.& \eqref{eq:definition_fibres} \\[4pt] \hline
        $c^\pm(M)$ &Deligne's periods of the motive $M$. & \eqref{eq:Deligne's_period_definition}\\[4pt] \hline
        $L(M,s)$ & The $L$-function associated to the motive $M$.  & \eqref{eq:definition_Lfunction}\\[4pt] \hline
        
	\end{tabularx}
}
\newpage

\setcounter{page}{1}
\section{Introduction} \label{sect:Introduction}
\vskip-10pt
In this work, we analyse the arithmetic properties of Calabi--Yau fourfolds. The motivation of our analysis is two-fold. From the physics perspective the Hodge theoretic decomposition of the cohomology classes of the Calabi--Yau fourfolds is an important ingredient in determining the low energy effective theory of a Calabi--Yau fourfold compactifications in the context of string theory, M-theory or F-theory. It has been noted --- see, e.g., refs.~\cite{Moore:1998pn,Moore:1998zu,Candelas:2007mb,Candelas:2019llw,Kachru:2020abh,Candelas:2021tqt,Bonisch:2022slo} --- that studying the arithmetic properties of compactification spaces is a powerful tool to gain interesting Hodge theoretic information that effects the resulting low energy effective description of the analysed compactification scenario. From the mathematical point of view, while there are already quite a few results on the arithmetic properties of K3~surfaces (see appendix \ref{app:K3_modularity} and, for instance, the reviews \cite{Yui2003a,Yui2012a} and references therein) and Calabi--Yau threefolds (for example refs. \cite{Candelas:2007mb,Candelas:2019llw,Candelas:2021tqt,Candelas:2023yrg,Schimmrigk:2020dfl}), the arithmetic of Calabi--Yau fourfolds remains rather unexplored. 

For Calabi--Yau threefolds~$Y$, arithmetic considerations have proven to be a powerful method to find \textit{attractor points} in its complex structure moduli space \cite{Candelas:2007mb,Candelas:2019llw}. That is to say, to find loci in the complex structure moduli space where a homology three-cycle in $H_3(Y,\mathbb{Z})$ is Poincar\'e dual to a three-form cohomology class in $H^3(Y,\mathbb{Z}) \cap \left( H^{(3,0)}(Y,\IC) \oplus H^{(0,3)}(Y,\IC) \right)$. An attractor point is said to be of \textit{rank two} if there exist two independent such cycles. Hodge theoretically this implies that the Hodge structure splits, meaning that there exist two subspaces $V,V' \subset H^3(Y, \IQ)$ with $V$ of Hodge type $(3,0)+(0,3)$ and $V'$ of type $(2,1)+(1,2)$, such that the third de Rham cohomology group can be written as a direct sum
\begin{align} \label{eq:Hodge_Split}
H^3(Y,\IQ) \= V \oplus V'~.
\end{align}
Such loci in the moduli space of Calabi--Yau threefolds give rise to additional non-perturbative BPS states in the spectrum of the type~II string compactifications \cite{Moore:1998pn,Moore:1998zu}.

Similar considerations are also important in the context of string compactifications with background fluxes (e.g. refs. \cite{Candelas:2023yrg,Kachru:2020abh,Kachru:2020sio,Candelas:2019llw,Schimmrigk:2020dfl}), which in type~II string theories are described in terms of the effective superpotential \cite{Gukov:1999ya,Taylor:1999ii}
$$
W \= \int_Y \Omega \wedge G \ ,
$$
where $\Omega$ is the holomorphic $(3,0)$-form of the Calabi--Yau threefold and the background flux $G$ is a three-form cohomology class. Analogous to rank-two attractor points, for some families, there exist special points in moduli space, where the middle cohomology splits similarly to \eqref{eq:Hodge_Split}, but with one of the subspaces being two-dimensional of Hodge type $(1,2)+(2,1)$. At these points the background flux $G$ can be taken to be of this Hodge type and the low-energy effective theory admits supersymmetric vacua.

We search for analogous points in the complex structure moduli space of Calabi--Yau fourfolds $X$, where a homology four-cycle in $H_4(X,\mathbb{Z})$ is Poincar\'e dual to a four-form cohomology class in $H^4(X,\mathbb{Z}) \cap \left( H^{(4,0)}(X,\IC) \oplus H^{(2,2)}(X,\IC) \oplus H^{(0,4)}(X,\IC) \right)$. In such a case, we can find an integral four-form cohomology class $G$ of Hodge type $(4,0)+(2,2)+(0,4)$. That is, 
\begin{align}
G \in H^4(X,\mathbb{Z}) \cap \left( H^{(4,0)}(X,\IC) \oplus H^{(2,2)}(X,\IC) \oplus H^{(0,4)}(X,\IC) \right)~.
\end{align}
In physics, such loci are relevant for instance in M-theory and F-theory compactifications on a Calabi--fourfold $X$ with four-form background fluxes. Specifically, the three-dimensional low-energy effective $\mathcal{N}=2$ supergravity action for M-theory compactifications on a Calabi--Yau fourfold $X$ with background fluxes is derived in refs.~\cite{Haack:2001jz,Berg:2002es}. In particular, the background four-form flux~$G$ induces the superpotential~\cite{Gukov:1999ya}
\begin{equation} \label{eq:W}
W(\varphi) = \int_X \Omega(\varphi) \wedge G \ .
\end{equation}
Here $\Omega(\varphi)$ is the holomorphic $(4,0)$-form of the Calabi--Yau fourfold which varies as a function of the complex structure moduli $\varphi$. The background flux $G$ is a quantised four-form cohomology class, which obeys the quantisation condition \cite{Witten:1996md}
\begin{equation} \label{eq:Gquant}
\left[ G - \frac{c_2(X)}{2} \right] \, \in \, H^4(X,\mathbb{Z}) \ ,
\end{equation}
where $c_2(X)$ is the second Chern class of the fourfold $X$. In the associated three-dimensional low energy effective $\mathcal{N}=2$ supergravity theory the flux-induced superpotential $W(\varphi)$ yields the contribution $V_W(\varphi)$ to the scalar potential \cite{Haack:2001jz,Berg:2002es,deWit:2003ja}
\begin{equation} \label{eq:scalpot}
V_W(\varphi,\bar\varphi) = e^{-K(\varphi,\bar \varphi)} \left( g^{i\bar\jmath} \cD_i W(\varphi) \cD_{\bar\jmath} \overline{W(\varphi)} - 4 |W(\varphi)|^2 \right) \ ,
\quad K(\varphi,\bar \varphi) = \log \int \overline{\Omega(\varphi)} \wedge \Omega(\varphi) \ ,
\end{equation} 
where $K(\varphi,\bar \varphi)$ is the K\"ahler potential of the complex structure moduli space $\cM_{\IC S}$ of $X$, $g^{i\bar\jmath}$ its inverse K\"ahler metric, and $\cD_i$ the K\"ahler covariant derivative. The quantised fluxes $G$ of Hodge type $(4,0)+(2,2)+(0,4)$ minimise the potential $V_W(\varphi,\bar\varphi)$ at the negative value~$- 4\, e^{-K} |W(\varphi)|^2$. Nevertheless, $G$~fluxes of this Hodge type give rise to supersymmetric Minkowski vacua \cite{Becker:1996gj,Haack:2001jz}, because the three-dimensional low energy effective $\mathcal{N}=2$ supergravity theory has a no-scale structure due to an additional contribution to the scalar potential that compensates the negative value of $V_W(\varphi,\bar\varphi)$ at its minimum \cite{Haack:2001jz}.

Calabi--Yau fourfolds for which a homology cycle of the required type exists can be found in cases where the Hodge structure splits over $\IQ$ so that there exists a subspace $V \subset H^4(X, \IQ)$ with $V$ being of Hodge type $(4,0)+(2,2)+(0,4)$. For Calabi--Yau fourfolds with one complex structure parameter, this implies the existence of a two-dimensional subspace $V' \subset H^4(X, \IQ)$ of type $(3,1)+(1,3)$. If a Calabi--Yau fourfold has such a subspace $V'$, we call the corresponding points in the complex structure moduli space \textit{attractive K3 (\AK{}) points}. This is owing to the fact that the arithmetic properties of such fourfolds are closely related to the arithmetic properties of attractive (also called singular) K3~surfaces.\footnote{In mathematics literature attractive K3~surfaces are usually called singular. This just means that they have Picard number~$\rho(S) = 20$. Such surfaces are still smooth manifolds. However, in this paper we use the terminology \textit{attractive} proposed by Moore~\cite{Moore:1998pn}.}

We search for attractive K3 points by applying and developing arithmetic techniques. Namely, for a Calabi--Yau fourfold $X$ defined over the field of rational numbers $\mathbb{Q}$, one can define a local zeta function $\zeta_p(X,T)$ for each prime $p$. This is the generating function for the number of solutions of the associated Calabi--Yau fourfolds $X/\mathbb{F}_{p^n}$ defined over the finite fields $\mathbb{F}_{p^n}$ for all $n$. As a consequence of the Weil conjectures \cite{Weil1949a} --- later proven by Dwork \cite{Dwork1960a}, Grothendieck \cite{Grothendieck1995a}, and Deligne \cite{Deligne1974a,Deligne1980a} --- the local zeta functions of Calabi--Yau fourfolds are rational functions in the formal variable~$T$. For the type of Calabi--Yau fourfolds discussed in this work, the local zeta functions take the general form\footnote{In this work, we analyse compact Calabi--Yau fourfolds $X$ for which the third Betti number $b_3(X)$ vanishes such that the numerator of the local zeta function $\zeta_p(X,T)$ becomes one, and for which the Picard group of $X$ is generated by divisors defined over the field $\mathbb{Q}$.}
\begin{equation} \label{eq:zeta_simplified}
\zeta_p(X,T) = \frac{1}{(1-T) (1 - pT)^{b_2(X)} R^{(p)}_4(X,T) (1-p^3 T)^{b_2(x)}(1 - p^4T)} \ ,
\end{equation}
where $b_2(X)$ is the second Betti number of $X$, and  $R^{(p)}_4(X,T)$ is a polynomial in $T$ of degree $b_4(X)$, which is the fourth Betti number of $X$. 

In refs.~\cite{Candelas:2021tqt,multiparameter_zeta} methods are developed for computing the zeta functions of Calabi--Yau threefolds numerically. These are based on the fact that the polynomials $R_n^{(p)}(X,T)$ can be expressed as characteristic polynomials of the so-called Frobenius map acting on the $n$'th cohomology of $X$. The action of the Frobenius map can be explicitly written down as a matrix by using periods of the manifold. For a generic Calabi--Yau fourfold the dimension of the middle cohomology is too large to be able to use the fourfold analogues of these methods to compute the polynomials $R^{(p)}_4(X,T)$ in practice. To make progress, we study subspaces $H^4_I(X,\IC)$ of the middle cohomology $H^4(X,\IC)$ of the Calabi--Yau fourfold $X$ that are generated by the action of the Picard--Fuchs differential ideal $I$ on the holomorphic $(4,0)$-form $\Omega$, i.e., 
\begin{align}
H_I^4(X,\IC) \defineas \langle \partial \Omega \rangle_{\partial \in I}~.  
\end{align}
We make the assumption that the Frobenius map has a well-defined action on these subspaces $H_I^4(X,\IC)$. In particular, this assumption means that the Frobenius map has a well-defined action on the horizontal piece $H^4_H(X,\IC)$ of the middle cohomology, by which we mean the subspace generated by the derivatives of the holomorphic $(4,0)$-form $\Omega$ with respect to all complex structure moduli. In this paper, we mainly focus on principal differential ideals~$I$ generated by the logarithmic derivative $\theta \defineas \varphi \partial_\varphi$ with respect to a single complex structure modulus $\varphi$.

We are able to make extensive consistency checks on this assumption: By concentrating on these lower-dimensional subspaces, we are able to compute the corresponding characteristic polynomials $R_I^{(p)}(X,T)$. According to our assumption, these should appear as factors of the full polynomial $R^{(p)}_4(X,T)$, so the Weil conjectures governing the behaviour of the zeta function have several implications for the properties of the polynomials $R_I^{(p)}(X,T)$. In all the cases that we have studied, all of these properties are satisfied, lending credence to the assumption we have made. Further evidence is obtained, for a particular manifold, by studying the modularity properties of the polynomial when it factorises further.

Of further interest is that many of the modularity considerations, such as the computation of Deligne's periods, depend on the rational structure of the horizontal piece of the middle cohomology (or in the multiparameter case, the relevant subspace $H_I^4(X,\IC)$). It is not clear that the decomposition of $H^4(X,\IC)$ into the horizontal part and the remainder $H^4(X,\IC) = H_H^4(X,\IC) \oplus H_\perp^4(X,\IC)$ is defined over $\IQ$. However, mirror symmetry arguments can be used to find an integral basis of periods, which gives rise to a natural $\IQ$ structure on $H^4_H(X,\IC)$. We can use this structure to compute Deligne's periods, and find that they take the conjectured form. We view this as an indication that the $\IQ$-structure given by mirror symmetry argument is in fact the correct $\IQ$ structure to use from the point-of-view of arithmetic geometry as well.

The subspace $H_I^4(X,\IC)$ may even split further over $\IQ$. Such splittings are expected to be reflected as persistent factorisations of the polynomial $R^{(p)}_I(X,T)$ over $\IQ$. Conversely, we examine such factorisations of $R^{(p)}_I(X,T)$ over $\IQ$ in order to look for rational splittings of $H_I^4(X,\IC)$, which are, in turn, relevant for the construction of M-theory fluxes of a particular Hodge type.

In this paper, we discuss Calabi--Yau fourfolds of Hodge types $(1,1,1,1,1)$ and $(1,1,2,1,1)$.\footnote{In this notation the vector $(i_0,\dots,i_4)$ collects the dimensions $i_k \defineas \dim_\IC H_I^4(X,\IC) \cap H^{(4-k,k)}(X,\IC)$.} These correspond to Calabi--Yau fourfolds with a single complex structure modulus, i.e., with $h^{3,1}=1$. Furthermore, they are governed --- in the former case --- by a five-dimensional or --- in the latter case --- by a six dimensional Hodge structure, which respectively involves a single $(2,2)$-form cohomology class or two linearly-independent $(2,2)$-form cohomology classes.\footnote{In the physics context, the relevance of these two Hodge structures is discussed, for instance, in refs.~\cite{Honma:2013hma,Gerhardus:2015sla,Gerhardus:2016iot}.} Because the degree of the polynomial $R_H^{(p)}(X,T)$ reflects the dimensionality of the Hodge structure, it is either of degree five or six, respectively, in these cases. 

For all examined Calabi--Yau fourfolds of Hodge type $(1,1,1,1,1)$, the polynomial $R_H^{(p)}(X,T)$ has always a linear factor $(1\pm p^2 T)$ when factorised over $\IQ$, for any prime $p$. This is a direct consequence of the form the zeta function takes for Calabi--Yau fourfolds, and does not indicate existence of an attractor or an \AK{} point in complex structure moduli space. However, among the examined Calabi--Yau fourfolds of this type, we find that the one-parameter Calabi--Yau fourfold $\text{HV}/\mathbb{Z}_6$ --- which is a certain $\mathbb{Z}_6$ orbifold of the Hulek--Verrill fourfold \cite{Hulek2005a, Candelas:2021lkc} --- has a point in the complex structure moduli space, where the polynomial $R_H^{(p)}(\text{HV}/\mathbb{Z}_6,T)$ further factorises over~$\IQ$~to
\begin{equation} \label{eq:RHfac}
R_H^{(p)}(\text{HV}/\mathbb{Z}_6,T) = (1 \pm p^2 T) (1 - c_p T + p^4 T^2) (1 - b_p p T + p^4 T^2)\ .
\end{equation}
We propose that this factorisation indeed furnishes an \AK{} point in the complex structure moduli space. By studying the geometric structure of the Calabi--Yau fourfold $\text{HV}/\mathbb{Z}_6$ in detail, we give geometric and arithmetic evidence for this proposal. In particular, we argue that, in agreement with Serre's modularity conjecture \cite{Serre1975a,Serre1987a}, the factor $(1- b_p p T + p^4T^2)$ in eq.~\eqref{eq:RHfac} corresponds to a weight-3 modular form, which we are able to explicitly identify. We test the identification by comparing the coefficients $b_p$ to Fourier coefficients of the modular form, and find a perfect agreement to at least 130 first primes.

By numerical analytic continuation, we are able to explicitly see a splitting of the Hodge structure analogous to \eqref{eq:Hodge_Split}, with a two-dimensional subspace of Hodge type $(3,1)+(1,3)$ generated by $\cD \Pi$ and $\overline{\cD \Pi}$. In such a situation Deligne's conjecture \cite{Deligne1979a} can be used to obtain a prediction of relation between the period vector $\cD \Pi$ and the critical $L$-function values $L_3(1)$ and $L_3(2)$, associated to the weight-three modular form. We also compare the Calabi--Yau periods to periods and quasi-periods of the modular form along the lines of ref.~\cite{Bonisch:2022mgw}. Analogously to the threefold cases studied in refs.~\cite{Candelas:2019llw,Kachru:2020abh,Yang:2019kib,Yang:2020lhd,Yang:2020sfu,Candelas:2023yrg}, and in accordance with the conjecture, we find that, to at least 100 digits, the following beautiful identity~holds
\begin{align}
\cD \Pi(1) \= -\frac{3}{28} \left(2 \; \frac{L_3(1)}{\pi^2} \left(
\begin{array}{r}
	12 \\
	5 \\
	20 \\
	-50 \\
	10 \\
\end{array}
\right) + \ii \frac{L_3(2)}{\pi^3} \left(
\begin{array}{r}
	0 \\
	5 \\
	24 \\
	-80 \\
	20 \\
\end{array}
\right) \right)~.
\end{align}
Geometric considerations provide yet more evidence in favour of modularity: we argue that the factor $(1- b_p p T + p^4T^2)$ in eq.~\eqref{eq:RHfac} appears also (up to a Tate twist amounting to a rescaling $T \mapsto T/p$) in the local zeta function of a an attractive K3~surface. This modular surface turns out to be embedded into the fourfold $\text{HV}_1/\mathbb{Z}_6$ corresponding to the \AK{} point $\varphi=1$ in the complex structure moduli space $\cM_{\IC S}(\HV/\IZ_6)$, making it natural to speculate that existence of this surface is the geometric origin of modularity in this~case. The modularity of this K3 surface was also already noted in ref.~\cite{Bonisch:2020qmm}.

For the examined Calabi--Yau fourfolds of Hodge type $(1,1,2,1,1)$, we find for some choices of complex structure moduli sporadic factorisations of the $R_H^{(p)}(X,T)$ over $\mathbb{Q}$. However, as these factorisations are not universal for (almost) all primes $p$, we do not find any evidence for attractor or \AK{} points in these examples. 

The structure of this paper is as follows: In section \ref{sect:Deformation_method}, we recall some of the central properties of the local zeta function $\zeta_p(X,T)$, and introduce the deformation method for computing the factors $R_I^{(p)}(X,T)$ in the rational expression for $\zeta_p(X,T)$. We give a detailed algorithm for computing the polynomials $R_H^4(X,T)$ for one-parameter Calabi--Yau fourfolds of Hodge type $(1,1,1,1,1)$, delegating the discussion of the type $(1,1,2,1,1)$ manifolds to the appendix \ref{app:(1,1,2,1,1)}. In the following section \ref{sect:Hulek--Verrill_Fourfolds}, we introduce Hulek--Verrill fourfolds their $\IZ_6$ quotients, and discuss their geometry in some detail before moving on to their arithmetic properties. Using this example, we also explore the relation of mirror symmetry and Deligne's conjecture, and speculate on the possible geometric origin of the modularity properties that we observe for the Hulek--Verrill manifolds. 

In section \ref{sect:physics}, the arithmetic considerations of the previous section are related to the physics of flux compactifications of M-theory. We discuss how the factorisations of the polynomials $R_I^{(p)}(X,T)$ can be used to find non-trivial geometries that support fluxes of different Hodge types, and how these give rise to different types of low-energy effective theories. We also discuss how number theory conjectures relate some central physical quantities in these cases to number theoretically interesting objects, such as $L$-functions and their critical values. We end with a brief summary in section \ref{sect:Conclusions} and speculate on interesting directions for future research. 

Matters that would otherwise disrupt the narrative are relegated to appendices. In appendices \ref{app:fourfolds_cohomology} and \ref{app:modularity} we give brief reviews of the middle cohomology of Calabi--Yau fourfolds and modularity, respectively. In appendix \ref{app:(1,1,2,1,1)}, we add to the discussion in section \ref{sect:Deformation_method} by presenting the algorithm for computing the polynomials $R_H^4(X,T)$ appearing in the local zeta functions of Calabi--Yau fourfolds of Hodge type $(1,1,2,1,1)$. A short review of the periods and quasi-periods of modular forms is given in appendix \ref{app:Modular_forms_for_Gamma_0_15}, where we also give computational details for the periods and quasi-periods of the modular form associated to the Hulek--Verrill fourfold. We present a few additional examples of fourfolds whose zeta functions we have studied in appendix~\ref{app:non-modular_examples}, and argue that based on the data we obtained, it is unlikely that these families contain modular members. Finally, in appendix~\ref{app:periods_motives} some additional details on periods and motives are given to complement the discussion of Deligne's periods in section~\ref{sect:Delignes_periods}.

\newpage
\section{The Zeta Function of Fourfolds from the Deformation Method} \label{sect:Deformation_method}
\vskip-10pt
In this section, we give a brief review of the deformation method we use to compute the polynomials $R_I^{(p)}(X,T)$ appearing in the zeta functions of the fourfolds we investigate. For the most part the method for fourfolds works analogously to the case of threefolds discussed in refs.~\cite{Candelas:2021tqt,multiparameter_zeta,Kuusela:2022hga}, although there are certain subtleties related to the different structure of the middle cohomology of fourfolds, which we will highlight in the following.

\subsection{The local zeta function of a Calabi--Yau fourfold} \label{sect:Weil_conjectures}
\vskip-10pt
Let us consider complex $d$-dimensional Calabi--Yau manifolds which are, at least locally, defined as zero loci of polynomials whose coefficients are in $\IQ$. We say that such a Calabi--Yau manifold is \textit{defined over $\IQ$}, and denote this by $X/\IQ$. By multiplying out the denominators, we can take the defining polynomials to have coefficients in $\IZ$, and further, by considering the natural projection $\IZ \hookrightarrow \IF_p$ of $\IZ$ into the finite field $\IF_p \equiv \IZ/p\IZ$, we can define the corresponding manifolds $X/\IF_p$ over finite fields. In other words, we obtain $X/\IF_p$ by considering the defining equations of $X$ modulo $p$. The set of points on the manifold $X/\IF_p$ that lie in $\IF_p$ is denoted by $X(\IF_p)$, and the number of such points by $N_p(X) \defineas \# X(\IF_p)$.

The local zeta function $\zeta_p(X,T)$ associated to a Calabi--Yau manifold $X$ can be defined as a generating function for the numbers of points $N_{p^n}(X) \defineas \# X(\IF_{p^n})$:
\begin{align} \label{eq:zeta_definition}
\zeta_p(X, T) \defineas \exp \left( \sum_{n=1}^\infty \frac{N_{p^n}(X) T^n}{n} \right).
\end{align}
The behaviour of this function is described by the Weil conjectures, originally stated by Weil \cite{Weil1949a}, and later proven by Dwork \cite{Dwork1960a}, Grothendieck \cite{Grothendieck1995a}, and Deligne \cite{Deligne1974a,Deligne1980a}. These can be stated as:
\begin{enumerate}
	\item \textbf{Rationality}: $\zeta_p(X,T)$ is a rational function of $T$ of the form
	\begin{align}
	\zeta_p(X,T) \= \frac{R_1(X,T)R_{3}(X,T)\dots R_{2d-1}(X,T)}{R_{0}(X,T)R_{2}(X,T)\dots R_{2d}(X,T)}~,
	\end{align}
	where $R_{i}(X,T)$ is a polynomial in $T$ with integer coefficients whose degree is given by the Betti number $b_i(X)$.
	\item \textbf{Functional equation}: $\zeta_p(X,T)$ satisfies
	\begin{align} \label{eq:Weil_conjectures_functional_equation}
	\zeta_p\left(X,p^{-d} T^{-1}\right) \= \pm p^{\frac{d}{2}\chi} T^\chi \zeta_p(X,T)~,
	\end{align}
	where $\chi$ is the Euler characteristic of $X$.
	\item \textbf{Riemann hypothesis}: The polynomials $R_i(X,T)$ factorise over $\IC$ as 
	\begin{align}
	R_i(X,T) \= \prod_{j=1}^{b_i}(1-\lambda_{ij}(X)T)~,
	\end{align}
	where the $\lambda_{ij}(X)$ are algebraic integers of absolute value $p^{i/2}$.
\end{enumerate}
By these statements, the local zeta function of a generic Calabi--Yau fourfold $X$ is given by
\begin{align} \label{eq:zeta_fourfold_generic}
\zeta_p(X,T) \= \frac{R_3^{(p)}(X,T)R_5^{(p)}(X,T)}{(1-T)R_2^{(p)}(X,T)R_4^{(p)}(X,T)R_6^{(p)}(X,T)(1-p^4T)}~,
\end{align}
where $R^{(p)}_i(X,T)$ are prime-dependent polynomials of degree $b_i(X)$ in $T$. If we concentrate on fourfolds for which the third Betti number $b_3(X)$ is vanishing, the numerator becomes one. If, in addition, the Picard group of $X$ is generated by divisors defined over the field $\IF_p$, then 
\begin{align}
R_2^{(p)}(X,T) \= (1-pT)^{b_2(X)} \qquad \text{and} \qquad R_6^{(p)}(X,T) \= (1-p^3 T)^{b_2(X)}~,
\end{align}
and we get to the form \eqref{eq:zeta_simplified} mentioned in the introduction.

It can be shown that for Calabi--Yau fourfolds defined over $\IQ$, the Lefschetz fixed-point theorem implies that the number of points in $\IF_{p^n}$ on the manifold $X$ is given by
\begin{align} \label{eq:Lefschetz}
N_{p^n}(X) \= \sum_{m=0}^{8} (-1)^m \, \Tr \left(\Fr_{p^n} \big| H^m(X) \right)~,
\end{align}
where $\Fr_{p^n} \big| H^m(X)$ denotes the action on the $p$-adic cohomology $H^m(X)$, such as the Monsky-Washnitzer cohomoly \cite{vanderPut1986a} (for a brief explanation, see ref.~\cite{Candelas:2021tqt} and references therein) of the Frobenius map
\begin{align} \label{eq:definition_Fr}
\Fr_{p^n}|H^m(X) : H^m(X) \to H^m(X)~.
\end{align}
This is the map induced by the action of the Frobenius map 
\begin{align} \label{eq:Frobenius_map_definition}
\Frob_{p^n}: (x_1,\dots,x_k) \mapsto (x_1^{p^n},\dots,x_k^{p^n})
\end{align}
on the ambient space where the fourfold $X$ is embedded. From this it follows that if we denote by $\mtU_p^i(X)^{-1}$ the matrix representing the Frobenius action on $H^i(X)$, the polynomials $R_i(X,T)$ are the characteristic polynomials
\begin{align} \label{eq:P_i_definition}
R_i^{(p)}(X,T) \= \det(\II - T \Fr_{p}^{-1}|H^{i}_H(X)) \= \det(\II - T \mtU_p^i(X))~.
\end{align}
For the purposes of this paper, we are interested in the action of the Frobenius map on the middle cohomology $H^4(X)$, or even more specifically, on the subspaces $H_I^4(X)$ generated by the action of differential operators in the Picard--Fuchs differential ideal~$I$ on the holmorophic $(4,0)$-form $\Omega$ of $X$. That is to say, on the spaces $H_I^4(X)$ we define as
\begin{align} \label{eq:H_I_definition}
H_I^4(X,\IC) \defineas \langle \partial \Omega \rangle_{\partial \in I}~.
\end{align}
In particular, for Calabi--Yau manifolds $X$ with a single complex structure modulus~$\varphi$, taking the Picard--Fuchs ideal~$I$ to be generated by the logarithmic derivative $\theta = \varphi \partial_\varphi$, i.e., $I = \langle \theta \rangle$, the subspace $H_I^4(X,\IC)$ is the horizontal subspace of the middle cohomology. More generally, the \textit{horizontal cohomology}~$H_H^4(X,\IC)$ is defined as the part of the middle cohomology~$H^4(X,\IC)$ generated from the action of the Picard--Fuchs differential ideal~$J$ of the entire complex structure moduli space on the holomorphic $(4,0)$-form~$\Omega$. It is given by
\begin{align} \label{eq:H_H_definition}
H_H^4(X,\IC) \defineas \langle \partial \Omega \rangle_{\partial \in J}~,
\end{align}
where $\partial$ runs over all the partial derivatives with respect to the holomorphic coordinates of the complex structure moduli space. For more details see appendix~\ref{app:fourfolds_cohomology}. \footnote{We thank Sheldon Katz, David P. Roberts, and Masha Vlasenko for exchanging ideas on mathematically defining horizontal subspaces of complex and $p$-adic middle cohomology groups.} 

Let us remark that from a physical perspective we expect that $H^4(X,\IC)$ can be given a rational structure by constructing a generating set of D4-branes, whose cycle classes are by definition Poincar\'e dual to rational cohomology elements. However, it is less clear whether it is always possible for a Calabi--Yau fourfold~$X$ to construct a set of D4-branes, whose Poincar\'e dual rational cohomology elements generates the horizontal cohomology~$H_H^4(X,\IC)$, which would give a physics-motivated definition of $H_H^4(X,\IQ)$. See Appendix~A in ref.~\cite{Intriligator:2012ue} for a related discussion on this issue.

We assume that, for the studied Calabi--Yau fourfolds, the action of the Frobenius map on $H^4(X)$ is reducible in such a way that it has a well-defined action on $H_I^4(X)$. To explain this assumption in more detail, we recall (see for example ref.~\cite{Goresky2004a}) that there exist so-called étale cohomologies $H^4(X,\IQ_\ell)$, which are vector spaces of dimension $b^4(X)$ over the $\ell$-adic numbers ($\ell \neq p$). The absolute Galois group $\text{Gal}(\overline{\IQ}/\IQ)$ acts on this space via a representation 
\begin{align}
	\rho: \text{Gal}(\overline{\IQ}/\IQ) \to \GL(H^4(X,\IQ_\ell))~.
\end{align}
This representation contains Frobenius elements $\Frob_{p^n}$ that correspond to (the lift of, see e.g. \cite{Goresky2004a}) the Frobenius map \eqref{eq:Frobenius_map_definition}. Our assumption then means that there exists a subspace $H_I^4(X,\IQ_\ell) \subset H^4(X,\IQ_\ell)$ corresponding to the cohomology $H_I^4(X,\IC)$ defined in eq. \eqref{eq:H_I_definition} furnishes a subrepresentation of the group generated by $\Frob_{p^n}$. Note that this would for instance follow if the Hodge structure would \textit{split} into the direct sum\footnote{There is a small subtlety associated to how the Galois representation split. See footnote \ref{foot:Galois_splitting}.}
\begin{align}
	H^4(X,\IQ) \= H_I^4(X,\IQ_\ell) \oplus H_\perp^4(X,\IQ)~.
\end{align} 
However, as discussed above, existence of such splitting is not a priori clear. For more details on the relation of the Hodge structure of the middle cohomology and the Galois representation $\rho$, see section \ref{sect:splitting_of_Hodge_structure}. 

In practical terms, this means that we can find a matrix $\mtU_p(X)^{-1}$ representing the Frobenius action on $H^4_I(X)$, which yields the characteristic polynomial
\begin{align} \label{eq:R_H_definition}
R_I^{(p)}(X,T) \= \det(\II - T \Fr_{p}^{-1}|H^{4}_I(X)) \= \det(\II - T \mtU_p(X))~.
\end{align}
As a consequence, the characteristic polynomial $R_4^{(p)}(X,T)$ corresponding to the middle cohomology factorises as
\begin{align}
R_4^{(p)}(X,T) \= R_I^{(p)}(X,T) R_\perp^{(p)}(X,T)~,
\end{align}
where $R_\perp^{(p)}(X,T)$ is the remaining factor. When $H_I^4(X,\IC) = H_H^4(X,\IC)$, which is true for the most cases studied in this paper, we denote the polynomial $R_I^{(p)}(X,T)$ by $R^{(p)}_H(X,T)$, too.

While we have no proof of this assumption, we are able to perform numerous consistency checks on the results obtained under the assumption. By computing the zeta function to a high $p$-adic precision, we are able to confirm that the polynomials $R_I^{(p)}(X,T)$ we compute are compatible with the Weil conjectures. Even more intricate checks can be performed by studying the manifold which we identify as an attractive K3 point in section \ref{sect:Hulek--Verrill_Fourfolds}. The polynomials $R_I^{(p)}(X,T)$ computed for this manifold not only satisfy the expected modularity properties, but these properties together with Deligne's conjecture can be used to express the period vector corresponding to $H^4_I(X,\IQ)$ in terms of certain $L$-function values. We are able to identify the likely geometrical origin of the modularity. We view the latter two as particularly strong checks, as the periods and the complex geometry of the manifold are completely independent of the assumption made on the behaviour of the Frobenius map. 

\subsection{Frobenius map and the deformation method}
\vskip-10pt
To explicitly find the matrix $\mtU_p(X)$, we will use the deformation method, first used by Dwork \cite{Dwork1962a,Dwork1964a}, and later discussed in mathematics and physics literature (see for example refs.~\cite{Lauder2004a,Lauder2004b,Candelas:2007mb,Candelas:2021tqt} for discussion relevant for zeta functions of Calabi--Yau manifolds). The idea of this method is to first find the matrix $\mtU_p(X)$ for a manifold $X$ for which it has a known expression. Then, by studying how the matrix changes as we move in the complex structure moduli space $\cM_{\IC S}(X)$ of $X$, we can deduce the matrix $\mtU_p(X')$ for any other manifold $X'$ that belongs to the same moduli space as $X$. 

In this paper, we study one-dimensional families of Calabi--Yau fourfolds, and choose the complex structure moduli space coordinate $\varphi$ such that the large complex structure point is at $\varphi=0$.\footnote{Note that this choice does not entirely fix the coordinate $\varphi$. However, additional constraints are imposed when the periods near the large complex structure point are required to take the Frobenius form \eqref{eq:frobenius_basis_definition}. The correct choice of coordinate is discussed in some detail in \cite{Candelas:2024vzf} in the threefold case.} For simplicity, we denote $\mtU_p(X_\varphi)$ by $\mtU_p(\varphi)$ if $X_\varphi$ is the manifold corresponding to the point $\varphi \in \cM_{\IC S}(X)$.

We consider the vector bundle $\cH$ whose base is given by the moduli space $\cM_{\IC S}$ and whose fibres are the cohomology groups $H^4(X_{\varphi},\IC)$. On this bundle, there is the natural Gauss--Manin connection $\nabla$ corresponding to the Picard--Fuchs equation of the family of manifolds. For our purposes, it will be enough to study these connections along the vector fields given by the logarithmic derivatives $\theta \defineas \varphi \partial_\varphi$.

The key insight that allows us to understand how the matrix $\mtU_p(\varphi)$ varies as $\varphi$ varies in the moduli space is that the covariant derivative $\nabla_\theta$ and the Frobenius map satisfy a compatibility condition, which states that for any section $v \in \Gamma(\cH)$
\begin{align} \label{eq:GM_Frob_compatibility}
\nabla_\theta \Fr_p v \= p \, \Fr \nabla_\theta v~.
\end{align}
In addition, the connection and the Frobenius map satisfy a Leibniz rule and a linearity relation, such that for any function $f: \cM_{\IC S} \to \IC$,
\begin{align}
\begin{split}
\nabla_\theta \left(f(\varphi) v \right) &\= (\theta f)(\varphi) \, v + f(\varphi) \nabla_\theta v~,\\
\Fr \left( f(\varphi) v \right) &\= f(\varphi^p) \, \Fr(v)~. 
\end{split}
\end{align}
These relations can be used to derive a differential equation \eqref{eq:U_differential_eq} for the matrix $\mtU_p(\varphi)$ \cite{Candelas:2021tqt}. As we will shortly see, this can be used to express the variation of $\mtU_p(\varphi)$ in terms of the periods of the manifold $X_\varphi$, owing to the relation of the Gauss--Manin connection to the Picard--Fuchs equations.\footnote{For more details on the Gauss--Manin connection in this context, see for example ref. \cite{Candelas:2021tqt}.} In addition to the above relations, there is also a useful compatibility condition of the Frobenius map with the wedge product
\begin{align} \label{eq:Wedge_compatibility}
\int_{X_\varphi} \Fr \, \xi \wedge \Fr \, \eta \= p^4 \, \Fr \int_{X_\varphi} \xi \wedge \eta~.
\end{align}
Since there exists a differential equation for the matrix $\mtU_p(\varphi)$, to completely fix it, we only need its value at some fixed point $\varphi_0$ to be used as the initial data for the differential equation. In his investigation of the quintic threefold, Dwork \cite{Dwork1962a} used the Fermat quintic whose corresponding matrix $\mtU_p(0)$ was originally given by Weil \cite{Weil1949a}. In refs.~\cite{Candelas:2021tqt,multiparameter_zeta}, the large complex structure point was used, and it was shown that it is possible to find an universal expression for the matrix $\mtU_p(0)$, which only depends on a few manifold-dependent constants.\footnote{In many cases these constants can even be expressed in terms of the topological data of the manifold \cite{Candelas:2021tqt}, although there are cases in which such a formula is not known and the coefficients are computed numerically \cite{multiparameter_zeta}.} As we are looking to compute the zeta functions for several different families of fourfolds, it is fruitful to emulate this approach and use the large complex structure limit of the relation \eqref{eq:GM_Frob_compatibility} together with some other conditions explained in the following section to find the form that the matrix $\mtU_p(0)$ takes. 

In this paper, we study both manifolds whose horizontal cohomology $H_H(X,\IC)$ is of Hodge type $(1,1,1,1,1)$ and those with $H_H(X,\IC)$ of type $(1,1,2,1,1)$ (see appendix \ref{app:fourfolds_cohomology}). These two cases must be treated differently. The case $(1,1,1,1,1)$ proceeds largely analogously to the threefold case treated in ref.~\cite{Candelas:2021tqt} --- we find that the conditions imposed by eqs.~\eqref{eq:GM_Frob_compatibility} and \eqref{eq:Wedge_compatibility} fix the form of the matrix $\mtU_p(0)$ up to two constants. These can be fixed numerically by requiring that the matrix $\mtU_p(\varphi)$ takes on a rational form.

In the case of manifolds of type $(1,1,2,1,1)$, the periods are not linearly independent in the large complex structure limit. Consequently, the matrix $\mtU_p(\varphi)$ is, at least a priori, not regular in the limit $\varphi \to 0$. This presents additional difficulties, rendering some conditions we are able to use in the case $(1,1,1,1,1)$ trivial. To make progress, we can use the requirement that to take the rational form, the matrix $\mtU(\varphi)$ cannot contain logarithms of $\varphi$. Therefore the derivation in the case $(1,1,2,1,1)$ differs significantly from that for the case $(1,1,1,1,1)$, so we delegate it to the appendix~\ref{app:(1,1,2,1,1)}.

Before we evaluate the matrix $\mtU_p(\varphi)$, there is an important subtlety that has to be addressed. The fact that the Frobenius map properly acts on a $p$-adic cohomology implies that the matrix $\mtU_p(\varphi)$ should be thought of as a matrix of power series with coefficients that are $p$-adic numbers. In practice this just means that we must find an embedding of the finite field $\IF_p$ to the field of $p$-adic numbers $\IQ_p$. A convenient choice is to use the Teichmüller representatives (see for example ref.~\cite{Candelas:2000fq} for a brief physicist-oriented review) $\varphi = \Teich(x)$ of integers $x \in \IF_p$. \footnote{Here and in the following, we often think of the finite field $\IF_p$ to be represented by integers $\!\!\!\! \mod p$, that is~$\IZ/p\IZ$.} Then every element $x \in \IF_p$ is identified with an element of $\IQ_p$ via the embedding given by identifying the integer $x$ with its Teichmüller representative $\Teich_p(x)$. Therefore, when we say we evaluate the matrix $\mtU_p(\varphi)$ at $\varphi = x \in \IF_p$, we mean that we compute $\mtU_p(\Teich_p(x))$.

The Teichmüller representatives have several convenient properties, one of them being that they are eigenvalues of the Frobenius map, that is
\begin{align} \label{eq:Teichmueller_property}
\Teich_p(x)^p \= \Teich_p(x)~.
\end{align}
This relation leads to the final subtlety that needs to be accounted for while computing the zeta functions using the deformation method. We will show below, in eq.~\eqref{eq:U_solution}, that the matrix $\mtU_p(\varphi)$ can be written in a form
\begin{align}
\mtU_p(\varphi) \= \mtE(\varphi^{p})^{-1} \mtV_p(0) \mtE(\varphi)~,
\end{align}
which almost looks like a conjugation of a constant matrix. Indeed, if we were to carelessly substitute in $\varphi = \Teich_p(x)$ in $\mtE(\varphi^{p})^{-1}$, and use the property \eqref{eq:Teichmueller_property}, we would (falsely) deduce that $\mtU_p(\varphi)$ is just a conjugate of a constant matrix. However, this cannot be correct as it would imply that the characteristic polynomial $R_I^{(p)}(X,T)$ does not vary when we move in the moduli space. 

The resolution to this ostensible paradox is that the Teichmüller representatives have $p$-adic norm $|\Teich_p(x)|_p = 1$, and the series in the matrix $\mtE(\varphi)$ only converge inside an `open' disc $|\varphi|_p<1$.\footnote{The topology induced by the $p$-adic norm makes the space $\IQ_p$ of $p$-adic numbers into a totally disconnected space, so we use the word ``open'' only to describe the defining inequality being strict.} However, the matrix $\mtU_p(\varphi)$ can be shown to be convergent for $|\varphi|_p < 1+ \delta$ for some $\delta >0$. In fact, we will see below that, modulo $p^n$, $\mtU_p(\varphi)$ can be expressed as a matrix of rational functions.\footnote{The quantities we are interested in can be shown to be integers smaller than $p^m$ for some integer $m$ (see section~\ref{sect:R_H}). Therefore, to compute these, it is enough to find the values of the entries of the matrix $\mtU_p(\varphi)$ modulo $p^n$ for some $n>m$.} Thus, while it is not correct to try to evaluate the individual matrices appearing at the above product at values $\varphi = \Teich_p(x)$, it is permissible to evaluate the product of matrices first and then substitute in the value  $\varphi = \Teich(x)$. An alternative way of phrasing this is that by finding the rational expression for $\mtU_p(\varphi)$, we are analytically continuing it from inside the region $|\varphi|_p < 1$ to the whole of $|\varphi|_p < 1 + \delta$.
\subsection{Practical evaluation for manifolds of type \texorpdfstring{$(1,1,1,1,1)$}{(1,1,1,1,1)}} \label{sect:practical_zeta_function}
\vskip-10pt
Consider a family of Calabi--Yau fourfolds with one complex structure modulus $\varphi$. The Picard--Fuchs equation of such a family is given by
\begin{align} \label{eq:PF_equation}
\cL \defineas \sum_{i=0}^b S_i \,  \theta^i \qquad \text{with } b \= \dim H^4_H(X,\IC)~,
\end{align}
where $\theta \defineas \varphi \partial_\varphi$ is the logarithmic derivative.

Here we consider the case where the horizontal cohomology $H_H(X,\IC)$ is of type $(1,1,1,1,1)$, leaving the derivation of the case $(1,1,2,1,1)$ to appendix \ref{app:(1,1,2,1,1)}. In this case $b=5$, and near the large complex structure (maximal unipotent monodromy) point $\varphi=0$, we can find a basis of solutions $\varpi = (\varpi^0,\dots,\varpi^5)$ with
\begin{align} \label{eq:frobenius_basis_definition}
\varpi^i \= \sum_{m=0}^i \frac{\log^m \varphi}{m!} f_{i-m}(\varphi)~,
\end{align}
where $f_i(\varphi)$ are holomorphic functions of $\varphi$ satisfying the condition $f_i(0) = \delta_{i,0}$.

We call the vector $\varpi$ formed out of these solutions the \textit{period vector} in the \textit{arithmetic Frobenius basis}. Picking a basis of solutions of the Picard--Fuchs equation is equivalent to choosing a particular (constant) basis $\langle v_i \rangle$ of $H^4_H(X,\IC)$ in which the holomorphic three-form can be expanded as
\begin{align} 
\Omega \= \sum_{i=0}^b \varpi^i v_i~.
\end{align}
We denote by $\mtE$ the change-of-basis matrix from the constant basis $\langle v_i \rangle$ to the basis $\langle \theta^i \varpi \rangle$.
\begin{align} \label{eq:E_matrix_definition}
\left[\mtE\right]^{ij} \defineas \theta^i \varpi^j~.
\end{align}
As in the case of threefolds \cite{Candelas:2021tqt}, we expect $\mtU_p(\varphi)$ to be a matrix of rational functions of $\varphi$.\footnote{In fact, for manifolds of Hodge type $(1,1,1,1,1)$, a simple computation analogous to that of ref. \cite{Candelas:2021tqt} shows that the logarithms appearing in the expansions of the periods near the large complex structure point drop out of the expression for $\mtU_p(\varphi)$.} Therefore, to speed up numerical computations, it is useful to define the following logarithm-free quantities
\begin{align} \label{eq:omega_tilde_definition}
\wt{\theta^k \varpi^i} \defineas \left. \theta^k \varpi^i \right|_{\log \varphi \to 0} \= \sum_{j=0}^i \binom{k}{i-j} \theta^{k+j-i} f_j
(\varphi)~.
\end{align}

Using these functions, we can define the logarithm-free change-of-basis matrix, which will be used in evaluating the matrix $\mtU_p(\varphi)$:
\begin{align} \label{eq:E_tilde_definition}
\big[ \, \wt \mtE \, \big]^{ij} \defineas \wt{\theta^i \varpi^j}~.
\end{align}

To invert the matrix $\mtE$, we define a matrix $\mtW$ related to the wedge products of the holomorphic four-form and its logarithmic derivatives
\begin{align} \label{eq:wedge_product}
W^{ij}\defineas \int_X \theta^i \Omega \wedge \theta^j \Omega \= (\theta^i \varpi)^T \sigma \,\theta^j \varpi~,
\end{align}
where $\sigma$ is the matrix representing this product in the Frobenius basis, up to an overall rational factor which is not important for the discussion in this paper, as the quantities we consider are defined up to a rational scale. We fix the overall factor by demanding that the intersection form derived from this matrix using the mirror map agrees with the intersection form computed geometrically using the Hirzebruch-Riemann-Roch theorem. A direct computation reveals that the correct contant of normalisation is given, in the basis we are using, by the intersection number
\begin{align}
\kappa \defineas \int_{\wt X} D^4~.
\end{align}
This is calculated in the mirror geometry with $D \in H^2(\widetilde{X},\IZ)$ the generator of the second cohomology of the mirror Calabi--Yau fourfold $\wt X$ of $X$. With this normalisation $\sigma$ becomes
\begin{align} \label{eq:sigma_matrix}
\sigma \= \frac{\kappa}{(2\pi\ii)^4}\left(
\begin{array}{ccccc}
0 & 0 & 0 & 0 & 1 \\
0 & 0 & 0 & -1 & 0 \\
0 & 0 & 1 & 0 & 0 \\
0 & -1 & 0 & 0 & 0 \\
1 & 0 & 0 & 0 & 0 \\
\end{array}
\right).
\end{align}
We can then express the matrix $\mtW$ as
\begin{align} \label{eq:W_matrix_definition}
\mtW \= \mtE^T \sigma \mtE \= \wt \mtE^T \sigma \wt \mtE~.
\end{align}
By either explicit computation or using the Picard--Fuchs equation to derive the conditions satisfied by this matrix, one can show that its entries are rational functions of $\varphi$. This makes it relatively simple to invert $\mtW$. We utilise this to write the inverse of the matrix $\wt \mtE$ in a form that is convenient to compute in practice:
\begin{align}
\wt \mtE^{-1} \= \mtW^{-1} \sigma \wt E^T~.
\end{align}
\subsubsection*{The large complex structure matrix $\mtU_p(0)$}
\vskip-5pt
Writing the compatibility relation \eqref{eq:GM_Frob_compatibility} in matrix form, we see that the matrix $\mtU_p(\varphi)$, representing the action of the inverse Frobenius map on $H_I^4(X_\varphi)$, satisfies the differential equation
\begin{align} \label{eq:U_differential_eq}
\theta \mtU_p(\varphi) \= \mtU_p(\varphi) \mtB(\varphi) - p \mtB(\varphi^p) \mtU_p(\varphi)~,
\end{align}
where $\mtB(\varphi)$ is the matrix defining the Gauss--Manin connection:
\begin{align}
\theta \mtE(\varphi) \= \mtE(\varphi) \mtB(\varphi)~.
\end{align}
Writing the Picard--Fuchs equation \eqref{eq:PF_equation} in the first-order form, we identify the matrix $\mtB(\varphi)$ as
\begin{align}
\begin{split}
&\hskip110pt \mtB(\varphi) \= \epsilon + \mtS~, \\[5pt]
&\epsilon \= \left(
\begin{array}{ccccc}
0 & 0 & 0 & 0 & 0 \\
1 & 0 & 0 & 0 & 0 \\
0 & 1 & 0 & 0 & 0 \\
0 & 0 & 1 & 0 & 0 \\
0 & 0 & 0 & 1 & 0 \\
\end{array}
\right), \qquad \mtS \= \left(
\begin{array}{ccccc}
0 & 0 & 0 & 0 & -\frac{S_0}{S_5} \\
0 & 0 & 0 & 0 & -\frac{S_1}{S_5} \\
0 & 0 & 0 & 0 & -\frac{S_2}{S_5} \\
0 & 0 & 0 & 0 & -\frac{S_3}{S_5} \\
0 & 0 & 0 & 0 & -\frac{S_4}{S_5} \\
\end{array}
\right).
\end{split}
\end{align}
The differential equation \eqref{eq:U_differential_eq} implies that $\mtU_p(\varphi)$ takes the form\footnote{Note that unlike in the case of threefolds, for fourfolds in general $V(0) \neq U(0)$. In fact, we show in appendix \ref{app:(1,1,2,1,1)} that the equality $V(0) = U(0)$ does not holds for manifolds of Hodge type $(1,1,2,1,1)$.}
\begin{align} \label{eq:U_solution}
\mtU_p(\varphi) \= \mtE(\varphi^{p})^{-1} \mtV_p(0) \mtE(\varphi)~.
\end{align}
We can make progress by studying three conditions: the compatibility conditions \eqref{eq:GM_Frob_compatibility} and \eqref{eq:Wedge_compatibility}, and the assumption, following refs. \cite{Lauder2004a,Lauder2004b,Candelas:2021tqt}, that the matrix $\mtU(\varphi) \! \mod p^n$ is a matrix of rational functions of $\varphi$.

First, we can take the limit $\varphi \to 0$ of the condition \eqref{eq:U_differential_eq}. In this limit, $\varphi \to 0$, $B(\varphi) \to \epsilon$, so the identity \eqref{eq:U_differential_eq} becomes
\begin{align} \label{eq:U(0)_commutation}
0 \= \mtU_p(0) \epsilon - p \epsilon \mtU_p(0)~.
\end{align}
Noting that $\mtE(\varphi) = \varphi^\epsilon \wt\mtE(\varphi)$, this condition implies that $\mtU_p(0) = \mtV_p(0)$, and that this is given by
\begin{align} \label{eq:matrix_U0}
\mtU_p(0) \= \mtV_p(0) \= u \left(
\begin{array}{ccccc}
1 & 0 & 0 & 0 & 0 \\
\alpha  & 1 & 0 & 0 & 0 \\
\beta  & \alpha  & 1 & 0 & 0 \\
\gamma  & \beta  & \alpha  & 1 & 0 \\
\delta  & \gamma  & \beta  & \alpha  & 1 \\
\end{array}
\right) \diag(1,p,p^2,p^3,p^4)~,
\end{align}
where $\alpha$, $\beta$, $\gamma$ and $\delta$ are as-of-yet unidentified constants. To partially fix these constants, we now use the compatibility condition \eqref{eq:Wedge_compatibility}, which can be written in the matrix form as
\begin{align}
\mtV_p(0) \sigma \mtV_p(0)^T \= p^4 \sigma~.
\end{align}
Note that here we must use the matrix $\mtV_p(0)$ and not $\mtU_p(0)$, as it is $\mtV_p(0)$ that gives the action of the inverse Frobenius map in the constant basis $\langle v_i \rangle$ (which is the basis where $\sigma$ represents the wedge product), whereas the matrix $\mtU_p(0)$ gives the action in the basis spanned by~$\theta^i \Omega$.

A straightforward computation shows that this imposes
\begin{align} \label{eq:U(0)_constant_relations}
u \= \pm 1~, \quad \beta \= \frac{\alpha^2}{2}~, \qquad \delta \= \frac{1}{8}(8 \alpha \gamma - \alpha^4)~,
\end{align}
leaving us with two unidentified constants, $\alpha$ and $\gamma$. The choice of the sign of $u = \pm 1$ gives only an overall sign that could be absorbed in the definition of the variable $T$ appearing in the characteristic polynomial \eqref{eq:R_H_definition}. We can thus choose $u = 1$ to obtain a unique solution that only depends on the values of the constants $\alpha$ and $\gamma$.
\subsubsection*{The matrix $\mtU_p(\varphi)$}
\vskip-5pt
Finally, to fix these constants, we require that the elements of the matrix $\mtU_p(\varphi)$ are rational functions, at least to the $p$-adic accuracy of $p^{6}$, which is sufficient for computing the polynomials $R_H^{(p)}(X,T)$ (see the discussion below in section \ref{sect:R_H}).\footnote{Although we also compute the polynomials $R_H^{(p)}(X,T)$ to a much higher $p$-adic accuracy to ensure as a consistency check that the properties expected of these polynomials by Weil conjectures are indeed satisfied.} For the matrix $\mtU_p(\varphi)$ to take the rational form, the terms containing logarithms in $\mtE(\varphi)$ must drop out of the expression \eqref{eq:U_solution} for $\mtU_p(\varphi)$. This indeed follows straightforwardly by recalling that $\mtE(\varphi) = \varphi^\epsilon \wt\mtE(\varphi)$ from which, together with the commutation relation \eqref{eq:U(0)_commutation} it follows that the matrix $\mtU_p(\varphi)$ can be in fact written as
\begin{align}
\wt{\mtE}(\varphi^{p})^{-1} \mtU_p(0) \wt{\mtE}(\varphi)~.
\end{align}
This form is also useful for practical computations, as the series appearing in the matrix $\wt \mtE(\varphi)$ take on a much simpler form than those in $\mtE(\varphi)$.

A conjectural formula for the denominator in the threefold case is given in ref. \cite{multiparameter_zeta}. We assume that an analogous expression holds for fourfolds, so that in the cases we study it would be given by
\begin{align} \label{eq:U(varphi)_denominator}
P_n(\varphi^p) \= \Delta(\varphi^p)^{n-4} \cW(\varphi^p)~,
\end{align} 
where $\Delta(\varphi)$ is the discriminant of the Picard--Fuchs operator and $\cW(\varphi)$ the denominator of the matrix $W^{-1}(\varphi)$ (see eq. \eqref{eq:W_matrix_definition}). We find that this form seems to hold in the fourfold case as well, so that the matrix $\mtU_p(\varphi)$ takes the form
\begin{align} \label{eq:U_rational_form}
\mtU_p(\varphi)=\frac{\cU(\varphi)}{P_n(\varphi^p)} + \cO(p^{n})~,
\end{align}
with $\cU(\varphi)$ a matrix of polynomials in $\varphi$, when the coefficients $\alpha$ and $\gamma$ are chosen appropriately. In fact, requiring the above form fixes these coefficients uniquely for each manifold. In all cases we have studied, the coordinates can always be chosen such that $\alpha=0$. We do not know, however, a closed-form formula for $\gamma$. In the case of one-parameter threefolds, the analogous coefficient was (conjecturally) identified as $\gamma \= 6 \chi(X) \zeta_p(3)$, where $\zeta_p(3)$ is the $p$-adic zeta function (see for example ref. \cite{Koblitz1984a}). It is tempting to assume that there exists a similar formula for fourfolds. However, we have been unable to find one that would reproduce the numerical values we find for~$\gamma$.
\subsection{Computing the polynomials \texorpdfstring{$R_H^{(p)}(X,T)$}{associated to the horizontal cohomology}} \label{sect:R_H}
\vskip-10pt
The Weil conjectures state that $|\lambda_{i}|=p^2$ for any eigenvalue $\lambda_{i}$ of the matrix $\mtU_p(X)$. This simple fact has useful consequences that we can use to significantly simplify the computations required to find the polynomials $R_H^{(p)}(X,T)$.  Since the coefficients of the characteristic polynomial $R_H^{(p)}(X,T)$ are real, in the case $(1,1,1,1,1)$ the eigenvalues $\lambda_i$ necessarily appear in complex conjugate pairs, except the remaining fifth eigenvalue which necessarily needs to be real. Thus, relabelling the eigenvalues if necessary, we can write them as $\lambda_1= \overline{\lambda_2} = p^2e^{\ii \Theta_1}$, $\lambda_{3}= \overline{\lambda_4} = p^2e^{\ii \Theta_2}$ and $\lambda_{5}=\epsilon p^2$, with $\epsilon = \pm 1$. Writing the polynomial $R_H^{(p)}(X,T)$ using these expressions, we find that
\begin{align} \label{eq:R_H_(1,1,1,1,1)_lambda}
R_H^{(p)}(X,T) = \prod_{i=1}^5 (1{-}\lambda_i T) = \left(1{+}\epsilon p^2 T \right) \left(1{-}2 p^2 \cos(\Theta_1) T{+}p^4 T^2\right) \left(1{-}2 p^2 \cos(\Theta_2) T{+}p^4 T^2\right).
\end{align}
In particular, over $\IQ$, $R_H^{(p)}(X,T)$ always factorises into a linear factor and a quartic.\footnote{We thank Duco van Straten for pointing out to us this simple but important fact.} In addition, from this form it is clear that the polynomial satisfies the functional equation 
\begin{align} \label{eq:functional_eq_R_H}
R^{(p)}_H(X,p^{-4}T^{-1}) \= \epsilon p^{-2b} T^{-b} R_H^{(p)}(X,T)~,
\end{align}
where, as before, $b \defineas \dim_\IC(H_H^4(X,\IC)) = 5$. Consequently, we can write the polynomial $R_H^{(p)}(X,T)$ also~as
\begin{align} \label{eq:R_H_ab}
R_H^{(p)}(X,T) \= 1 + a_p T + b_p p T^2 + \epsilon b_p p^3 T^3 + \epsilon a_p p^6 T^4 + \epsilon p^{10} T^5~,
\end{align}
where $a_p$ and $b_p$ are integers. Therefore, it is enough to compute the coefficients $a_p$ and $b_p$, and the sign $\epsilon$ to completely fix the polynomial. Using the equation \eqref{eq:R_H_definition}, these can be obtained from the matrix $\mtU_p(X)$ via the relations
\begin{align} \label{coefficients_ap_bp}
	a_p \= -\Tr(\mtU_p(X))~,\quad p \, b_p \= \frac{1}{2}\left(\Tr(\mtU_p(X))^2-\Tr(\mtU_p(X)^2)\right) \ .
\end{align}
The sign $\epsilon$ can be found by computing the coefficient of $T^3$ in the polynomial $R_H^{(p)}(X,T)$. Denoting by $c_p$ the combination $c_p \defineas \epsilon b_p$, we have
\begin{align} \label{coefficient_cp}
p^3 c_p \= - \frac{1}{6} \left( \Tr(\mtU_p(X))^3 - 3 \Tr(\mtU_p(X)^2)\Tr(\mtU_p(X)) + 2\Tr(\mtU_p(X)^3) \right)~.
\end{align}
Crucially, comparing the expression \eqref{eq:R_H_(1,1,1,1,1)_lambda} to eq. \eqref{eq:R_H_ab} allows us to derive bounds for the magnitude of the coefficients $a_p$, $b_p$ and $c_p$ that we can use to deduce the $p$-adic accuracy to which they need to be computed to obtain the exact value.
\begin{align}
	\begin{split}
		a_p&\=-\sum_{i=1}^{5} \lambda_{i} \= -p^2(2\cos(\Theta_1)+2\cos(\Theta_2) - \epsilon)~, \\
		p \, b_p&\=\frac{1}{2}a_p^2-\frac{1}{2}\sum_{i=1}^{5}\lambda_{i}^2 \= \frac{1}{2}\left(a_p^2-p^4(2\cos(2\Theta_1)+2\cos(2\Theta_2)+1) \right) \ ,\\
        p^3 \, c_p & \= -\frac{1}{3}a_p^3+pa_pb_p-\frac{1}{3}\sum_{i=1}^5 \lambda_i^3=-\frac{1}{3}a_p^3+pa_pb_p-\frac{1}{3}p^6\left(2\cos(3\Theta_1)+2\cos(3\Theta_2)+\epsilon \right)
	\end{split}
\end{align}
From these we can deduce bounds on $a_p$, $b_p$ and $c_p$, which are given by
\begin{align}
	\begin{split}
		|a_p|\leq 5 p^2~, \; \qquad \; \left|b_p-\frac{a_p^2}{2p}\right| \; \leq \; \frac{5}{2}p^3 \; < \; 3 p^3~,\qquad \left|c_p+\frac{a_p^3}{3p^3}-\frac{a_pb_p}{p^2}\right|\leq \frac{5}{3}p^3<2p^3
	\end{split}
\end{align} 
These bound show that we are able to obtain exact values for $a_p$, $b_p$ and $c_p$ working $\!\!\!\! \mod p^4$ for primes $p \geq 5$.
\newpage

\section{Modular Example: The \texorpdfstring{$\IZ_6$}{}-Symmetric Hulek--Verrill Fourfold} \label{sect:Hulek--Verrill_Fourfolds}
\vskip-10pt
In this section we study the arithmetic properties of the family of Hulek--Verrill Calabi--Yau fourfolds $\HV_{(\varphi_1,\dots,\varphi_6)}$ with six complex structure parameters $\varphi_i$, $i=1,\dots,6$. This is the higher-dimensional analogue of the Hulek--Verrill Calabi--Yau threefold introduced in ref.~\cite{Hulek2005a}. This is an interesting example to study as, as shown in  ref.~\cite{Hulek2005a}, the modularity of threefolds can be seen to be related in many cases to the fibration structure of the manifold. Since the fourfold displays similar fibration structure, but with a K3 fibre instead of an elliptic fibre, we might hope to see modularity in the fourfold case as well. Indeed, in ref.~\cite{Bonisch:2020qmm}, it was already noted, based in part on Bessel function identities \cite{Zhou:2017vhw}, that the K3 fibre appearing in this fibration has modular properties.

We focus on a one-dimensional subfamily of Hulek--Verrill Calabi--Yau fourfolds, which admit a $\IZ_6$-group action. As opposed to the lower-dimensional smooth $\IZ_5$-quotients of Hulek--Verrill threefolds examined in refs.~\cite{Candelas:2019llw,Candelas:2021lkc}, the $\IZ_6$-quotients of the Hulek--Verrill Calabi--Yau fourfold $\HV$ exhibit orbifold singularities, because the $\IZ_6$-group does not act freely on the fourfolds~$\HV$.

After discussing the geometry of the Hulek--Verrill Calabi--Yau fourfolds, their $\IZ_6$-invariant subfamilies and their mirror geometries, we compute the polynomials $R_H^{(p)}(\HV/\IZ_6,T)$ appearing in the local zeta functions of the one-parameter subfamilies. Since the generic Calabi--Yau $\IZ_6$-orbifold $\HV/\IZ_6$ is singular, it is not clear that these polynomials have a consistent arithmetic interpretation as factors in the local zeta functions $\zeta_p(\HV/\IZ_6,T)$ . However, we can alternatively consider the subspace \begin{align}
H_{\langle \theta_1 + \dots + \theta_6 \rangle}^4(X,\IC) = \langle (\theta_1+\dots+\theta_6)^n \Omega \rangle_{n=0}^5 \subset H^4(\HV_\varphi,\IC)
\end{align}
on the smooth covering space $\HV_\varphi \defineas \HV_{(\varphi,\dots,\varphi)}$. According to the assumption we have made, the Frobenius map should have a well-defined action on this space, and the explicit computation of the corresponding characteristic polynomial $R_{\langle \theta_1 + \dots + \theta_6 \rangle}(\HV_\varphi,T)$ shows that this is exactly the polynomial $R_H^{(p)}(\HV_\varphi/\IZ_6,T)$ corresponding to the singular quotient. Therefore, even though the manifold $\HV_\varphi/\IZ_6$ is singular, the polynomials should also appear as factors of the zeta function of the covering space. In fact, the whole discussion of arithmetic geometry in this section applies mutatis mutandis to the smooth covering space.

We find that, for a the value $\varphi=1$ of the complex structure modulus, $R_H^{(p)}(\HV_1/\IZ_6,T)$ factorises, for every prime we have computed, into a linear factor and two quadratics. This indicates that the Hodge structure of the corresponding manifold splits, in the sense that there exists a two-dimensional subspace $\Lambda \subset H^4_H(\HV/\IZ_6,\IQ)$, which we verify numerically. 

The splitting of the Hodge structure further implies that the manifold has modularity properties, by which we mean that the coefficients of the local zeta functions, specifically the coefficients of one of the quadratic factors, are Fourier coefficients of a weight-three modular form. In this way, the manifold is uniquely associated to a modular form and the $L$-function given by the Hecke $L$-series of this modular form. The periods of the modular manifold, and consequently some physical quantities, can be expressed in terms of the critical values of these $L$-functions. This relation is given by Deligne's conjecture, which we review in some detail. We also discuss its implications, and numerically verify the conjecture to a high accuracy in the case we study. We also speculate on the possible geometric origin of the modularity, related to an attractive K3~surface appearing in the fourfold, and the fact that this surface is associated to the same modular form as the fourfold itself. 

\subsection{The geometry of the Hulek--Verrill fourfold}
\vskip-10pt
Analogously to the Hulek--Verrill Calabi--Yau threefold \cite{Hulek2005a}, we can express a smooth Hulek--Verrill Calabi--Yau fourfold as a complete intersection in a six-dimensional ambient Fano toric variety by using the formulation of Batyrev--Borisov for complete intersection Calabi--Yau varieties~\cite{Borisov1993a,Batyrev:1994pg}. We construct the six-dimensional ambient Fano toric variety $\mathbb{P}_\Delta$ in terms of the lattice polyhedron $\Delta \subset M \otimes_\mathbb{Z} \mathbb{R}$ in the six-dimensional lattice $M=\langle e_1, \ldots, e_6 \rangle$ given by the convex hull of the lattice points
\begin{equation} \label{eq:polyDelta}
	\Delta \= \operatorname{Conv}\left(\left\{e_1,-e_1,\ldots,e_6,-e_6\right\} \cup \left\{ e_i - e_j \right\}_{i\ne j} \right) \subset M \otimes_\mathbb{Z} \mathbb{R} \ .
\end{equation}  
The latter elements of the convex hull are the simple roots of the $A_5$-sublattice of $M$.

Now the family of Hulek--Verrill Calabi--You fourfolds $\HV$ is defined as the nef-partition of $\Delta = \text{Mink}(\Delta_1,\Delta_2)$ in terms of the Minkowski sum of the lattice polyhedra
\begin{equation} \label{eq:NefParHV}
	\Delta_1 \= \operatorname{Conv}\left(\left\{0, e_1, \ldots, e_6\right\}\right) \subset M \otimes_\mathbb{Z} \mathbb{R}\ , \quad
	\Delta_2 \= \operatorname{Conv}\left(\left\{0, -e_1, \ldots, -e_6\right\}\right) \subset M \otimes_\mathbb{Z} \mathbb{R}\ .
\end{equation}
The complete intersection associated to this nef-partition is the fourfold analogue of the Hulek--Verrill threefold studied for instance in refs.~\cite{Hulek2005a,Candelas:2019llw,Candelas:2021lkc,Bonisch:2020qmm}. 

Using the combinatorial formulas of Batyrev--Borisov~\cite{Batyrev:1994pg} together with eq.~\eqref{eq:h22}, we readily calculate the Hodge numbers $h^{1,1}$, $h^{2,1}$, $h^{2,2}$, and the Euler characteristic~$\chi$ of the Calabi--Yau fourfold $\HV$
\begin{equation}
	h^{1,1}(\HV) = 106 \ , \quad h^{2,1}(\HV)  = 0 \ , \quad h^{3,1}(\HV)  = 6 \ , \quad 
	h^{2,2}(\HV)  = 492 \ , \quad \chi(\HV)  = 720 \ .
\end{equation}
Thus, the Calabi--Yau fourfold $\HV$ possesses six complex structure moduli. In the patch $(\mathbb{C}^*)^6 \subset \mathbb{P}_\Delta$, generic global sections of the divisors of the polyhedra~\eqref{eq:NefParHV} of the nef-partition yield the Calabi--Yau fourfold $\HV$ as the zero locus of the Laurent polynomials
\begin{equation} \label{eq:HVlocaleq}
	\HV_{\bm\varphi}: \quad
	1 + x_1 + \ldots +  x_6 \= 0 \ , \qquad 1 + \frac{\varphi_1}{x_1} + \ldots + \frac{\varphi_6}{x_6} \= 0 
	\quad \text{for} \ {\bm x}\in (\mathbb{C}^*)^6 \subset \mathbb{P}_\Delta \ .
\end{equation}
The complex coefficients ${\bm \varphi}= ( \varphi_1,\ldots,\varphi_6)$ parametrise (locally) the six-dimensional complex structure moduli space of the family of Calabi--Yau fourfolds $\HV_{\bm \varphi}$ --- as indicated in the following by the subscript ${\bm\varphi}$.

Upon calculating the Jacobian matrix from the intersection equations~\eqref{eq:HVlocaleq}, one finds that the Hulek--Verrill Calabi--Yau fourfold becomes singular in the patch $(\mathbb{C}^*)^6 \subset \mathbb{P}_\Delta$ at points $\bm \varphi$ where
\begin{equation} \label{eq:SingHV}
	\frac{\varphi_1}{x_1^2} \= \frac{\varphi_2}{x_2^2} \= \ldots \= \frac{\varphi_6}{x_6^2} \quad \text{for any} \ {\bm x}\in (\mathbb{C}^*)^6  \cap \HV_{\bm\varphi}  \ . 
\end{equation} 
For generic moduli ${\bm \varphi}$ there are no singularities in $(\mathbb{C}^*)^6  \cap \HV_{\bm\varphi}$, whereas for non-generic choices of ${\bm\varphi}$ --- such that the conditions~\eqref{eq:SingHV} can be fulfilled at some points ${\bm x}\in(\mathbb{C}^*)^6  \cap \HV_{\bm\varphi}$  --- the Hulek--Verrill Calabi--Yau fourfold becomes singular.

Let us consider the holomorphic map $\Phi: (\mathbb{C}^*)^6 \setminus \{ x_3 + \ldots + x_6=0\} \subset \mathbb{P}_\Delta \to (\mathbb{C}^*)^7, \, {\bm x} \mapsto {\bm y}$ given by the relations
\begin{equation}
	\begin{aligned}
		y_0&\=-1-x_1-x_2 \ , \quad y_1 \= x_1 \ , \quad y_2 \= x_2 \ ,  \\  
		y_3&\=-1-\frac{x_4}{x_3}-\frac{x_5}{x_3} - \frac{x_6}{x_3} \ , \quad y_4 \= \frac{x_4}{x_3} \ , \quad y_5 \= \frac{x_5}{x_3} \ , \quad y_6 \= \frac{x_6}{x_3} \ .
	\end{aligned}
\end{equation}    
The map $\Phi$ restricted to the locus~\eqref{eq:HVlocaleq} of the Hulek--Verrill Calabi--Yau fourfold defines in $(\mathbb{C}^*)^7$ a locus of co-dimension three. Parameterised over the affine complex coordinate $z \in \mathbb{C}$, we arrive at the local equations for an elliptic curve $\mathcal{E}_{\bm \varphi_1}$ and a polarised K3~surface $\text{K3}_{\bm\varphi_2}$, which are again lower-dimensional analogues of the Hulek--Verrill Calabi--Yau threefold, namely
\begin{equation} \label{eq:definition_fibres}
	\begin{aligned}
		\mathcal{E}_{(z,\varphi_1,\varphi_2)}:&&&1+y_0+y_1+y_2 \= 0 \ , && 1 + \frac{z}{y_0} + \frac{\varphi_1}{y_1} + \frac{\varphi_2}{y_2} \= 0 \  , \\
		\text{K3}_{\left(\frac{z}{\varphi_3},\frac{\varphi_4}{\varphi_3},\frac{\varphi_5}{\varphi_3},\frac{\varphi_6}{\varphi_3}\right)}:&&&1+y_3+y_4+y_5+y_6 \= 0 \ , 
		&& 1 + \frac{\frac{z}{\varphi_3}}{y_3} + \frac{\frac{\varphi_4}{\varphi_3}}{y_4} + \frac{\frac{\varphi_5}{\varphi_3}}{y_5} + \frac{\frac{\varphi_6}{\varphi_3}}{y_6} \=0 \  .
	\end{aligned}
\end{equation}
As in ref.~\cite{Hulek2005a} the construction of the map $\Phi$ restricted to the fourfold $\HV_{\bm\varphi}$ extends to a birational map to the double-fibred Calabi--Yau fourfold $\mathcal{E}\times_{\IP^1} \operatorname{K3}$, i.e., 
\begin{equation} \label{eq:HV_double_fibration}
	\HV_{\bm\varphi} \  \sim_\text{rat}\  \mathcal{E}_{{\bm \varphi_1}(z,{\bm\varphi})} \times_{\IP^1} \operatorname{K3}_{{\bm \varphi_2}(z,{\bm\varphi})} \ .
\end{equation}  
Here $z$ is the (affine) coordinate of the base $\mathbb{P}^1$ and $\mathcal{E}_{{\bm \varphi_1}(z,{\bm\varphi})}$ and $\operatorname{K3}_{{\bm \varphi_2}(z,{\bm\varphi})}$ are, respectively, the elliptic fibre and the K3 fibre of the Hulek--Verrill type, meaning that that are associated to root lattices $A_n$ \cite{Hulek2005a}. Their moduli ${\bm\varphi_1}$ and ${\bm\varphi_2}$ are determined by the moduli ${\bm\varphi}$ of the Hulek--Verrill fourfold $\HV_{\bm\varphi}$ and the point $z \in\mathbb{P}^1$ on the base locus.

Upon setting $\varphi \defineas \varphi_1 = \varphi_2 = \ldots =\varphi_6$, we observe that the resulting one-parameter subfamily of Calabi--Yau fourfolds~$\HV_{\varphi} \defineas \HV_{(\varphi,\ldots,\varphi)}$ possesses a $\IZ_6$ symmetry that cyclically permutes the affine coordinates $x_i$, $i=1,\ldots,6$.\footnote{The $\IZ_6$ symmetry requires also a suitable non-generic choice of K\"ahler parameters. As we are focusing here on the complex structure moduli space, we are not specific about the necessary choices in the K\"ahler moduli space of the Calabi--Yau fourfold $\HV_{\varphi}$.} On the ambient space~$\IP_\Delta$ the $\IZ_6$~action possesses --- with respect to the subgroup $\IZ_3$ --- a co-dimension four fixed-point locus at $x_1=x_3=x_5$, $x_2=x_4=x_6$. With respect to the subgroup $\IZ_2$ there is a co-dimension three fixed-point locus at $x_1=x_4$, $x_2=x_5$, $x_3=x_6$. These two fixed-point loci intersect on $\mathbb{P}_\Delta$ in the $\IZ_6$ fixed-point locus $x_1=x_2=\ldots=x_6$ of co-dimension five. 

For a generic choice of the modulus~$\varphi$ the $\IZ_6$~fixed-point locus of co-dimension five in the ambient space~$\IP_\Delta$ does not intersect the Calabi--Yau fourfold~$\HV_{\varphi}$. Comparing with eqs.~\eqref{eq:SingHV} we actually find that the Hulek--Verrill fourfold $\HV_\varphi$ only intersects the $\IZ_6$ fixed-point locus if it is singular.  Conversely, from eqs.~\eqref{eq:SingHV} we see that the Calabi--Yau fourfold $\HV_\varphi$ is singular if and only if the modulus $\varphi$ satisfies the equation
\begin{equation} \label{eq:SingHVquo}
	(4 \varphi - 1)(16\varphi -1)(36\varphi-1) = 0 \ .
\end{equation}
For the $\IZ_6$-symmetric Calabi--Yau fourfolds $\HV_\varphi$, we may consider the four-dimensional Calabi--Yau orbifold $\HV_\varphi/\IZ_6$. As discussed above, the $\IZ_6$-action is not freely acting and hence the quotient $\HV_\varphi/\IZ_6$ has orbifold singularities. As the Calabi--Yau fourfold $\HV_\varphi$ is smooth for generic values of the complex structure modulus~$\varphi$, the singularities of the quotient $\HV_\varphi/\IZ_6$ arise solely from the fixed-point locus of the $\IZ_6$~action. Locally, the orbifold singularities are of the form $\IC^4/\Gamma$, where the discrete Abelian group $\Gamma$ acts on $\IC^4$ via a complex four-dimensional representation. Specifically, the isolated orbifold singularity arising from the fixed-point locus of the subgroup $\IZ_3$ has the form
\begin{equation} \label{eq:orbsingZ3}
  \IC^4/\IZ_3 \simeq \IC^4/\!\!\sim  \quad \text{with} \quad
  (z_1, z_2, z_3, z_4) \sim 
  (\omega z_1, \omega^2 z_2, \omega z_3, \omega^2 z_4) \ .
\end{equation}
Here $\omega$ is a primitive third root of unity. In terms of the local coordinates ${\bm x} \in (\IC^*)^6$, the coordinates $z_i$ are defined as 
\begin{align}
\begin{split}
z_1 &\defineas x_1 + \omega^2 x_3 + \omega x_5~, \qquad z_2 \defineas x_1 + \omega x_3 + \omega^2 x_5~,\\
z_3 &\defineas x_2 + \omega^2 x_4 + \omega x_6~, \qquad z_4 \defineas x_2 + \omega x_4 + \omega^2 x_6~.
\end{split}
\end{align}
Furthermore, the fixed-point locus of the subgroup $\IZ_2$ yields a curve of $\IZ_2$~orbifold singularities, which are locally described as
\begin{equation} \label{eq:orbsingZ2}
  \IC^4/\IZ_2  \simeq
  \IC \times ( \IC^3/\!\!\sim) \quad \text{with} \quad
  (v_1, v_2, v_3) \sim (-v_1, -v_2, -v_3) \ .
\end{equation}
The invariant factor $\IC$ parameterises locally the curve of orbifold singularities, which are described by the second factor $\IC^2/\!\!\sim$ in terms of the coordinates 
\begin{align}
v_1 = x_1-x_4~, \qquad  v_2=x_2-x_5~, \qquad v_3=x_3-x_6~. 
\end{align}
As a consequence of the $\SU(4)$ holonomy of smooth Calabi--Yau fourfolds, it is a necessary condition for the existence of a crepant resolution of the orbifold singularities $\IC^4/\Gamma$ to a smooth Calabi--Yau fourfold that the discrete Abelian group $\Gamma$ acts on $\IC^4$ by multiplication with $\SU(4)$ matrices~\cite{Aspinwall:1994ev}. While this condition is fulfilled for the isolated $\IZ_3$~singularity~\eqref{eq:orbsingZ3}, it is not met for the curve of $\IZ_2$~singularities~\eqref{eq:orbsingZ2}. Hence, the $\IZ_6$~orbifold $\HV_\varphi/\IZ_6$ does not admit a smooth resolution preserving its trivial canonical class. Nevertheless, it would be interesting to study in detail string or M-theory on the singular Calabi--Yau orbifold~$\HV/\IZ_6$ with a single complex structure parameter. This, however, is beyond the scope of the present work. 

Let us close the discussion of the geometry of the Hulek--Verrill Calabi--Yau fourfold $\HV_{\bm\varphi}$ by introducing the mirror Hulek--Verrill Calabi--Yau fourfold $\MHV$, which we obtain from the Batyrev--Borisov mirror construction for complete intersections in toric varieties \cite{Batyrev:1994pg,Borisov1993a}.\footnote{Our construction of the mirror Hulek--Verrill Calabi--Yau fourfold parallels the description of the mirror Hulek--Verrill Calabi--Yau threefold in ref.~\cite{Candelas:2021lkc}.} The ambient toric variety $\mathbb{P}_\nabla$ of the mirror geometry is obtained from the lattice polyhedron $\nabla \subset N \otimes_{\mathbb{Z}} \mathbb{R}$, where $N=\langle e_1^*, \ldots, e_6^* \rangle$ is the lattice dual to the lattice $M$ and the lattice polyhedron $\nabla$ is dual to the polyhedron
$$
\nabla^* \= \operatorname{Conv}\left( \Delta_1, \Delta_2 \right) 
\subset M \otimes_{\mathbb{Z}} \mathbb{R} \ ,
$$ 
expressed in terms of the polyhedra~\eqref{eq:NefParHV}. The resulting polyhedron~$\nabla$ of the toric variety $\mathbb{P}_\nabla$ is a six-dimensional hypercube, which is explicitly  given by
\begin{equation}
	\nabla \= \operatorname{Conv}\left( 
	\left\{ \varepsilon_1 e_1^*+ \ldots + \varepsilon_6 e_6^*\, \middle| \,
	\varepsilon_1,\ldots,\varepsilon_6 \in \{ -1,+1\} \right\} \right) 
	\subset  N \otimes_\mathbb{Z} \mathbb{R} \ .
\end{equation}
The polyhedra~$\nabla_1$ and $\nabla_2$ of the nef-partition which is mirror dual to the nef-partition of the polyhedra of eq.~\eqref{eq:NefParHV} are
\begin{equation} \label{eq:NefParMHV}
	\begin{aligned}
		\nabla_1 &\= \operatorname{Conv}\left( 
		\left\{ \delta_1 e_1^*+ \ldots + \delta_6 e_6^*\, \middle| \,
		\delta_1,\ldots,\delta_6 \in \{ 0,1\} \right\} \right)  \subset N \otimes_\mathbb{Z} \mathbb{R}\ , \\[0.3ex]
		\nabla_2 &\= \operatorname{Conv}\left( 
		\left\{ -\delta_1 e_1^*- \ldots - \delta_6 e_6^*\, \middle| \,
		\delta_1,\ldots,\delta_6 \in \{ 0,1\} \right\} \right) 
		\subset N \otimes_\mathbb{Z} \mathbb{R}\ .
	\end{aligned}    
\end{equation}
These polyhedra obey $\nabla = \text{Mink}(\nabla_1,\nabla_2)$ and $\Delta^* = \operatorname{Conv} \left( \nabla_1, \nabla_2 \right) \subset N \otimes_\mathbb{Z}\mathbb{R}$ and hence furnish the nef-partition of the mirror Hulek--Verrill Calabi--Yau fourfold~$\MHV \subset \mathbb{P}_\nabla$. Note that the ambient toric variety $\mathbb{P}_\nabla$ is isomorphic to a product of six projective lines $\mathbb{P}^1$ and the nef-partition yields the mirror Hulek--Verrill Calabi--Yau fourfold~$\MHV$ as the complete intersection
\begin{equation} \label{eq:MirrorHV_iso}
	\MHV \; \simeq \; \cicy{\IP^1\\\IP^1\\\IP^1\\\IP^1\\\IP^1\\\IP^1}{1 & 1\\1 & 1\\1 & 1\\ 1 & 1\\ 1 & 1\\ 1 & 1} \ ,
\end{equation}
with the Hodge numbers and the Euler characteristic
\begin{equation}
	h^{1,1}(\MHV) = 6 \ , \quad h^{2,1}(\MHV) = 0 \ , \quad h^{3,1}(\MHV) = 106 \ , \quad 
	h^{2,2}(\MHV) = 492 \ , \quad \chi(\MHV) = 720 \ .
\end{equation}
The six-dimensional K\"ahler cone is spanned by the hyperplane classes $h_i$, $i=1,\ldots,6$, of the six projective lines, such that the (complexified) K\"ahler class 
\begin{align} \label{eq:Kahler_class_HV}
J \= \sum_{i=1}^6 t_i h_i
\end{align}
is parameterised by the six K\"ahler parameters~${\bm t}=(t_1,\ldots,t_6)$, and we denote the (symplectic) family of mirror Calabi--Yau fourfolds by $\MHV_{\bm t}$.

Choosing the complex structure of the mirror manifold suitably, the one-dimensional subfamily of symplectic mirror Calabi--Yau fourfolds~$\MHV_t \defineas \MHV_{(t,\ldots,t)}$ has again a $\IZ_6$~symmetry, which cyclically permutes the six projective lines of the projective ambient space $\mathbb{P}_\nabla \simeq \mathbb{P}^1 \times \ldots \times \mathbb{P}^1$. Taking the $\IZ_6$~orbifold of $\MHV_t$ yields the mirror Calabi--Yau orbifold~$\MHV_{t}/\IZ_6$ with a single K\"ahler parameter. The discussion of the singularity structure of the Calabi--Yau orbifold $\HV/\IZ_6$ --- which does not admit a smooth crepant resolution --- implies that the mirror Calabi--Yau orbifold~$\MHV_{t}/\IZ_6$ cannot be deformed to a smooth Calabi--Yau fourfold. 

\subsection{Periods of the Hulek--Verrill Calabi--Yau fourfold} \label{sect:periods}
\vskip-10pt
To any homology four-cycle $\Gamma \in H_4(\HV_{\bm \varphi},\mathbb{Z})$ we associate the integral period $\Pi_\Gamma({\bm\varphi})$ of the Hulek--Verrill Calabi--Yau fourfold, which is defined in terms of the nowhere vanishing holomorphic $(4,0)$-form $\Omega({\bm\varphi})$ of the fourfold $\HV_{\bm\varphi}$ as
\begin{equation} \label{eq:definiton_intergal_periods}
	\Pi_\Gamma({\bm\varphi}) \= \int_\Gamma \Omega({\bm\varphi}) \ ,
\end{equation}
which locally in the complex structure moduli space is a holomorphic function of the complex structure moduli ${\bm\varphi}$. The period of the homology four-cycle associated to the $T^4$ fiber of the Strominger--Yau--Zaslow fibration of the Calabi--Yau fourfold $\HV_{\bm\varphi}$ in the vicinity of the point of maximal unipotent monodromy ${\bm\varphi=0}$ is the fundamental period $\Pi_0(\bm\varphi)$ of $\HV_{\bm\varphi}$ \cite{Strominger:1996it}. We determine the fundamental period by the standard method of direct integration over the Strominger--Yau--Zaslow fiber $T^4$, which --- using the canonical normalization of the holomorphic $(4,0)$-form --- yields for the complete intersection Calabi--Yau fourfold~$\HV_{\bm\varphi}$ expressed in terms of the defining equations~\eqref{eq:HVlocaleq} the expression
\begin{equation} \label{eq:Fundamental_Period_Series}
	\begin{aligned}
		\Pi_{0}({\bm \varphi}) &\= \frac1{(2\pi\ii)^6} \oint_0 \frac{\dd x_1}{x_1} \cdots \oint_0 \frac{\dd x_6}{x_6} 
		\frac1{(1+x_1+\ldots+x_6)\left(1 + \frac{\varphi_1}{x_1} + \ldots + \frac{\varphi_6}{x_6} \right)} \\
		&\= \sum_{n_1,\ldots,n_6=0}^{+\infty} \binom{n_1 + \ldots + n_6}{n_1,\ldots,n_6}^2 \varphi_1^{n_1} \cdot\ldots\cdot \varphi_6^{n_6} \ ,
	\end{aligned}  
\end{equation}
with the multinomial coefficients $\binom{n_1 + \ldots + n_6}{n_1,\ldots,n_6} \defineas \frac{(n_1 + \ldots + n_6)!}{n_1! \cdot \ldots \cdot n_6!}$. 

Non-trivial integral periods $\Pi_\Gamma$ of other homology four-cycles $\Gamma \in H_4(\HV_{\bm \varphi},\mathbb{Z})$ exhibit logarithmic singularities at a point of maximally unipotent monodromy $\bm\varphi=0$.\footnote{This is also a large complex structure point, in a sense that it is mirror to a large volume point.} We determine the asymptotic behavior of such integral periods with the help of mirror symmetry as follows. The homology four-cycle class~$\Gamma \in H_4(\HV_{\bm \varphi},\mathbb{Z})$ of a non-vanishing period $\Pi_\Gamma$ maps under mirror symmetry to a K-theory class $\mathcal{E}_\Gamma \in K^0_\text{alg}(\MHV_{\bm t})$ together with its mirror period $\Pi_{\mathcal{E}_\Gamma}({\bm t})$, which is a function of the (complexified) K\"ahler moduli ${\bm t}$ of the mirror fourfold~$\MHV_{\bm t}$. The polynomial part of the mirror period~$\Pi_{\mathcal{E}_\Gamma}$ of the mirror Hulek--Verrill Calabi--Yau fourfold $\MHV_{\bm t}$ --- mapping to the asymptotic behavior of the integral period $\Pi_{\Gamma}$ of the Hulek--Verrill fourfold~$\HV_{\bm \varphi}$ --- is calculated as \cite{Libgober:1999aaa,Iritani:2007aaa,MR2553377,MR2483750,Halverson:2013qca,Hori:2013ika,Gerhardus:2016iot,CaboBizet:2014ovf}
\begin{equation} \label{eq:APeriodAsy}
	\Pi_{\mathcal{E}_\Gamma}^\text{asy}({\bm t}) \= \int_{\MHV_{\bm t}} e^J \, \Gamma_{\MHV} \,
	\operatorname{ch} \mathcal{E}_\Gamma^\lor \ .
\end{equation}
Here $J$ is the complexified K\"ahler class, the gamma class $\Gamma_{\MHV}$ a multiplicative characteristic class of the fourfold~$\MHV$ based on the series~$e^{\frac{z}4}\Gamma(1-\frac{z}{2\pi i})$, and $\operatorname{ch} \mathcal{E}_\Gamma^\lor$ the Chern character of the K-theory class $\mathcal{E}_\Gamma^\lor$ dual to $\mathcal{E}_\Gamma \in K^0_\text{alg}(\MHV_{\bm t})$. Recall from the equation \eqref{eq:Kahler_class_HV} that we can explicitly express the K\"ahler class~$J$ of the mirror Hulek--Verrill Calabi--Yau fourfold in terms of the hyperplane classes ${\bm h}=(h_1,\ldots,h_6)$. Thus the characteristic gamma class of the mirror Hulek--Verrill Calabi--Yau fourfold reads explicitly
\begin{equation}
	\begin{aligned}
		\Gamma_\MHV&\= 1+ \frac{1}{24} c_2(\MHV) - \frac{\zeta(3)}{(2 \pi \ii)^3} c_3(\MHV) + \frac{1}{5\,760}\left(7 c_2(\MHV)^2 - 4 c_4(\MHV)\right)  \\
		&\=1 + \frac1{12} s_2({\bm h}) +\frac{4\,\zeta(3)}{(2 \pi \ii)^3} s_3({\bm h}) +\frac{1}{1\,440} \left(7s_2({\bm h})^2 - 24 s_4({\bm h}) \right)\ ,
	\end{aligned}
\end{equation}
where, in the last line, the Chern classes $c_\ell(\MHV)$ are written as 
\begin{equation}
	c(\MHV) \= \sum_{\ell=0}^4 c_\ell(\MHV) \= \frac{\prod_{i=1}^6 (1+h_i)^2}{\left(1 + \sum_{i=1}^6 h_i \right)^2} 
	\= 1 + 2 s_2({\bm h}) -4 s_3({\bm h}) +24 s_4({\bm h}) \ .
\end{equation}
These, in turn, are given by the hyperplane classes ${\bm h}=(h_1,\ldots,h_6)$ of the ambient product space $\IP^1 \times \dots \times \IP^1$ where $s_\ell({\bm h})$ denotes the $\ell$'th elementary symmetric polynomial of six variables as a function of the hyperplane classes ${\bm h}$.

The K-theory classes of $\mathcal{E} \in K^0_\text{alg}(\MHV_{\bm t})$ describe the charges of B-branes of the Calabi--Yau fourfold~$\MHV_{\bm t}$ \cite{Witten:1998cd}. We now construct for B-branes on the Calabi--Yau fourfold~$\MHV_{\bm t}$ the explicit asymptotic mirror periods following ref.~\cite{Gerhardus:2016iot}:
\begin{itemize}
	\item  The zero-dimensional B-brane located at any position $p\in\MHV_{\bm t}$ is described by the skyscraper sheaf $\mathcal{O}_p$, which, independent of the location $p$, yields the asymptotic period
	\begin{equation}
		\Pi_{\mathcal{O}_p}^\text{asy}({\bm t}) \= 1 \ .
	\end{equation}
	\item As detailed in ref.~\cite{Gerhardus:2016iot}, we associate to each curve $C_i$ that is a generator of the Mori cone of the Calabi--Yau fourfold $\MHV_{\bm t}$ a two-dimensional B-brane, which is a sheaf~$\mathcal{C}_i$ supported on the curve $C_i$. The associated K-theory class gives rise to the asymptotic period
	\begin{equation}
		\Pi_{\mathcal{C}_i}^\text{asy}({\bm t}) \= t_i \ , \quad i\=1,\ldots, 6 \ .
	\end{equation}  
	\item Let us consider the four-dimensional B-branes that are given in terms of the structure sheaves $\mathcal{O}_{h_i \cap h_j}$ of the hyperplane divisors $h_i$ and $h_j$  for $1\le i < j \le 6$ intersected with the Calabi--Yau fourfold $\MHV_{\bm t}$. Using the projective resolution of the sheaf $\mathcal{O}_{h_i \cap h_j}$ in terms of the exact complex
	\begin{equation}
		\begin{CD}
			0 @>>> \mathcal{O}( - h_i - h_j ) @>>> \mathcal{O}(-h_i) \oplus \mathcal{O}(-h_j)  @>>> \mathcal{O} @>>> \mathcal{O}_{h_i \cap h_j} @>>> 0
		\end{CD} \ ,
	\end{equation}
    where $\mathcal{O}$ denotes the structure sheaf of the Calabi--Yau fourfold $\MHV_{\bm t}$, we obtain for the sheaf $\mathcal{O}_{h_i \cap h_j}$ the Chern character~$\operatorname{ch} \mathcal{O}_{h_i \cap h_j} =1 - \operatorname{ch}\mathcal{O}(-h_i) - \operatorname{ch}\mathcal{O}(-h_j)+ \operatorname{ch} \mathcal{O}(-h_i - h_j)$ and hence for the dual sheaf $\mathcal{O}_{h_i \cap h_j}^\lor$ the Chern character~$\operatorname{ch} \mathcal{O}_{h_i \cap h_j}^\lor = 1 - \operatorname{ch}\mathcal{O}(h_i) - \operatorname{ch}\mathcal{O}(h_j)+ \operatorname{ch} \mathcal{O}(h_i + h_j)$. Then the asymptotic mirror integral period of the K-theory class of the B-brane $\mathcal{O}_{h_i \cap h_j}$ becomes
	\begin{equation}
		\Pi_{\mathcal{O}_{h_i \cap h_j}}^\text{asy}({\bm t}) \= 2 s_2({\bm t}_{\widehat{ij}}) + 1 \ .
	\end{equation}
	Here $s_2({\bm t}_{\widehat{ij}})$ is the second elementary symmetric polynomial in four variables, as a function of the K\"ahler parameters~${\bm t}_{\widehat{ij}}=(t_1,\ldots,\widehat{t_i},\ldots,\widehat{t_j},\ldots,t_6)$, where $t_i$ and $t_j$ are omitted in this tuple. 
	\item We consider the six-dimensional B-branes given by the structure sheaves $\mathcal{O}_{h_i}$ of the hyperplane divisors~$h_i$, $i=1,\ldots,6$, which enjoy the projective resolution
	\begin{equation}
		\begin{CD}
			0 @>>> \mathcal{O}(-h_i) @>>> \mathcal{O} @>>> \mathcal{O}_{h_i} @>>> 0 
		\end{CD} \ .
	\end{equation}
	This short exact sequence implies $\operatorname{ch}  \mathcal{O}_{h_i} = 1 - \operatorname{ch} \mathcal{O}(-h_i)$ and $\operatorname{ch}  \mathcal{O}_{h_i}^\lor = 1 - \operatorname{ch} \mathcal{O}(h_i)$ so that the corresponding asymptotic mirror periods become
	\begin{equation}
		\Pi_{\mathcal{O}_{h_i}}^\text{asy}({\bm t}) \= - 2 s_3({\bm t}_{\widehat{i}}) - s_1({\bm t}_{\widehat{i}}) + \frac{80\,\zeta(3)}{(2\pi \ii )^3}  \ ,
	\end{equation}
	where the elementary symmetric polynomials $s_\ell({\bm t}_{\widehat{i}})$ of five variables are functions of the K\"ahler parameters~${\bm t}_{\widehat{i}}=(t_1,\ldots,\widehat{t_i},\ldots,t_6)$ with the K\"ahler parameter $t_i$ omitted.
	\item The structure sheaf $\mathcal{O} \simeq \mathcal{O}^\lor$ of the Calabi--Yau fourfold $\MHV_{\bm t}$ with the Chern character~$\operatorname{ch} \mathcal{O} = 1$ realises an eight-dimensional B-brane of the mirror Hulek--Verrill fourfold $\MHV_{\bm t}$. The associated asymptotic integral period becomes
	\begin{equation}
		\Pi^{\text{asy}}_{\mathcal{O}}(\bm t) 
		\= 2 s_4(\bm t) + s_2(\bm t) - \frac{80\,\zeta(3)}{(2 \pi \ii)^3} s_1(\bm t) + \frac{3}{8} \ 
	\end{equation}
	when written in terms of the elementary symmetric polynomials $s_\ell({\bm t})$ in six variables that are functions of the K\"ahler parameters~${\bm t}$. 
\end{itemize}
Mirror symmetry allows us to identify the asymptotic behavior of the integral periods $\Pi_\Gamma({\bm\varphi})$ of the Hulek--Verrill Calabi--Yau fourfold~$\HV_{\bm \varphi}$ using the asymptotic integral mirror periods of the mirror Hulek--Verrill Calabi--Yau fourfold~$\MHV_{\bm t}$. Namely, from the above list of periods we arrive at
\begin{equation} \label{eq:asyperHV}
	\begin{aligned}
		\Pi_{\mathcal{O}_p}^\text{asy}({\bm t}): \quad &\Pi_0({\bm \varphi}) = 1 + \ldots \ , \\
		\Pi_{\mathcal{C}_i}^\text{asy}({\bm t}): \quad  &\Pi^i_{1}({\bm \varphi}) = \frac{1}{2\pi\ii} \log \varphi_i + \ldots \ , \\
		\Pi_{\mathcal{O}_{h_i\cap h_j}}^\text{asy}({\bm t}): \quad &\Pi^{i,j}_{2}({\bm \varphi}) 
		= \frac{2}{(2\pi\ii)^2} s_2({\bmlog \bm\varphi}_{\widehat{ij}}) +1 + \ldots \ , \\
		\Pi_{\mathcal{O}_{h_i}}^\text{asy}({\bm t}): \quad &\Pi^i_3({\bm \varphi}) = 
		-\frac2{(2\pi\ii)^3} s_3({\bmlog \bm\varphi}_{\widehat{i}})
		- \frac1{2\pi\ii} s_1({\bmlog \bm\varphi}_{\widehat{i}})
		+ \frac{80\,\zeta(3)}{(2\pi \ii )^3}  +  \ldots  \ , \\
		\Pi_{\mathcal{O}}^\text{asy}({\bm t}): \quad &\Pi_{4}(\bm\varphi) =
		\frac2{(2\pi\ii)^4} s_4(\bmlog \bm\varphi) + \frac1{(2\pi\ii)^2} s_2(\bmlog\bm\varphi) - \frac{80\,\zeta(3)}{(2 \pi \ii)^4} s_1(\bmlog \bm\varphi) + \frac{3}{8} 
		+ \ldots \ .
	\end{aligned}   
\end{equation}
The subscript $d$ in $\Pi_{d}(\bm\varphi)$ indicates the (complex) dimension of the support of the B-brane on the mirror Hulek--Verrill Calabi--Yau fourfold $\MHV_{\bm t}$, from which the period originates. The elementary symmetric polynomials $s_\ell$ in these expressions are functions of the tuples 
\begin{align}
\begin{split}
\bmlog\bm\varphi &\= (\log \varphi_1,\ldots,\log\varphi_6)~,\\ 
\bmlog\bm\varphi_{\widehat{i}}  &\= (\log \varphi_1,\ldots,\widehat{\log\varphi_i},\ldots,\log\varphi_6)~,\\
\bmlog\bm\varphi_{\widehat{ij}}  &\= (\log \varphi_1,\ldots,\widehat{\log\varphi_i},\ldots,\widehat{\log\varphi_j},\ldots,\log\varphi_6)~.
\end{split}
\end{align}
The displayed terms in eq. \eqref{eq:asyperHV} are singular or non-vanishing in the limit $\bm\varphi \to 0$, whereas the remaining terms indicated by `\ldots' vanish in the limit $\bm\varphi\to 0$ and are of the orders $O(\varphi)$, $O(\varphi \log\varphi)$, $O(\varphi \log^2\varphi)$, etc. This asymptotic structure suffices to unambiguously determine the entire periods as solutions to the Picard--Fuchs differential ideal.

Finally, let us discuss a rational basis of periods for the complex structure submoduli space of the one-parameter family of the $\IZ_6$-symmetric Calabi--Yau fourfolds~$\HV_\varphi$. On the $\IZ_6$-symmetric subfamily of mirror Calabi--Yau fourfolds~$\MHV_t$, these rational periods are associated to $\IZ_6$-invariant B-brane configurations. Alternatively, the constructed periods can be viewed as B-branes on the mirror Calabi--Yau orbifold~$\MHV_t/\IZ_6$.\footnote{For our purposes, it is not necessary to consider B-branes localised at orbifold singularities, known as fractional branes \cite{Diaconescu:1997br,Diaconescu:1999dt}.} By mirror symmetry these periods correspond to integral periods on the Calabi--Yau orbifold~$\HV_\varphi/\IZ_6$.

The fundamental period $\Pi_0(\varphi)$ of $\HV_\varphi/\IZ_6$ is obtained by setting  $\varphi=\varphi_1 = \varphi_2 = \ldots = \varphi_6$ in the expression \eqref{eq:Fundamental_Period_Series} for the fundamental period of the covering space~$\HV_{\varphi}$. Doing this, we arrive at
\begin{equation} \label{eq:fundamental_period_series_quotient}
	\Pi_0(\varphi) \= 
	\sum_{n_1,\ldots,n_6=0}^{+\infty} \binom{n_1 + \ldots + n_6}{n_1,\ldots,n_6}^2 \varphi^{n_1+\ldots+n_6} \ .
\end{equation}
We determine the Picard--Fuchs operator $\mathcal{L}$ for the Calabi--Yau orbifold~$\HV_\varphi/\IZ_6$ with the single complex structure modulus~$\varphi$ by requiring that it is of order $5$ and annihilates the fundamental period~$\Pi_0(\varphi)$. In terms of the logarithmic derivatives $\theta \defineas \varphi \frac{\dd}{\dd \varphi}$, it reads
\begin{multline} \label{eq:PFHV}
	\cL \= \theta ^5-2 (2 \,\theta +1) \left(14\, \theta  (\theta +1) \left(\theta ^2+\theta
	+1\right)+3\right) \varphi -1\,152 (\theta +1)^2 (\theta +2)^2 (2 \theta +3)
	\varphi ^3
	\\+4 (\theta +1)^3 (196 \,\theta  (\theta +2)+255) \varphi ^2 \ .
\end{multline}
The discriminant~$\Delta$ of this Picard--Fuchs operator~$\mathcal{L}$ is given by
\begin{equation}
	\Delta \= (4 \varphi -1) (16 \varphi -1) (36 \varphi -1) \ ,
\end{equation}
which vanishes at those points in complex structure moduli space, where the $\IZ_6$-symmetric Calabi--Yau fourfold~$\HV_\varphi$ becomes singular according to eq.~\eqref{eq:SingHVquo}. Upon averaging over images of branes with respect to the $\IZ_6$~orbifold action, we obtain, in addition to the fundamental period~$\Pi_0(\varphi)$, further rational periods from the asymptotic expressions of the periods~\eqref{eq:asyperHV} of the covering space~$\HV_\varphi$. Specifically, we choose for the mirror Calabi--Yau orbifold~$\MHV_t/\IZ_6$ the asymptotic periods
\begin{equation}
\begin{aligned}
   \Pi_1^\text{asy}(t) &\= \frac16\sum_{i=1}^6 \Pi^\text{asy}_{\mathcal{C}_i}(t,\ldots,t) \ , \qquad&
   \Pi_2^\text{asy}(t) &\= \frac16 \sum_{i=1}^3 \Pi_{\mathcal{O}_{h_i\cap h_{i+3}}}^\text{asy}(t, \ldots, t) \ , \\
   \Pi_3^\text{asy}(t) &\= \frac16 \sum_{i=1}^6 \Pi_{\mathcal{O}_{h_i}}^\text{asy}(t, \ldots, t) \ , \qquad&
   \Pi_4^\text{asy}(t) &\= \frac16 \Pi_{\mathcal{O}}^\text{asy}(t, \ldots, t) \ ,
\end{aligned} 
\end{equation}
such that the rational periods of the Calabi--Yau orbifold~$\HV_\varphi/\IZ_6$ become
\begin{equation} \label{eq:one-parameter_periods}
	\begin{aligned}
		\Pi_0(\varphi) &\= 1 + \ldots \ , \\
		\Pi_{1}(\varphi) &\= \frac{1}{2\pi\ii} \log \varphi + \ldots \ , \\
		\Pi_{2}(\varphi) &\= \frac{6}{(2\pi\ii)^2} \log^2 \varphi + \frac12 +\ldots \ , \\
		\Pi_3(\varphi) &\=   -\frac{20}{(2\pi\ii)^3} \log^3\varphi
		- \frac5{2\pi\ii} \log \varphi
		+ \frac{80\,\zeta(3)}{(2\pi \ii )^3}  +  \ldots  \ , \\
		\Pi_{4}(\varphi) &\=
		\frac5{(2\pi\ii)^4} \log^4\varphi + \frac5{2(2\pi\ii)^2} \log^2 \varphi - \frac{80\,\zeta(3)}{(2 \pi \ii)^4} \log\varphi + \frac{1}{16} 
		+ \ldots \ .
	\end{aligned}   
\end{equation}
All these periods are annihilated by the Picard--Fuchs operator~\eqref{eq:PFHV}, which determines the term `\ldots' unambiguously. We gather the periods~\eqref{eq:one-parameter_periods} in the period vector $\Pi = (\Pi_0(\varphi),\dots, \Pi_4(\varphi))^T$, for which the monodromy around the large volume singularity can readily be evaluated to be
\begin{equation} \label{eq:MLV}
\mtM_0 \= \left(
\begin{array}{ccccc}
 1 & 0 & 0 & 0 & 0 \\
 1 & 1 & 0 & 0 & 0 \\
 6 & 12 & 1 & 0 & 0 \\
 -20 & -60 & -10 & 1 & 0 \\
 5 & 20 & 5 & -1 & 1 \\
\end{array} \right) \ .
\end{equation}
The monodromy matrices about the remaining zeros of the discriminant~\eqref{eq:SingHVquo} the point $\varphi=\infty$ are determined via analytic continuation of the periods and are given by
\begin{align} \label{eq:Mnum}
\begin{split}
&\mtM_{\frac1{36}} \= \left(
\begin{array}{ccccc}
 1 & 0 & 0 & 0 & -6 \\
 0 & 1 & 0 & 0 & 0 \\
 0 & 0 & 1 & 0 & -6 \\
 0 & 0 & 0 & 1 & 0 \\
 0 & 0 & 0 & 0 & -1 \\
\end{array}
\right), \quad
\mtM_{\frac1{16}} \=\left(
\begin{array}{ccccc}
 1 & 0 & 0 & 6 & 30 \\
 0 & 1 & 0 & -1 & -6 \\
 0 & 0 & 1 & 6 & 30 \\
 0 & 0 & 0 & 11 & 60 \\
 0 & 0 & 0 & 2 & 11 \\
\end{array}
\right),\\[10pt]
&\mtM_{\frac14} \= \left(
\begin{array}{ccccc}
 1 & 0 & -30 & -24 & -60 \\
 0 & 1 & 10 & 9 & 24 \\
 0 & 0 & -41 & -36 & -96 \\
 0 & 0 & -100 & -89 & -240 \\
 0 & 0 & -20 & -18 & -49 \\
\end{array}
\right), \quad 
\mtM_\infty \= \left(
\begin{array}{ccccc}
 1 & 120 & -90 & -36 & -60 \\
 -1 & 61 & -40 & -15 & -24 \\
 -6 & 252 & -161 & -60 & -96 \\
 20 & -660 & 410 & 151 & 240 \\
 -5 & 140 & -85 & -31 & -49 \\
\end{array}
\right) .
\end{split}
\end{align}
The monodromy matrix $M_\infty$ about $\varphi=\infty$ is determined by the relation
\begin{align}
M_\infty\=\left(\mtM_0 \mtM_{\frac1{36}} \mtM_{\frac1{16}} \mtM_{\frac14}\right)^{-1}~.
\end{align}
The intersection pairing $I$ for the set of B-branes associated to the period vector~$\Pi$ reads
\begin{equation} \label{eq:Imat}
  \mtI \= \begin{pmatrix}
    0 & 0 & 0 & 0 & 1 \\
    0 & 0 & 0 & 1 & 0 \\
    0 & 0 & \frac65 & 0 & 1 \\
    0 & 1 & 0 & -10 & 0\\
    1 & 0 & 1 & 0 & \frac13
  \end{pmatrix} \ .
\end{equation}
Up to an overall normalization, this intersection matrix is entirely determined by the requirement that the intersection pairing is invariant with respect to the monodromy action, i.e., $\mtI = \mtM_p \mtI \mtM_p^T$ for all monodromy matrices $\mtM_p$ with $p=0, \frac1{36}, \frac1{16}, \frac14, \infty$. The rational entries in the intersection matrix~$\mtI$ are a consequence of the singularities of the $\IZ_6$-orbifold $\MHV_{t}/\IZ_6$. These singularities lead to fractional B-brane charges \cite{Diaconescu:1997br,Diaconescu:1999dt}, which we do not keep track of in the presented description of the B-branes on the $\IZ_6$-quotient $\MHV_{t}/\IZ_6$. Nevertheless, for those B-branes $\mathcal{E}$ and $\mathcal{F}$, whose intersection pairing is not affected by the singularities and hence is insensitive to fractional brane charges, we can unambiguously compute the intersection index $I(\mathcal{E},\mathcal{F})$ geometrically with the Hirzebruch--Riemann--Roch index theorem on the covering space $\MHV_{t}$ of the orbifold $\MHV_{t}/\IZ_6$ as\footnote{Such a formula determines the intersection pairing for all B-branes in a freely acting and hence smooth Calabi--Yau orbifold because there are no fractional branes in this case (see ref.~\cite{Brunner:2001eg}).}
\begin{equation}
  I(\mathcal{E},\mathcal{F})
  \= \frac16 \int_{\MHV} \operatorname{Td}(\MHV_{t}) \operatorname{ch} \pi^*( \mathcal{E}\otimes\mathcal{F}^\lor) \ .
\end{equation}
Here $\pi^*$ is the pullback with respect to the map $\pi: \MHV_{t}\to \MHV_{t}/\IZ_6$, $\operatorname{Td}(\MHV_{t})$ is the Todd class of the covering space $\MHV_{t}$, and the prefactor $\frac16$ is the order of the orbifold group $\IZ_6$. This intersection pairing for the B-branes computes all integral entries of the intersection matrix~\eqref{eq:Imat}, whereas for the rational entries a more careful analysis --- taking into account the fractional brane contributions --- is required.

The period vector $\Pi$ is related to the Frobenius basis introduced in eq.~\eqref{eq:frobenius_basis_definition} by the change of basis $\Pi = \mtT \varpi$, given by the matrix
\begin{align} \label{eq:change_of_basis_integral_frobenius}
\begin{split}
	\mtT \= &\rho \nu^{-1}~, \qquad \nu \= \text{Diag}\left(1,2\pi\ii,(2\pi\ii)^2,\dots,(2\pi \ii)^4\right)~,\\
	\rho &\= \left(
	\begin{array}{ccccc}
		1 & 0 & 0 & 0 & 0 \\
		0 & 1 & 0 & 0 & 0 \\
		\frac{1}{2} & 0 & 12 & 0 & 0 \\
		\frac{10 \ii \zeta (3)}{\pi ^3} & -5 & 0 & -120 & 0 \\
		\frac{1}{16} & -\frac{10 \ii \zeta (3)}{\pi^3} & 5 & 0 & 120 \\
	\end{array}
	\right).
\end{split}
\end{align}
The intersection matrix $\sigma$ of the horizontal cohomology elements of the Calabi--Yau orbifold $\HV_{\varphi}/\IZ_6$ expressed in terms of the Frobenius basis is given in eq.~~\eqref{eq:wedge_product}. In the rational B-brane basis of the mirror Calabi--Yau orbifold~$\MHV_{t}/\IZ_6$ it transforms into the rational intersection matrix~$\Sigma$, which reads
\begin{align}
	\Sigma \= \left(\mtT^{-1} \right)^T\sigma \; \mtT^{-1} \= \left(
\begin{array}{ccccc}
 \frac{1}{2} & 0 & -\frac{5}{6} & 0 & 1 \\
 0 & 10 & 0 & 1 & 0 \\
 -\frac{5}{6} & 0 & \frac{5}{6} & 0 & 0 \\
 0 & 1 & 0 & 0 & 0 \\
 1 & 0 & 0 & 0 & 0 \\
\end{array}
\right)\ .
\end{align}
Note that the intersection matrix~$\Sigma$ of the wedge product is dual to the intersection pairing~$I$ given in eq.~\eqref{eq:Imat}, so that $\Sigma$ fulfills the relations
\begin{equation}
  \mtI = \Sigma^{-1} \ , \qquad \Sigma = \mtM_p^T \Sigma \mtM_p \quad \text{for} \quad
  p=0,~\frac1{36},~\frac1{16},~\frac14,~\infty \ ,
\end{equation}
in terms of the monodromy matrices~$\mtM_p$ given in eqs.~\eqref{eq:MLV} and \eqref{eq:Mnum}.

\subsection{Searching for persistent factorisations of \texorpdfstring{$R_H^{(p)}(\HV_\varphi/\IZ_6,T)$}{the polynomial corresponding to the horizontal cohomology}}
\label{sect:persistent_factorisation}
\vskip-10pt
Even though we have seen above that the $\IZ_6$ quotient of the $\IZ_6$-symmetric one-dimensional subfamily $\HV_{\varphi}$ is singular, we can still proceed to compute the polynomials $R_H^{(p)}(\HV_\varphi/\IZ_6,T)$ using the algorithm described in section \ref{sect:Deformation_method}. Although it is not a priori clear that this computation will give a sensible result, somewhat surprisingly, we find that the polynomials satisfy the properties expected of factors appearing in the local zeta functions of a smooth variety. 

This observation can be explained by noting that the polynomial $R_H^{(p)}(\HV_\varphi/\IZ_6,T)$ is in fact related to the zeta function of the smooth covering space $\HV_{\varphi}$ on the $\IZ_6$-symmetric locus where ${\varphi_1 = \dots = \varphi_6 = \varphi}$. Namely, the horizontal piece of the middle cohomology $H^4(\HV_\varphi/\IZ_6,\IC)$ of the quotient variety $\HV_\varphi /\IZ_6$ corresponds to the subspace $H_I^4(\HV_\varphi,\IC)$ defined in eq. \eqref{eq:H_I_definition}, generated by the ideal $I = \langle \theta_1 + \dots + \theta_6 \rangle$,
\begin{align}
H_{\langle \Theta \rangle}^4(\HV_\varphi, \IC) = \langle \Theta^n \Omega \rangle_{n=0}^5 \subset H_H^4(\HV_\varphi,\IC)~, \qquad \text{where} \quad \Theta = \theta_1 + \dots + \theta_6~.
\end{align}
According to our assumption, this subspace should give rise to a well-defined action of  of the Frobenius map. Under this assumption, the action can be computed completely analogously to the $(1,1,1,1,1)$ case discussed in section \ref{sect:Deformation_method}, and the characteristic polynomial $R_{\langle \Theta \rangle}(\HV_\varphi,T)$ is exactly the polynomial $R_H^{(p)}(\HV_\varphi/\IZ_6,T)$ corresponding to the horizontal part of the quotient variety. 

We find it most convenient to discuss the computation and the properties of the polynomials $R_H^{(p)}(\HV_\varphi/\IZ_6,T)$, and the corresponding subspaces $H_H^4(\HV_\varphi/\IZ_6,\IC)$ and $\langle \Theta^n \Omega \rangle_{n=0}^5$ using the language of the one-parameter case $\MHV_\varphi/\IZ_6$. However, the discussion below applies mutatis mutandis to the case of the covering manifold $\HV_{\varphi}$ if one replaces $\HV_\varphi/\IZ_6$ by the $\IZ_6$-symmetric manifold $\HV_{\varphi}$, $H_H^4(\HV_\varphi/\IZ_6)$ by the space $H_{\langle \Theta \rangle}^4(\HV_\varphi,\IC) \subset H_H^4(\HV_{\varphi},\IC)$, and $R_H(\HV_\varphi/\IZ_6,T)$ by the polynomial $R_{\langle \Theta \rangle}(\HV_\varphi,T)$.

To obtain the data needed for the algorithm to compute the polynomials $R_H^{(p)}(\HV_\varphi/\IZ_6,T)$, we compute the matrix $\mtW$ defined in eq. \eqref{eq:W_matrix_definition} by using the Picard--Fuchs equation to find the differential equation satisfied by it, or alternatively by explicitly evaluating the matrix in terms of the periods $\varpi_i$ and noting that the series that appear in the matrix entries truncate:
\begin{align} \notag
\mtW^{-1} = \left(
\begin{smallmatrix}
 -12 \varphi  (\varphi  (768 \varphi -85)+1) & -4 \varphi  (\varphi  (3456 \varphi -451)+7) & -12 \varphi  (\varphi  (1632 \varphi -281)+7) & -12 \varphi  (4 \varphi  (216
   \varphi -49)+7) & -\Delta \\
 -4 \varphi  (\varphi  (3456 \varphi -451)+7) & 4 \varphi  (\varphi  (2304 \varphi -451)+14) & 4 \varphi  (4 \varphi  (216 \varphi -49)+7) & \Delta & 0 \\
 -12 \varphi  (\varphi  (1632 \varphi -281)+7) & 4 \varphi  (4 \varphi  (216 \varphi -49)+7) & -\Delta & 0 & 0 \\
 -12 \varphi  (4 \varphi  (216 \varphi -49)+7) & \Delta & 0 & 0 & 0 \\
 -\Delta & 0 & 0 & 0 & 0 \\
\end{smallmatrix}
\right),
\end{align}
where $\Delta$ denotes the discriminant of the Picard--Fuchs equation. Consequently, the (conjectural) denominator \eqref{eq:U(varphi)_denominator} of the matrix $\mtU_p(\varphi)$ takes the form
\begin{align} \label{eq:U(varphi)_denominator_HV}
P_n(\varphi^p) \= \Delta(\varphi^p)^{n-4} \= \left((4 \varphi^p -1) (16 \varphi^p -1) (36 \varphi^p -1)\right)^{n-4}~.
\end{align} 
Computing the series solutions to the Picard--Fuchs equations to 6000 terms, and requiring that the matrix $\mtU_p(\varphi)$ be rational $\!\!\!\mod p^n$ with the denominator given by \eqref{eq:U(varphi)_denominator_HV} allows us to numerically obtain the values of the coefficients $\alpha$ and $\gamma$ that appear in the expression for $\mtU(0)$.\footnote{One can in principle also leave the denominator a priori unfixed, and use the rationality condition of $\mtU_p(\varphi)$ to solve for it.} Working to the $p$-adic accuracy of $p^9$, we find that $\alpha = 0$, and the coefficients $\gamma$ take the values the first few of which are included in \tref{tab:gamma_values_HV}.
\begin{table}[H]
	\renewcommand{\arraystretch}{1.3}
	\begin{center}
		\begin{tabular}{|l|l||l|l||l|l|}
			\hline
			\hfil $p$ & \hfil $\gamma + \cO\left(p^{9}\right)$ & \hfil $p$ & \hfil $\gamma + \cO\left(p^{9}\right)$ & \hfil $p$ & \hfil $\gamma + \cO\left(p^{9}\right)$\\ \hline \hline
			7 & 997787 & 23 &22194055873 & 43 & 4946872588338\\ \hline 
			11 & 6344877 & 29 &416835911016 & 47 & 5597431098007\\ \hline 
			13 & 347778509 & 31 &246702637383 & 53 & 339881724690\\ \hline 
			17 & 6679847451 & 37 &1533396234945 & 59 & 92128291575420\\ \hline 
			19 & 11041988586 & 41 &4341907833009 & 61 & 183276963102347\\ \hline 
		\end{tabular}
		\vskip10pt
		\capt{5.3in}{tab:gamma_values_HV}{The values of the prime-dependent constants $\gamma \!\mod p^{9}$ for the first few primes $p\geq 7$ for the one-parameter family $\HV_\varphi/\IZ_6$ of quotients of Hulek--Verrill fourfolds.}		
	\end{center}
 \vskip-30pt
\end{table}
With this data and series expansions of the periods to 6000 first terms, we are able to find the polynomials $R_H^{(p)}(\HV_\varphi/\IZ_6,T)$ for all primes $7 \leq p \leq 733$ and for all $\varphi \in \IF_p$. As discussed in \ref{sect:R_H}, the Weil conjectures imply that these polynomials always factorise over $\IQ$ into a linear polynomial and a quartic, which is indeed what we observe. However, the quartic factorises over $\IQ$ often further into two quadrics, which may themselves factorise further. We call these cases \textit{quadratic factorisations} to differentiate these from the generic case where we have only a linear factor. We display some examples in table \ref{tab:R_H_HV}.

\tablepreambleFourfold{1}
7 & 3 & -(p^2T-1) \left(p^4 T^2+58 T+1\right) \left(p^4 T^2-10 p T+1\right)
 \tabularnewline[.5pt] \hline 
7 & 5 & -(p^2T-1) \left(p^8 T^4+48 p^4 T^3+586 p T^2+48 T+1\right)
 \tabularnewline[.5pt] \hline 
7 & 6 & -(p^2T-1) \left(p^8 T^4+8 p^4 T^3+426 p T^2+8 T+1\right)
 \tabularnewline[.5pt] \hline 
11 & 1 & -(p^2T-1)^3 \left(p^2 T+1\right)^2
 \tabularnewline[.5pt] \hline 
11 & 5 & \left(p^2 T+1\right) \left(p^8 T^4-144 p^4 T^3+1946 p T^2-144 T+1\right)
 \tabularnewline[.5pt] \hline 
13 & 1 & \left(p^2 T-1\right)^2 \left(p^2 T+1\right) \left(p^4 T^2+310 T+1\right)
 \tabularnewline[.5pt] \hline 
17 & 1 & \left(p^2 T+1\right) \left(p^4 T^2+70 T+1\right) \left(p^4 T^2+14 p T+1\right)
 \tabularnewline[.5pt] \hline 
17 & 14 & \left(p^2 T+1\right) \left(p^8 T^4-24 p^5 T^3+4406 p T^2-24 p T+1\right)
 \tabularnewline[.5pt] \hline 
19 & 1 & \left(p^2 T+1\right) \left(p^4 T^2-338 T+1\right) \left(p^4 T^2+22 p T+1\right)
 \tabularnewline[.5pt] \hline  
19 & 5 & -(p^2T-1) \left(p^8 T^4+344 p^4 T^3+354 p T^2+344 T+1\right)
 \tabularnewline[.5pt] \hline 
\tablepostamble
\vskip-40pt
\begin{table}[H]
\begin{center}
\capt{5.9in}{tab:R_H_HV}{Some examples of the polynomials $R_H^{(p)}(\HV_\varphi/\IZ_6,T)$ corresponding to the horizontal cohomology of the manifold $\HV_\varphi/\IZ_6$ for the first few primes $p$ and some values of the complex structure modulus $\varphi \in \IF_p$.}
\end{center}
\vskip-30pt
\end{table}

We plot the number of factorisations for the 130 primes in the range $7 \leq p \leq 733$ in \fref{fig:Factorisations_HV}, which makes it clear that there is at least one factorisation for every prime. This suggests that there is what was termed a \textit{persistent factorisation} in \cite{Candelas:2019llw}, meaning that there exists a complex structure modulus $\varphi \in \IC$ such that the polynomial $R_H^{(p)}(\HV_\varphi/\IZ_6,T)$ of the corresponding manifold $\HV_\varphi/\IZ_6$ has a quadratic factorisation for all (apart from possibly finitely many) primes. 
\begin{figure}[H]
	\centering
	\begin{center}
		\includegraphics[width=14cm, height=5cm]{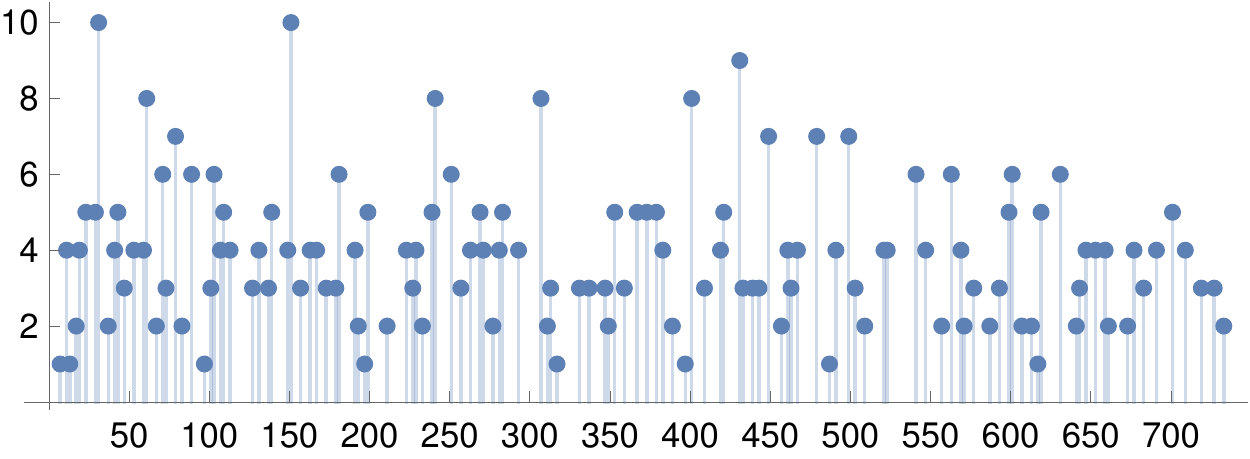}
	\end{center}
	\vskip-10pt
	\capt{6in}{fig:Factorisations_HV}{The number of quadratic factorisations of $R_H^{(p)}(\HV_1/\IZ_6,T)$ for the first 130 primes $p\geq 7$. The vertical axis gives the number of such factorisations for the prime indicated by the horizontal axis.}	
    \vskip-20pt
\end{figure}
We can look for a manifold for which there exists a persistent factorisation using a procedure analogous to that used in ref. \cite{Candelas:2019llw}: We look for integers $k_0,k_1 \in \IZ$, with $k_1 \neq 0$, such that 
\begin{align} \label{eq:linear_factorisation_ansatz}
k_0 + k_1 \varphi \= 0 \mod p~, \qquad \text{or} \qquad k_1 \= 0 \mod p~.
\end{align}
when $R_H^{(p)}(\HV_\varphi/\IZ_6,T)$ has a quadratic factorisation or $\Delta = 0 \!\! \mod p$, the latter condition implying the existence of a singularity or an apparent singularity. Existence of such a pair of integers would indicate that the complex structure modulus $\varphi = -k_0/k_1$ corresponds to a manifold $\HV_\varphi/\IZ_6$ that has a persistent factorisation, as the solution to the first equation in \eqref{eq:linear_factorisation_ansatz} gives a representative of $ -k_0/k_1$ in $\IF_p$. The second condition in \eqref{eq:linear_factorisation_ansatz} is included to ensure that we do not miss the cases where $-k_0/k_1$ is not defined in the field $\IF_p$, which happens when $k_1$ is not invertible in $\IF_p$. Similarly, the cases  where $\Delta = 0 \mod p$ are included to make sure that the so-called \textit{bad primes} $p$ for which $\HV_\varphi$ is develops further singularities over $\IF_p$ do not cause us to miss a point with a persistent factorisation.

Searching over all  $-1000 \leq k_0 \leq 1000$ and $0 < k_1 \leq 1000$, we find four distinct pairs of solutions, corresponding to the following rational values of $\varphi$
\begin{align} \label{eq:linear_factorisations_points}
\varphi = \frac{1}{4}~, \qquad \varphi=\frac{1}{16}~, \qquad \varphi \= \frac{1}{36}~, \qquad \varphi \= 1~.
\end{align}
The first three just correspond to the singularities where $\Delta = 0$, whereas the last value $\varphi=1$ seems to correspond to a genuine persistent factorisation, as apart from the bad prime $p=7$, the polynomial $R_H^{(p)}(\HV_1/\IZ_6,T)$ has a quadratic factorisation for all primes studied. 

As we have several primes for which the polynomials $R_H^{(p)}(\HV_\varphi/\IZ_6,T)$ only factorise for a single value of $\varphi$, it may be tempting to immediately conclude that there cannot be further persistent factorisations. However, there may be values of $\varphi$ such that at these point the value of the representative in $\IF_p$ of $\varphi$ coincides with $\varphi=1$ or corresponds to a singular manifold. Since in our search we did not find such points where $\varphi$ satisfies the linear equation \eqref{eq:linear_factorisation_ansatz} and would therefore correspond to an element of $\IQ$, we next look for points where $\varphi$ belongs to a quadratic field extension, that is $\varphi \in \IQ(\alpha)$ with $\alpha \notin \IQ$, $\alpha^2 \in \IQ$. To do this, we simply search for coefficients $k_0,k_1,k_2 \in \IZ$ such that 
\begin{align} 
p(\varphi) \defineas k_0 + k_1 \varphi + k_2 \varphi^2 = 0 \mod p~, \qquad \text{or} \qquad p(\varphi) = 0 \text{ has no solutions over $\IF_p$}
\end{align}
when $R_H^{(p)}(\HV_\varphi/\IZ_6,T)$ has a quadratic factorisation or $\Delta = 0 \!\! \mod p$. The second condition above makes again sure that the cases where a representative of $\varphi$ does not exist in the field $\IF_p$ are taken into account. We search over the values $-1000 \leq k_0, k_1 \leq 1000$ and $0 < k_2 \leq 1000$, but the only solutions we find are given by the rational values in eq. \eqref{eq:linear_factorisations_points}. Additional arguments for non-existence of further persistent factorisations can be made along the lines of appendix \ref{app:non-modular_examples}.
\subsection{Splitting of the Hodge structure at the point \texorpdfstring{$\varphi = 1$}{where the modulus equals 1}} \label{sect:splitting_of_Hodge_structure}
\vskip-10pt
The persistent factorisations --- the evidence of existence of which we have observed above --- are intimately linked to a \textit{splitting of the Hodge structure}. By this we mean the following: Using the rational periods $\Pi_i$ given by mirror symmetry, we assign a $\IQ$-structure on the horizontal cohomology which we therefore denote by $H^4_H(\HV_1/\IZ_6,\IQ)$. Then the horizontal cohomology is said to split if there exists a two-dimensional subspace $\Lambda \subset H^4_H(\HV_1/\IZ_6,\IQ)$ of Hodge type $(4,0)+(0,4)$ or $(3,1)+(1,3)$ and the remainder $\Xi$. That is,
\begin{align}
H_H^4(\HV_1/\IZ_6,\IQ) \= \Lambda \oplus \Xi~,
\end{align}
where
\begin{align}
\begin{split}
&\Lambda \= \Lambda_{\text{attr}} \subseteq \left(H^{(4,0)}(\HV_1/\IZ_6,\IC) \oplus H^{(0,4)}(\HV_1/\IZ_6,\IC)\right) \cap H^4_H(\HV_1/\IZ_6,\IQ)~,\\[5pt]
&\text{or}\\[5pt]
&\Lambda \= \Lambda_{\AK{}} \subseteq \left(H^{(3,1)}(\HV_1/\IZ_6,\IC) \oplus H^{(1,3)}(\HV_1/\IZ_6,\IC)\right) \cap H^4_H(\HV_1/\IZ_6,\IQ)~.
\end{split}
\end{align}
Recall that we have assumed that the polynomial $R_4^{(p)}(\HV_\varphi/\IZ_6,T)$ corresponding to the full middle cohomology $H^4(\HV_\varphi/\IZ_6)$ factorises into a polynomial $R_H^{(p)}(\HV_\varphi/\IZ_6,T)$ associated to the horizontal part, and the remainder $R_\perp^{(p)}(\HV_\varphi/\IZ_6,T)$,
\begin{align}
R_4^{(p)}(\HV_\varphi/\IZ_6,T) \= R_H^{(p)}(\HV_\varphi/\IZ_6,T) R_\perp^{(p)}(\HV_\varphi/\IZ_6,T)~.
\end{align}
Then the quadratic factorisation we have found implies the existence of a quadratic factor in the full polynomial $R_4^{(p)}(\HV_\varphi/\IZ_6,T)$. Therefore, it is natural to assume that the persistent factorisation we have observed corresponds in fact to a splitting of the Hodge structure of the full middle cohomology $H^4(\HV_1/\IZ_6,\IQ)$.

We call a point $\varphi$ in the complex structure moduli space corresponding to a Calabi--Yau fourfold $X_\varphi$ with a subspace of type $\Lambda_\text{attr}$ a \textit{rank-two attractor point}. This is motivated by the comparison to the threefold case, where such points are attractors of the flow in the complex structure moduli space (see for example refs. \cite{Moore:1998pn,Moore:1998zu,Candelas:2019llw,Candelas:2021mwz}) given by the attractor mechanism \cite{Ferrara:1995ih} of the four-dimensional $\cN=2$ supergravity (for a review, see for instance ref. \cite{Pioline:2006ni}). Similarly, we term points corresponding to manifolds with a subspace $\Lambda_{\AK{}}$ \textit{attractive K3 (\AK{}) points}, as two-dimensional spaces of Hodge type type $(1,2)+(2,1)$ are so-called Tate twists of two-dimensional subspaces of Hodge type $(2,0)+(0,2)$, which appear as transcendental lattices of attractive K3~surfaces (see also appendix \ref{app:K3_modularity}). These points can also be thought of as analogues of the \textit{supersymmetric flux vacuum points} of threefolds, where existence of a subspace of Hodge type $(1,2)+(2,1)$ implies that the corresponding Calabi--Yau threefold supports supersymmetric flux vacua \cite{Candelas:2019llw,Kachru:2020sio,Kachru:2020abh,Candelas:2023yrg}.

The connection between the splitting of the Hodge structure of a Calabi--Yau fourfold $X$ and the existence of a persistent factorisation arises from the fact that the Frobenius map $\Fr_p$ can be, roughly speaking, viewed as an element of the absolute Galois group $\text{Gal}(\overline{\IQ}/\IQ)$ of automorphisms of the algebraic closure $\overline{\IQ}$ that leave $\IQ$ pointwise fixed, which have a natural action on the (étale) cohomology $H^4(X)$, furnishing a $b_4(X)$-dimensional representation $\rho$ of $\text{Gal}(\overline{\IQ}/\IQ)$.\footnote{$\Fr_p$ defines an element of $\text{Gal}(\overline{\IF_p}/\IF_p)$, and there exists a natural projection map $\pi: \text{Gal}(\overline{\IQ}_p/\IQ_p) \to \text{Gal}(\overline{\IF_p}/\IF_p)$, with $\IQ_p$ the field of $p$-adic numbers, and the inclusion map $r: \text{Gal}(\overline{\IQ}_p/\IQ_p) \to \text{Gal}(\overline{\IQ}/\IQ)$. If the Galois representation $\rho$ is unramified at $p$, one can use these maps to define a canonical lift of $\Fr_p$, $\rho(\Fr_p)$ as an element of the representation. For ramified primes the situation is slightly more complicated (see for example ref. \cite{Yang:2020sfu}).} While we need not concern ourselves with the details of this group or the representations, there are several number theoretic results and conjectures that connect the structure of the middle cohomology to the properties of the Frobenius map and thus the zeta function, one prominent example being the Weil conjectures reviewed in \ref{sect:Weil_conjectures}. In particular, assuming the Hodge conjecture, existence of a two-dimensional subspace of a type $\Lambda_\text{attr}$ or $\Lambda_{\AK{}}$ would imply that the representation of $\text{Gal}(\overline{\IQ}/\IQ)$ splits into a two-dimensional representation $\rho_\Lambda$ acting on $\Lambda$ and the remainder \cite{Kachru:2020sio}\footnote{\label{foot:Galois_splitting}Strictly speaking, the two-dimensional subspace corresponding to $\Lambda_{\AK{}}$ is only known to furnish a representation of $\text{Gal}(\overline{\IQ}/\IE)$ with $\IE \supset \IQ$ some field extension of $\IQ$. However, we assume that $\IE = \IQ$ for the case of the Hulek--Verrill fourfold. We view the fact that we find that the manifold $\MHV_1/\IZ_6$ has the properties that would be expected in the case $\IE = \IQ$ as a partial a posteriori justification of this assumption. This is related to the assumption of existence of a motive, roughly speaking an algebraically-defined piece of cohomology, $M_{\AK{}}$ corresponding to $\Lambda_{\AK{}}$ we make in section \ref{sect:Delignes_periods}. For some discussion on possible subtleties, see e.g. ref. \cite{Kachru:2020sio} or \cite{Kuusela:2022hga}.}:
\begin{align}
\rho \= \rho_\Lambda \oplus \rho_\Xi~.
\end{align}
Consequently, the Frobenius map would also split into $\Fr_{\Lambda,p} \oplus \Fr_{\Xi,p}$, where $\Fr_{\Lambda,p}$ acts on $\Lambda$ and $\Fr_{\Xi,p}$ on the remainder $\Xi$. Explicitly, this would imply that the matrix $\mtU_p(\varphi)$ representing the action of the Frobenius map would, in a suitable basis, take on a block diagonal form, with one $2 \times 2$ block. If such a split exists, the degree-two piece that appears in $R_H^{(p)}(\HV_1/\IZ_6,T)$ could be attributed to the action of $\Fr_p$ on $\Lambda$, or equivalently to the $2\times 2$ block in the matrix $\mtU_p(\varphi)$. Therefore, with some prescience, let us denote one of the quadratic factors appearing in the polynomial $R^{(p)}_H(\HV_1/\IZ_6,T)$ by $R_{\AK{}}^{(p)}(\HV_1/\IZ_6,T)$ so that 
\begin{align}
R_H^{(p)}(\HV_1/\IZ_6,T) \= R_{\AK{}}^{(p)}(\HV_1/\IZ_6,T) R_\Xi^{(p)}(\HV_1/\IZ_6,T)~.
\end{align}
We observe that we can always choose $R_{\AK{}}^{(p)}(\HV_1/\IZ_6,T)$ to be a factor that is of the form
\begin{align}
R_{\AK{}}^{(p)}(\HV_1/\IZ_6,T) \= 1 - b_p p T + p^4 T^2~,
\end{align}
which uniquely fixes the factor.

Given that we have in fact observed that the second factor $R_\Xi^{(p)}(\HV_1/\IZ_6,T)$ of degree three always factorises further into a linear factor and a quadratic, it would be tempting to assume that there are, in fact, two sublattices. Recall, however, that it was argued in section \ref{sect:Weil_conjectures} that the functional equation arising from the Weil conjectures is enough to guarantee that there always exists a linear factor. Thus it is not a priori clear that the second factor should correspond to a two-dimensional subspace. Indeed, below we will find numerical evidence indicating that no such piece exists.

\subsubsection*{Verifying the existence of a subspace $\Lambda_{\AK{}}$}
\vskip-5pt
We can analytically continue the periods collected into the vector $\Pi = \mtT \varpi$ by numerically integrating the Picard--Fuchs equation to find the value of the period vector and its derivatives at the point $\varphi=1$. We choose a path that circles the singularities in the upper half plane (see \fref{fig:moduli_space}), although any other choice of path only differs by the action of the monodromies.
\begin{figure}[H] 
    \centering

\begin{tikzpicture}
    \usetikzlibrary{decorations.pathmorphing}

    \tikzset{snaking/.style={decorate, decoration=snake}}

    \filldraw[fill=gray!30,draw=none] (0,0) circle (1);

	  \path [draw=black,snaking]
    (-2,0) -- (0,0);

    \path [draw=black,snaking]
    (1,0) -- (11,0);
	
	\draw [line width=1pt] (0,0) -- (0.5,0.5) -- (9.5,0.5) -- (10,0);		

	\draw [line width=1pt] (-2.5,-2.5) -- (-2.5,2.5) -- (11.5,2.5) -- (11.5,-2.5) -- cycle;		 
	
	\foreach \Point in {(0,0), (1,0), (2,0), (5,0)}{
		\node at \Point {\textcolor{red}{\textbullet}};}

  	\foreach \Point in {(10,0)}{
		\node at \Point {\textcolor{black}{\textbullet}};}
\end{tikzpicture}
\vskip1pt
\place{1.6}{1}{$\varphi{=}1/36$}
\place{3.23}{1}{$\varphi{=}1/4$}
\place{5.27}{1}{$\varphi{=}1$}
\capt{6in}{fig:moduli_space}{A schematic representation of the complex structure moduli space of the family $\HV_\varphi/\IZ_6$. The red dots denote the singularities of the Picard--Fuchs operator at $\varphi=0,1/36,1/16,1/4$, the shaded region denotes the large complex structure region where the series expressions such as \eqref{eq:fundamental_period_series_quotient} converge. The black dot denotes the point $\varphi=1$ which we have argued corresponds to a manifold with a persistent factorisation. To obtain the numerical expression for the period vector at this point, we numerically integrate the Picard--Fuchs equation along the path indicated with the black line.}	
\vskip-15pt
\end{figure}
Doing this, we find that, to the numerical accuracy of at least 100 digits 
\begin{align} \label{eq:ReImDPi}
\Re \, \cD \Pi \= c_1 F~, \qquad \Im \, \cD \Pi \= c_2 H~,
\end{align}
where $F$ and $H$ are the integral vectors 
\begin{align} \label{eq:F_H_definition}
F \= (12, 5, 20, -50, 10)^T~, \qquad H \= (0, 5, 24, -80, 20)^T~,
\end{align}
and $c_1$ and $c_2$ are constants that are given by
\begin{align} \label{eq:c1_c2_numerical}
\begin{split}
c_1 &\= -0.011783350037859430679992523295962605881917251716994...\ ,\\[5pt]
c_2 &\= -0.003042447897277474018298818684307367137030444536707...\ .
\end{split}
\end{align}
Here $\cD$ denotes the Kähler covariant derivative defined by
\begin{align}
\cD \defineas \partial_\varphi - (\partial_\varphi K)~, \qquad \text{with} \qquad K \defineas \log \left(-\ii \Pi^\dag \Sigma \Pi \right)~.
\end{align}
This is defined such that $\cD \Omega$, which corresponds to the vector $\cD \Pi$, belongs to $H_H^{(3,1)}(\HV_\varphi,\IC)$. As $\Re \, \cD \Omega$ and $\Im \, \cD \Omega$ (or equivalently $\cD \Omega$ and $\overline{\cD \Omega}$) span the space $H_H^{(3,1)}(\HV_\varphi,\IC) \oplus H_H^{(1,3)}(\HV_\varphi,\IC)$, the two-dimensional subspace corresponding to the $\IQ$-span of the integral vectors $F$ and $H$,
\begin{align}
\Lambda_{\AK{}} \defineas \langle \Re \, \cD \Omega /c_1 , \Im \, \cD \Omega /c_2 \rangle_\IQ \subset H_H^4(\HV_1/\IZ_6,\IQ)
\end{align}
is of the Hodge type $(1,3)+(3,1)$, and we can identify $\varphi=1$, associated to the manifold $\HV_1 /\IZ_6$, as an \AK{} point.

We can use the same technique to investigate whether there exists a two-dimensional subspace $\Lambda_{\text{attr}}$ of Hodge type $(4,0)+(0,4)$, corresponding to the vectors $\Re \, \Pi$ and $\Im \, \Pi$. We find that 
\begin{align}
\begin{split}
&\Re \, \Pi \= c_3 \, (1, 1, 8, -10, 2)^T~, \qquad \text{where}\\[3pt]
&c_3 \= -0.156167172440226442754946776771084997380089550669633... \ .
\end{split}
\end{align}
However, based on a similar numerical computation, it seems that $\Im \, \Pi \notin c \IQ$ for any real coefficient $c$. For instance, considering the ratio $\Im \, \Pi_0(1) / \Im \, \Pi_1(1)$, the rational number $p/q$ with smallest denominator that satisfies
\begin{align}
\left| \frac{\Im \, \Pi_1(1)}{\Im \, \Pi_0(1)} - \frac{p}{q} \right| < 10^{-100}
\end{align}
has height\footnote{The \textit{height} of an irreducible rational number $p/q$ is defined as $|p|+|q|$.} of order $\approx 10^{50}$. In addition, if we increase the accuracy to which we work, we find a different rational number. This fact, together with the fact that the other rational numbers that appear in the vectors $\Pi, \cD \Pi$ have much smaller heights, give a relatively strong indication that the ratio $\Im \, \Pi_0(1) / \Im \, \Pi_1(1)$ is not rational, and therefore there is no two-dimensional subspace $\Lambda_{\text{attr}}$ of Hodge type $(4,0)+(0,4)$.
\subsubsection*{Identifying modular forms}
\vskip-10pt
By Serre's modularity conjecture \cite{Serre1975a,Serre1987a}, proven by Khare, Wintenberger, and Kisin \cite{Khare2009a,Khare2009b,Kisin2009a}, a~two-dimensional representation --- such as the one associated to a subspace $\Lambda_{\AK{}}$ --- of the absolute Galois group $\text{Gal}(\overline{\IQ}/\IQ)$ is attached to a modular form. Under this correspondence the eigenvalues of the Frobenius element $\rho(\Frob_p)$ are related to the modular form coefficients. In the case of $\HV_1/\IZ_6$, assuming that $\Lambda_{\AK{}}$ gives rise to such a representation (and not a representation of $\text{Gal}(\overline{\IQ}/\IE)$ with $\IE \supset \IQ$ a field extension of $\IQ$), we expect that the coefficients $b_p$ appearing in the factor 
\begin{align}
R_{\AK{}}^{(p)}(\HV_1/\IZ_6,T) \= 1 - b_p p T + p^4 T^2~,
\end{align}
are the Fourier coefficients of a modular form. This phenomenon is known as \textit{(arithmetic) modularity} (for further discussion, see appendix \ref{app:modularity}). We can further predict the weight of the modular form by noting that the Hodge type of $\Lambda_{\AK{}}$ is $(3,1)+(1,3)$, this is related by a so-called Tate twist (see section \ref{sect:Delignes_periods} and appendix \ref{app:periods_motives}) to a space of Hodge type $(2,0)+(0,2)$. Such spaces appear as transcendental lattices of attractive K3~surfaces, which have been proven \cite{Livne1995a} to be related to weight-3 modular forms (see~appendix~\ref{app:K3_modularity}). The effect of the Tate twist on $R_{\AK{}}^{(p)}(\HV_1/\IZ_6,T)$ is to rescale $T$ by $T \mapsto T/p$, giving a polynomial 
\begin{align}
R_{\AK{}}^{(p)}(\HV_1/\IZ_6,T/p) \= 1 - b_p T + p^2 T^2~,
\end{align}
which is a form that appears in the local zeta functions of an attractive K3~surface. This form also explains, why we expect the coefficient $b_p$, rather than the combination $b_p p$ that appears in $R_{\AK{}}^{(p)}(\HV_1/\IZ_6,T)$, to be a Fourier coefficient of the associated modular form.

Since we have computed the polynomials $R_H^{(p)}(\MHV_1/\IZ_6,T)$ for primes $p$ up to $p \leq 733$, we can easily read off the corresponding values of $b_p$, the first few of which in table \ref{tab:b_n_coefficients}. By comparing these coefficients to the millions of modular forms listed on LMFDB, we find that among those forms, there is exactly one form, with the label \textbf{15.3.d.b}, whose Fourier coefficients agree with the $b_p$.
\begin{table}[H]
	\begin{center}
		\begin{tabular}{|c|c|c|c|c|c|c|c|c|c|c|c|c|c|c|c|c|c|c|}
			\hline
			$p$    & 7 & 11 & 13 & 17 & 19 & 23 & 29 & 31 & 37 & 41 & 43 & 47 & 53 & 59 & 61 & 67 & 71 & 73 \\ \hline
			$b_p$  & - & 0  & 0  & -14 & -22 & 34 & 0 & 2 & 0 & 0 & 0 & -14 & -86 & 0 & -118 & 0 & 0 & 0     \\ \hline
		\end{tabular}
		\vskip10pt
		\capt{5.6in}{tab:b_n_coefficients}{The coefficients $b_p$ appearing in the polynomials $R_{\AK{}}^{(p)}(\HV_1/\IZ_6,T)$ for the first few primes $p$. The coefficient $b_7$ is not included, as the manifold $\HV_1/\IZ_6$ develops an additional singularity over $\IF_7$.}
	\end{center}
 \vskip-20pt
\end{table}
This is the modular form
\begin{align} \label{eq:modular_form}
	f \in S_3\left(\Gamma_0(15),\chi_{-15} \right)~, \qquad f(\tau) \= \left(\eta(3 \tau)\eta(5 \tau)\right)^3 + (\eta(\tau)\eta(15\tau))^3~,
\end{align}
where $\chi_{-15}$ is the Dirichler character given by the Kronecker symbol $\left( \frac{-15}{\cdot}\right)$. For a brief review of some relevant notation and results concerning modular forms, see appendices~\ref{app:modularity} and \ref{app:Modular_forms_for_Gamma_0_15}.

The fact that we identify a unique modular form associated to the polynomials $R_{\AK{}}^{(p)}(\HV_1/\IZ_6,T)$ provides strong evidence that we have identified the correct modular form, and that the assumptions we have made are consistent. However, we can go even further: In the following section \ref{sect:Delignes_periods}, we show that the $L$-function values associated to the modular form \textbf{15.3.d.b} appear in the numerical expressions for the derivatives of integral periods $\Pi_i$ in the way predicted by Deligne's conjecture, providing yet another highly non-trivial consistency check. In section \ref{sect:Geometric_origin_of_splitting} we find a natural geometric interpretation for the modular form, by noting that the manifolds $\HV_{\bm \varphi}$ are birational to a K3 fibrations, where, for $\varphi=1$, an attractive K3 corresponding to the modular form \textbf{15.3.d.b} appears as a fibre over $\IP^1$. This can be interpreted as giving rise to the modular form and the Tate twist. The modularity of the attractive K3 surfaces that appear in the fibration and their relation to the modular form \eqref{eq:modular_form} was noted already in ref.~\cite{Bonisch:2020qmm}.
\subsection{Deligne's periods and \texorpdfstring{$L$}{L}-function values} \label{sect:Delignes_periods}
\vskip-10pt
We show in this section, that, analogously to refs. \cite{Candelas:2019llw,Candelas:2023yrg,Yang:2020sfu,Yang:2019kib,Yang:2020lhd,Bonisch:2022mgw}, the value of the vector $\cD \Pi$ at the \AK{} point can be expressed in terms of critical values of the $L$-function of the modular form  \textbf{15.3.d.b} associated to the local zeta function of the modular Calabi--Yau manifold. This striking correspondence, which is used in section \ref{sect:physics} to write certain physical quantities in terms of the $L$-function values, can be explained by appealing to Deligne's conjecture \cite{Deligne1979a} (for a physicist-friendly introduction, see also refs. \cite{Yang:2020lhd,Yang:2020sfu,Candelas:2023yrg}, for example), which predicts a relationship between periods and $L$-function values. In this section, we first briefly review the conjecture. Then, by explicitly computing the periods and $L$-function values, we numerically verify Deligne's conjecture in the case of $\HV_{1}/\IZ_6$.

\subsubsection*{Motives and their realisations}
\vskip-5pt
Deligne's conjecture is most conveniently formulated in the language of \textit{motives}, which can be thought of as generalisations of cohomology theories, in the following sense: It is well-known that there is no well-defined algebraically-defined $\IQ$ cohomology (see e.g. ref. \cite{Milne2013a}). However, different cohomology theories, such as the étale cohomology, de Rham cohomology, and crystalline cohomology share many of the same properties as if they arose from such a cohomology theory. Motives are essentially a way of explaining this underlying common structure, keeping track of some properties that are independent of the choice of a `good' cohomology theory.

In particular, given a smooth projective variety $X$ defined over $\IQ$, one can associate a motive to its middle cohomology. This motive is usually denoted (for fourfolds) by $h^4(X)$. Since this is essentially the only motive we will discuss, we shall denote it simply by $M \defineas h^4(X)$. However, we will not need the full machinery of motives for the purposes of the present discussion. Instead, it is enough to think of two concrete \textit{realisations} of the motive in question. These classical realisations are (see for example ref. \cite{Yang:2020lhd} or the appendices of ref. \cite{Kachru:2020sio} for details):
\begin{enumerate}
	\item The Betti realisation is in our case just the singular cohomology $H^4_{\text{sing.}}(X,\IQ)$ which we denote by $H^4_\BB(X,\IQ)$ to emphasise the motivic point-of-view we take.
	\item The de Rham realisation is given by the algebraic de Rham cohomology defined as the hypercohomology $\IH(X,\Omega^\bullet)$ of the sheaf $\Omega^{\bullet}$ of algebraic differential forms. This is a finite-dimensional vector space over $\IQ$, so we denote it by $H^4_\dR(X,\IQ)$.
\end{enumerate}
For concreteness, in this section, we only discuss the motive $M$, its submotives we introduce below, and their realisations. However, the generalisation to other motives is obtained simply by replacing the cohomologies $H^4(X)$ appearing here with the relevant cohomology $H^i(X)$ associated to any other motive $N$. For more details on motives, we refer an interested reader to an accessible summary ref. \cite{Milne2013a} and references therein. In physics, Calabi--Yau motives have also been investigated using methods complementary to ours, for example in~refs.~\cite{Kadir:2010dh,Schimmrigk:2006dy}.

Recall that as $X$ is a complex Kähler manifold the singular cohomology admits a Hodge structure 
\begin{align}
	H^4_\BB(X,\IQ) \otimes_\IQ \IC \= \bigoplus_{p+q=4} H^{p,q}(X,\IC)~,
\end{align}
which also gives a natural Hodge filtration $F^p H_\BB(X,\IQ) \= \bigoplus_{s \geq p} H^{s,4-s}(X,\IC)$. On the other hand, the de Rham cohomology has a filtration $F^p H_\dR(X,\IQ)$ inherited from the Hodge filtration of the complex $\Omega^\bullet$ (see ref. \cite{Voisin2007a}, for instance). 

The Betti and de Rham realisations have a canonical \textit{comparison isomorphism} between them. This is the map
\begin{align}
I_\infty: H^4_\BB(X,\IQ) \otimes_\IQ \IC \to H^4_{\dR}(X,\IQ) \otimes_\IQ \IC~,
\end{align}
which is given by de Rham's theorem (see e.g. ref. \cite{Griffiths1978a}), in the following sense: Recall that from de Rham's theorem it follows that there exists a non-degenerate bilinear pairing between the singular homology (denoted here $H_\BB^4(X,\IQ)^\vee$ to keep in line with the `motivic' notation) and the de Rham cohomology 
\begin{align}
\langle *,* \rangle : H_\BB^4(X,\IQ)^\vee \otimes_\IQ \IC &\times H_\dR^4(X,\IQ) \otimes_\IQ \IC \to \IC~, \qquad \langle \gamma, \eta \rangle \= \int_\gamma \eta~.
\end{align}
Note that we can define a linear map $f_\eta \in \text{Hom}(H_\BB^4(X,\IQ)^\vee,\IC) = H_\BB^4(X,\IQ) \otimes_\IQ \IC$ by taking
\begin{align}
f_\eta(\gamma) \= \int_\gamma \eta~, \qquad \text{for any } \gamma \in H_\BB^4(X,\IQ)^\vee~,
\end{align}
so we can define the inverse map $I_\infty^{-1}$ as
\begin{align}
I_\infty^{-1}(\eta) \= f_\eta~.
\end{align}
This isomorphism can be used to define a natural action of complex conjugation on $H_\BB^4(X,\IQ)$. To do this, simply note that $I_\infty^{-1}$ sends the natural complex conjugation on $H^4_{\dR}(X,\IQ) \otimes_\IQ \IC$, which acts simply as $\II \otimes c$, $c$ being the ordinary complex conjugation on $\IC$, to an involution $F_\infty \otimes c$. 

Using this canonical isomorphism, one can also define two (generically different) $\IQ$-structures on $H_\dR(X,\IQ) \otimes \IC$, one being the canonical $\IQ$-structure of $H^4_{\dR}(X,\IQ)$, and the other induced from $H^4_{\BB}(X,\IQ)$ via the comparison isomorphism. To be explicit, we can choose a basis $\Upsilon_i$ for $H^4_\dR(X,\IQ)$, and a basis $A^i$ for $H^4_\BB(X,\IQ)^\vee$. Let then $\wt A_i$ be a basis of $H^4_\BB(X,\IQ)$ dual to $A^i$ in the sense that $\wt A_j(A^i) \= \delta^i_j$, and denote by $I_\infty(A^i) \defineas \alpha_i \in H^4_\dR(X,\IQ)$ the dual basis of $\wt A_j$ under the pairing $\langle *, *\rangle$.\footnote{Note that from the definitions above it follows that $\langle A_i, I_\infty(\wt A^j)\rangle = \delta_i^j$.} With these definitions, we can write
\begin{align} \label{eq:I_infty_determinant_explicit}
	\Upsilon_i \= \langle A^j, \Upsilon_i \rangle \, \alpha_j \= \left( \int_{A^j} \Upsilon_i \right) \alpha_j \defineas [\mtI_\infty]_i^j \alpha_j~,
\end{align}
where the last expression defines the matrix $\mtI_\infty$ corresponding to the isomorphism $I_\infty$ in the chosen basis. This matrix essentially gives the difference between the two different $\IQ$-structures. Note that the matrix $\mtI_\infty$ is only well-defined up to a choice of the rational bases of $H^4_\BB(X,\IQ)$ and $H^4_\dR(X,\IQ)$.

The fact that the two $\IQ$-structures which are related by $\mtI_\infty$ are generically different can be used to define \textit{periods}, in the sense of Kontsevich and Zagier \cite{Kontsevich2001a}. Prominent examples of such a periods are Deligne's periods \cite{Deligne1979a}, to whose definition we now turn. See also appendix \ref{app:periods_motives} for a simple introductory example of this notion of a period.

\subsubsection*{Definition of Deligne's period}
\vskip-5pt
\textit{Deligne's periods} $c^\pm(M)$ are defined as follows: Define $M_{\text{B}}^{\pm}$ as subspaces of the Betti realisation $H^{4}_{\text{B}}(X,\IQ)$ of the motive $M$ as the positive and negative eigenspaces of the complex conjugation map $F_{\infty}$.\footnote{In fact, this map can in fact be identified as the Frobenius map $\text{Fr}_p$ at the `infinite prime'.} That is, we define $M^\pm_{\text{B}}$ so that
\begin{align}
	F_\infty \big|_{M^\pm_{\text{B}}} = \pm \II~.
\end{align}
We also define two subspaces $M_\dR^\pm$ of the de Rham realisation of $M$ by
\begin{align}
	M^{\pm}_{\text{dR}} \= H_{\text{dR}}(X,\IQ)/F^{\mp}~,
\end{align}
where $F^\pm \defineas F^pH_{\dR}^4(X,\IQ) \subset H^4_\dR(X,\IQ)$ for $p$ such that
\begin{align}
	\dim_{\IQ} F^\pm \= \dim_{\IQ} M^\pm_{\text{B}}~.
\end{align}
With these definitions, it can be shown that the comparison isomorphism $I_\infty$ induces the isomorphisms $I_\infty^\pm: M^{\pm}_\BB \otimes_\IQ \IC \to M^\pm_\dR \otimes_\IQ \IC$.

We can express the maps $I_\infty^\pm$ in terms of matrices $\mtI_\infty^\pm$ in a way completely analogous to eq.~\eqref{eq:I_infty_determinant_explicit}. Explicitly, if we choose the bases $A^{\pm,i}$ for $(M^{\pm}_\BB)^\vee$, and $\Upsilon^\pm_i$ for $M^{\pm}_\dR$, and let $\beta^{\pm,i}$ be the basis of $M^{\pm}_\dR$ consisting of Poincaré duals of $A^{\pm,i}$, then 
\begin{align}
	\left[\mtI_\infty^\pm \right]^i_{j} \= \int_{A^{\pm,i}} \Upsilon^\pm_j \= \frac{1}{(2\pi \ii)^4} \int_X \Upsilon^\pm_j \wedge \beta^{\pm,i}~.
\end{align}
Deligne's periods can then be expressed as the determinants of these matrices:
\begin{align} \label{eq:Deligne's_period_definition}
	c^\pm(M) \defineas \det(\mtI_\infty^\pm)~.
\end{align}
The determinants are well-defined up to an overall rational factor, which depends on the choices of the bases above.

\subsubsection*{Deligne's conjecture}
\vskip-5pt
Given the motive $M$ corresponding to the middle cohomology of a fourfold $X$, one can associate to it a \textit{motivic $L$-function},\footnote{The relation of the $L$-function to the motive $M$ can be explained by considering the étale realisation of $M$.} which is defined by using the polynomials $R_4^{(p)}(X,T)$ appearing in the expression \eqref{eq:zeta_fourfold_generic} for the local zeta function $\zeta_p(X,T)$\footnote{Here we are ignoring the slight subtlety introduced by \textit{ramification} (see for example refs. \cite{Yang:2020sfu,Frenkel:2005pa}). This will not affect our discussion.}
\begin{align} \label{eq:definition_Lfunction}
	L(M,s) \=\!\!' \;\; \prod_{p \text{ good}}\frac{1}{R_4^{(p)}(X,p^{-s})}~.
\end{align}
Here $='$ denotes equality up to finitely many factors, and the product is over the \textit{primes of good reduction} (see appendix \ref{app:modularity} for definitions). 

Deligne's conjecture concerns the relationship between the so-called \textit{critical motives} and the values of these motivic $L$-functions. We leave the precise definition of a critical motive to appendix \ref{app:periods_motives}, but note here that in the cases we are interested in, the condition for a motive to be critical can be stated in terms of its Hodge structure as the requirement that for every pair $(p,q)$ such that $h^{p,q}(H^4_\BB(X,\IQ)) \neq 0$, either $p < 0$ and $q \geq 0$ or $p \geq 0$ and $q < 0$.\footnote{For motives with a Hodge structure of even weight there is also a condition on the space $H^{w/2,w/2}$. However, this space will be trivial in the case we study later.}

Deligne's conjecture can now be stated as the prediction that, for a critical motive $M$, the motivic $L$-function value $L(M,0)$ is a rational multiple of the period $c^+(M)$, that is
\begin{align}
\frac{L(M,0)}{c^+(M)} \in \IQ~.
\end{align}
\subsubsection*{The motive $M^{\AK{}}$ corresponding to the subspace $\Lambda^{\AK{}}$}
\vskip-5pt
We assume, that, as suggested by the zeta function factorisation and Hodge-like conjectures (see refs. \cite{Bonisch:2020qmm,Kachru:2020sio}, for instance), at the \AK{} points the motive $M$ splits into a direct sum of submotives
\begin{align} \label{eq:M_SFV_submotive}
M \= M^{\AK{}} \oplus M'~,
\end{align}
where $M_{\AK{}}$ is of Hodge type $(3,1)+(3,1)$ associated to the subspace $\Lambda_{\AK{}}$, and $M'$ is of type $(4,0)+(2,2)+(0,4)$. We are particularly interested in the motive $M^{\AK{}}$, as being a rank-two motive, its properties are better-understood than those of its higher-rank counterparts, and it is a subject to various intriguing conjectures. To the motive $M^{\AK{}}$ we can associate a motivic $L$-function, which is defined in terms of the polynomials $R_{\AK{}}^{(p)}(M^{\AK{}},T)$ which we have computed (for the first 130 primes) in the previous section:
\begin{align}
L(M^{\AK{}},s) \=\!\!' \;\; \prod_{p \text{ good}} \frac{1}{R_{\AK{}}^{(p)}(M^{\AK{}},p^{-s})}~.
\end{align}
In the case of the manifold $\MHV_1/\IZ_6$, we have (conjecturally) identified the weight-3 modular form \textbf{15.3.d.b} as the modular form whose Fourier coefficients determine the polynomials $R_H^{(p)}(X,T)$. Therefore we expect that the motivic $L$-function $L(M^{\AK{}},s)$ is exactly the $L$-function with the LMFDB label \textbf{2-15-15.14-c2-0-1} associated to the modular form \textbf{15.3.d.b} (for definitions of modular forms and their associated $L$-functions, see appendix \ref{app:modularity}). 

From the definition of a critical motive given above, it follows that the motive $M^{\AK{}}$, being of Hodge type $(3,1)+(1,3)$ is not critical. Thus, at first glance it might seem that Deligne's conjecture would not be able to tell us anything useful about $M^{\AK{}}$. However, a motive of Hodge type $(1,-1)+(-1,1)$, or $(0,-2)+(-2,0)$ would be critical. 

It turns out that $M^{\AK{}}$ can be trivially related to a motive of Hodge type $(1,-1)+(-1,1)$ or $(0,-2)+(-2,0)$ by studying its \textit{Tate twists}. The Tate twist $M(m)$ of a motive $M$ is defined as a tensor product with the \textit{Tate motive} $\IQ(m)$
\begin{align}
M(m) \defineas M \otimes_\IQ \IQ(m)~.
\end{align}
For instance, the Betti realisation of the Tate motive is $(2\pi \ii)^m \IQ$, so the Tate twist $M(m)$ of $M$ has the Betti realisation $H^{4}_\BB(X,\IQ) \otimes_\IQ (2\pi\ii)^m \IQ$.\footnote{Some further details on these properties and the definition of the Tate motive can be found in appendix \ref{app:periods_motives}.}

Taking the Tate twist has three effects that are of central importance to us:
\begin{enumerate}
	\item The Betti realisation $(2\pi \ii)^m \IQ$ can be given a pure Hodge structure of type $(-m,-m)$. In particular, the Tate twists $M^{\AK{}}(2)$ and $M^{\AK{}}(3)$ have Hodge types $(1,-1)+(-1,1)$ and $(0,-2)+(-2,0)$, respectively, and are thus critical.
	\item Twisting introduces multiples of $2\pi\ii$; if $\alpha \in H^4_{\text{B}}(X,\IQ)$, then $(2\pi \ii) \alpha \in H^4_{\text{B}}(X,\IQ) \otimes_\IQ (2\pi \ii) \IQ$. When computing Deligne's periods, this elementary observation has the effect of multiplying the periods by powers of $(2\pi \ii)$, and changing the eigenvalue of the complex conjugation map $F_\infty$, thus exchanging the periods $c^{\pm}$. The cases we are interested in are
	\begin{align}
    \begin{split}
		c^{\pm}(M^{\AK{}}(2)) &\= (2\pi \ii)^2 \, c^\pm(M^{\AK{}})~,\\
		c^{\pm}(M^{\AK{}}(3)) &\= (2\pi \ii)^3 \, c^\mp(M^{\AK{}})~. 
    \end{split}
	\end{align}
	\item The twist by $\IQ(m)$ affects the polynomials $R^{(p)}_{\AK{}}(M,T)$ appearing in the zeta function by rescaling the argument $T$ by $T \mapsto p^{-m} T$ so that 
	\begin{align}
		R^{(p)}_{\AK{}}(M^{\AK{}}(m),T) \= R^{(p)}_{\AK{}}(M^{\AK{}},p^{-m} T)~.
	\end{align}
	Recalling that these polynomials appear in the definition of the motivic $L$-function, we have
	\begin{align}
    \begin{split}
		L(M^{\AK{}}(m),s) &\= \prod_{p} \frac{1}{R^{(p)}_{\AK{}}(M^{\AK{}}(m),p^{-s})}\\
        &\= \prod_{p} \frac{1}{R^{(p)}_{\AK{}}(M^{\AK{}},p^{-(s+m)})} \= L(M^{\AK{}},s+m)~.
    \end{split}
	\end{align}		
	Then, comparing to the definition of the weight-3 $L$-function $L_3(s)$ with LMFDB label \textbf{2-15-15.14-c2-0-1}, which is associated to the modular form \textbf{15.3.d.b}, we see that
	\begin{align}
		L_3(s) \= L(M^{\AK{}}(1),s) \= L(M^{\AK{}},s+1)~.
	\end{align}
	In particular, the $L$-function values that appear in Deligne's conjecture are
	\begin{align}
		L(M^{\AK{}}(2),0) \= L_3(1)~, \qquad L(M^{\AK{}}(3),0) \= L_3(2)~.
	\end{align}	
\end{enumerate}
The explicit values of $L_3(1)$ and $L_3(2)$ can be computed using Dokchitser's algorithm \cite{Dokchitser2004a}, which has been implemented in the computer algebra system Sage \cite{sagemath}: 
\begin{align} \label{eq:L-function_values_numeric}
\begin{split}
L_3(1) & \= 0.54271934916842485520176378379613163082128362217397375229323... \ ,\\
L_3(2) & \= 0.88045982535822981044968910894132568513898932226249847773441...
\end{split}
\end{align} 

These statements explain our previous assertion that the Tate twisted motive $M(m)$ is trivially related to the original $M$, as both Deligne's periods and the $L$-functions of the twist are related to those of $M$ via simple relations. In the case of $M^{\AK{}}$, Deligne's conjecture predicts two highly non-trivial relations
\begin{align} \label{eq:Deligne's_conjecture_ratios}
\begin{split}
\frac{c^+(M^{\AK{}}(2))}{L(M^{\AK{}}(2),0)} &\= (2\pi \ii)^2 \, \frac{c^+(M^{\AK{}})}{L(M^{\AK{}},1)} \= (2\pi \ii)^2 \,\frac{c^+(M^{\AK{}})}{L_3(1)} \in \IQ~, \\[5pt]
\frac{c^+(M^{\AK{}}(3))}{L(M^{\AK{}}(3),0)} &\= (2\pi \ii)^3 \, \frac{c^-(M^{\AK{}})}{L(M^{\AK{}},2)} \= (2\pi \ii)^3 \,\frac{c^-(M^{\AK{}})}{L_3(2)} \in \IQ~.
\end{split}
\end{align}
\subsubsection*{Explicit evaluation of Deligne's periods of $M^{\AK{}}$}
\vskip-5pt
To test whether Deligne's conjecture holds in the case of the motive $M^{\AK{}}$, we evaluate the ratios of Deligne's periods to $L$-function values numerically. We have already given the relevant $L$-function values in eq. \eqref{eq:L-function_values_numeric}, so it remains to compute the periods $c^+(M^{\AK{}}(2))$ and $c^+(M^{\AK{}}(3))$, which we will do by relating these to the rational periods given in eq. \eqref{eq:one-parameter_periods}. If Deligne's conjecture holds for the motive $M^{\AK{}}$, we would expect the ratios \eqref{eq:Deligne's_conjecture_ratios} to be rational numbers of low height.

For simplicity, in this section, we work under the assumption that there exists a motive $M_H$ whose de Rham realisation is generated by $\{\theta^i \Omega\}_{i=0}^4$, which corresponds to the horizontal cohomology $H^4_H(\HV_\varphi/\IZ_6)$ of Hulek--Verrill manifolds. However, strictly speaking, this assumption is not necessary. By carefully going through the arguments presented below, it is clear that we need to only assume existence of the motive $M^\AK{}$ introduced in eq. \eqref{eq:M_SFV_submotive}, and the assumptions we made in section \ref{sect:Deformation_method} to be able to compute the polynomials $R_H^{(p)}(\HV_\varphi/\IZ_6,T)$ related to the local zeta functions.

First, we construct the map $I_\infty$. The de Rham realisation $H^4_{H,\dR}(X,\IQ)$ is $\langle \Omega, \theta \Omega, \dots, \theta^5 \Omega \rangle_\IQ$. The mirror map, given by the matrix $\mtT$ in eq. \eqref{eq:change_of_basis_integral_frobenius}, gives the relation between this basis, and a basis of $H_{H,\BB}^4(X,\IQ)$, thus giving us essentially the comparison isomorphism. Using these bases, the matrix $\mtI_\infty$ corresponding to the comparison isomorphism $I_\infty$ can be written as\footnote{Note that the matrix $\mtE(\varphi)$ appears here transposed, as we have defined $\mtE$ using the ``transposed'' conventions of~ref.~\cite{Candelas:2021tqt}}
\begin{align}
\mtI_\infty \= (2\pi \ii)^4 \mtT \mtE(\varphi)^T~.
\end{align}
Recall from earlier that the map $I_\infty$ has the property that it maps the involution $F_\infty \otimes c$ acting on $H^4_\BB(X,\IQ) \otimes_\IQ \IC$ to $\II \otimes c$ acting on $H^4_{\dR}(X,\IQ) \otimes_\IQ \IC$. Noting that $\II \otimes c$ acts trivially on the basis $\{\Omega,\dots,\theta^4 \Omega\}$, this implies that
\begin{align}
\mtT \mtE(\varphi)^T\= \mtF_\infty \overline{\mtT \mtE(\varphi)^T}~,
\end{align}
from which $\mtF_\infty$ can be immediately solved, as the matrices $\mtE(\varphi)$ and $\mtT$ are invertible. To find explicitly the matrix $\mtF_\infty$ for the manifold $\HV_1/\IZ_6$, we analytically continue the periods as in section~\ref{sect:splitting_of_Hodge_structure}. Doing this, and numerically evaluating the matrix $\mtF_\infty$, we find that at least to an accuracy of 100 digits, we have
\begin{align}
\mtF_{\infty} \= \left(
\begin{array}{ccccc}
 1 & 0 & -30 & -24 & -60 \\
 0 & -1 & -10 & -9 & -24 \\
 0 & 0 & -41 & -36 & -96 \\
 0 & 0 & 100 & 89 & 240 \\
 0 & 0 & -20 & -18 & -49 \\
\end{array}
\right).
\end{align}
As we know a priori that the matrix $\mtF_\infty$ should have rational entries, the above expression is very likely actually exact.

Simple computation then shows that the vectors $F$ and $H$ introduced in \eqref{eq:F_H_definition} are eigenvectors of $F_\infty$ with eigenvalues $1$ and $-1$, respectively. Denoting these by $V_{+}$ and $V_-$ for clarity, we can write the eigenspaces $M_\BB^{\AK{},\pm}$ as their $\IQ$-spans 
\begin{align}
M_\BB^{\AK{},+} \= \langle V_+ \rangle_{\IQ}~, \qquad M_\BB^{\AK{},-} \= \langle V_- \rangle_{\IQ}~.
\end{align}
The spaces $F^{\AK{},\pm}$ are both by definition given by the 1-dimensional subspace in the Hodge filtration of the de Rham realisation, that is
\begin{align}
F^{\pm} \= \langle \cD \Pi \rangle_\IQ~.
\end{align}
The periods $c^{\pm}(M^{\AK{}})$ are then given by determinants of the maps $I^{\AK{},\pm}: \langle V_+ \rangle \otimes \IC \to \langle \cD \Pi \rangle \otimes \IC$. The spaces appearing here are one-dimensional, so the determinants are simply given by
\begin{align}
c^{\pm}(M^{\AK{}}) \= \int_{\HV_1/\IZ_6} \Gamma^{\pm} \wedge \cD \Omega \= V^{\pm} \, \Sigma \, \cD \Pi~,
\end{align}
where $\Gamma^{\pm}$ denote the four-forms corresponding to the vectors $V^{\pm}$. Plugging in the numerical expression for $\cD \Pi$ from eq. \eqref{eq:ReImDPi} and the critical values of the $L$-functions we have evaluated in eq. \eqref{eq:L-function_values_numeric}, we find that
\begin{align}\label{eq:Deligne_period_ratios}
\frac{c^{+}(M^{\AK{}})}{L_3(1)/(2\pi \ii)^2} \= -4 \in \IQ~, \qquad \frac{c^{-}(M^{\AK{}})}{L_3(2)/(2\pi \ii)^3} \= 60 \in \IQ~,
\end{align}
to the accuracy of at least 100 digits, which is the numerical accuracy we have used to make all of our computations, giving a strong evidence that Deligne's conjecture is satisfied in this case. 

We can also view Deligne's conjecture as predicting the form of the vector $\cD \Pi(1)$, thus explaining the appearance of the $L$-functions. In fact, using the ratios \eqref{eq:Deligne_period_ratios}, we can identify the coefficients $c_1$ and $c_2$ \eqref{eq:c1_c2_numerical} that determine the vector $\cD \Pi$ via eq. \eqref{eq:ReImDPi} as rational multiples of $L$-function values. Explicitly,
\begin{align}
c_1 \=  -\frac{1}{56} \, \frac{L_3(1)}{\pi^2}~, \qquad c_2 \= -\frac{3}{112} \frac{L_3(2)}{\pi^3}~,
\end{align}
so we can write the covariant derivative $\cD \Pi$ of the period vector as
\begin{align} \label{eq:DPi_L-functions}
\cD \Pi(1) \=  -\frac{3}{28} \left(2 \; \frac{L_3(1)}{\pi^2} \left(
\begin{array}{r}
	12 \\
	5 \\
	20 \\
	-50 \\
	10 \\
\end{array}
\right) + \ii \frac{L_3(2)}{\pi^3} \left(
\begin{array}{r}
	0 \\
	5 \\
	24 \\
	-80 \\
	20 \\
\end{array}
\right) \right)~.
\end{align}
\subsection{Calabi--Yau periods and modular form periods}
\vskip-10pt
Complementary to the approach taken above, which uses Deligne's periods to express the periods of the Calabi--Yau manifold associated to the motive $M^\AK$ in terms critical of $L$-function values, is the approach taken in \cite{Bonisch:2022mgw}, where the Calabi--Yau periods are expressed in terms of periods (and quasi-periods) of modular forms. We review these notions in some detail in appendix \ref{app:Modular_forms_for_Gamma_0_15}, where most of the details of the explicit computations are also relegated. 

Before stating the results of the computations, let us comment on the relation of the two methods of expressing Calabi--Yau periods in terms of number theoretic quantities. Since we have reviewed the construction of Deligne's periods in detail above, and the modular form periods are discussed in detail in \cite{Bonisch:2022mgw} and reviewed in appendix \ref{app:Modular_forms_for_Gamma_0_15}, we will keep these very brief. Recall that $H_H^4(X,\IQ)$ has a Hodge filtration 
\begin{align}
	H^4_H(X,\IQ) = F^0 H_H^4 \supset F^1 H_H^4 \supset \cdots \supset F^4H_H^4 \supset \emptyset
\end{align}
and decomposition
\begin{align} \label{eq:defintion_Hodge_decomposition}
	H^4_H(X,\IQ) \otimes_\IQ \IC \= \bigoplus_{p+q=4} H_H^{p,q}(X,\IC)~, \qquad \text{with} \qquad H_H^{q,p}(X,\IC) \= \overline{H_H^{p,q}(X,\IC)}~. 
\end{align}
 The filtration is algebraic as we can realise the subspaces $F^p H_H^4$ as
\begin{align} \label{eq:Hodge_filtration_realisation}
	F^4H_H^4 \= \langle \Omega \rangle~, \quad F^3H_H^4 \= \langle \Omega, \theta \Omega \rangle~, \quad \dots~, \quad F^0 H_H^4 \= \langle \Omega, \dots, \theta^4 \Omega \rangle~,
\end{align}
whereas, due to the complex conjugation in definition \eqref{eq:defintion_Hodge_decomposition}, the Hodge decomposition is in general~not.

These differences between these two constructions are analogous to the those between expressing the Calabi--Yau periods in terms of Deligne's periods or periods of modular forms: Since the approach based on Deligne's conjecture relies on the split of the period vector into its real and imaginary parts, Deligne's periods can be thought of as describing the `difference' (or `incommensurability') of the rational structure given by $\langle \cD \Omega, \overline{\cD \Omega} \rangle_\IQ$, and that given by the Betti realisation of the motive $M^\AK$. In contrast, the modular form periods allow us to study the algebraic filtration, and thus these periods describe incommensurability between the Betti realisation and $\langle \partial \Omega, \partial^3 \Omega \rangle_\IQ$ projected to the subspace $H^{3,1}(X,\IC) \otimes H^{1,3}(X,\IC)$ corresponding to the submotive $M^\AK$. In general the $\IQ$-structures defined by the all three construction are different, thus leading to different expressions for the same periods. In terms of the period polynomials, these two $\IQ$-structures are related to the isomorphisms \eqref{eq:Eichler-Shimura_isomorphism} and \eqref{eq:isoSH1}.

\subsubsection*{The Calabi--Yau periods and periods of the modular form \textbf{15.3.d.b}}
\vskip-5pt
Since we are interested in the submotive $M^\AK$, we need to be slightly careful in treating the Hodge filtration. We still consider the filtration given by eq. \eqref{eq:Hodge_filtration_realisation}, but we project to the subspace $H^{3,1}(X,\IC) \otimes H^{1,3}(X,\IC)$. For this purpose we define the operator $\cP_\AK$ inductively by
\begin{align}
	\cP_\AK [\Omega] &\defineas \Omega~,\\
	\cP_\AK [\partial^n \Omega] &\defineas \partial_\varphi^n \Omega - \sum_{\substack{i < n\\i \notin \{1,3\}}} \frac{\int_X \partial^n \Omega \wedge \overline{\cP_\AK [\partial^i \Omega]}}{\int_X \cP_\AK [\partial^i \Omega] \wedge \overline{\cP_\AK [\partial^i \Omega]}} \; \cP_\AK [\partial^i \Omega]~.
\end{align}
Then we can study the filtration 
\begin{align}
	\langle \cP_\AK[\partial \Omega], \cP_\AK[ \partial^3 \Omega ] \rangle \supset \langle \cP_\AK[\partial \Omega] \rangle~,
\end{align}
and we expect to be able to express the corresponding projected period vectors 
\begin{align}
	\cP_\AK[\Pi](1) \= \cD \Pi(1) \qquad \text{and} \qquad \cP_\AK[\partial^3 \Pi](1)
\end{align} 
in terms of the periods and quasi-periods of the modular form \textbf{15.3.d.b} in eq.~\eqref{eq:modular_form}. The details of the computation of these periods can be found in appendix \ref{app:Modular_forms_for_Gamma_0_15}. Then numerical computations indicate that
\begin{align}
	\begin{split}
		\cD \Pi(1) &\= \frac{3}{7}  \left(\frac{\omega_f^-}{(2\pi \ii)^3} \left(
		\begin{array}{r}
			12 \\
			5 \\
			20 \\
			-50 \\
			10 \\
		\end{array}
		\right) - \frac{\omega_f^+}{(2\pi\ii)^3} \left(
		\begin{array}{r}
			0 \\
			5 \\
			24 \\
			-80 \\
			20 \\
		\end{array}
		\right) \right)~,\\
		\cP_\AK[\partial^3\Pi](1) &\= \frac{2488}{225} \cD\Pi(1) + \frac{32}{525}\left( \frac{\eta_F^-}{(2\pi \ii)^3} \left(
		\begin{array}{r}
			12 \\
			5 \\
			20 \\
			-50 \\
			10 \\
		\end{array}
		\right) - \frac{\eta_F^+}{(2\pi\ii)^3} \left(
		\begin{array}{r}
			0 \\
			5 \\
			24 \\
			-80 \\
			20 \\
		\end{array}
		\right) \right)~,
	\end{split}	
\end{align}
verifying the expectation.

\subsection{Possible geometric origin of the splitting} \label{sect:Geometric_origin_of_splitting}
\vskip-10pt
We have above given extensive evidence for the modularity of the Hulek--Verrill Calabi--Yau fourfold $\HV_1/\IZ_6$ by studying various cohomology theories.  However, \textit{Hodge-like conjectures} (see for example refs. \cite{Candelas:2019llw,Bonisch:2022mgw} for a brief physicist-oriented summary) suggest that there should be a geometric interpretation of the modularity. Roughly speaking, the Hodge-like conjectures can be thought of as generalisations of the original Hodge conjecture. 

From the original conjecture it follows that if $X$ is a compact Kähler manifold, then the Hodge classes of $X$, that is the elements of $H^{2k}(X,\IQ) \cap H^{(k,k)}(X,\IC)$, arise as rational linear combinations of classes of closed algebraic subsets of $X$. Generalising this, Hodge-like conjectures essentially state that any subspace $V \subseteq H^k(X,\IQ)$ compatible with the Hodge decomposition of $H^k(X,\IQ)$ (such as $\Lambda_{\AK{}}$, which is of Hodge type $(3,1)+(1,3)$) arises from geometry (i.e. it defines what is called a \textit{geometric motive}, see for instance ref. \cite{Bonisch:2022slo}).

Hulek and Verrill have identified, for several resolutions of singular Hulek--Verrill threefolds, cycles responsible for modularity by investigating a double elliptic fibration over $\IP^1$ \cite{Hulek2005a}, analogous to the double fibration $\mathcal{E}\times_{\IP^1} \operatorname{K3}$ that the fourfolds are birational to (see eq. \eqref{eq:HV_double_fibration}). The cycles they find arise from an elliptic curve that is fibred over a singular elliptic fibre of Kodaira type $I_n$, $n>1$. Two cycles of Hodge type $(1,2)$ and $(2,1)$ are then be obtained by taking one of the real 1-cycles on the elliptic curve and one of the $\IP^1$ components of the singular fibre. Similar geometric structure can also be observed in smooth Hulek--Verrill threefolds exactly when they are modular \cite{Candelas:2023yrg}. The elliptic curves forming part of the cycles were identified in ref. \cite{Candelas:2023yrg} as being related to $L$-functions appearing in Deligne's periods, and several physical quantities, such as the axiodilaton or the F-theory fibre in flux vacuum compactifications of IIB/F-theory on modular threefolds.

The fact that the observations we have made regarding the modularity of the Hulek--Verrill Calabi--Yau fourfolds are closely analogous to the threefold case motivates us to study the double fibration \eqref{eq:HV_double_fibration} to search for a possible geometric origin of modularity. In particular, we are interested in studying the K3 fibres over singular elliptic fibres of type $I_n$, $n>1$, as these could provide natural cycles responsible for modularity.

If we specialise to the case of the $\IZ_6$-symmetric one-parameter subfamily $\HV_\varphi$ of Hulek--Verrill fourfolds by setting $\varphi_i = \varphi$, $i=1,\dots,6$ in eq. \eqref{eq:definition_fibres}, the elliptic curve in the fibration has the Weierstrass model given by
\begin{equation}
\begin{aligned}
y^{2}&\=\left(x+\frac{2\left(z^{2}+1+2\varphi^2\right)-\left(z+1+2\varphi\right)^{2}}{8}\right)^{2}x-\frac{A(\varphi,z)}{64}x~,\\[5pt]
\text{where }\quad A(\varphi,z)&\=\prod_{\epsilon_i=\pm1}\left(z-\left(1+\epsilon_1\sqrt{\varphi}+\epsilon_2\sqrt{\varphi}\right)^{2}\right)~.
\end{aligned}
\end{equation}
As in eq. \eqref{eq:HV_double_fibration}, $z$ describes the affine coordinate of the $\mathbb{P}^1$. From this expression, one can immediately read off the coefficients $f$ and $g$ in the form
\begin{align}
y^2 \= x^3 + f x^2 + g~,
\end{align}
and the discriminant
\begin{align}
\Delta = (z-1)^2 z^2 \varphi^4 \left(z^2-2 z (4 \varphi +1)+(1-4 \varphi )^2\right)~.
\end{align}
The discriminant has singularities at
\begin{align}
z \in \left\{0,1,1+4\varphi \pm 4\sqrt{\varphi} \defineas z_\pm, \infty \right\}~.
\end{align}
Generically, the singular elliptic fibres above these points are of type $I_2$ above $z=0$ and $z=1$, and of type $I_1$ elsewhere. However, for certain values of $\varphi$, the singularities at $z_\pm$ can coincide with the other singularities, enhancing them to $I_3$. It is easy to check that this happens exactly when $\varphi=0,1/4,1$. Among these, the fourfold is only non-singular for $\varphi=1$, which is exactly the case where we observe a persistent quadratic factorisation of $R_H^{(p)}(\HV_\varphi/\IZ_6,T)$. The singular elliptic fibres in the case $\varphi=1$ are listed in table \ref{tab:fibred_product_singular_fibres_varphi=1}.
\begin{table}[H]
	\renewcommand{\arraystretch}{1.3}
	\begin{center}
		\begin{tabular}{|c|c|c|c|c|}
			\hline
			$z$ & $0$ & $1$ & $9$ & $\infty$ \\ \hline
			sing. & $I_2$ & $I_3$ & $I_1$ & $I_6$  \\ \hline 
		\end{tabular}
	\vskip5pt
	\capt{6in}{tab:fibred_product_singular_fibres_varphi=1}{The singular elliptic fibres appearing in the fibred product $\cE_{z,1,1} \times_{\IP^1} \text{K3}_{z,1,1,1}$ when $z \in \IP^1$ is varied. The fibre over $z = \infty$ is obtained by resolving the singularities of the fibred product \eqref{eq:HV_double_fibration}, similarly as in ref. \cite{Hulek2005a}.}	
	\end{center}	
    \vskip-30pt
\end{table}
\subsubsection*{K3~surfaces of Hulek--Verrill type and their modularity}
\vskip-5pt
The family of K3~surfaces $S_{\varphi}$, given by a toric compactification of the sublocus of $\IP^3\setminus\{X_i \= 0\}$ which is given by\footnote{Here we have scaled $\varphi$ to set $z=1$ compared to the fibre in eq. \eqref{eq:definition_fibres}.}
\begin{align}
	\left( X_3 + X_4 + X_5 + X_6 \right) \left( \frac{1}{X_3} + \frac{1}{X_4} + \frac{1}{X_5}  + \frac{1}{X_6} \right) &\= 1/\varphi
\end{align} 
has been previously studied by Verrill in \cite{Verrill1996a}. In particular, by explicitly constructing the generators of the Picard group, they show that the Picard number of a generic member of this family is $\rho = 19$. Thus we expect to find a Picard--Fuchs equation of degree 3, so that, for a generic member $S_\varphi$, we can decompose the middle cohomology as (see appendix \ref{app:K3_modularity} for more details)
\begin{align}
H^2(S_\varphi,\IQ) \= H^2_H(S_\varphi,\IQ) \oplus H_\perp^2(S_\varphi,\IQ)~,
\end{align}
where, as for fourfolds, $H^2_H(S_\varphi,\IQ)$ denotes the part of the middle cohomology generated by $\langle \Omega, \theta \Omega, \theta^2 \Omega \rangle$, and the spaces appearing in this decomposition are related to the Néron-Severi group and the transcendental lattice as 
\begin{align}
H^2_H(S_\varphi,\IQ) \simeq T(S_\varphi) \otimes \IQ~, \qquad \text{and} \qquad H^2_\perp(S_\varphi,\IQ) \simeq \text{NS}(S_\varphi) \otimes \IQ~.	
\end{align}
In particular, for the purposes of computing the zeta function of the family, we can concentrate on the three-dimensional subspace $H^2_H(S_\varphi,\IQ)$, as it is known that for K3~surfaces with $\rho \geq 19$ the polynomial $R_2^{(p)}(S_\varphi,T)$ factorises into a degree-3 factor corresponding to the subspace $H^2_H(S_\varphi,\IQ)$ and the remainder (see appendix~\ref{app:K3_modularity} for more details).

The fundamental period associated to this family takes the form analogous to the fourfold fundamental period:\footnote{The coordinate $\varphi$ we use here corresponds to $\nu$ of ref. \cite{Verrill1996a}. In addition, the fundamental period considered in ref. \cite{Verrill1996a} differs from ours by scaling $\Omega \mapsto -\varphi \Omega$.}
\begin{align}
	\varpi_{\text{K3}}^0(\varphi) \= \sum_{n_1,\ldots,n_4=0}^{+\infty} \binom{n_1 + \ldots + n_4}{n_1,\ldots,n_4}^2 \varphi^{n_1+\ldots+n_4}~.
\end{align} 
The Picard--Fuchs operator annihilating this fundamental period is given by
\begin{align} \label{eq:K3_PF-operator}
	\cL_{\text{K3}} \= \theta^3+64 \varphi^2 (\theta +1)^3 -2 \varphi (2 \theta +1) (5 \theta  (\theta
	+1)+2)~,
\end{align}
which agrees with the result in ref. \cite{Verrill1996a} derived using the Griffiths-Dwork method.

We can also introduce a period vector $\Pi_\text{K3}$ in a rational basis. We could again work with the gamma class to find an integral basis, in a manner similar to \ref{sect:periods}, but since we only need the period vector in a rational basis, it is enough to note that for K3~surfaces, the gamma class contains only rational terms, which can be absorbed by a rational change of basis. Therefore we can simply take
\begin{align}
\Pi_\text{K3} \= \nu_\text{K3}^{-1} \varpi_\text{K3}~, \qquad \text{with} \qquad \nu_\text{K3} \= \diag\left(1,2\pi\ii,(2\pi\ii)^{2} \right)~.
\end{align}
Using the deformation method with the differential operator \eqref{eq:K3_PF-operator}, we can compute the degree-3 factor $R_H^{(p)}(S_\varphi,T)$ of the polynomial $R_2^{(p)}(S_\varphi,T)$ corresponding to $H^2(S_\varphi)$, which determines the local zeta function of the K3~surface (see appendix \ref{app:K3_modularity}). This is done essentially using the prescription given in section \ref{sect:practical_zeta_function} for computing the zeta function of fourfolds of type $(1,1,1,1,1)$, with only minimal obvious modifications to take into account that $H^2_H(S_\varphi,\IQ)$ is three-dimensional.

Due to the Weil conjectures, the degree-three polynomial $R_H^{(p)}(T)$ always factorises into a linear piece, together with a quadratic one:
\begin{align} \label{eq:R_H_K3}
R_H^{(p)}(T) \= (1\pm p T) (1-b_p T + p^2 T^2)~.
\end{align}
As for fourfolds, generically this does not imply the existence of a two-dimensional subspace 
\begin{align}
\Lambda_{\text{attr}} \subset \left(H^{(2,0)}(S_\varphi,\IC) \oplus H^{(0,2)}(S_\varphi,\IC)\right) \cap H^2(S_\varphi,\IQ)~.
\end{align}
Indeed, the existence of such a space would imply that $\dim T(S_\varphi) = 2$, which would be in contradiction of the result of Verrill's that $\dim T(S_\varphi) = 3$ for a generic member of the family. Consequently, we do not expect that over generic point the coefficients $b_p$ are coefficients of a weight-three modular form.\footnote{However, see ref. \cite{Yui2012a} for discussion on modularity results for non-attractive K3~surfaces.} This expectation is corroborated by that fact that we have computed the coefficients $b_p$ for several generic values of the modulus, such as $\varphi=1/2$, and do not find any matching modular forms on LMFDB.

However, at $\varphi=1$, corresponding to a K3 fibre over a singular elliptic curve of Kodaira type $I_3$ in the fibration \eqref{eq:HV_double_fibration}, we find that the coefficients $b_p$ (see \tref{tab:b_n_coefficients_K3}) are exactly those given by the modular form \textbf{15.3.d.b} which is also the modular form associated to the Hulek--Verrill fourfold~$\HV_1/\IZ_6$.
\begin{table}[H]
	\begin{center}
		\begin{tabular}{|c|c|c|c|c|c|c|c|c|c|c|c|c|c|c|c|c|c|c|}
			\hline
			$p$    & 7 & 11 & 13 & 17 & 19 & 23 & 29 & 31 & 37 & 41 & 43 & 47 & 53 & 59 & 61 & 67 & 71 & 73 \\ \hline
			$b_p$  & 0 & 0  & 0  & -14 & -22 & 34 & 0 & 2 & 0 & 0 & 0 & -14 & -86 & 0 & -118 & 0 & 0 & 0     \\ \hline
		\end{tabular}
		\vskip10pt
		\capt{5.5in}{tab:b_n_coefficients_K3}{The coefficients $b_p$, for the primes $7 \geq p \leq 73$, appearing in the polynomials $R_H^{(p)}(S_1,T)$. Note the perfect agreement with the coefficients $b_p$ displayed in table \ref{tab:b_n_coefficients} and the Fourier coefficients of the modular form \textbf{15.3.d.b}.}
	\end{center}
    \vskip-20pt
\end{table}
This is a strong indication that the the K3 fibre $S_1$ is attractive, as attractive K3 manifolds are known to be modular \cite{Livne1995a} (see appendix \ref{app:K3_modularity} for a review of modularity results for K3~surfaces). Using numerical integration of the Picard--Fuchs equation in a manner completely analogous to that used in section \ref{sect:splitting_of_Hodge_structure}, we can numerically verify the existence of a two-dimensional subspace $\Lambda_\text{attr}$, showing that the K3~surface $S_1$ is indeed attractive. Working to the accuracy of at least 100 digits, we find that
\begin{align}
\Pi \= \frac{1}{4} \left( \frac{L_3(2)}{(2\pi \ii)^2} A - \frac{L_3(1)}{2\pi \ii} B \right)~,
\end{align}
where $L_3(1)$ and $L_3(2)$ are the critical values of the $L$-function associated to the modular form \textbf{15.3.d.b}, which we have given in eq. \eqref{eq:L-function_values_numeric}, and $A$ and $B$ are the integral vectors
\begin{align}
A \= (24,24,7)^T~, \qquad B \= (24,8,1)^T~.
\end{align}
The appearance of the $L$-function values can again be explained by Deligne's conjecture, following essentially the same arguments as in section \ref{sect:Delignes_periods}.

The observations above give a very strong indication that the K3~surface $S_1$ is attractive, and therefore modular, and that it is associated to the same modular form as the fourfold. Given that $S_1$ appears as a fibre over the $I_3$ type degenerate elliptic fibre in the double fibration \eqref{eq:HV_double_fibration} that is birational to the Hulek--Verrill fourfold $\HV_{(1,\dots,1)}$, it is tempting to assume that existence of this fibre is the geometric reason for modularity of $\HV_1/\IZ_6$. Indeed, there is a natural pair of cycles, arising from taking one of the $\IP^1$ components in the fibre $I_3$, together with the two transcendental cycles of the attractive K3 fibre, that could give rise to the subspace $\Lambda_{\AK{}}$ of Hodge type $(3,1)+(1,3)$, with the $\IP^1$ being essentially responsible for the Tate twist. In the threefold case, it was shown by Hulek and Verrill \cite{Hulek2006a} that an analogous mechanism, with an elliptic curve in place of the K3~surface, accounts for the modularity in various cases. We give a brief overview of their argument in appendix \ref{app:modularity}.

\newpage

\section{Physics of Modular Calabi--Yau Fourfolds} \label{sect:physics}
\vskip-10pt
F-theory compactifications on Calabi--Yau fourfolds give rise to four-dimensional low-energy effective theories with four supercharges, which is the minimal amount of supersymmetry in four space-time dimensions. This class of Calabi--Yau fourfold compactifications has extensively been studied in string phenomenology over the years. For an enlightening review, see for instance ref.~\cite{Weigand:2010wm}.

In order to construct phenomenologically interesting effective four-dimensional theories, an important ingredient is the four-form background flux~$G$. This flux generates a scalar potential in the four-dimensional effective theory, which lifts flat directions and possibly breaks supersymmetry spontaneously. Consistency of the F-theory compactification imposes a quantisation condition on the background flux~$G$ \cite{Witten:1996md}, which requires the $G$-flux to be an integral or half-integral four-form cohomology class according to eq.~\eqref{eq:Gquant}. The properties of the four-dimensional low-energy effective scalar potential are governed by the flux-induced superpotential~\eqref{eq:W} \cite{Gukov:1999ya}, which depends on the complex structure moduli of the Calabi--Yau fourfold and hence on the Hodge type of the integral or half-integral four-form flux $G$. The arithmetic methods presented in this work offer a powerful tool to examine the Hodge type of the flux~$G$ as a function of the complex structure moduli, which allows us to characterise the vacuum structure of the four-dimensional low energy effective theory, at a given point in the moduli space.

F-theory requires the Calabi--Yau fourfold to be elliptically fibred and there are typically further conditions imposed on the background flux~$G$, see for instance refs.~\cite{Donagi:2009ra,Collinucci:2010gz,Braun:2011zm,Marsano:2011hv,Krause:2011xj} and references therein. We do not consider these additional constraints here, but instead we analyse the four-form flux~$G$ in the context of M-theory compactifications, where the four-form flux~$G$ is only required to fulfill the flux quantisation condition~\eqref{eq:Gquant}.  As the minimal supersymmetric four-dimensional theory arising from F-theory compactifications on Calabi--Yau fourfolds, the low energy effective theory of M-theory compactifications on Calabi--Yau fourfolds has four supercharges \cite{Becker:1996gj}. The resulting three-dimensional low energy effective $\mathcal{N}=2$ supergravity theory is derived in refs.~\cite{Haack:2001jz,Berg:2002es}. The four-form flux~$G$ induces the superpotential~\eqref{eq:W}, which yields in the effective $\mathcal{N}=2$ supergravity theory the contribution to the scalar potential~\eqref{eq:scalpot}. This in turn has an expression in terms of the superpotential~\eqref{eq:W} and the (semi-classical) K\"ahler potential~\eqref{eq:scalpot} of the complex structure moduli fields coupling to the $G$-flux.\footnote{If the flux~$G$ is not primitive, there is also a flux-induced twisted superpotential, which couples to the K\"ahler moduli~\cite{Haack:2001jz,Berg:2002es,Intriligator:2012ue}.} Therefore, we discuss now the physical implications of the arithmetic properties of the Calabi--Yau fourfold in this described M-theory context.

As detailed in section~\ref{sect:splitting_of_Hodge_structure}, determining persistent factorisations of the polynomials~$R_H^{(p)}(X_{\bm \varphi},T)$ appearing in the local zeta function~$\zeta_p(X_{\bm \varphi},T)$ of the Calabi--Yau fourfold~$X$ for all but finitely many primes~$p$ offers a powerful arithmetic tool to discover points $\bm \varphi$ in the complex structure moduli space of $X$ such that the Hodge structure of the horizontal four-form cohomology~$H_H^4(X_{\bm \varphi},\IC)$ of the corresponding manifold $X_{\bm \varphi}$ splits over $\IQ$. These splittings allow us to choose a four-form flux~$G$ that is of a non-generic Hodge type. In the following, we classify possible four-form fluxes $G$ according to their Hodge type, and discuss the resulting properties of the low-energy effective supergravity action as a consequence of the flux-induced superpotential~\eqref{eq:W} and scalar potential~\eqref{eq:scalpot}:
\begin{itemize}
\item \textbf{Hodge Type $G\in H^{4,0}(X_{\bm \varphi})\oplus H^{2,2}(X_{\bm \varphi}) \oplus H^{0,4}(X_{\bm \varphi})$}:
The $G$-flux of this Hodge type yields vanishing covariant derivatives $\mathcal{D}_iW(\bm \varphi)$ of the superpotential $W(\bm \varphi)$ with respect to the complex structure moduli $\bm \varphi$. As a result the scalar potential~$V_W(\bm \varphi,\bar{\bm \varphi})$ of the effective three-dimensional $\mathcal{N}=2$ supergravity theory is minimised at the negative value $-4 e^{-K(\bm \varphi,\bar{\bm \varphi})}|W(\bm \varphi)|^2$. Due to the no-scale structure of the entire scale potential resulting from M-theory compactifications on Calabi--Yau fourfolds \cite{Haack:2001jz}, such a four-form flux~$G$ yields a supersymmetric Minkowski vacuum. Compactifying M-theory on the modular Hulek--Verrill Calabi--Yau orbifold $\HV/\IZ_6$ of section~\ref{sect:Hulek--Verrill_Fourfolds} yields a one-parameter example that admits a $G$-flux of this type.
\item \textbf{Hodge Type $G\in H^{2,2}(X_{\bm \varphi})$}:
For a $G$-flux of this Hodge type both the superpotential~$W(\bm \varphi)$ and its covariant derivatives $\mathcal{D}_iW(\bm \varphi)$ vanish. As a result the contribution~$V_W(\bm \varphi,\bar{\bm \varphi})$ to the scalar potential and the scalar potential itself vanishes \cite{Haack:2001jz}, which describes, in the effective three-dimensional $\mathcal{N}=2$ supergravity theory, a supersymmetric Minkowski vacuum \cite{Becker:1996gj,Haack:2001jz,Berg:2002es}. The Hessian~$\mathcal{D}_i\mathcal{D}_j W(\bm \varphi)$ yields a complex mass matrix for the three-dimensional $\mathcal{N}=2$ chiral multiplets associated to the complex structure moduli $\bm \varphi$ of the Calabi--Yau fourfold~$X_{\bm \varphi}$. The rank of this matrix determines the number of stabilised complex structure moduli. Assuming the Hodge conjecture, the existence of a $G$-flux of this Hodge type implies that the $G$-flux is a rational multiple of an algebraic cycle class. In the context of M-theory compactifications on Calabi--Yau fourfolds, the choice of consistent $G$-fluxes in terms of algebraic cycle classes plays an important role in the low energy effective three-dimensional $\mathcal{N}=2$ supergravity action arising from extremal transitions in Calabi--Yau fourfolds \cite{Intriligator:2012ue,Jockers:2016bwi}. 

For one-parameter Picard--Fuchs systems for which there is both an attractor point and an attractive K3 point in the complex structure moduli space --- i.e., for which the Hodge structure splits simultaneously into $H^{4,0}(X_{\bm \varphi})\oplus H^{0,4}(X_{\bm \varphi})$ and $H^{3,1}(X_{\bm \varphi})\oplus H^{1,3}(X_{\bm \varphi})$ --- the $G$-flux can be chosen to be of Hodge type $(2,2)$ and hence algebraic. Given the automatic factorisations that follow from the Weil conjectures, it is not immediately clear from the form of the local zeta functions $\zeta_p(X_{\bm \varphi},T)$ whether a point in moduli space is of this type. However, such a point must be among those where we have a persistent quadratic factorisation, which is already a very restrictive condition. Therefore, one can in principle examine all such points explicitly, in order to determine whether the Hodge structure splits in this anticipated form.
\item \textbf{Hodge Type $G\in H^{4,0}(X_{\bm \varphi}) \oplus H^{0,4}(X_{\bm \varphi})$}: 
For a $G$-flux of this Hodge type, the Calabi--Yau fourfold must have an attractor point, at which the piece $H^{4,0}(X_{\bm \varphi})\oplus H^{0,4}(X_{\bm \varphi})$ splits off in Hodge structure. The part $V_W(\bm \varphi,\bar{\bm \varphi})$ of the scalar potential of the low energy effective three-dimensional supergravity action has a minimum with the negative value $-4 e^{-K(\bm \varphi,\bar{\bm \varphi})} | W(\bm \varphi)|^2$, which --- again to due to the no-scale structure of the low energy effective supergravity theory --- results in a supersymmetric Minkowski vacuum \cite{Becker:1996gj,Haack:2001jz,Berg:2002es}. As such an attractor point is presumably associated to a modular form of weight five, the non-vanishing cosmological constant may be expressed in terms of critical $L$-function values associated to the weight-five modular form.
\item \textbf{Hodge Type $G\in H^{3,1}(X_{\bm \varphi})\oplus H^{2,2}(X_{\bm \varphi}) \oplus H^{1,3}(X_{\bm \varphi})$}: For such a $G$-flux the superpotential vanishes, while some of its covariant derivatives $\mathcal{D}_iW(\bm \varphi)$ are non-vanishing. As a result, the flux-induced contribution to the scalar potential $V_W(\bm \varphi,\bar{\bm \varphi})$ arising from the complex structure moduli~$\bm \varphi$ is not minimised. This indicates that the effective three-dimensional $\mathcal{N}=2$ supergravity theory is not in a stable vacuum state. 
\item \textbf{Hodge Type $G\in H^{3,1}(X_{\bm \varphi}) \oplus H^{1,3}(X_{\bm \varphi})$}: For such a $G$-flux the superpotential vanishes as in the previous case, while again some covariant derivatives $\mathcal{D}_iW(\bm \varphi)$ are non-vanishing. As for the previous $G$-flux, we do not expect that the effective three-dimensional $\mathcal{N}=2$ supergravity description has a stable vacuum configuration at this point in moduli space. However, as the splitting of the Hodge structure into a $H^{3,1}(X_{\bm \varphi}) \oplus H^{1,3}(X_{\bm \varphi})$ piece possibly admits a description in terms of Siegel modular forms, the covariant derivatives $\mathcal{D}_iW(\bm \varphi)$ at such a point in the complex structure moduli space may enjoy an expression in terms of $L$-function values of these Siegel modular forms. In the special case that the splitting yields a $H^{3,1}(X_{\bm \varphi}) \oplus H^{1,3}(X_{\bm \varphi})$ piece of dimension $1+1$, we have an attractive K3 point in the complex structure moduli space, which is expected to be related to a modular form of weight three, instead of a Siegel modular form arising from splittings of the Hodge structure into higher-dimensional $H^{3,1}(X_{\bm \varphi}) \oplus H^{1,3}(X_{\bm \varphi})$ pieces.
\end{itemize}

\newpage 

\section{Conclusions} \label{sect:Conclusions}
\vskip-10pt
\subsection{Summary}
\vskip-10pt
In this paper, we analyse the arithmetic properties of Calabi--Yau fourfolds, focusing on examples with a single complex structure parameter. We generalise the deformation methods for Calabi--Yau threefolds developed in refs.~\cite{Candelas:2021tqt,multiparameter_zeta} to Calabi--Yau fourfolds~$X$ in order to compute a factor of their local zeta functions~$\zeta_p(X,T)$. These deformation methods probe the horizontal part of the middle cohomology both for Calabi--Yau threefolds and for Calabi--Yau fourfolds. While the local zeta functions of Calabi--Yau threefolds are essentially determined by the polynomial associated to the middle cohomology --- which for threefolds coincides with the horizontal cohomology --- the local zeta functions for Calabi--Yau fourfolds are typically not entirely determined in terms of the horizontal piece of the middle cohomology. It is this factor that we focus on in this paper, leaving the determination of the entire zeta function for a future work.

More specifically, under the assumption that the action of the Frobenius map is well-defined on the horizontal part $H_H^4(X,\IC)$ of the middle cohomology of the Calabi--Yau fourfold~$X$ --- or more generally on the cohomology $H_I^4(X,\IC)$ generated by a Picard--Fuchs differential ideal $I$ --- we are able to compute the corresponding polynomials~$R_H^{(p)}(X,T)$. These polynomials furnish a factor of the polynomials~$R_4^{(p)}(X,T)$ in the local zeta functions~$\zeta_p(X,T)$ of the Calabi--Yau fourfold~$X$, which are attributed to the entire middle cohomology~$H^4(X,\IC)$.

A further persistent factorisation (which means that the polynomial factorises over $\IQ$ for all but finitely many primes $p$) of the polynomials $R_H^{(p)}(X_{\bm \varphi},T)$ of the local zeta functions is related, by a local-to-global-type argument, to a splitting of the Hodge structure of the horizontal cohomology over~$\IQ$. Inspired by the terminology for Calabi--Yau threefolds, we call a point $\bm \varphi \in \cM_{\IC S}(X)$ in the complex structure moduli space of the Calabi--Yau fourfold~$X$ either a rank-two attractor point or an attractive K3~(AK3)~point. The type of the point depends on whether the rational subspace~$\Lambda \subset H_H^4(X_{\bm \varphi},\IQ)$ appearing in the split of the horizontal cohomology of the corresponding manifold $X_{\bm \varphi}$ is of Hodge type~$(4,0)+(0,4)$ or of Hodge type~$(3,1)+(1,3)$.

More generally, we list all in principle possible rational Hodge theoretic splittings of the horizontal cohomology in the context of M-theory compactifications on Calabi--Yau fourfolds. In such compactifications a quantised four-form background flux needs to be specified, which corresponds to a (half-)integral middle-dimensional cohomology class of the compactification space. Depending on the Hodge type of such quantised background fluxes, we spell out the resulting interactions in the three-dimensional low-energy effective $\mathcal{N}=2$ supergravity theory. We find that if the Hodge theoretic splitting gives rise to modular forms then the interactions of the background fluxes in the supergravity action are given in terms of critical $L$-function values of these modular forms.

For all families of Calabi--Yau fourfolds of Hodge type~$(1,1,1,1,1)$ studied in this work, we find that $R_H(X,T)$ always factorises into a linear factor $(1\pm p^2 T)$ and a remainder of degree four. The appearance of the linear factor is a consequence of the functional equation that the local zeta functions~$\zeta_p(X,T)$ satisfy. Hence, such a linear factor does not indicate a rational splitting of the Hodge structure. However, we identify a point of persistent factorisation in the $\IZ_6$-invariant sublocus of the complex structure moduli space of the family of Hulek--Verrill Calabi--Yau fourfolds. In this case we are able to explicitly identify a two-dimensional rational subspace $\Lambda_{\AK{}}$ of the middle cohomology of Hodge type~$(3,1)+(1,3)$, which verifies the point where the persistent factorisation occurs as an \AK{} point.

Furthermore, we study the modular properties of the Hulek--Verrill Calabi--Yau fourfold at the $\IZ_6$ invariant attractive K3 point. The modularity conjectures predict that the coefficients of the degree-two factor in $R_H^{(p)}(X,T)$ corresponding to the rational subspace $\Lambda_{\AK{}}$ are Fourier coefficients of a modular form. We identify the weight-3 modular form \textbf{15.3.d.b}, whose Fourier coefficients agree with the polynomial coefficients, at least for the first $130$ primes. We establish that the coefficients of this modular form appear also in the local zeta functions of the K3~surface, which arises as a fibre in the double-fibred fourfold birational to the modular Hulek--Verrill Calabi--Yau fourfold. Finally, with the help of mirror symmetry, we determine a natural $\IQ$~structure of $H^4_H(X,\IC)$, which allows us to compute the Deligne's periods related to the rational subspace $\Lambda_{\AK{}}$. We find, as expected by Deligne's conjecture, that --- up to an irrelevant rational prefactor --- these periods are given by the critical $L$-function values associated to the modular form \textbf{15.3.d.b}. This explains the observed relation of periods of the modular manifold to $L$-function values, and intriguingly indicates that the $\IQ$-structure given by mirror symmetry coincides with the $\IQ$-structure that is natural from the arithmetic point-of-view. This occurs to us as a surprising observation because a priori it is not clear to us that the horizontal part of the middle cohomology inherits a well-defined $\IQ$-structure from $H^4(X,\IQ)$. One can also define a distinct $\IQ$-structure given by the projection of the Hodge filtration of $H_H^4(X,\IC)$ to the subspace $\Lambda_{\AK{}}$. This $\IQ$-structure can be used to express the periods of the modular Calabi--Yau fourfold in terms of the periods and quasi-periods of the weight-three modular form.

We also investigate Calabi--Yau fourfolds of Hodge type $(1,1,2,1,1)$. The examined examples show sporadic factorisations for some primes, but none of them indicates a persistent factorisation of the local zeta functions as in the case of the Hulek--Verrill fourfold. However, we leave a systematic search for persistent factorisations in this class of Calabi--Yau fourfolds to future work.

\subsection{Outlook}
\vskip-10pt
With this work, we initiate the study of modularity of Calabi--Yau fourfolds in connection to physics, and lay out this connection explicitly in the context M-theory compactifications. We analyse a limited number of explicit examples of Calabi--Yau fourfolds with the introduced arithmetic techniques. Among these examples we find a modular manifold within the family of Hulek--Verrill Calabi--Yau fourfolds. However, to be able to apply the presented arithmetic techniques, we make two important assumptions. Firstly, we assert that the Frobenius map has a well-defined action on the subspaces $H^4_I(X) \subset H^4(X)$ defined in eq.~\eqref{eq:H_I_definition} associated to a Picard-Fuchs differential ideal $I$. Secondly, we assume that a persistent factorisation of the factors $R^{(p)}_I(X,T)$ corresponds to a rational splitting of the Hodge structure of $H_I^4(X,\IC)$, which then carries over to a rational splitting of the Hodge structure of the full middle cohomology $H^4(X,\IC)$. While we perform highly non-trivial consistency checks on the examples analysed in this paper, the presented examples are somewhat limited in scope. Therefore, additional studies of arithmetic properties of Calabi--Yau fourfolds are necessary to further test and possibly refine the assumptions made here.

We test the first assumption for particular examples of one-parameter fourfolds of both Hodge type~$(1,1,1,1,1)$ and $(1,1,2,1,1)$, and for the principal differential ideal $I$ associated to the one-parameter $\IZ_6$-symmetric family of Hulek--Verrill Calabi--Yau fourfolds within the six-dimensional complex structure moduli space of the generic Hulek--Verrill fourfolds. The self-consistency of our results offers convincing evidence that the assumption is legitimate for at least these examples. It is certainly interesting to see whether the presented deformation method can be consistently applied to a wider variety of differential ideals $I$. Establishing the deformation method for a broader class of Calabi--Yau fourfolds promises useful applications in various physical contexts, such as the F-theory limits discussed recently in ref.~\cite{Kim:2022jvv}. It would also be useful to find examples of Calabi--Yau fourfolds for which the full local zeta functions can be determined independently. Then we could directly compare these results with the polynomials~$R_I^{(p)}(X,T)$ computed here, which would offer another check on the assumptions made.

Given the first assumption, the second assumption follows more naturally. Nevertheless, it is not a priori obvious that $H^4(X,\IQ)$ splits over the rational numbers into $H_I^4(X,\IQ)$ and a rational complement, which would guarantee existence of a split of the entire middle cohomology $H^4(X,\IQ)$ given a split of $H_I^4(X,\IQ)$. This is intimately connected to the question of the existence of the putative motive~$M^{\AK{}}$. In similar situations, one often applies the Hodge conjecture to argue for the existence of the motive. In our case, use of such an argument hinges on the rational split of $H^4(X,\IQ)$ described above, which is the content of the second assumption.

We propose to employ mirror symmetry to define a natural $\IQ$-structure on the subspace $H^4_I(X,\IC)$ of $H^4(X,\IC)$, as illustrated for the Hulek--Verrill Calabi--Yau fourfold in this work. In this example, we observe that the $\IQ$-structure on $H^4_I(X,\IC)$ obtained in this way behaves as if the cohomology group $H^4(X,\IQ)$ split over the rational numbers into $H_I^4(X,\IQ)$ and a rational complement. By studying the geometry of the Hulek--Verrill Calabi--Yau fourfold in greater detail some of these assumptions can possibly be established more rigorously in this case. For instance, it seems possible to prove the connection between the modularity of the Hulek--Verrill Calabi--Yau fourfold and the existence of the attractive K3 ~fibre, analogously to what was done in the Hulek--Verrill Calabi--Yau threefold case in ref.~\cite{Hulek2005a}.

From the M-theory perspective, it is natural to consider compactifications on Calabi--Yau orbifolds. This motivates the investigation of arithmetic properties of singular Calabi--Yau fourfolds with the presented methods. While many properties of local zeta functions $\zeta_p(X,T)$ of a variety $X$ are mathematically proven only under the assumption of smoothness of $X$, the relationship between persistent factorisations of the polynomials~$R_I^{(p)}(X,T)$ appearing in $\zeta_p(X,T)$ and the rational splitting of the horizontal cohomology groups possibly extends to Calabi--Yau fourfolds with singularities, for at least some differential ideals $I$. For the singular Hulek--Verrill orbifold~$\HV/\IZ_6$ the presented variation of Hodge structure gives evidence for a generalisation of our findings to at least certain classes of singular Calabi--Yau fourfolds. A systematic investigation of the presented methods to a larger class of compactification spaces promises interesting applications to M- and F-theory compactifications with background fluxes.

Apart from putting the applied methods on a more rigorous footing, the present work could be expanded on in the future by identifying more examples of modular Calabi--Yau fourfolds. 
For Picard--Fuchs differential ideals~$I$ of Calabi--Yau fourfolds with a low number of generators a systematic search for persistent factorisations of the polynomials $R^{(p)}_I(X,T)$ may not only yield new examples of families of Calabi--Yau fourfolds with attractive K3 points, but may also provide examples of attractive Calabi--Yau fourfolds, which we did not explicitly identify here. 

While Calabi--Yau manifolds with dimensions beyond fourfolds do not play such a prominent role in the context of compactifications,\footnote{However, for F-theory compactifications on elliptically-fibred Calabi--Yau fivefolds see, for instance, ref.~\cite{Schafer-Nameki:2016cfr}.} the presented techniques promise to be applicable for higher-dimensional Calabi--Yau manifolds as well. The Hulek--Verrill Calabi--Yau manifolds turn out to be modular in dimensions 1, 2, 3, and 4. It would be interesting to investigate whether this pattern continues for Hulek--Verrill fivefolds and beyond. For instance, we expect that the Hulek--Verrill fivefold is birational to a fibred product $\HV^3 \times_{\IP^1} \cE$, where the fibres $\HV^3$ and $\cE$ denote a threefold and an elliptic curve of the Hulek--Verrill type, respectively. It is then tempting to speculate that there is a modular Hulek--Verrill fivefold with a two-dimensional subspace $\Lambda$ of Hodge type $(4,1)+(1,4)$, which arises from a modular Hulek--Verrill threefold fibre in the described double fibration. 

The study of higher-dimensional Calabi--Yau manifolds may also turn out to have utility in the analysis of lower-dimensional geometries. For instance, in the present work we come across a surprising way of finding modular K3~surfaces. Since the Hulek--Verrill fourfolds are birational to K3 fibrations, we could (conjecturally) identify an attractive K3~surface by studying the local zeta functions of Hulek--Verrill fourfolds. This effectively allows us to bypass the difficulty that the polynomial $R_H(S,T)$ associated to the horizontal cohomology $H^2_H(S)$ of K3~surfaces~$S$ with Picard number~$\rho(S)=19$ always contains a linear factor, and as such cannot be used to identify attractive K3~surfaces.

\bigskip
\section*{Acknowledgements}
\vskip-10pt
It is a pleasure to acknowledge the many fruitful conversations regarding arithmetic and geometry of Calabi--Yau manifolds with Philip Candelas, Xenia de la Ossa, and Joseph McGovern. We thank Michael Haack for interesting comments and for pointing out an error regarding the physical interpretation of solutions that appeared in an earlier version of this work. We also wish to thank Duco van Straten for instructive comments regarding the zeta function and Weil conjectures, and for his interesting lectures on $p$-adic cohomology at JGU Mainz. We are grateful to Sheldon Katz and Manfred Lehn for informative discussion on the geometries studied in this paper. We thank Mateo Galdeano, Miroslava Mosso Rojas, and Maik Sarve for interesting conversations and comments. We would like to thank David P. Roberts and Masha Vlasenko for helpful discussions on the motivic aspects of this work. This work is supported by the Cluster of Excellence Precision Physics, Fundamental Interactions, and Structure of Matter (PRISMA+, EXC 2118/1) within the German Excellence Strategy (Project-ID 390831469).
\newpage

\appendix

\section{The Middle Cohomology of Calabi--Yau Fourfolds} \label{app:fourfolds_cohomology}
\vskip-10pt
The middle cohomology $H^4(X,\mathbb{C})$ of Calabi--Yau fourfolds~$X$ holds the key to finding modular Calabi--Yau fourfolds, analogously to modular K3~surfaces and Calabi--Yau threefolds respectively studied in the physics context in refs.~\cite{Moore:1998pn,Schimmrigk:2006dy,Yang:2020gwr} and refs.~\cite{Candelas:2019llw,Kachru:2020abh,Kachru:2020sio,Candelas:2023yrg,Schimmrigk:2020dfl}. In comparison to Calabi--Yau threefolds, the structure of the middle cohomology of Calabi--Yau fourfolds $H^4(X,\mathbb{C})$ is more complicated. For instance, the middle cohomology of Calabi--Yau fourfolds --- as well as of polarised K3~surfaces --- is not entirely generated by varying the holomorphic volume-form over the complex structure moduli space. Here we collect some properties of the middle cohomology of Calabi--Yau fourfolds relevant for us.

In terms of the Hodge numbers $h^{p,q} = \dim_\mathbb{C} H^{p,q}(X,\IC)$, the Hodge diamond of a Calabi--Yau fourfold~$X$ with $SU(4)$~holonomy (and not a subgroup thereof) reads
\begin{align}
	h^{p,q}\left( X \right) \= \begin{matrix}
		& & & & 1 & & & & \\
		& & & 0 &  & 0 & & & \\
		& & 0 &  & h^{1,1}  &  & 0 & & \\ 
		& 0 &  & h^{2,1} &   & h^{2,1} &  & 0 & \\ 
		1 &  & h^{3,1} & & h^{2,2} & & h^{3,1} &  & 1 \\
		& 0 &  & h^{2,1} &   & h^{2,1} &  & 0 & \\ 
		& & 0 &  & h^{1,1}  &  & 0 & & \\
		& & & 0 &  & 0 & & & \\
		& & & & 1 & & & & \\ 
	\end{matrix} \ ,
\end{align}
where the Hodge numbers $h^{p,q}$ obey the relations \cite{Sethi:1996es}
\begin{align} \label{eq:h22}
\begin{split}
	0 &\= 44 + 4 h^{1,1} - 2 h^{2,1} + 4 h^{3,1} - h^{2,2} \\
    \chi(X)&\= 6(8+h^{1,1}+h^{3,1}-h^{2,1})
\end{split}
\end{align}
with $\chi(X)$ denoting the Euler characteristic of $X$. In particular, the middle cohomology $H^4(X,\mathbb{C})$ enjoys the Hodge decomposition
\begin{align}
H^4(X,\mathbb{C}) \= H^4(X,\mathbb{Z})\otimes_{\mathbb{Z}} \mathbb{C} \= \bigoplus_{p+q=4}  H^{p,q}(X,\IC)  \ ,
\end{align}
and the decreasing filtration of weight four
\begin{equation}
	0 \subset F^4H^4 \subset F^3H^4 \subset \ldots  \subset F^0H^4 \= H^4(X,\mathbb{C}) \ ,
\end{equation}
with the filtered pieces $F^kH^4 = \bigoplus_{p\ge k} H^{p,4-p}(X,\IC)$ that obey $H^{p,4-p}(X,\IC) = F^p H^4 \cap \overline{F^{4-p}H^4}$. 

Of particular importance to us is the primary horizontal subspace of the middle cohomology $H^4(X,\mathbb{C})$ introduced and defined in refs.~\cite{Greene:1993vm,Becker:1996ay} as follows:\footnote{The used definition of the primary horizontal subspace slightly differs from the definition of the horizontal four-form cohomology given in Appendix~A of ref.~\cite{Intriligator:2012ue}.} Let us consider a family $\mathcal{X}$ of Calabi--Yau fourfolds over an open set $\mathcal{U}$ of the complex structure moduli space $\cM_{\IC S}$, which is given by the fibration:
$$
\begin{CD}
	X_\varphi @>>> \mathcal{X} \\
	@. @VV{\pi}V \\
	@. \mathcal{U}
\end{CD} 
$$
For any $\varphi\in\mathcal{U} \subset \cM_{\IC S}$ the inverse image $X_\varphi = \pi^{-1}(\varphi)$ of the projection $\pi$ yields the Calabi--Yau fourfold $X_\varphi$ as a fiber of the fibration~$\mathcal{X}$. Furthermore, the filtered spaces $F^kH^4$ furnish the fibers of holomorphic bundles over $\mathcal{U}$, and in particular the holomorphic $(4,0)$-form $\Omega(\varphi)$ of the Calabi--Yau fourfold $X_\varphi$ becomes a holomorphic section of the line bundle associated to the one-dimensional filtered piece $F^4H^4$. Then the primary horizontal subspace~$H^4_H(X_\varphi,\mathbb{C})$ is spanned by all cohomology classes that realise the holomorphic $(4,0)$-form $\Omega(\varphi)$ for some Calabi--Yau fourfold $X_\varphi$ at the point $\varphi\in\mathcal{U}$ in complex structure moduli space. Analogously to the (topological) cohomology group $H^4(X_\varphi,\mathbb{C})$, the primary horizontal subspace $H^4_H(X_\varphi,\mathbb{C})$ is by construction independent of the choice of complex structure $\varphi \in \mathcal{U}$, at least for the considered connected component in the complex structure moduli space. Therefore, for ease of notation we often drop in the following the complex structure coordinate $\varphi$ for the cohomology group $H^4(X,\mathbb{C})$ and for its primary horizontal subspace $H^4_H(X,\mathbb{C})$ --- even if we refer to a family $\mathcal{X}$ of Calabi--Yau fourfolds. 

Given a Calabi--Yau fourfold $X_\varphi$ with holomorphic $(4,0)$-form $\Omega(\varphi)$, we can construct the primary horizontal subspace as the span of $\Omega(\varphi)$ and all its higher order derivatives, i.e., 
\begin{equation}
	H^4_H(X,\mathbb{C}) \= \left\langle \, \partial \Omega(\varphi) \, \right\rangle_{\partial \in J} \ ,
\end{equation}
where $J=\langle \partial_{\varphi^1},\dots, \partial_{\varphi^{h^{3,1}}}\rangle$ is the Picard--Fuchs differential ideal generated by the derivatives $\partial_{\varphi^i}$ with respect the the holomorphic coordinates $\varphi=(\varphi_1,\varphi_2,\ldots,\varphi_{h^{3,1}}) \in \mathcal{U}$. Note that while the above span arises from a countable infinite dimensional set, the primary horizontal subspace $H^4_H(X,\mathbb{C})$ becomes finite dimensional due to the relations among the partial derivatives~$\partial\Omega$, which are captured by the Picard--Fuchs differential ideal. 

As all primary horizontal four-form classes must be primitive for any $\varphi\in\mathcal{U}$ and because there is a space of $h^{1,1}$-dimensional imprimitive four-form cohomology classes \cite{Intriligator:2012ue}, the dimensionality of the primary horizontal four-form classes is bounded by
$$
\dim_{\mathbb{C}} H^4_H(X,\mathbb{C}) \le b_4 - h^{1,1} = 2 + 2 h^{3,1} + h^{2,2} - h^{1,1} \ .
$$
Upon setting $h^{2,2}_H = \dim_\mathbb{C} (H^{2,2}(X,\IC) \cap H^4_H(X,\mathbb{C}))$ this implies the inequality
$$
h^{2,2}_H + h^{1,1} \le h^{2,2}  \ .
$$  

By specifying the tuple $(h^{4,0},h^{3,1}, h^{2,2}_H,h^{3,1},h^{0,4})$, we refer to the Hodge type of the Calabi--Yau fourfold~$X$. In this work, we consider Calabi--Yau fourfolds $X$ of Hodge type~$(1,1,1,1,1)$ and $(1,1,2,1,1)$, which means that we focus on Calabi--Yau fourfolds $X$ with one complex structure modulus with either $\dim_\mathbb{C} H^4_H(X,\mathbb{C})= 5$ or  $\dim_\mathbb{C} H^4_H(X,\mathbb{C}) = 6$, see, for instance, refs.~\cite{Honma:2013hma,Gerhardus:2016iot}. In terms of the complex structure modulus $\varphi\in \mathcal{U}\subset \cM_{\IC S}$ a basis of the primary horizontal cohomology classes for a Calabi--Yau fourfold of Hodge type $(1,1,1,1,1)$ generates $H_H^4(X,\IC)$ according to
\begin{equation}
	H_H^4(X,\IC) \= \left\langle \Omega(\varphi), \partial_\varphi \Omega(\varphi), \partial_\varphi^2\Omega(\varphi), \partial_\varphi^3 \Omega(\varphi), \partial_\varphi^4 \Omega(\varphi) \right\rangle_\IC \ ,
\end{equation}
whereas for Calabi--Yau fourfolds of Hodge type $(1,1,2,1,1)$, this is generated by
\begin{equation}
	H_H^4(X,\IC) \= \left\langle \Omega(\varphi), \partial_\varphi \Omega(\varphi), \partial_\varphi^2\Omega(\varphi), \partial_\varphi^3 \Omega(\varphi), \partial_\varphi^4 \Omega(\varphi), \partial_\varphi^5 \Omega(\varphi) \right\rangle_\IC \ .
\end{equation}
For the former type of Calabi--Yau fourfolds $\partial_\varphi^5\Omega(\varphi)$ and for the latter type of Calabi--Yau fourfolds $\partial_\varphi^6\Omega(\varphi)$ become linearly dependent in terms of the above given bases, such that the complex structure moduli spaces is governed by a Picard--Fuchs operator of order five and of order six, respectively.

More generally, instead of looking at the entire horizontal cohomology $H_H^4(X_\varphi,\IC)$, one may consider subspaces $H_I^4(X_\varphi,\IC)$ thereof generated by a Picard--Fuchs differential ideal $I \subset J$. An example of this is discussed in section \ref{sect:Hulek--Verrill_Fourfolds} in the context of the family of Hulek--Verrill Calabi--Yau fourfolds, with the relevant definitions spelled out in section \ref{sect:Weil_conjectures}.

\newpage
\section{Modularity} \label{app:modularity}
\vskip-10pt
An interesting property of the zeta function is that under certain conditions, it is possible to associate uniquely a modular form to it. We very briefly review some aspects of this correspondence, making no attempt at rigour or completeness. Good mathematical reviews on the subject are refs.~\cite{Yui2003a,Hulek2006a,Yui2012a}. Some aspects have also been reviewed in physics literature, see for example refs.~\cite{Candelas:2019llw,Kachru:2020abh,Kachru:2020sio,Kuusela:2022hga}.

The prototypical, and mathematically best-understood example is that of elliptic curves. The middle cohomology of an elliptic curve $\cE$ is two-dimensional, and of Hodge type $(1,0)+(0,1)$:
\begin{align}
H^1(\cE,\IQ) \= H^{(1,0)}(\cE,\IQ) \oplus H^{(0,1)}(\cE,\IQ)~.
\end{align}
Thus, by Weil conjectures, the zeta function of an elliptic curve has the form
\begin{align}
\zeta_p(\cE,T) \= \frac{1-c_p T + p T^2}{(1-T)(1-p^2T)}~.
\end{align}
Given the set of polynomials $\{R_1^{(p)}(\cE,T) = 1-c_pT+pT^2 \; | \; p \text{ a prime} \}$ appearing in the zeta function, it is possible to associate a \textit{Hasse-Weil $L$-function} to it. This is defined, up to finitely many factors, as
\begin{align}
L(\cE,s) \= \prod_{p \text{ good}} \frac{1}{R_1^{(p)}(\cE,p^{-s})}~,
\end{align}
where the product runs over the \textit{primes of good reduction} $p$ (\textit{good primes} for short) for which the variety $\cE/\IF_p$ is smooth. The finitely many factors not included here correspond to the \textit{primes of bad reduction} $p$ (\textit{bad primes} for short) for which $\cE/\IF_p$ is singular.

In particular, the $L$-function corresponding to the two-dimensional middle cohomology of the elliptic curve $\cE$ is given by
\begin{align} \label{eq:Hasse-Weil_L-function}
L(\cE,s) \= \prod_{p \text{ good}} \frac{1}{1-c_p p^{-s} + p^{1-2s}} \prod_{p \text{ bad}} \frac{1}{1\pm p^{-s}}~,
\end{align}
The sign chosen in the second product depends on the type of singularity the curve has (see ref.~\cite{Silverman2009a}). In particular we see that this $L$-function contains all the non-trivial information encoded in the zeta function. We will shortly see that these can also be related to modular forms by what is known as the \textit{modularity theorem}.

We define the \textit{slash operator with the character} $\chi$ is defined as
\begin{equation} \label{eq:slash}
	\left( \sla{f}k\chi\gamma\right)(\tau) \= \chi(d)^{-1} \left(c \tau + d\right)^{-k} f(\gamma \tau) \ , \qquad
	\gamma \= \begin{pmatrix} a & b \\ c & d \end{pmatrix} \in \Gamma \subseteq \SL(2,\IZ) \ .
\end{equation}
In terms of this operator, we can write the transformation property satisfied by a \textit{modular form of weight} $k$ with a Dirichlet character $\chi: \mathbb{Z} \to \mathbb{C}$ for a group $\Gamma $ simply as 
\begin{align}
	f = \sla{f}k\chi\gamma~.
\end{align}

We consider in particular the congruence subgroups
\begin{align}
	\Gamma(N) &\defineas \left\{ \left(
	\begin{array}{ccccc}
		a & b \\
		c & d
	\end{array}
	\right) \in \SL(2,\IZ) \;\; \bigg| \; \; a,d = 1 \!\! \mod N~, b,c = 0  \!\! \mod N \right\}~,\\
	\Gamma_0(N) &\defineas 
	\left\{ \left(
	\begin{array}{ccccc}
		a & b \\
		c & d
	\end{array}
	\right) \in \SL(2,\IZ) \;\; \bigg| \; \; c = 0  \!\! \mod N \right\}~.
\end{align}
Choosing $a=d=b=1$, $c=0$ in the formula \eqref{eq:slash} and using the transformation property of modular forms (with character) for $\SL(2,\IZ)$ (and many important subgroups $\Gamma$, such as all groups $\Gamma_0(N)$) shows that these modular forms are periodic, and thus have a Fourier expansion 
\begin{align}
f(\tau) \= \sum_{i=0} a_i q^i~, \qquad q \= \me^{2\pi \ii \tau}~.
\end{align}
A modular form for $\Gamma \subset \SL(2,\IZ)$ is called a \textit{cusp form} if $a_0 = 0$, and the space of such weight-$k$ cusp forms with character $\chi$ is denoted as $S_k(\Gamma,\chi)$.

Particularly interesting cases of cusp forms are \textit{normalised Hecke eigenforms} for $\Gamma_0(N)$ (or \textit{Hecke forms} for short). To define these, we first define \textit{Hecke operators} $T_m$ whose action on a modular form $f$ of weight $k$ can be written as
\begin{align} \label{eq:Hecke_operator_definition}
	T_m f  \= m^{k-1} \sum_{\gamma \in \Gamma_0(N)\setminus M_m} \sla{f}k\chi\gamma~, 
\end{align}
where $M_m$ is the set of $2 \times 2$ integral matrices with determinant $m$ satisfying the condition $c = 0 \!\! \mod N$. It can be shown that $T_m$ and $T_n$ commute for all $m,n$ (see for example ref.~\cite{Zagier2008a}). A modular form that is a simultaneous eigenvector of the operators $T_m$ for all $m$, normalised such that $a_1=1$ is called a Hecke form. It can be shown that for these forms
\begin{align} \label{eq:Hecke_eigenproperties}
T_m f = a_m f~, \qquad a_m a_n \= \sum_{r \mid (m,n)} r^{k-1} a_{mn/r^2}~.
\end{align}
To a Hecke form, one can associate the \textit{Hecke $L$-series}
\begin{align}
L(f,s) \= \sum_{n=1}^\infty \frac{a_n}{n^s}~.
\end{align} 
The properties in eq. \eqref{eq:Hecke_eigenproperties} can be used to show that this can be written as an Euler product
\begin{align}
L(f,s) \= \prod_{p \text{ prime}} \left( 1 + \frac{a_p}{p^s} + \frac{a_{p^2}}{p^{2s}} + \dots \right) \= \prod_{p \text{ prime}} \frac{1}{1 - a_p p^{-s} + p^{k-1-2s}}~.
\end{align}
Note that for $k=2$, this $L$-series and the Hasse-Weil $L$-function \eqref{eq:Hasse-Weil_L-function} have essentially the same form, up to the factor corresponding to the bad primes. This turns out to not be a coincidence: the modularity theorem \cite{Wiles:1995ig,Breuil2001a,Diamond1996a,Taylor1995a} can be stated as follows:\footnote{Strictly speaking this is a consequence of the modularity theorem, but for this paper, we only need this formulation.} Let $\cE/\IQ$ be an elliptic curve over $\IQ$. Then there exists a Hecke form (for a suitable subgroup $\Gamma \subset \SL(2,\IZ)$) of weight two, such that
\begin{align}
L(f,s) \= L(\cE,s)~,
\end{align}
In fact, the Fourier coefficients $a_p$ of the modular form $f$ agree with the coefficients $c_p$ appearing in the zeta functions, possibly apart from the bad primes:
\begin{align}
a_p \= c_p \qquad \text{for } p \text{ a prime of good reduction}.
\end{align}
Note that this form of modularity is intimately linked to the fact that the middle cohomology $H^1(\cE,\IC)$ is two-dimensional. In fact, using more abstract language, one can view the coefficients $a_p$ as arising from the action of the so-called absolute Galois group $\text{Gal}(\overline{\IQ},\IQ)$ on the middle cohomology. This group is topologically generated by the Frobenius map, explaining the relation to the matrix $\mtU_p(\varphi)$ we use in section \ref{sect:Deformation_method} to find the zeta function. 

It also turns out that the above modularity for elliptic curves generalises to higher-dimensional Calabi--Yau manifolds if the representation of the absolute Galois group $\text{Gal}(\overline{\IQ},\IQ)$ splits into a direct sum of representations, of which at least one is two-dimensional. Then, according to Serre's modularity conjecture \cite{Serre1975a,Serre1987a} (subsequently proved by Khare and Wintenberger \cite{Khare2009a,Khare2009b}) --- subject to a few technical assumptions --- one can associate to the two-dimensional piece a modular form. For instance, the following generalisations of the modularity theorem for higher-dimensional Calabi--Yau manifolds hold: 
\begin{enumerate}
	\item Attractive K3~surfaces $S$ are weight-3 modular with the modular forms associated to 2-dimensional Galois representations on the transcendental lattice $T(S)$.
	\item Rigid Calabi--Yau threefolds are weight-4 modular with the modular forms associated to 2-dimensional Galois representations on $H^3(Y,\IQ)$.
\end{enumerate}

In addition, there are results for non-rigid Calabi--Yau threefolds, where one can rigorously prove that the Galois representation on the middle cohomology of dimension $\geq 4$ factorises, and identify the corresponding modular forms. One very interesting family\footnote{This family and its quotients have subsequently been studied extensively in the physics literature in connection to the attractor mechanism, supersymmetric flux vacua, and field theory amplitudes \cite{Candelas:2019llw,Candelas:2021mwz,Candelas:2021tqt,Candelas:2023yrg,Candelas:2021lkc,Bonisch:2020qmm}.} of such a Calabi--Yau threefolds $Z$ was studied by Hulek and Verrill \cite{Hulek2005a}. By identifying a set of birational maps\footnote{Here we only sketch the main ideas of their argument, leaving any details to ref.~\cite{Hulek2005a}.}
\begin{align}
\phi: \cE \times \IP^1 \to Z~,
\end{align}
where $\cE$ is an elliptic curve, they were able to obtain cohomology homomorphisms
\begin{align}
	H^3(Z) \to H^1(E) \times H^2(\IP^1)~, 
\end{align}
where $E$ is an elliptic surface birational to $\cE \times \IP^1$. In a situation where the varieties as well as the maps are all defined over $\IQ$, this gives rise to an exact sequence of $\text{Gal}(\overline{\IQ},\IQ)$ modules \cite{Hulek2006a}, which can be used to show that the $\text{Gal}(\overline{\IQ},\IQ)$ representation contains two-dimensional pieces, essentially corresponding to the elliptic curves $\cE$. This also gives a geometric reason for the modularity in this case. Analogous relations are, in the general case, subject to numerous conjectures, such as so-called Hodge- and Tate-like conjectures, which roughly speaking state that there should be a geometric reason for the splitting of the Galois representation.
\subsection{Modularity of K3~surfaces} \label{app:K3_modularity}
\vskip-10pt
As seen in section \ref{sect:Hulek--Verrill_Fourfolds}, the modularity of certain K3~surfaces is intimately linked to the modularity of Hulek--Verrill fourfolds. We therefore find it useful to give a brief review of these properties. In addition, K3 manifolds display some properties analogous to those which distinguish the case of Calabi--Yau fourfolds from the threefolds. In this way, it is useful to draw analogies between the arithmetic geometry of K3~surfaces and the more involved case of fourfolds. For a more exhaustive discussion on the geometry of K3 manifolds, we refer the reader to refs. \cite{Huybrechts2016a,Aspinwall:1996mn}.
\subsubsection*{The middle cohomology of K3~surfaces}
\vskip-5pt
The Hodge diamond of a K3~surface $S$ is given by
\begin{align}
h^{p,q}\left( S \right) \= 
\begin{matrix}
&  & 1 &  & \\ 
& 0 &   & 0 &  \\ 
1 & & 20 & & 1\\
& 0 &   & 0 & \\ 
&  & 1  &  & 
\end{matrix}~.
\end{align}
The Néron-Severi group can be defined as
\begin{align}
\text{NS}(S) \defineas H^{2}(S,\IZ) \cap H^{1,1}(S,\IC)~,
\end{align}
which is isomorphic to the group of holomorphic line bundles on $S$. The rank of this group is defined to be the Picard number $\rho(S)$:
\begin{align}
\rho(S) \defineas \text{rk}_{\IZ}(\text{NS}(S))~.
\end{align}
The \textit{transcendental lattice} $T(S)$ is defined as the smallest subspace of $H^2(S,\IZ)$ that contains $H^{(2,0)}(S,\IC)$. For complex K3~surfaces, $T(S)$ is the orthogonal complement of the Néron-Severi group \cite{Huybrechts2016a}
\begin{align}
T(S) \= \text{NS}(S)^{\perp}~.
\end{align}
A K3~surface is called \textit{attractive} or \textit{singular}\footnote{Note that in this context ``singular'' simply means that the surface has the maximal Picard number $\rho(S)=20$. These are in fact smooth manifolds. Moore has proposed alternative terminology ``attractive K3~surface'' for these owing to their appearance as fixed points in the attractor flows of IIA/B supergravities compactified on the K3 \cite{Moore:1998pn,Moore:1998zu}. One sometimes encounters this terminology in physics papers, and it is the terminology we adopt here.} if it has a maximal Picard number $\rho(S)=20$ so that the middle cohomology of the surface takes the form
\begin{align}
H^2(S,\IC) \= T(S) \oplus \text{NS}(S)~.
\end{align}
\subsubsection*{Periods and Picard--Fuchs equations}
\vskip-10pt
Analogously to other Calabi--Yau varieties, and Riemann surfaces, one can define the periods of the K3~surfaces $S$ as integrals of the holomorphic two-form over cycles $\gamma \in H_2(S,\IZ)$. The corresponding \textit{period point} is
\begin{align}
	\left(\int_{\gamma_1} \Omega, \dots, \int_{\gamma_{22}} \Omega \right) \= \left(\int \Omega \wedge [\gamma_1], \dots, \int \Omega  \wedge [\gamma_{22}] \right) \in \IP^{21}~,
\end{align}
where $\gamma_i$ give a basis of $H^2(S,\IZ)$, and $[\gamma_i]$ are the corresponding cohomology classes. The period point is only defined up to overall scaling, as the two-form $\Omega$ is only defined up to such a scaling.

Note that by definition the elements of the Néron-Severi group correspond to forms $\gamma \in H^{(1,1)}(X,\IC)$. The corresponding integral vanishes, so by a change of basis, we can take the non-zero components of the period point to live in $\IP^{21 - \rho}$. We identify this point with the period vector $\varpi$. 

It can also be shown that if $[\gamma] \in \text{NS}(S)$, then also
\begin{align}
	\int_\gamma \theta^i \Omega \= 0~,
\end{align}
where $\theta^i$ schematically denotes any number of logarithmic derivatives in the algebraic moduli of the surface. Therefore, we can decompose the middle cohomology over $\IQ$ as
\begin{align}
	H^2(S,\IQ) \= H^2_H(S,\IQ) \oplus H_\perp^2(S,\IQ)~,
\end{align}
where the piece $H^2_H(S,\IQ)$ is spanned by the two-form $\Omega$ and its derivatives. The summand $H_\perp^2(S,\IQ) \supseteq NS(S) \otimes \IQ$ is its orthogonal complement under the intersection pairing defined by the integral above.

The Picard--Fuchs equations are derived similarly to other Calabi--Yau manifolds by considering $\Omega$ and its derivatives $\theta^i \Omega$. By definition these live in $H_H^2(S,\IC)$, which has dimension $\leq 22-\rho(S)$. Therefore we can have at most $22-\rho$ of the $\theta^i \Omega$ being independent and thus $\Omega$ satisfies a (partial) differential equation of degree $\leq 21-\rho(S)$. For instance, consider the case $\rho \geq 19$. In this case $H_H^2(S,\IC) = \langle \Omega, \theta \Omega, \theta^2 \Omega \rangle$, and $\Omega$ satisfies a third-degree differential equation.
\subsubsection*{Modularity of K3~surfaces with $\rho \geq 19$}
\vskip-10pt
From the Weil conjectures it follows that the zeta function of a K3~surface takes the form
\begin{align}
\zeta_p(S,T) \= \frac{1}{(1-T)R_2^{(p)}(S,T)(1-p^2T)}~,
\end{align}
where $R_2^{(p)}(S,T)$ is a degree-22 polynomial associated to the middle cohomology of the K3~surface. 

As discussed above, for families of K3~surfaces with $\rho=19$, the middle cohomology splits over $\IQ$ into a three-dimensional piece $H_H(S,\IQ)$ associated to the transcendental lattice, and the remaining 19-dimensional piece associated to the Néron-Severi group. Thus the polynomial $R(T)$ factorises correspondingly (see for example ref.~\cite{Yui2012a}) as
\begin{align}
R_2^{(p)}(S,T) \= R_H^{(p)}(S,T) R_{\perp}^{(p)}(S,T)~.
\end{align}
The behaviour of $ R_{\perp}^{(p)}(S,T)$ is understood as a consequence of Tate's conjecture whose validity for K3~surfaces has been proven \cite{Yui2012a}. In particular, if the divisors generating $\text{NS}(S)$ are defined over $\IQ$, then the Frobenius map acts on them as a multiplication by $p$, and consequently 
\begin{align}
R_{\perp}^{(p)}(S,T) \= (1-pT)^{\rho(S)}~.
\end{align}
The piece $H_H(S,\IQ)$ is spanned by the holomorphic two-form $\Omega$ and its derivatives, and is thus analogous to the horizontal cohomology of the fourfold of type $(1,1,1,1,1)$. By carefully following through the arguments in section \ref{sect:Deformation_method}, one can see that the polynomials $R_H^{(p)}(S,T)$ can be computed using the deformation method, which proceeds completely analogously, just with the matrix $\mtU_p(0)$ taking the form
\begin{align}
	\mtU_p(0) \= \left(
	\begin{array}{ccc}
		1 & 0 & 0 \\
		\alpha  & p & 0 \\
		\frac{\alpha ^2}{2} & \alpha  p & p^2 \\
	\end{array}
	\right)~.
\end{align}
In the case of K3~surfaces of Hulek--Verrill type, we find that $\alpha = 0$, at least to the $p$-adic accuracy we need to work to find the polynomials $R_H^{(p)}(S,T)$.

Attractive K3~surfaces have $\rho = 20$, and thus we would expect further factorisation of the degree-three polynomial $R_H^{(p)}(S,T)$ into a linear piece and a quadratic piece, which could be associated to an ordinary modular form. In fact, it has been shown by Livné \cite{Livne1995a} that in these cases there is a 2-dimensional Galois representation acting on the transcendental lattice $T(S)$ and this representation is modular in the sense that there exists a weight-3 modular form $f$ such that
\begin{align}
L(T(S) \otimes \IQ_\ell,s) \= L(f,s)~.
\end{align}
In other words, in these cases, there is a quadratic factor $R_T^{(p)}(T)$ of $R_H^{(p)}(S,T)$ with
\begin{align}
R_T^{(p)}(T) \= 1 - a_p T + p^2 T~,
\end{align}
and the coefficients $a_p$ are Fourier coefficients of a weight-3 modular form $f$.
\newpage

\section{Zeta Function for Fourfolds of Type \texorpdfstring{$(1,1,2,1,1)$}{(1,1,2,1,1)}} \label{app:(1,1,2,1,1)}
\vskip-10pt
In the case where the horizontal cohomology is of the type $(1,1,2,1,1)$, the Picard--Fuchs equation is again of the form \eqref{eq:PF_equation} but with $b=6$. This has two holomorphic solutions $\varpi_0^1, \varpi_0^2$, which we choose so that in the limit $\varphi \to 0$, these are given by
\begin{align}
 \varpi_0^1(\varphi) \= 1 + \cO(\varphi)~, \quad \text{and} \quad \varpi_0^2(\varphi) \= \varphi(1+\cO(\varphi))~.
\end{align}
We take the period vector to be given by $\varpi = (\varpi_0^1,\varpi_1,\varpi_2,\varpi_3,\varpi_4,\varpi_0^2)$. The logarithm-free vector $\wt \varphi$ and the matrices $\mtE(\varphi)$ and $\wt \mtE(\varphi)$ can again be defined using the definitions given in eqs. \eqref{eq:omega_tilde_definition}, \eqref{eq:E_matrix_definition}, and \eqref{eq:E_tilde_definition}, respectively. Similarly, $W$ is given by eq. \eqref{eq:wedge_product}, with the only difference to the case $(1,1,1,1,1)$ being that the matrix $\sigma$ takes the form
\begin{align} \label{eq:sigma_matrix_(1,1,2,1,1)}
\sigma \= \frac{\kappa}{(2\pi\ii)^4}\left(
\begin{array}{cccccc}
0 & 0 & 0 & 0 & 1 & 0 \\
0 & 0 & 0 & -1 & 0 & 0\\
0 & 0 & 1 & 0 & 0  & 0\\
0 & -1 & 0 & 0 & 0 & 0\\
1 & 0 & 0 & 0 & 0  & 0\\
0 & 0 & 0 & 0 & 0  & \sigma_{55}\\
\end{array}
\right), \qquad \kappa \defineas \int_{\wt X} D^4~,
\end{align}
where $\sigma_{55}$ denotes a manifold-specific constant, and, as in the case of manifolds of Hodge type $(1,1,1,1,1)$, $D \in H^2(\widetilde{X},\IZ)$ is the generator of the second cohomology of the mirror Calabi--Yau fourfold $\wt X$ of~$X$.

In the case $(1,1,2,1,1)$, the matrix $\mtE(\varphi)$ becomes singular in the limit $\varphi \to 0$, due to the inclusion of the period $\varpi_0^2$, which vanishes in this limit. This also implies that in this case $\mtV_p(0) \neq \mtU_p(0)$, meaning that this case requires slightly more careful treatment. In particular, the conditions arising from the relations \eqref{eq:U_differential_eq} turn out to be significantly less constraining, even when the limit $\varphi \to 0$ is treated carefully, so we need to look for alternative constraints in this case. Such constraints are provided by the observation that $\mtV_p(0) \neq \mtU_p(0)$ and the above observations imply that the logarithms appearing in $\mtE(\varphi)$ do not automatically drop out from the expression $\mtU_p(\varphi) = \mtE(\varphi^p)^{-1}\mtV_p(0) \mtE(\varphi)$. However, in order to have any hope of obtaining a rational expression of the form \eqref{eq:U_rational_form} for $\mtU(\varphi)$, logarithmic terms should not appear. Therefore we impose in this case as an additional constraint that the logarithmic terms cancel.\footnote{One could have also derived the form of the matrix $\mtU_p(0)$ in the case $(1,1,1,1,1)$, as well as in the threefold cases considered in refs.~\cite{Candelas:2021tqt,multiparameter_zeta} by imposing this requirement. In this sense the requirement that the logarithms drop out is just a natural generalisation of the earlier results.} This restricts the matrix $\mtV_p(0)$ to be of the form
\begin{align}
\mtV_p(0) \= u \left(
\begin{array}{cccccc}
1 & 0 & 0 & 0 & 0 & u_{0,5} \\
\alpha  & p & 0 & 0 & 0 & u_{1,5} \\
\beta  & \alpha  p & p^2 & 0 & 0 & u_{2,5} \\
\gamma  & \beta  p & \alpha  p^2 & p^3 & 0 & u_{3,5} \\
\delta  & \gamma  p & \beta  p^2 & \alpha  p^3 & p^4 & p^2 \epsilon \\
\eta & 0 & 0 & 0 & 0 & \mu \\
\end{array}
\right)~.
\end{align}
The compatibility condition \eqref{eq:Wedge_compatibility} has four solutions, imposing $u = \pm 1$, and 
\begin{align} \label{eq:Wedge_compatibility_solutions_(1,1,2,1,1)}
u_{i,j} = 0~, \qquad \beta = \frac{\alpha^2}{2}~, \qquad \delta = \frac{1}{8}(8 \alpha \gamma - \alpha^4 - 4 \sigma_{55} \epsilon^2)~, \qquad \eta = \pm \sigma_{55} \epsilon~, \qquad \mu = \mp p^2~,
\end{align}
where the opposite signs are to be chosen in the last two relations. In all cases we have tested, only the solution with $\mu$ having positive sign will result in a rational matrix $\mtU_p(\varphi)$. Furthermore, the requirement for $\mtU_p(\varphi)$ to have the form \eqref{eq:U_rational_form} with $n=13$, gives that
\begin{align}
\alpha = 0 + \cO(p^{13})~, \qquad \epsilon = 0 + \cO(p^{13})~,
\end{align}
so the relations \eqref{eq:Wedge_compatibility_solutions_(1,1,2,1,1)} above are in fact completely analogous to the case $(1,1,1,1,1)$.

\subsection{Computing the polynomials \texorpdfstring{$R_H^{(p)}(X,T)$}{associated to the horizontal cohomology}} \label{sect:R_H_(1,1,2,1,1)}
\vskip-10pt
In the case $(1,1,2,1,1)$, the eigenvalues $\lambda_i$ either all appear in complex conjugate pairs so that $\lambda_1= \overline{\lambda_2} = p^2e^{\ii \Theta_1}$, $\lambda_{3}= \overline{\lambda_4} = p^2e^{\ii \Theta_2}$, and $\lambda_5= \overline{\lambda_6} = p^2e^{\ii \Theta_3}$, or then four of the eigenvalues come in pairs, and the last two eigenvalues are $\pm p^2$, so that $\lambda_1= \overline{\lambda_2} = p^2e^{\ii \Theta_1}$, $\lambda_{3}= \overline{\lambda_4} = p^2e^{\ii \Theta_2}$, and $\lambda_5= - \lambda_6 = p^2$.

In the former case the polynomial $R_H^{(p)}(X,T)$ does not generically factorise over $\IQ$, taking the form
\begin{align} \label{eq:R_H_(1,1,2,1,1)_lambda1}
R_H^{(p)}(X,T) = \left(1{-}2 p^2 \cos(\Theta_1) T{+}p^4 T^2\right) \left(1{-}2 p^2 \cos(\Theta_2) T{+}p^4 T^2\right) \left(1{-}2 p^2 \cos(\Theta_3) T{+}p^4 T^2\right),
\end{align}
which also implies that the functional equation \eqref{eq:functional_eq_R_H} is satisfied with $\epsilon = 1$. 

In the latter case $R_H^{(p)}(X,T)$ factorises over $\IQ$ into two linear factors, and a generically irreducible degree-4 piece:
\begin{align} \label{eq:R_H_(1,1,3,1,1)_lambda2}
R_H^{(p)}(X,T) = (1 {-} p^2 T)(1 {+} p^2 T)\left(1{-}2 p^2 \cos(\Theta_1) T{+}p^4 T^2\right) \left(1{-}2 p^2 \cos(\Theta_2) T{+}p^4 T^2\right).
\end{align}
From this, it can be easily seen that eq. \eqref{eq:functional_eq_R_H} is now satisfied satisfied with $\epsilon = -1$.

In both cases, the functional equation satisfied by the polynomials implies that they can be written in the form
\begin{align}
R_H^{(p)}(X,T) \= 1 + a_p T + b_p p T^2 + c_p p^3 T^3 + \epsilon b_p p^5 T^4 + \epsilon a_p p^{8} T^5 + \epsilon p^{12} T^6~,
\end{align}
where $c_p = 0$ if the functional equation \eqref{eq:functional_eq_R_H} is satisfied with $\epsilon = -1$. Thus, to completely fix the polynomial $R_H^{(p)}(X,T)$ in the case $(1,1,2,1,1)$, we need to find the values of $a_p$, $b_p$, $c_p$, and $\epsilon$, the last of which can be fixed by solving the value of $d_p \defineas b_p \epsilon$. 
As in the other case, these coefficients can be expressed in terms of traces of powers of the matrix $\mtU$. The equations \eqref{coefficients_ap_bp} and \eqref{coefficient_cp} hold for the $(1,1,2,1,1)$ case as well, and for $d_p$ we derive the relation
\begin{align}
    p^5d_p=\frac{1}{24}\left(\Tr(\mtU)^4-6\Tr(\mtU^2)\Tr(\mtU)^2+3\Tr(\mtU^2)^2+8\Tr(\mtU^3)\Tr(\mtU)-6\Tr(\mtU^4)\right)~.
\end{align}
Thus, from the structure of the eigenvalues, we can again deduce bounds on the values of the coefficients in analogy to the $(1,1,1,1,1)$ case. We obtain
\begin{align} \notag
    \begin{split}
		a_p&\=-\sum_{i=1}^{6} \lambda_{i} \= -2p^2(\cos(\Theta_1)+\cos(\Theta_2) + \cos(\Theta_3))~, \\
		p \, b_p&\=\frac{1}{2}a_p^2-\frac{1}{2}\sum_{i=1}^{6}\lambda_{i}^2 \= \frac{1}{2}a_p^2-p^4(\cos(2\Theta_1)+\cos(2\Theta_2)+\cos(2\Theta_3)) \ ,\\
        p^3 \, c_p & \= -\frac{1}{3}a_p^3+pa_pb_p-\frac{1}{3}\sum_{i=1}^6 \lambda_i^3=-\frac{1}{3}a_p^3+pa_pb_p-\frac{2}{3}p^6\left(\cos(3\Theta_1)+\cos(3\Theta_2)+\cos(3\Theta_3) \right)~, \\
        p^5 d_p&=\frac{1}{4}a_p^4-pb_pa_p^2+\frac{1}{2}p^2b_p^2+p^2c_pa_p -\frac{1}{4}\sum_{i=1}^6 \lambda_i^4 \\
        &\=\frac{1}{4}a_p^4-pb_pa_p^2+\frac{1}{2}p^2b_p^2+p^2c_pa_p - \frac{1}{2}p^8(\cos(4\Theta_1)+\cos(4\Theta_2)+\cos(4\Theta_3)) ~.
	\end{split}
\end{align}
From these expressions we can immediately read off the following bounds
\begin{align}
	\begin{split}
		&|a_p|\leq 6 p^2~, \; \qquad \; \left|b_p-\frac{a_p^2}{2p}\right| \;  \leq \; 3 p^3~,\qquad \left|c_p+\frac{a_p^3}{3p^3}-\frac{a_pb_p}{p^2}\right|\leq 2p^3 \\
        &\left|d_p-\frac{a_p^4}{4p^5}+\frac{a_p^2b_p}{p^4}-\frac{b_p^2}{2p^3}-\frac{a_pc_p}{p^3}\right|\leq \frac{3}{2}p^3 < 2 p^3 \ .
	\end{split}
\end{align} 
As in the other case, we can deduce from these bounds that working mod $p^4$ for primes $p \geq 7$ gives the exact values for the coefficients $a_p$, $b_p$, $c_p$, and $d_p$.

In the case $\lambda_5=-\lambda_6=p^2$, the relations above are slightly modified to give
\begin{align}
    \begin{split}
		a_p&\=-\sum_{i=1}^{6} \lambda_{i} \= -2p^2(\cos(\Theta_1)+\cos(\Theta_2))~, \\
		p \, b_p&\=\frac{1}{2}a_p^2-\frac{1}{2}\sum_{i=1}^{6}\lambda_{i}^2 \= \frac{1}{2}a_p^2-p^4(\cos(2\Theta_1)+\cos(2\Theta_2)+1) \ ,\\
        p^3 \, c_p & \= -\frac{1}{3}a_p^3+pa_pb_p-\frac{1}{3}\sum_{i=1}^6 \lambda_i^3 \= -\frac{1}{3}a_p^3+pa_pb_p-\frac{2}{3}p^6\left(\cos(3\Theta_1)+\cos(3\Theta_2) \right)~, \\
        p^5 d_p&\=\frac{1}{4}a_p^4-pb_pa_p^2+\frac{1}{2}p^2b_p^2+p^2c_pa_p -\frac{1}{4}\sum_{i=1}^6 \lambda_i^4 \\
        &\= \frac{1}{4}a_p^4-pb_pa_p^2+\frac{1}{2}p^2b_p^2+p^2c_pa_p - \frac{1}{2}p^8(\cos(4\Theta_1)+\cos(4\Theta_2)+1) ~.
	\end{split}
\end{align}
which makes the bounds on the coefficients stricter, but only marginally. In particular the conclusion that it is enough to compute the coefficients mod $p^4$ for primes $p \geq 7$ to obtain the exact values for the coefficients $a_p$, $b_p$, $c_p$, and $d_p$ still applies.
\newpage

\section{Weight Three Modular Forms of the Congruence Subgroup $\Gamma_0(15)$} \label{app:Modular_forms_for_Gamma_0_15}
\vskip-10pt
In this appendix we calculate the periods and quasi-periods of the weight three modular forms of the congruence subgroup $\Gamma_0(15)$ of $\operatorname{SL}(2,\mathbb{Z})$ with the Dirichlet character $\chi_{-15}$. The performed calculations follow closely the analysis presented in Appendix~A of ref.~\cite{Bonisch:2022mgw} and in ref.~\cite{Bonisch:2023}, to which we refer the reader for more details on the performed computations.\footnote{We would like to thank the referee of this article for suggesting that we compare our original computations to the periods and quasi-periods as calculated in ref.~\cite{Bonisch:2022mgw}.} Many computations pertaining to modular forms performed in the appendix are made using algorithms implemented in the computer algebra system Sage \cite{sagemath}.

\subsection{The Congruence Subgroup $\Gamma_0(15)$ and its Modular Curve}
\vskip-10pt
\begin{figure}[H]
	\centering
	\begin{center}
		\includegraphics[scale=0.65]{figures/Fdomain2.pdf}
	\end{center}
	\place{0.69}{0.25}{$-\frac{1}{2}$}
	\place{5.61}{0.25}{$\frac{1}{2}$}
		
	\place{1.17}{0.25}{$-\frac{2}{5}$}
	\place{5.13}{0.25}{$\frac{2}{5}$}
	
	\place{1.49}{0.25}{$-\frac{1}{3}$}
	\place{4.81}{0.25}{$\frac{1}{3}$}
	
	\place{2.2}{0.25}{-$\frac{1}{5}$}		
	\place{4.17}{0.25}{$\frac{1}{5}$}
	
	\place{3.22}{0.25}{$0$}
	
	\place{0.9}{1}{$-\frac{1}{2} {+} \ii \frac{\sqrt{3}}{30}$}	
	\place{5.15}{1}{$\frac{1}{2} {+} \ii \frac{\sqrt{3}}{30}$}
	
	\place{2.44}{1}{$-\frac{1}{10} {+} \ii \frac{\sqrt{3}}{30}$}
	\place{3.52}{1}{$\frac{1}{10} {+} \ii \frac{\sqrt{3}}{30}$}
											
	\vskip10pt
	\capt{6in}{fig:Fdomain}{Shown is the fundamental domain of the level $15$ congruence subgroup $\Gamma_0(15)$. Boundary segments that are drawn in like colors are identified by a M\"obius transformation of an element of the discrete group $\Gamma_0(15)$.}	
\end{figure}
The level $15$ congruence subgroup $\Gamma_0(15)$ of $\operatorname{SL}(2,\mathbb{Z})$ is given by
\begin{equation}
  \Gamma_0(15) = \left\{ \begin{pmatrix} a & b \\ c & d \end{pmatrix}\in \operatorname{SL}(2,\mathbb{Z}) \, \middle| \, c = 0 \mod 15 \right\} \ ,
\end{equation}
which acts in the usual way by M\"obius transformations on the upper half plane $\mathcal{H} = \left\{ \tau \in \mathbb{C}\,\middle| \,\operatorname{Im}\tau >0 \right\}$. The fundamental domain in the upper half plane $\mathcal{H}$ is depicted in Figure~\ref{fig:Fdomain}. The boundary segments of this fundamental domain are identified with these group elements of $\Gamma_0(15)$
\begin{equation} \label{eq:GenGamma015first}
\begin{aligned}
   \gamma_0 &= \begin{pmatrix} 1 & 1 \\ 0 & 1 \end{pmatrix} \ , &\quad
   \gamma_1 &= \begin{pmatrix} -2 & -1 \\ 15 & 7 \end{pmatrix} \ , &\quad
   \gamma_2 &= \begin{pmatrix} 11 & 4 \\30 & 11 \end{pmatrix} \ , \\
   \gamma_3 &= \begin{pmatrix} 4 & 1 \\ 15 & 4 \end{pmatrix} \ , &\quad
   \gamma_4 &= \begin{pmatrix} 1 & 0 \\ 15 & 1 \end{pmatrix} \ , &\quad
   \gamma_5 &= \begin{pmatrix} 7 & -1 \\ 15 & -2 \end{pmatrix} \ .
\end{aligned}   
\end{equation}
These group elements also generate the modular subgroup $\Gamma_0(15)/\{ \pm \mtI \}$. The cusps $\frac13$, $-\frac13$ and the cusps $\frac15$, $-\frac15$, $\frac25$, $-\frac25$ are mutually equivalent, and the four inequivalent cusps $ \ii \infty$, $0$, $-\frac13$, and $-\frac25$ are the fixed points of the respective parabolic elements
\begin{equation} \label{eq:parelem}
    \gamma_0 = \begin{pmatrix} 1 & 1 \\ 0 & 1 \end{pmatrix} \ , \quad   
    \gamma_4 = \begin{pmatrix} 1 & 0 \\ 15 & 1 \end{pmatrix} \ , \quad   
    \gamma_3^{-1} \gamma_2 = \begin{pmatrix} 14 & 5 \\ -45 & -16 \end{pmatrix} \ , \quad   
    \gamma_1^{-1}\gamma_3^{-1}\gamma_5^{-1}\gamma_2 =  \begin{pmatrix} -31 & 12 \\ -75 & 29 \end{pmatrix} \ . 
\end{equation}
Furthermore, there is one non-trivial relation among the generators~\eqref{eq:GenGamma015first}
\begin{equation}
   \gamma_0^{-1} \gamma_5 \gamma_4 \gamma_1 = \mtI \ ,
\end{equation}
which allows us to remove for instance $\gamma_5$ from the list of generators~\eqref{eq:GenGamma015first}. Thus, in the following we consider the five independent generators for the modular subgroup $\Gamma_0(15)/\{ \pm \mtI \}$ given by
\begin{equation} \label{eq:G15gen}
  \Gamma_0(15)/\{ \pm \mtI \} = \left\langle \gamma_0, \ldots, \gamma_4 \right\rangle \ .
\end{equation}  

The modular curve $\mathcal{C} = \mathcal{H} / \Gamma_0(15)$ of $\Gamma_0(15)$ is of genus one, which we represent by elliptic curve given in the Tate form
\begin{equation} \label{eq:TateElliptic}
     y^2 + 3 x y + 8 y  = x^3 + 2 x^2 - 16 x - 33  \ .
\end{equation}
The affine coordinates $(x,y)$ parametrizing~$\mathcal{C}$ are meromorphic modular functions of $\Gamma_0(15)$ that map the cusp at $\ii \infty$ to the zero point $O$ of the elliptic curve~$\mathcal{C}$. Hence, the meromorphic modular functions $x(q)$ and $y(q)$ have poles of order $2$ and $3$ at $q=e^{2\pi i \tau} = 0$, respectively. As a consequence the field of modular functions $M^\text{mero}_0(\Gamma_0(15))$ of the modular group $\Gamma_0(15)$ are generated by $x(q)$ and $y(q)$ modulo the relation arising from the Tate equation~\eqref{eq:TateElliptic}, namely
\begin{equation}
    M^\text{mero}_0(\Gamma_0(15)) \simeq \mathbb{C}(x,y)/ \left\langle y^2 + 3 x y + 8 y  - x^3 - 2 x^2 + 16 x + 33 \right\rangle \ .
\end{equation}   

In order to describe the field of modular meromorphic functions $M^\text{mero}_0(\Gamma_0(15))$ explicitly, we first study the space of holomorphic modular forms $M_4(\Gamma_0(15))$ of weight $4$ of $\Gamma_0(15)$. The eight-dimensional vector space $M_4(\Gamma_0(15))$ possess (up to normalization) a unique holomorphic modular form of weight $4$ with maximal vanishing order $8$ at the cusp $\ii \infty$. This modular form can be expressed as
\begin{equation}
  h(q) = \frac{\eta(\tau)\,\eta(15\tau)^{15}}{\eta(3\tau)^3\, \eta(5\tau)^5} \ , \qquad q = e^{2\pi i \tau} \ ,
\end{equation}  
in terms of the Dedekind $\eta(\tau)$ function. The modular form $h(q)$ enjoys the Fourier expansion
\begin{equation} \label{eq:Defh}
  h(q) = q^8 - q^9 - q^{10} + 3\,q^{11} - 3\,q^{12} + 3\,q^{13} + 4\,q^{14} - 13\,q^{15} + 9\,q^{16} + 7\,q^{17} - 9\,q^{18} 
  + \ldots \ .
\end{equation}
The modular forms of weight $4$ with vanishing order $6$ and $5$ at the cusp $\ii \infty$, characterized by their Fourier expansions
\begin{equation}
\begin{aligned}
  f(q)  &= q^6 + 2q^9 + 5q^{12} + 10q^{15} + 20q^{18} + 26q^{21} + 45q^{24} + 60q^{27} + 85q^{30} + 100q^{33} 
  + \ldots  \ , \\
  g(q) &= q^5 + 9q^{10} + 27q^{15} + 73q^{20} + 126q^{25} + 243q^{30} + 344q^{35} + 585q^{40} + 729q^{45} 
  + \ldots \ ,
\end{aligned}
\end{equation}
yield the generators $x(q)$ and $y(q)$ of the field of modular meromorphic functions as
\begin{equation} \label{eq:GenMero}
\begin{aligned}
  x(q) &= \frac{f(q)}{h(q)} = q^{-2} + q^{-1} + 2q + 4q^2 + 2q^4 - q^5 + 3q^6 - 4q^7 + 2q^8 - 3q^9 + q^{10} 
  + \ldots \ , \\
  y(q) &= \frac{g(q)}{h(q)} = q^{-3} + q^{-1} + q^2 + 6q^3 + 6q^4 + 2q^5 + 7q^6 - 3q^7 - 7q^8 - 21q^9 + 3q^{10} 
  + \ldots \ . 
\end{aligned}  
\end{equation}
It is straight forward to check that these meromorphic modular forms fulfill the Tate equation~\eqref{eq:TateElliptic}.

\subsection{Weight Three Modular Forms with Character}
\vskip-10pt
In order to set the stage to compute the periods and quasi-periods of weight three modular forms of $\Gamma_0(15)$ with the Dirichlet character $\chi_{-15} \defineas \left( \frac{-15}{\cdot} \right)$, we construct in this subsection the finite dimensional vector space of modular forms $\mathbb{S}_3(\Gamma_0(15),\chi_{-15})$. 

As in ref.~\cite{Bonisch:2022mgw}, we define the vector spaces $\mathbb{S}_k(\Gamma,\chi)$ of modular forms of weight $k$ of a modular subgroup $\Gamma \subseteq \operatorname{SL}(2,\mathbb{Z})$ with a Dirichlet character $\chi: \mathbb{Z} \to \mathbb{C}$ as follows. The vector space of meromorphic modular forms with character $\chi$ are defined as
\begin{equation}
  M^\text{mero}_k(\Gamma,\chi) \= \left\{ F: \overline{\mathcal{H}} \to \overline{\mathbb{C}} \, \middle| \, \text{$F$ meromorphic and } \sla{F}k\chi\gamma = F  \
  \text{for any $\gamma\in \Gamma$} \right\} \ ,
\end{equation}
where $\overline{\mathcal{H}}$ is the union of the upper half plane $\mathcal{H}$ with the cusps of the modular subgroup $\Gamma$ and $\overline{\mathbb{C}} = \mathbb{C} \cup \{\infty\}$. The slash operator with the character $\chi$ is defined in eq.~\eqref{eq:slash}. We define
\begin{align}
	S_k^\text{mero}(\Gamma,\chi) \defineas \{F \in M^\text{mero}_k(\Gamma,\chi) | \; \text{$F$ is locally $(k{-}1)$-st derivative} \}~.
\end{align}
In particular, the constant terms in the Fourier expansions about the cusps vanish for such modular forms. These spaces allow us to define $\mathbb{S}_k(\Gamma,\chi)$ as
\begin{equation} \label{eq:DefDoubleS}
    \mathbb{S}_k(\Gamma,\chi) \; \defineas \; S^\text{mero}_k(\Gamma,\chi) / (D^{k-1} M^\text{mero}_{2-k}(\Gamma,\chi) ) \ , 
\end{equation}
in terms of the derivative $D \defineas \frac1{2\pi i} \frac{d}{d\tau} = q \frac{d}{dq}$. Thus, the elements of vector space $\mathbb{S}_k(\Gamma,\chi)$ of meromorphic modular forms with character $\chi$ which are $(k{-}1)$-st local derivatives modulo \mbox{$(k{-}1)$-st} global derivatives. The fact that $D^{k-1} M^\text{mero}_{2-k}(\Gamma,\chi)  \subset S^\text{mero}_k(\Gamma,\chi)$ is a consequence of Bol's identity \cite{MR33411,Bonisch:2022mgw}, which yields for any meromorphic function $f: \mathcal{H} \to \overline{\mathbb{C}}$ and for any $\gamma \in \operatorname{SL}(2,\mathbb{R})$  the non-trivial relation
\begin{equation} \label{eq:Bol}
  D^{k-1}(\sla{f}{2-k}\chi\gamma) \= (D^{k-1}f)\sla{\vphantom{f}}{k}\chi\gamma \ .
\end{equation}

Furthermore, we have the isomorphism \cite{Bonisch:2022mgw}
\begin{equation} \label{eq:DoubleSiso}
   \mathbb{S}_k(\Gamma,\chi) \simeq S_k(\Gamma,\chi) \oplus S_k(\Gamma,\chi)^\lor \ ,
\end{equation}   
where $S_k(\Gamma,\chi)$ is the (finite-dimensional) vector space of holomorphic cusp forms with character $\chi$ and $S_k(\Gamma,\chi)^\lor$ its dual with respect to the Eichler pairing $\{ \cdot , \cdot \}: S^\text{mero}_k(\Gamma,\chi) \times S^\text{mero}_k(\Gamma,\chi) \to \mathbb{C}$ given by
\begin{equation} \label{eq:Eichler_pairing_definition}
  \left\{ F, G \right\} = (2\pi i)^k \sum_{\tau\in\Gamma \backslash \overline{\mathcal{H}}} \operatorname{Res}_\tau (\widetilde F G ) \ .
\end{equation}  
Here $\widetilde F$ is an \textit{Eichler integral} of the meromorphic modular form $F$ with character $\chi$, that is any meromorphic function $\wt F : \overline{\mathcal{H}} \to \overline{\mathbb{C}}$ such that $D^{k-1}\wt F = F$. A particularly useful expression for Eichler integrals is given by
\begin{equation} \label{eq:Eichler}
  \widetilde F_{\tau_0}(\tau) = \frac{(2\pi i)^{k-1}}{(k-2)!} \int_{\tau_0}^\tau dz \, (\tau -z)^{k-2} F(z) \ ,
\end{equation}  
for some base point $\tau_0 \in \overline{\mathcal{H}}$. 

The vector space of holomorphic cusp forms with character $\chi$, $S_k(\Gamma,\chi)$, decomposes into eigenforms with respect to the Hecke operators $T_n$ which are defined as in eq.~\eqref{eq:Hecke_operator_definition}. Via the pairing \eqref{eq:Eichler_pairing_definition} that is equivariant with respect to $T_n$, this also induces a decomposition of meromorphic modular forms in $S_k(\Gamma,\chi)^\lor$ into eigenspaces with the same eigenvalues as in $S_k(\Gamma,\chi)$. Note that the meromorphic modular forms with character in the dual space $S_k(\Gamma,\chi)^\lor$ are only eigenforms of the Hecke operators $T_n$ modulo global total derivatives in $D^{k-1} M^\text{mero}_{2-k}(\Gamma,\chi)$. Altogether, this allows us to decompose the entire finite-dimensional vector space $\mathbb{S}_k(\Gamma,\chi)$ into a basis of cusps forms and meromorphic modular forms with character that are eigenforms with respect to the Hecke operators $T_n$. 

We are now ready to construct a Hecke eigenbasis for the for us relevant vector space of meromorphic modular forms with character $\mathbb{S}_3(\Gamma_0(15),\chi_{-15})$. The space of cusp forms $S_3(\Gamma_0(15),\chi_{-15})$ is two-dimensional and it is spanned by two linearly independent newforms $f_{\pm1}$ that are unambigiously distinguished by their Hecke eigenvalues $\pm 1$ of the Hecke operator $T_2$. They respectively enjoy the Fourier expansion 
\begin{equation} \label{eq:felem}
\begin{aligned}
  f_{+1}(q) &= q + q^2 - 3q^3 - 3q^4 + 5q^5 - 3q^6 - 7q^8 + 9q^9 + 5q^{10} + 9q^{12} - 15q^{15} + 5q^{16} 
    + \ldots \ , \\
  f_{-1}(q) &= q - q^2 + 3q^3 - 3q^4 - 5q^5 - 3q^6 + 7q^8 + 9q^9 + 5q^{10} - 9q^{12} - 15q^{15} + 5q^{16} 
    + \ldots \ .
\end{aligned}  
\end{equation}  
In order to complete these holomorphic modular forms to a basis of $\mathbb{S}_3(\Gamma_0(15),\chi_{-15})$ we take advantage of the isomorphism \cite{Bonisch:2022mgw}
\begin{equation} \label{eq:Siso}
  \mathbb{S}_3(\Gamma_0(15),\chi_{-15}) \simeq \mathbb{S}^{[\infty,-7]}_3(\Gamma_0(15),\chi_{-15}) \ ,
\end{equation}
where $\mathbb{S}^{[\infty,-7]}_3(\Gamma_0(15),\chi_{-15})$ is defined as in eq.~\eqref{eq:DefDoubleS} but restricted to meromorphic forms with character of weight three with poles only at the cusp $\ii \infty$ with maximal pole order $7$. Therefore --- as opposed to $\mathbb{S}_3(\Gamma_0(15),\chi_{-15})$ --- the vector space $\mathbb{S}^{[\infty,-7]}_3(\Gamma_0(15),\chi_{-15})$ arises as a quotient of the finite dimensional vector spaces $S^{[\infty,-7]}_3(\Gamma_0(15),\chi_{-15})$ and $D^2M^{[\infty,-7]}_{-1}(\Gamma_0(15),\chi_{-15})$.

Let us consider a meromorphic form $F \in S^{[\infty,-7]}_3(\Gamma_0(15),\chi_{-15})$. By definition $F$ has only a pole at the cusp $\ii \infty$ maximally of order $7$. Multiplying $F$ with the weight $4$ holomorphic modular form $h$ defined in eq.~\eqref{eq:Defh} yields a holomorphic cusp form $k(q) = F(q)h(q)$ of weight seven with character $\chi$ because $h(q)$ has maximal vanishing order $8$ at $\ii \infty$. Conversely, given a cusp form $k \in S_7(\Gamma_0(15),\chi_{-15})$ divided by the holomorphic modular form $h(q)$ gives rise to a meromorphic modular form with character $F(q) = \frac{k(q)}{h(q)}$. Since $h(q)$ has maximal vanishing order at the cusp $\ii \infty$ and hence no other zeros, the Fourier expansion about any cusp other than $\ii \infty$ has no constant term. However, it may have a constant term in the Fourier expansion about the cusp $\ii \infty$. As the vector space $S_7(\Gamma_0(15),\chi_{-15})$ of holomorphic cups forms with character $\chi$ is ten dimensional, we obtain ten linearly independent meromorphic modular forms $F(q)$. However, due to the appearance of the constant term in the Fourier expansion about the cusp $\ii \infty$ of some of the modular forms $F(q)$, only a nine dimensional subspace span the vector space $S^{[\infty,-7]}_3(\Gamma_0(15),\chi_{-15})$, and thus we find
\begin{equation}
   \dim S^{[\infty,-7]}_3(\Gamma_0(15),\chi_{-15}) =  \dim S_7(\Gamma_0(15),\chi_{-15}) - 1 = 9 \ .
\end{equation}   
Similarly, we can construct a basis of $D^2M^\text{mero}_{-1}(\Gamma_0(15),\chi_{-15})$ with the ansatz
\begin{equation}
  F(q) = D^2 \left( \frac{l(q)}{h(q)} \right) \ ,
\end{equation}
where $l(q)$ is a holomorphic modular form of weight three with character $\chi_{-15}$. By construction any choice of $l(q)$ yields a meromorphic modular form in $S^\text{mero}_3(\Gamma_0(15),\chi_{-15})$. However, the order of the pole at $\ii \infty$ is possibly eight and hence bigger than $7$. Restricting to those linearly independent meromorphic modular forms with character $F(q)$ with a pole of order $7$ or smaller, we find
\begin{equation}
  D^2M^\text{mero}_{-1}(\Gamma_0(15),\chi_{-15}) = 5 \ .
\end{equation}  
This proves with eqs.~\eqref{eq:DefDoubleS} and \eqref{eq:Siso} that
\begin{equation} \label{eq:S3dim}
  \dim  \mathbb{S}_3(\Gamma_0(15),\chi_{-15}) = \dim \mathbb{S}^{[\infty,-7]}_3(\Gamma_0(15),\chi_{-15}) = 4 \ ,
\end{equation}
which is in agreement with eq.~\eqref{eq:DoubleSiso} and $\dim S_3(\Gamma_0(15),\chi_{-15}) = 2$. Furthermore, the described construction allows us to complete the holomorphic cusp forms~\eqref{eq:felem} to a basis of $\mathbb{S}_3(\Gamma_0(15),\chi_{-15})$ by adding meromorphic representatives, which we choose to be the meromorphic modular forms $F_{\pm 1}$ with character $\chi_{-15}$ which have the Fourier expansions
\begin{equation}
\begin{aligned}
  F_{+1}(q) &= q^{-2} + \frac14q^{-1} - 3q^3 - \frac54q^5 - 3q^6 - 12q^7 + \frac94q^9 - 45q^{10} + \frac{47}4q^{11} + 36q^{12}  
     + \ldots \ , \\
  F_{-1}(q) &= q^{-2} - \frac14 q^{-1} - 3q^3 + \frac54q^5 + 3q^6 - 12q^7 - \frac94 q^9 - 45q^{10} - \frac{47}4q^{11} + 36q^{12} 
     + \ldots \ .
\end{aligned}
\end{equation}
It can be checked that the meromorphic modular forms $F_{+1}$ and $F_{-1}$ are Hecke eigenforms with eigenvalues $+1$ and $-1$ with respect to the Hecke operator $T_2$ in $\mathbb{S}_3(\Gamma_0(15),\chi_{-15})$. Moreover, these representatives are chosen such that they have a pole of order $2$ at $\infty$ normalized to one and such that the coefficient of $q$ vanishes.

Finally, let us record that the constructed basis of $\mathbb{S}_3(\Gamma_0(15),\chi_{-15})$ can also be expressed in terms of the generators $x(q)$ and $y(q)$ of the field of meromorphic functions given in eq.~\eqref{eq:GenMero} and the Eisenstein series $G_{3,\chi_{-15}}(q)$ of weight three with the character $\chi_{-15}$, which is defined as \cite{MR2385372}
\begin{equation}
\begin{aligned}
  G_{3,\chi_{-15}}(q) &= c_3(\chi_{-15}) + \sum_{n=1}^{+\infty} \left( \sum_{d|n} \chi_{-15}(d) d^2 \right) q^n \\
  &= -8 + q + 5q^2 + q^3 + 21q^4 + q^5 + 5q^6 - 48q^7 + 85q^8 + q^9 + 5q^{10}  
  + \ldots \ .
\end{aligned}  
\end{equation}
Here the constant $c_3(\chi_{-15})= \frac12 L(\chi_{-15},-2)=-8$ is given by the Dirichlet series $L(\chi_{-15},s) = \sum_{n=1}^{+\infty} \chi_{-15}(n) n^{-s}$ via analytic continuation to the critical value $-2$. We compute 
\begin{equation} \label{eq:DoubleS3basis}
\begin{aligned}
  f_{+1}(q) &= -\frac{5x + y + 2}{8x^2 + 23x + 15y + 2} \,  G_{3,\chi_{-15}}(q) \ , \\
  F_{+1}(q) &=- \frac{4x^3 + 7x^2 + 5xy - 26x + 6y - 16}{4(8x^2 + 23x + 15y + 2)} \,  G_{3,\chi_{-15}}(q) \ , \\
  f_{-1}(q) &=  -\frac{3x + y + 16}{8x^2 + 23x + 15y + 2} \,  G_{3,\chi_{-15}}(q) \ , \\
  F_{-1}(q) &= -\frac{4x^3 + x^2 + 3xy - 38x + 10y - 8}{8x^2 + 23x + 15y + 2}\,  G_{3,\chi_{-15}}(q) \ .
\end{aligned}
\end{equation}

\subsection{The weight-three parabolic cohomology of $\Gamma_0(15)$ with character $\chi_{-15}$}
\vskip-10pt
To a modular form $F(\tau)$ of weight $k$ of a congruence subgroup $\Gamma \subset \operatorname{SL}(2,\mathbb{Z})$ with character $\chi$, we can assign to it an Eichler integral $\widetilde F(\tau)$ which is generically not modular. For any group element $\gamma \in \Gamma$ we can measure  the failure of modularity with the function $r_F(\gamma)$ given by
\begin{equation} \label{eq:periodmap}
  r_F(\gamma)(\tau) \= \left(\widetilde F\sla{\vphantom{F}}{2-k}{\chi}\gamma\right)(\tau) - \widetilde F(\tau) \ . 
\end{equation}  
Bol's identity~\eqref{eq:Bol} implies for a modular form $F$ that $D^{k-1} r_F(\gamma)(\tau) =0$, which in turn shows that $r_F(\gamma)(\tau) \in V_{k-2}(\IC)$ is a polynomial in $\tau$ of degree $k-2$. Here $V_{k-2}(\IK)$ denotes the vector space of polynomials in $\tau$ of degree $k-2$ with coefficients in the field $\IK$, and for brevity we denote this usually simply by $V_{k-2}$. The polynomials $r_F(\gamma)$ are called \textit{period polynomials} and satisfy the cocycle relation
\begin{equation} \label{eq:r_rels}
  r_F(\gamma_1 \gamma_2) \= r_F(\gamma_1)\sla{\vphantom{F}}{2-k}{\chi}\gamma_2 + r_F(\gamma_2) \ , \qquad \gamma_1,\gamma_2 \in \Gamma \ .
\end{equation}  
Different choices of Eichler integrals $\widetilde F$ and $\widetilde F'$ give rise to distinct period polynomials $r_F$ and $r_F'$, which differ by a degree $k-2$ polynomial
\begin{equation}
r_F(\gamma) - r_F'(\gamma) \= \sla{p}{2-k}\chi\gamma - p \in V_{k-2}(\mathbb{C}) \quad \text{for all} \ \gamma\in \Gamma \ .
\end{equation}

Following ref.~\cite{Bonisch:2022mgw}, these observations now motivate the definition of coboundaries
\begin{equation} \label{eq:cobdry}
  B^1(\Gamma,V_{k-2}^\chi) \= \left\{ r: \Gamma \to V_{k-2}^\chi, \gamma \mapsto \sla{p}{2-k}\chi\gamma - p \right\}_{p \in V_{k-2}} \ ,
\end{equation}
and the parabolic cocycles
\begin{multline}
  Z_\text{par}^1(\Gamma,V_{k-2}^\chi) = \left\{ r: \Gamma \to V_{k-2}^\chi \, \middle|
    \, r(\gamma_1\gamma_2) = \sla{r(\gamma_1)}{2-k}{\chi}\gamma_2 + r(\gamma_2) \quad  \forall \ \gamma_1,\gamma_2\in \Gamma\right. \\
    \text{\ and\ } \left.r(\gamma) \in W_{k-2}^\chi(\gamma)  \quad \forall \ \text{parabolic}\ \gamma \in \Gamma
    \right\} \ ,
\end{multline} 
with the vector space of polynomials $W_{k-2}^{\chi}(\gamma)$ given by
\begin{equation}
  W_{k-2}^{\chi}(\gamma) = \left\{ \sla{p}{2-k}\chi\gamma - p \right\}_{p \in V_{k-2}}  \subset V_{k-2} \ ,
\end{equation}
which --- due to $B^1(\Gamma,V_{k-2}) \subset Z_\text{para}^1(\Gamma,V_{k-2})$ --- gives rise to the parabolic cohomology group
\begin{equation}
  H^1_\text{para}(\Gamma,V_{k-2})  = Z_\text{para}^1(\Gamma,V_{k-2}) / B^1(\Gamma,V_{k-2}) \ .
\end{equation} 
The utility of this cohomology is that every modular form $f \in S_k(\Gamma,\chi)$ corresponds to a unique element $[r_f] \in H^1_{\text{para}}(\Gamma,V_{k-2}(\IC))$ given by the image of the period polynomial. By construction, this representative is independent of the choice of the Eichler integral. Further, the theorems proven by Eichler \cite{Eichler1957a} and Shimura \cite{Shimura1971a} show that every cohomology element corresponds to a modular form $f \in S_k(\Gamma,\chi)$ or its complex conjugate $\overline{f} \in \overline{S_k(\Gamma,\chi)}$. Namely, the map $f \mapsto [r_f]$, $\overline{f} \mapsto [\overline{r_f}]$ provides an isomorphism
\begin{align} \label{eq:Eichler-Shimura_isomorphism}
	H^1_{\text{para}}(\Gamma,V_{k-2}(\IC)) \simeq S_k(\Gamma,\chi) \oplus  \overline{S_k(\Gamma,\chi)}~.
\end{align}
Apart from this isomorphism, in ref.~\cite{Bonisch:2022mgw} it is shown that over the field of complex numbers $\mathbb{C}$ the vector space $\mathbb{S}_k(\Gamma,\chi)$ of meromorphic modular forms with character $\chi$ is isomorphic to the parabolic cohomology group $H^1_\text{par}(\Gamma,V_{k-2})$ via the period map \eqref{eq:periodmap}
\begin{equation} \label{eq:isoSH1}
     \mathbb{S}_k(\Gamma,\chi) \xrightarrow{\sim} H^1_\text{para}(\Gamma,V_{k-2}) \, , \ F \mapsto [r_F] \ .
\end{equation}
Note that the cocycle representative $r_F$ is dependent on the chosen Eichler integral $\wt F$. Upon choosing a different integral, the cocycle representative $r_F$ changes by a coboundary and hence the cohomology class $[r_F]$ of $H^1_\text{par}(\Gamma,V_{k-2})$ is independent of the chosen Eichler integral. Moreover, $(k{-}1)$-st global derivatives representing zero classes in $\mathbb{S}_k(\Gamma,\chi)$ are mapped to coboundaries in $H^1_\text{par}(\Gamma,V_{k-2})$.  Thus the above isomorphism is well-defined.

Due to the isomormorphism \eqref{eq:isoSH1}, the Hecke operators $T_n$ on $\mathbb{S}_k(\Gamma,\chi)$ act on the parabolic cohomology group $H^1_\text{para}(\Gamma,V_{k-2})$ as well, which is explicitly discussed in ref.~\cite{Bonisch:2022mgw}. This means that, the Hecke operators decompose the cohomology group $H^1_\text{para}(\Gamma,V_{k-2})$ further into Hecke eigenspaces. Moreover, as shown ref.~\cite{Bonisch:2022mgw}, the diagonal matrix $\epsilon=\operatorname{diag}(-1, 1)$ yields an involution $r \mapsto \sla{r}{2-k}\chi\epsilon$ on the cocycles $Z_\text{par}^1(\Gamma,V_{k-2})$ via the action 
\begin{equation} \label{eq:invol}
  (\sla{r}{2-k}\chi\epsilon)(\gamma) \= r(\epsilon \gamma \epsilon)\sla{\vphantom{r}}{2-k}\chi\epsilon \ ,
\end{equation}  
where the slash operator acts as in \eqref{eq:slash}. 

This involution commutes with the action of the Hecke operators and induces a decomposition of the parabolic cohomology group $H^1_\text{par}(\Gamma,V_{k-2}(\IC))$ into positive and negative eigenspaces, i.e.,
\begin{equation}
    H^1_\text{para}(\Gamma,V_{k-2}(\IC)) \=  H^1_{+,\text{para}}(\Gamma,V_{k-2}(\IC))  \oplus H^1_{-,\text{para}}(\Gamma,V_{k-2}(\IC)) \ .
\end{equation}
Thus, we can decompose the parabolic cohomology group $H^1_\text{par}(\Gamma,V_{k-2})$ simultaneously into positive and negative eigenspaces with respect to the involution induced by the matrix $\epsilon$ and into Hecke eigenspaces with respect to the action of the Hecke operators. If $f$ is a modular form with character $f \in \IS_k(\Gamma,\chi)$, this fact can be used to show that
\begin{align} \label{eq:periods_definition} 
	[r_f] \in \omega^+_f H^1_{+,\text{para}}(\Gamma_0(N),V_{k-2}(\IQ(f))) \oplus \omega^-_f H^1_{-,\text{para}}(\Gamma_0(N),V_{k-2}(\IQ(f)))~,
\end{align}
where $\IQ(f)$ is the field generated by the Hecke eigenvalues of $f$.
In other words, we can write
\begin{align}
	[r_f] \= \omega_f^+ [r^+] + \omega_f^-[r^-]~, \qquad [r^\pm] \in H^1_{\pm,\text{para}}(\Gamma_0(N),V_{k-2}(\IQ(f)))~,
\end{align}
and this expression defines the \textit{periods} $\omega_f^{\pm}$ of $f$, which are only well-defined up to multiplication by the elements of $\IQ(f)$. Note that the isomorphism \eqref{eq:Eichler-Shimura_isomorphism} guarantees that the periods $\omega_f^\pm$ always exist and are non-vanishing. If $F$ is the meromorphic partner of $f$, the periods $\omega_F^{\pm}$ of $F$, which we often denote by $\eta^\pm_F$ for clarity, are called the \textit{quasi-periods of $f$}.

With all the relevant ingredients at hand, we can now compute the for us relevant parabolic cohomology group $H^1_\text{para}(\Gamma_0(15),V_{1})$ of weight three. Due to the relations~\eqref{eq:r_rels} we can unambiguously specify a representative of a cohomology element by stating the five tuple of period polynomials 
\begin{equation}
   \left( r(\gamma_0), r(\gamma_1), \ldots, r(\gamma_4) \right) \in V_{1}^5\ ,
\end{equation}
in terms of the five generators~\eqref{eq:G15gen}. For weight three these polynomials are of the linear form
\begin{equation} \label{eq:linpols}
  r(\gamma_\ell)=\alpha_\ell \, \tau + \beta_\ell \ , \qquad \ell=0,\ldots,4 \ .
\end{equation}
These period polynomials $r$ are constrained by the parabolic cocycle conditions, namely $r(\gamma) \in W_1^{\chi_{-15}}(\gamma)$ for any parabolic element $\gamma \in \Gamma_0(15)$. For each inequivalent cusp we get one independent constraint, which we calculate from the inequivalent parabolic elements~\eqref{eq:parelem}, which yields on the coefficients $\alpha_\ell$ and $\beta_\ell$ the four constraints 
\begin{equation} \label{eq:linpolsrels}
    \alpha_0 = 0 \ , \quad \beta_4 \= 0 \ ,  \quad \alpha_2 - \alpha_3 - 3 \beta_2 + 3\beta_3 =0 \ , \quad
    \alpha_1 + 3 \alpha_2 + \alpha_4 + 5 \beta_0 - 8 \beta_2 - 2 \beta_3  = 0 \ .
\end{equation} 
Thus any set of linear polynomials~\eqref{eq:linpols} obeying the constraints~\eqref{eq:linpolsrels} describes a cocycle $Z^1_\text{para}(\Gamma_0(15),V_{1})$. The coboundaries $B^1(\Gamma_0(15),V_{1})$ in turn are given by
\begin{equation}
     B^1(\Gamma_0(15),V_{1}) \= \left\langle \xi_+ \,, \ \xi_- \right\rangle \ ,
\end{equation}     
specified in terms of the two five tuples
\begin{equation}
    \xi_+ = \left(  0,\,  -15\tau - 8,\, -30\tau-12,\, 15\tau+3,\, 15\tau \right) \ , \quad
    \xi_- = \left(  1,\,   \tau +1,\, -12\tau-4,\,  3\tau+1,\, 0 \right) \ .      
\end{equation}
The coboundaries $\xi_+$ and $\xi_-$ arise from their defining polynomial $p=1$ and $p=\tau$, and they are eigenvectors with eigenvalues $+1$ and $-1$ with respect to the involution~\eqref{eq:invol}. 

Altogether, we thus find that --- taking into account the four constraints~\eqref{eq:linpolsrels} --- the cocycles $Z^1_\text{para}(\Gamma_0(15),V_{1})$ are parametrized by $6$ parameters modulo the two linearly independent coboundaries $\xi_\pm$. Therefore, the parabolic cohomology group $H^1_\text{para}(\Gamma_0(15),V_{1})$  is of dimension~$4$, which is in accord with the isomorphism \eqref{eq:isoSH1} and the dimension~\eqref{eq:S3dim} of $\mathbb{S}_3(\Gamma_0(15),\chi_{-15})$. We calculate a basis for $H^1_\text{para}(\Gamma_0(15),V_{1})$ given by
\begin{equation}
     H^1_\text{para}(\Gamma_0(15),V_{1}) 
     = \left\langle [r^+_{+1}], [r^-_{+1}], [r^+_{-1}], [r^-_{-1}]  \right\rangle \ ,
\end{equation}
Here the four cocycle representatives $r^\pm_{+1},r^\pm_{-1}  \in Z^1_{\pm,\text{para}}(\Gamma_0(15),V_{1})$ are chosen as eigenvectors with respect to the involution~\eqref{eq:invol} and the Hecke operator $T_2$. That is to say the superscript $\pm$ denotes the eigenvalue $\pm 1$ of the involution~\eqref{eq:invol} and the subscript ${\pm1}$ displays the eigenvalue with respect to the Hecke operator $T_2$. The calculated cohomology representatives are explicitly given by the five tuples
\begin{equation} \label{eq:Cohreps}
\begin{aligned}
  r^+_{+1} &=\left( 0,\, 1,\, 30\tau+12,\, -15\tau-3,\, 0 \right) \ , & \quad  r^-_{+1} &= \left( 0,\, 2\tau+1,\, 0,\, 3\tau+1\,,0 \right) \ , \\
  r^+_{-1} &= \left( 0,\, 1,\, 10\tau+ 4,\,-5\tau -1,\,  0 \right) \ ,  & \quad  r^-_{-1} &=  \left( 0,\, 2\tau+1,\, -4\tau-\tfrac43,\, \tau+\tfrac13,\,  0 \right) \ .
\end{aligned}
\end{equation} 
Finally, we can now use the isomorphism~\eqref{eq:isoSH1} to expand the meromorphic modular forms with character~\eqref{eq:DoubleS3basis} of weight three into the constructed cohomology elements of $H^1_\text{para}(\Gamma_0(15),V_{1})$. For the holomorphic modular Hecke eigenforms $f_{\pm 1}$ and for the meromorphic Hecke eigenforms $F_{\pm 1}$ of weight three with character $\chi_{-15}$, we arrive for the period polynomials at the expansions
\begin{equation}
  [r_{f_{\pm 1}}] = \omega^+_{f_{\pm 1}} [ r^+_{\pm 1}] + \omega^-_{f_{\pm 1}} [ r^-_{\pm 1}] \ , \qquad
  [r_{F_{\pm 1}}] = \eta^+_{F_{\pm 1}} [ r^+_{\pm 1}] + \eta^-_{F_{\pm 1}} [ r^-_{\pm 1}] \ .
\end{equation}
Note that the since the Hecke eigenvalues of $f_{\pm 1}$ are in the field of rational numbers $\IQ$, by eq.~\eqref{eq:periods_definition} the cohomology representatives~\eqref{eq:Cohreps} are also constructed over $\IQ$, whereas the representatives $r_{f_{\pm 1}}$ and $r_{F_{\pm 1}}$ via their Eichler integrals are defined over the field of complex numbers~$\mathbb{C}$. Therefore, the periods and quasi-periods are complex numbers, which capture the transcendental nature of the embedding of the meromorphic modular forms with character of $\mathbb{S}_3(\Gamma_0(15),\chi_{-15})$ into the parabolic cohomology group $H^1_\text{para}(\Gamma_0(15),V_{k-2})$. 

Let us now numerically compute the periods $\omega^\pm_{f_{\pm 1}}$ and quasi-periods $\eta^\pm_{F_{\pm 1}}$ explicitly. For a given meromorphic modular form~$F$ with character this is done in the following way: First, we determine a Fourier expansion about the cusp $\ii \infty$ to sufficiently high order in $q$ of an Eichler integral $\widetilde F(q)$ of the modular form $F$. For the Eichler integral $\widetilde{F}(q)$ we simply take 
\begin{equation}
  \widetilde F(q) = \sum_{j \ne 0} \frac{a_j}{j^2} q^j \ ,
\end{equation}
for the meromorphic modular form $F(q)= \sum_{j \ne 0} a_j q^j$ (with vanishing constant term). Then for a list of $n\ge 2$ suitable numerical values $\tau_1,\ldots\tau_n \in \mathcal{H}$ we evaluate numerically these Fourier expansions up to the given order with the numerical values $r_F(\gamma_\ell)(\tau)$ explicitly. Using linear regression the calculated numerical data determines a linear function $\bar\alpha_\ell \tau + \bar\beta_\ell$, which approximates the period polynomial $r_F(\gamma_\ell)(\tau) = \alpha_\ell \tau + \bar\beta_\ell$ to a certain numerical precision (which depends on the chosen numerical values $\tau_1,\ldots,\tau_n$ and on the used order of the Fourier expansion~$\widetilde F(q)$). This procedure is repeated for all generators $\gamma_\ell$, $\ell =0,\ldots, 4$, in order to obtain a numerical approximation of the cohomology class $[r_F]$. For the meromorphic representatives~\eqref{eq:DoubleS3basis} we arrive at the numerically approximations expressed in terms of the period matrices
\begin{equation}
\begin{aligned}
  \begin{pmatrix}  \omega^+_{f_{+1}} &  \omega^-_{f_{+1}} \\ \eta^+_{F_{+1}} & \eta^-_{F_{+1}} \end{pmatrix}
  &=  \begin{pmatrix} 1.7609196507164596209\ldots & \ii \,6.8200124812342295488\ldots \\
 0.46706675860507249461\ldots & -\ii\,1.8089417776445254950\ldots  \end{pmatrix} \ , \\
   \begin{pmatrix}  \omega^+_{f_{-1}} &  \omega^-_{f_{-1}} \\ \eta^+_{F_{-1}} & \eta^-_{F_{-1}} \end{pmatrix}
  &=  \begin{pmatrix} 2.3625216251503139776\ldots & \ii\,9.1500069092620476587\ldots \\
 -0.6266358133628561690\ldots & \ii\,2.4269500692914810463\ldots  \end{pmatrix} \ .
\end{aligned}
\end{equation}  
The positive periods $\omega^+_{f_{\pm 1}}$ and quasi-periods $\eta^+_{F_{\pm 1}}$  are real, and the negative periods $\omega^-_{f_{\pm 1}}$ and quasi-periods $\eta^-_{F_{\pm 1}}$ are imaginary, which reflects that the eigenvalues of the Hecke operators of $f_{\pm}$ are rational, and thus real \cite{Bonisch:2022mgw}. We observe that the periods and quasi-periods fulfil the relations
\begin{equation}
   \omega^+_{f_{\pm1}} \eta^-_{F_{\pm1}} +  \omega^-_{f_{\pm1}} \eta^+_{F_{\pm1}} = 0 \ .
\end{equation}  
We also wish to point out explicitly the connection between the L-functions values that appear as Deligne's periods and the periods of the holomorphic modular form $f_{+1}$. A generalisation \cite{Pasol2013a} of a theorem of Manin's \cite{Manin1973a} can be used to show that the period polynomial for a holomorphic modular form $f \in S_k(\Gamma,\chi)$ can be written as 
\begin{align}
	r_f(\gamma)(\tau) = (2\pi \ii)^{k-2} L(f,1) p_1(\tau) + (2\pi \ii)^{k-3} L(f,2) p_2(\tau)~,
\end{align}
where $p_i(\tau)$ are polynomials in with coefficients in $\IQ(f)$. This shows that the periods of a holomorphic modular forms are proportional to the critical $L$-function values. In particular, for the weight-$3$ holomorphic modular form $f_{+1}$ of $\Gamma_0(15)$ it explains the relations
\begin{align}
	\omega_{f_{+1}}^+ \= 2 L_3(2)~, \qquad \omega_{f_{+1}}^- \= 2 (2\pi \ii) L_3(1)~.
\end{align}
This also helps to explain why we expect periods of a modular form to appear in the periods of a Calabi--Yau manifolds, as in section~\ref{sect:Delignes_periods} we have already seen that they should be given in terms of the L-function values.

\newpage
\section{Three Further Examples} \label{app:non-modular_examples}
\vskip-10pt
To contrast with the modular example of the Hulek--Verrill fourfold, we briefly present two cases where the zeta function data indicates that rank-two attractor points or \AK{} points do not exist. These are the family $\IP^7[2,2,4]^\vee$ of mirror manifolds of the complete intersection $\IP^7[2,2,4]$, which is of Hodge type $(1,1,1,1,1)$, and the mirror $X_{1,4}^\vee$ of a Grassmannian subvariety $X_{1,4} \subset \text{Gr}(2,5)$ of the Hodge type $(1,1,2,1,1)$. In particular, we will see that the number of quadratic factorisations of their local zeta functions is much lower, leading us to conclude that there are likely no modular manifolds in the moduli space. The examples also work to illustrate in practice the method for computing the zeta functions discussed in section~\ref{sect:Deformation_method}.

In addition, we discuss the family $\IP^5[6]^\vee$ of mirror manifolds of the sextic fourfolds $\IP^5[6]$. The polynomials $R_H^{(p)}(\IP^5[6]^\vee,T)$ for this family have frequent factorisation, making it tempting to assume that an attractor or an \AK{} point exists. However, a search for persistent factorisations yields no results, leaving the question of modularity open for this family.

\subsection{Hodge type \texorpdfstring{$(1,1,1,1,1)$}{(1,1,1,1,1)}: the mirror of the complete intersection \texorpdfstring{$\IP^7[2,2,4]$}{of two quartics and a quadric in the 7-dimensional projective space}}
\label{sect:P7224}
\vskip-10pt
As a first non-modular example, let us consider the mirror $\IP^7[2,2,4]^\vee$ of the complete intersection $\IP^7[2,2,4]$ of two quadrics and a quartic in projective space.

The topological data of this family of manifolds is given by
\begin{align}
	h^{1,1} \= 1~, \quad h^{2,1} \= 0~, \quad h^{2,2} \= 1100~, \quad h^{3,1} \=263 ~, \quad \chi \=1632 \ ,
\end{align}
whereas its periods can be computed as solutions of the Picard-Fuchs operator
 \begin{align}
 	\cL \= \theta^5 - 32(1+2\theta)^3(1+4\theta)(3+4\theta)\varphi~,
 \end{align}
indicating that this family of manifolds is of type $(1,1,1,1,1)$. The fundamental period $\varpi^0$ can be computed as the normalised holomorphic solution to this differential operator and reads
\begin{align}
	\varpi^0(\varphi)\=\sum_{n=0}^{\infty} \frac{((2n)!)^2(3n)!}{(n!)^9}\varphi^n \ .
\end{align}
Using the period vector, we can compute the logarithm-free period matrix $\widetilde{\mtE}$ which in turn gives the matrix $\mtW=\widetilde{\mtE}^T\sigma\widetilde{\mtE}$. Its inverse is given by 
\begin{align}
	\mtW^{-1} \= \left(
	\begin{array}{rrrrr}
		-192\varphi & -896 \varphi & -3840\varphi & -6144\varphi & 1-4096\varphi \\
		-896\varphi & 1792\varphi & 2048\varphi & -(1-4096\varphi) & 0\\
		-3840\varphi & 2048\varphi & 1-4096\varphi & 0 & 0 \\
		-6144\varphi & -(1-4096\varphi) & 0 & 0 & 0 \\
		1-4096\varphi & 0 & 0 & 0 & 0
	\end{array}
	\right).
\end{align}
As described in section \ref{sect:practical_zeta_function}, the constants $\alpha$ and $\gamma$ appearing in the matrix $\mtV_p(0)$ can be fixed by requiring all entries of $\mtU_p(\varphi)$ to be rational functions of $\varphi$ up to $p$-adic accuracy of $p^{13}$. In this case we have $\cW= 1$, so the denominator to the $p$-adic accuracy $n$ is given by
\begin{align}
P_n(\varphi^p) \= \Delta(\varphi^p)^{n-5} \= (1-2^{12} \varphi^p)^{n-5} \ .
\end{align}
For $\IP^7[2,2,4]^\vee$ we find for all primes $7\leq p\leq 317$ that 
\begin{equation}
    \alpha\= 0+ \mathcal{O}(p^{13})~,
\end{equation}
whereas the values for $\gamma \! \mod p^{7}$ are computed numerically for all primes $7\leq p \leq 317$. For the first primes these are listed in \tref{tab:gamma_values_p7[2,2,4]}.
\begin{table}[H]
	\renewcommand{\arraystretch}{1.3}
	\begin{center}
        \begin{tabular}{|l|l||l|l||l|l||l|l|}
			\hline
			\hfil $p$ & \hfil $\gamma + \cO\left(p^{7}\right)$ & \hfil $p$ & \hfil $\gamma + \cO\left(p^{7}\right)$ & \hfil $p$ & \hfil $\gamma + \cO\left(p^{7}\right)$& \hfil $p$ & \hfil $\gamma + \cO\left(p^{7}\right)$\\ \hline \hline
			7 & 507983 & 23 &2256856830 & 43 & 46222905083& 67& 5161332788111\\ \hline 
			11 & 14056691 & 29 &15950284055 & 47 & 378140704441& 71& 4752693726602\\ \hline 
			13 & 33071441 & 31 &22233202046 & 53 & 488630690470& 73& 75828749708\\ \hline 
			17 & 16045858 & 37 &46843843747 & 59 & 1734718936212& 79& 7276878958204\\ \hline 
			19 & 632536980 & 41 &115754335762 & 61 & 1353458876413& 83& 4201927721268\\ \hline 
		\end{tabular}
		\vskip10pt
		\capt{\textwidth}{tab:gamma_values_p7[2,2,4]}{The values of the prime-dependent constant $\gamma \! \mod p^{7}$ for the first $20$ primes for $\IP^7[2,2,4]^\vee$.}		
	\end{center}
	\vskip-30pt
\end{table}
With these values of $\gamma$, it is possible to compute the matrices $\mtU_p(\varphi)$ to the $p$-adic accuracy required to find their characteristic polynomials $R_H^{(p)}(\IP^7[2,2,4]^\vee,T)$. We can then investigate for which values of $\varphi \in \IF_p$ these factorise. In particular, we would expect that if there is a rank-two attractor point or an \AK{} point for a rational value $\varphi \in \IQ$, then we should find at least one factorisation of the form \eqref{eq:RHfac} for almost every prime $p$.\footnote{If the manifold corresponding to the point $\varphi$ in the complex structure moduli is singular over $\IF_p$ (or has an apparent singularity), meaning that $\Delta(\varphi) = 0 \mod p$, then we would not expect to see a factorisation as we do not compute the zeta functions corresponding to these points. Also, if $\varphi \in \IQ$ is of the form $\varphi = n/m$ with $n,m$ mutually prime, for primes $p \mid m$, there is no representative for $\varphi$ in $\IF_p$, so we would expect not to see the corresponding factorisation.} However, the data presented in figure \ref{fig:Factorisations_P7224}, strongly indicates that no such points exist for this family of manifolds, given several primes for which no factorisations exist at all. This should be contrasted with the modular example of section \ref{sect:Hulek--Verrill_Fourfolds}, where for every prime studied there exists at least one factorisation.
\begin{figure}[H]
	\centering
	\begin{center}
		\includegraphics[scale=0.9]{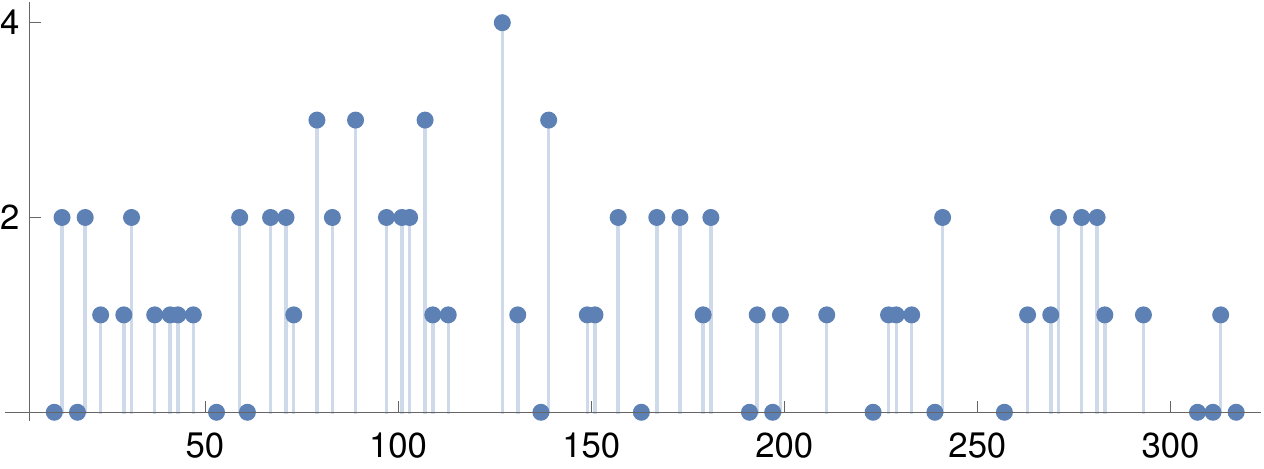}
	\end{center}
	\vskip-10pt
	\capt{6in}{fig:Factorisations_P7224}{The number of factorisations where the polynomial $R_H^{(p)}(\IP^7[2,2,4]^\vee,T)$ factorises over $\IQ$ to a product of one linear factor that is always present by the Weil conjectures, and two quadratic factors. The vertical axis gives the number of such factorisations for the prime indicated by the horizontal axis. Note that the numerous primes where no factorisation exist likely indicate that there are no persistent factorisations, and therefore no rank-two attractor or \AK{} points.}	
\end{figure}
To provide some more evidence in favour of non-existence of persistent factorisations, we can perform a search, similar to that done in section \ref{sect:persistent_factorisation}, looking for integers $k_0,k_1 \in \IZ$, with $k_1 \neq 0$, such that 
\begin{align}
k_0 + k_1 \varphi \= 0 \mod p~, \qquad \text{or} \qquad k_1 \= 0 \mod p
\end{align}
when $R_H^{(p)}(\IP^7[2,2,4]^\vee,T)$ has a quadratic factorisation or $\Delta = 0 \!\! \mod p$, the latter condition indicating the existence of a singularity or an apparent singularity. We find no such integers $k_0,k_1$, at least within the bounds $-1000 \leq k_0 \leq 1000$ and $0 < k_1 \leq 1000$.

This of course does not conclusively rule out existence of persistent factorisations. However, consider the following simple argument, ignoring for the moment the possibility of bad primes for which $\Delta(\varphi) = 0 \mod p$: Suppose we have a persistent factorisation at a rational value $\varphi=r/s$, but primes $p_1,\dots,p_n$ for which there are no factorisations. For this to be consistent we must have that $\varphi$ is not defined in the fields $\IF_{p_i}$, implying that $p_i \mid s$ so that
\begin{align}
s \= N p_1 \dots p_n~, \qquad N \in \IZ~.
\end{align}
Using the data displayed in the figure \ref{fig:Factorisations_P7224}, this implies that
\begin{align}
s>210595453066825643554741744147>10^{29}~,
\end{align}
which seems, at least naïvely, unlikely. A similar argument was used in ref. \cite{Candelas:2019llw} to argue against the existence of further attractor points on the family of $\IZ_5$ quotients of the Hulek--Verrill threefolds.

We could also take into account the possibility of existence of bad primes. Assume that $\varphi=r/s$ where $r$ and $s$ are coprime integers, and $\Delta(\varphi)=0 \mod p$ for primes $p=p_1,\dots,p_n$. Then we have a system of congruences
\begin{align}
2^{12} r \= s \mod p_1~, \qquad 2^{12} r \= s \mod p_2~, \qquad \dots \qquad  2^{12} r \= s \mod p_n~.
\end{align}
The Chinese remainder theorem then implies that
\begin{align}
s \= 2^{12} r + N p_1 \dots p_n~, \qquad N \in \IZ~.
\end{align}
Since $r$ and $s$ are assumed to be coprime, the only solution with $N=0$ is $r=1$, $s=2^{12}$, but we know that $\varphi=1/2^{12}$ corresponds to a conifold, ruling out this solution. Otherwise, for large $n$, either $r$ or $s$ needs to be large, making the existence of an attractor or an \AK{} point seem, at least naïvely, unlikely.

Similar arguments can also be utilised for providing evidence against existence of persistent factorisations for values of $\varphi$ which are defined over a quadratic field extension $\IQ(\alpha)$ \cite{Candelas:2019llw}: Assume that there is a quadratic equation 
\begin{align} 
p(\varphi) \defineas k_0 + k_1 \varphi + k_2 \varphi^2 = 0 \mod p~,
\end{align}
corresponding to an attractor or attractive K3 point $\varphi \in \IQ(\alpha)$. This equation has exactly two roots in $\IF_p$ if the discriminant $\Delta \defineas k_1^2 - 4 k_0 k_2$ is a square $\!\!\! \mod p$, corresponding to the fact that then $\alpha$ has two representatives in $\IF_p$. However, if $\Delta = 0 \mod p$, then these representatives coincide and the equation has only one root. Finally, if $\Delta$ is not a square $\!\!\! \mod p$, then there are no solutions to the above equation. 

If we again ignore the singularities and the cases where the polynomial does not have any roots, to be consistent with the data displayed in \fref{fig:Factorisations_P7224}, we would need to have that $\Delta$ is divisible by all primes where there exists only a single factorisation, which would imply that $\Delta > 10^{50}$.

Furthermore, it is a well-known fact that if we consider a large number of primes, $p_\text{min} \leq p \leq p_{\text{max}}$ with $p_{\text{max}} \to \infty$, the ratio of primes where a quadratic equation has two roots to all primes approaches $1/2$.  However, in the data displayed in \fref{fig:Factorisations_P7224}, we only observe two or more factorisations at the rate of $\approx 37.7\%$. 

This last argument generalises for field extensions of degree $n$ \cite{Candelas:2019llw}: by the Chebotarëv density theorem a polynomial of degree $n$ has exactly $n$ roots in fields $\IF_p$ with the frequency given by the order of the Galois group of the polynomial. Since the Galois group is always a subgroup of the symmetric group $S_n$, we would expect to see $n$ factorisations with a frequency $\geq 1/n!$. For instance, since we observe 3 or more factorisations with frequency $5/66 < 1/9$, we have an indication that a persistent factorisation with $\varphi$ in a cubic field extension of $\IQ$ might not exist.

\subsection{Hodge type \texorpdfstring{$(1,1,2,1,1)$}{(1,1,2,1,1)}: the mirror of \texorpdfstring{$X_{1,4} \subset \text{Gr}(2,5)$}{a complete intersection in a Grassmannian variety}}
\vskip-10pt
Relatively simple example of manifold of type $(1,1,2,1,1)$ can be constructed as the mirror of a complete intersection of hypersurfaces in Grassmannian varieties. We consider the manifold $X_{1,4}$ defined as the intersection of a hypersurface of degree $1$ with another hypersurface of degree $4$ inside the Grassmannian variety $\text{Gr}(2,5)$. The mirror $X_{1,4}^\vee$ of this manifold is of Hodge type $(1,1,2,1,1)$. This geometry has been studied in context of mirror symmetry in refs. \cite{Honma:2013hma,Gerhardus:2016iot}, for instance.

The relevant topological data of the manifold $X_{1,4}$ is given by
\begin{align}
h^{1,1} \= 1~, \quad h^{2,1} \= 0~, \quad h^{2,2} \= 1244~, \quad h^{3,1} \= 299~, \quad \chi \= 1848~.
\end{align}
The periods of the manifold $X_{1,4}^\vee$ satisfy the Picard--Fuchs equation $\cL\varpi=0$ for
\begin{align}
\begin{split}
\cL \=& (\theta-1)\theta^5 - 8 \varphi (2\theta+1)(4 \theta+1)(4 \theta+3)(11\theta^2+11\theta+3)\theta \\&- 64 \varphi^2 (2 \theta+1)(2\theta+3)(4\theta+1)(4\theta+5)(4\theta+7)~.
\end{split}
\end{align}
This differential equation has six solutions, indicating that the manifold is of type $(1,1,2,1,1)$, so we take the basis of these periods to be
\begin{align}
(\varpi_0^1,\varpi_1,\varpi_2,\varpi_3,\varpi_4,\varpi_0^2)~,
\end{align}
where $\varpi_0^i$ are the holomorphic periods. Using these periods and their derivatives, we construct the period matrix $\mtE(\varphi)$ as explained in section \ref{sect:Deformation_method}. In this case, the matrix $\sigma$
representing the wedge product \eqref{eq:wedge_product} is 
\begin{align}
\sigma \= \frac{20}{(2\pi \ii)^4} \left(
\begin{array}{cccccc}
0 & 0 & 0 & 0 & 1 & 0 \\
0 & 0 & 0 & -1 & 0 & 0\\
0 & 0 & 1 & 0 & 0  & 0\\
0 & -1 & 0 & 0 & 0 & 0\\
1 & 0 & 0 & 0 & 0  & 0\\
0 & 0 & 0 & 0 & 0  & 576\\
\end{array}
\right).
\end{align}
Using this, we can compute the matrix $\mtW$, which we refrain from giving here due to its size. It suffices to note that the denominator of $\mtW^{-1}$ is $\varphi$. Therefore, in this case, the denominator of the matrix $\mtU_p(\varphi) \mod p^n$ is
\begin{align} \label{eq:U_denominator}
P_n(\varphi) \= \varphi^p \Delta(\varphi^p)^{n-5} \=  \varphi^p (1-2816\varphi^p-65536\varphi^{2p})^{n-5}~.
\end{align}
By requiring that the matrix $\mtU_p(\varphi) \mod p^n$ is a matrix of rational entries with this denominator, we can fix the constants $\alpha$ and $\gamma$ appearing in $\mtU_p(0)$. We find that for all primes $7 \leq p \leq 131$
\begin{align}
\alpha \= 0 + \cO(p^{13})~, \qquad \epsilon = 0 + \cO(p^{13})~.
\end{align}
The values $\gamma \! \mod p^{7}$ are tabulated for the first few primes $p$ in \tref{tab:gamma_Values_X14}.

\begin{table}[H]
	\renewcommand{\arraystretch}{1.3}
	\begin{center}
		        \begin{tabular}{|l|l||l|l||l|l|l|l|}
			\hline
			\hfil $p$ & \hfil $\gamma + \cO\left(p^{7}\right)$ & \hfil $p$ & \hfil $\gamma + \cO\left(p^{7}\right)$ & \hfil $p$ & \hfil $\gamma + \cO\left(p^{7}\right)$& \hfil $p$ & \hfil $\gamma + \cO\left(p^{7}\right)$\\ \hline \hline
			7 & 808794 & 23 & 366372704 & 43 & 155628750954 & 67 & 1195806619330 \\ \hline 
			11 & 14509231 & 29 & 7433645255 & 47 & 304410385699 & 71 & 2840782556320 \\ \hline 
			13 & 56460703& 31 & 24965870894 & 53 & 643696656237 & 73 & 7434442033297 \\ \hline 
			17 & 83098482 & 37 & 3385241296 & 59 & 1590159024861 & 79 & 13071775373740 \\ \hline 
    		19 & 579825565 & 41 & 138567186762 & 61 & 1502565873047 & 83 & 3851767077829 \\ \hline 
		\end{tabular}
		\vskip10pt
		\capt{5.3in}{tab:gamma_Values_X14}{The values of the prime-dependent constants $\gamma \mod p^{7}$ for the first $20$ primes for the mirror $X_{1,4}^\vee$ of the complete intersection $X_{1,4}$ in the Grassmannian variety $\text{Gr}(2,5)$.}		
	\end{center}
\end{table}
With this data, we are able to find the polynomials $R_H^{(p)}(X_{1,4}^\vee,T)$ and study their factorisation properties. In this case, we do see a frequent appearance of a quadratic factor. However, almost all of these cases seem to follow directly from the Weil conjectures in a manner discussed in appendix \ref{sect:R_H_(1,1,2,1,1)}. Namely, in these cases the quadratic factor factorises further into two linear factors so that $R_H^{(p)}(X_{1,4}^\vee,T)$ takes on the form
\begin{align} \label{eq:R_H_factorisation_(1,1,2,1,1)}
R_H^{(p)}(X_{1,4}^\vee,T) \= (1-p^2 T)(1+p^2 T) R_4(X_{1,4}^\vee,T)~,
\end{align}
where $R_4(X_{1,4}^\vee,T)$ is a polynomial of degree 4 that is generically irreducible, and the functional equation \eqref{eq:functional_eq_R_H} is satisfied with $\epsilon = -1$. As such we do not expect these factorisations to correspond to splitting of the Hodge structure, which is indeed what we observe. Therefore, to search for \AK{} points, we look for additional quadratic factorisations apart from these linear ones. Tabulating the number of such factorisations in figure \ref{fig:Factorisations_X14}, we see that it is unlikely for any attractor of \AK{} points to exist.
\begin{figure}[H]
	\centering
	\begin{center}
        \includegraphics[scale=0.9]{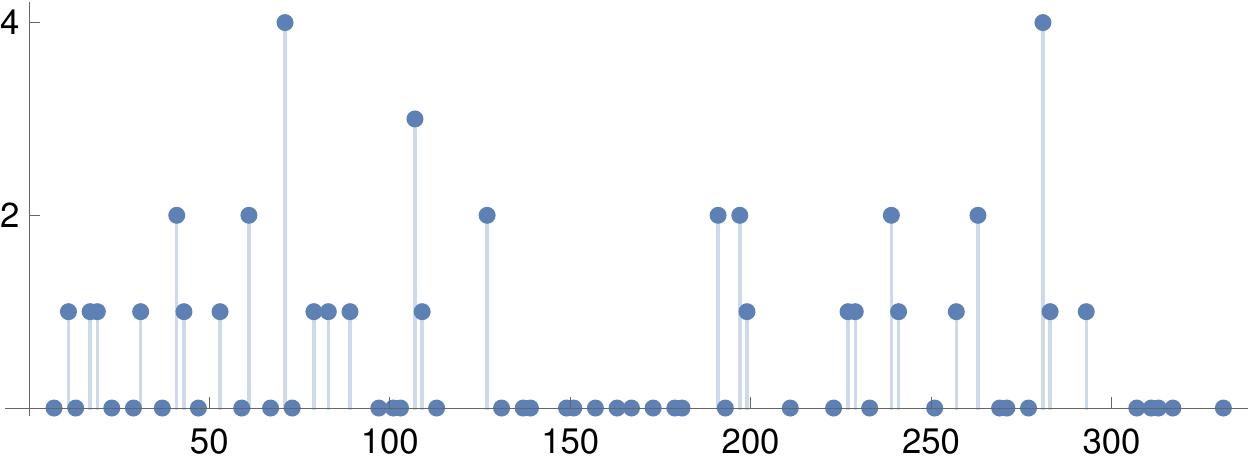}
	\end{center}
	\vskip-10pt
	\capt{6in}{fig:Factorisations_X14}{The number of factorisations where the polynomial $R_H^{p}(X_{1,4}^\vee,T)$ has a quadratic factor that is not a product of two linear factors that always appear when the functional equation \eqref{eq:Weil_conjectures_functional_equation} is satisfied with the negative sign. The vertical axis gives the number of such factorisations for the prime indicated by the horizontal axis. Note that numerous primes where no factorisation exist indicate that there are no persistent factorisation, and therefore no rank-two attractor or \AK{} points.}	
\end{figure}

\subsection{Additional example: the mirror of the sextic fourfold \texorpdfstring{$\IP^5[6]$}{}}
\vskip-10pt
We end the this appendix with an example where frequent quadratic factorisations do occur, but where we have not found any persistent factorisations, leaving open the question whether the example has attractor or \AK{} points outside of the Fermat point which cannot be studied using the choice of coordinates made here. The modularity of Fermat sextic has been studied in physics context for instance in ref. \cite{Schimmrigk:2008mp}.

The mirror family of sextic fourfolds can be realised as the vanishing locus of the polynomial
\begin{align}
\sum_{i=1}^6 X_i^6 - \varphi^{-6} \prod_{i=1}^6 X_i \= 0~.
\end{align}
Its non-trivial Hodge numbers and its Euler characteristic are given by
\begin{align}
h^{1,1} \= 1~, \quad h^{2,1} \= 0~, \quad h^{2,2} \=1752 ~, \quad h^{3,1} \=426 ~, \quad \chi \= 2610~.
\end{align}
The Picard--Fuchs operator can then be easily derived for example by using the Griffiths-Dwork reduction and is given by
\begin{align} \label{eq:PF_equation_sextic}
\cL \= \theta^5 - 6 \varphi \prod_{i=1}^5 (6 \theta+i) \defineas \sum_{i=0}^5 S_i \,  \theta^i ~.
\end{align}
From this we see that the manifold is of type $(1,1,1,1,1)$. The fundamental period $\varpi^0$ given by
\begin{align}
\varpi^0 \= \sum_{n=0}^\infty \frac{(6n)!}{(n!)^6} \, \varphi^n~.
\end{align}
In analogy to \ref{sect:P7224}, we can use the period vectors to form the logarithm-free period matrices $\wt \mtE$ defined in eq. \eqref{eq:E_matrix_definition} and the product matrix $\mtW = \wt \mtE^T \sigma \wt \mtE$ as given in eq. \eqref{eq:W_matrix_definition}, whose inverse is given by
\begin{align}
\mtW^{-1} \= \left(
\begin{array}{rrrrr}
-1440 \varphi  & -8424 \varphi  & -40176 \varphi  & -69984 \varphi  & 1-46656
\varphi  \\
-8424 \varphi  & 16848 \varphi  & 23328 \varphi  & 46656 \varphi -1 & 0 \\
-40176 \varphi  & 23328 \varphi  & 1-46656 \varphi  & 0 & 0 \\
-69984 \varphi  & 46656 \varphi -1 & 0 & 0 & 0 \\
1-46656 \varphi  & 0 & 0 & 0 & 0 \\
\end{array}
\right).
\end{align}
We have $\cW= 1$ for the mirror sextic fourfold, so the denominator to the $p$-adic accuracy $n$ is given by
\begin{align}
P_n(\varphi^p) \= \Delta(\varphi^p)^{n-5} \= (1-6^6 \varphi^p)^{n-5} \ .
\end{align}
Requiring this form allows us to solve for the constants $\alpha$ and $\gamma$. We find that for all primes $7 \leq p \leq 317$
\begin{align}
\alpha \= 0 + \cO(p^{13})~.
\end{align}
We tabulate the values of $\gamma$ for the first few primes below in \tref{tab:gamma_Values_Sextic}, although we have computed these up to $p=317$ and more are easily computable.
\begin{table}[H]
	\renewcommand{\arraystretch}{1.3}
	\begin{center}
        \begin{tabular}{|l|l||l|l||l|l|l|l|l|l|}
			\hline
			\hfil $p$ & \hfil $\gamma + \cO\left(p^{7}\right)$ & \hfil $p$ & \hfil $\gamma + \cO\left(p^{7}\right)$ & \hfil $p$ & \hfil $\gamma + \cO\left(p^{7}\right)$& \hfil $p$ & \hfil $\gamma + \cO\left(p^{7}\right)$& \hfil $p$ & \hfil $\gamma + \cO\left(p^{7}\right)$\\ \hline \hline
			7 & 177674 & 23 & 16413787 & 43 & 40547832 & 67 & 59855068 & 89 & 14407046764\\ \hline 
			11 & 3648271 & 29 & 21240596 & 47 & 45374641 & 71 & 115190198 & 97 & 2141051657\\ \hline 
			13 & 1933360 & 31 & 26067405 & 53 & 50201450 & 73 & 57156047 & 101 & 175321584405\\ \hline 
			17 & 6760169 & 37 & 30894214 & 59 & 55028259 & 79 & 1475260917 & 103 & 248074561120\\ \hline 
    		19 & 11586978 & 41 & 35721023 & 61 & 59855068 & 83 & 4834460747 & 107 & 47445553655\\ \hline 
		\end{tabular}
	\vskip10pt
	\capt{\textwidth}{tab:gamma_Values_Sextic}{The values of the prime-dependent constants $\gamma \! \mod p^{7}$ for the first $25$ primes for the mirror of the sextic fourfold.}		
	\end{center}
	\vskip-30pt
\end{table}
As for the previous examples, we compute from these the matrix $\mtU_p(\varphi)$ to necessary $p$-adic accuracy and its characteristic polynomial $R_H^{(p)}(\IP^5[6]^\vee,T)$. Figure \ref{fig:Factorisations_Sextic} again shows the number of factorisations for all primes $7\leq p \leq 317$ which indicates at least one factorisation for all primes. From our previous discussion, this could be seen a sign of a persistent factorisation for a mirror sextic fourfold. However, searching for persistent factorisations as described in section \ref{sect:persistent_factorisation} does not give any solutions. Thus, for this family of fourfolds, the question whether a point of persistent factorisation exists or not remains unanswered outside of the Fermat point $\varphi=\infty$.
\begin{figure}[H]
	\centering
	\begin{center}
		\includegraphics[scale=0.9]{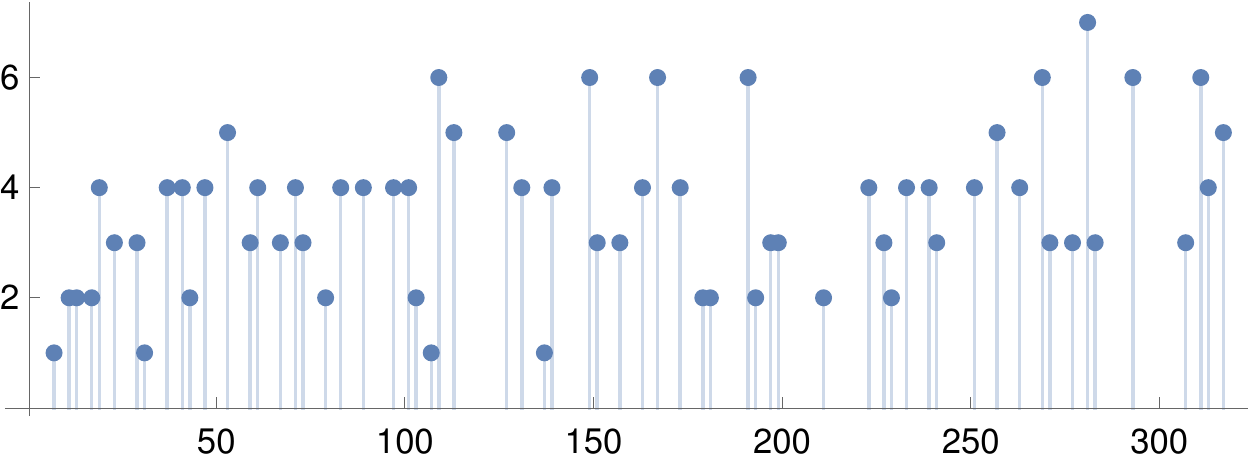}
	\end{center}
	\vskip-10pt
	\capt{6in}{fig:Factorisations_Sextic}{The number of factorisations where the polynomial $R_H^{p}(\IP^5[6]^\vee,T)$ factorises into a linear factor and two quadratic polynomials. The vertical axis gives the number of such factorisations for the prime indicated by the horizontal axis.}	
\end{figure}

\newpage

\section{Some Details on Periods and Motives} \label{app:periods_motives}
\vskip-10pt
In this appendix, we briefly review some elementary aspects of the theory of periods and give additional details on the construction of Deligne's periods. We will only focus on aspects relevant to the present work, and as such we do not consider the most general or rigorous definitions. Instead, we refer the reader to the excellent article of Kontsevich and Zagier \cite{Kontsevich2001a}, and a textbook treatment of ref. \cite{Huber2017a}. 

\subsection{Periods}
\vskip-10pt
Kontsevich and Zagier \cite{Kontsevich2001a} define (``naïve'') \textit{periods} as complex numbers whose real and imaginary parts are absolutely convergent integrals of rational differential forms over domains defined by polynomial inequalities with rational coefficients. This definition can be rephrased in terms of integrals of differential forms: periods are integrals
\begin{align}
I \= \int_\gamma \omega~,
\end{align}
where $\omega$ is a closed algebraic $n$-form on an quasi-projective variety $X$ vanishing on a subvariety $Y \subset X$, and $\gamma$ is a singular $n$-chain on $X(\IC)$ with boundary contained in $Y(\IC)$. In addition $\omega$ and $Y$ are taken to be defined over $\overline{\IQ}$.

To illustrate the above definition and to show the connection to the computation done in section \ref{sect:Delignes_periods}, we consider the simple example of the period $2\pi \ii$ in detail following ref. \cite{Brown2016a}. That this is a (``naïve'') period is immediately clear from the integral representation
\begin{align}
2 \pi \ii \= \oint \frac{\dd z}{z}~.
\end{align}
However, we can also express this as more formally in terms of algebraic forms. Consider the affine variety $\IC^*$, the complex plane with the origin removed. The singular homology $H_1^{\text{sing.}}(\IC^*,\IQ)$, that is the dual of the Betti cohomology, is one-dimensional. It is generated by the class of the cycle $\gamma$ corresponding to a circle around the origin:
\begin{align}
H_\BB^1(\IC^*,\IQ) \= \IQ [\gamma]~.
\end{align}
The algebraic de Rham cohomology group corresponds to the complex
\begin{align}
0 \longrightarrow \IQ\left[z,z^{-1}\right] \longrightarrow \IQ\left[z,z^{-1}\right] \dd z \longrightarrow 0~.
\end{align}
Since $\dd z /z = \dd \log z$, and $\log z \notin \IQ\left[z,z^{-1}\right]$, it follows that
\begin{align}
H_\dR^1(\IC^*, \IQ) \= \IQ \left[ \frac{\dd z}{z}\right]~.
\end{align}
Then one can express the period $2\pi \ii$ as
\begin{align} \label{eq:period_form}
2 \pi \ii \= \int_\gamma \frac{\dd z}{z}~. 
\end{align}
To connect this all to the discussion in section \ref{sect:Delignes_periods}, we note that the information contained in the integral can be also expressed in terms of the comparison isomorphism. Namely, the comparison isomorphism gives a map
\begin{align}
I_\infty: \; H^1_{\dR}(X,\IQ) \otimes \IC \to H^1_\BB(X,\IQ) \otimes \IC~,
\end{align}
defined so that if $\omega \in H^1_\dR(X,\IQ)$, then $I_\infty(\omega)$ is the map $f \in \text{Hom}(H_1^{\text{sing.}}(X,\IQ) \otimes \IC,\IC) = H_1^{\text{sing.}}(X,\IQ)^\vee = H^1_\BB(X,\IQ) \otimes \IC$ defined by
\begin{align}
f(\gamma) \= \int_\gamma \omega~,
\end{align}
for any $\gamma \in H_1^{\text{sing.}}(X,\IQ)$. In the present case, given \eqref{eq:period_form}, under the comparison isomorphism 
\begin{align}
\left[ \frac{\dd z}{z}\right] \otimes 1 \mapsto \left[ \gamma^\vee \right] \otimes 2\pi \ii~.
\end{align}
In this way, we can view $H_\BB^1(\IC^*,\IQ) \otimes \IC $ as having two $\IQ$-structures, one being the canonical $\IQ$-structure of $H_\BB^1(\IC^*,\IQ)$, and the other induced from the $\IQ$-structure of $H^1_\dR(\IC^*,\IQ)$ via the comparison isomorphism. The `difference' between these is the period $2\pi\ii$. Essentially the same logic is used in section \ref{sect:Delignes_periods} to give Deligne's periods in terms of the mirror map.

\subsection{The Tate motive}
\vskip-10pt
As the Tate motive $\IQ(1)$ plays a central role in this work, we gather its realisations and the most important properties below. The three canonical realisations are the following:\cite{Yang:2020lhd}
\begin{enumerate}
	\item The Betti realisation $\IQ(1)_\BB$ is given by $(2 \pi \ii) \IQ$. This has a natural Hodge structure of type $(-1,-1)$.  
	\item The de Rham realisation is simply given by $\IQ(1)_\dR = \IQ$. It has a natural Hodge filtration given by $F^0 \IQ(1)_\dR = 0$, $F^{-1} \IQ(1)_\dR = \IC$.
	\item The $\ell$-adic realisation $\IQ(1)_\ell$ is given by the inverse limit $\mathop{\lim_{\longleftarrow}}_{n} \mu_{\ell^n}(\overline{\IQ}) \otimes_{\IZ_\ell} \IQ_\ell$, where $\mu_{\ell^n}$ denotes the $\ell^n$'th root of unity.
\end{enumerate}
While it may at first seem strange to have a Hodge structure of type $(-1,-1)$, this is in fact completely natural in the setting of abstract theory of Hodge structures (for a review, see for instance ref. \cite{Filippini2015a}). In this context, a \text{(pure, rational) Hodge structure} of weight $n$ is defined as a vector space $H_\IQ$ together with a decomposition of the complexification $H_\IQ \otimes \IC$:
\begin{align}
H_\IQ \otimes \IC \= \bigoplus_{p+q=n} H^{p,q}~, \qquad \text{with} \qquad \overline{H^{p,q}} \= H^{q,p} \subseteq H_\IQ \otimes \IC~.
\end{align}
It is easy to verify that if we take $H_\IQ = (2\pi \ii) \IQ$ and $H_\IQ \otimes \IC = \IC = H^{-1,-1}$, this is a pure rational Hodge structure of weight $-2$. In fact, this is the unique Hodge structure of weight $w=-2$, up to an isomorphism.

The pure Hodge structure of weight $n$ can be equivalently defined in terms of \textit{the Hodge filtration}. This is defined as a decreasing filtration $H_\IQ \otimes \IC = F^0 H \supset F^1 H \supset \cdots \supset F^n H = \{0\}$ on $H_\IQ \otimes \IC$ such that $F^p H \oplus \overline{F^{n-p+1} H } =H_\IQ \otimes \IC$. It is easy to check that the Hodge filtration given above for $\IQ(1)_\dR$ gives a pure Hodge structure of weight $-2$. 

The central property of the $\ell$-adic realisation of the Tate motive we have used in the discussion in section \ref{sect:Delignes_periods} is that tensoring a motive $M$ by $\IQ(1)_\ell$ rescales the formal variable $T$ appearing in the characteristic polynomial $R^{(p)}(M\otimes\IQ(1)_\ell,T)$ by $T \mapsto p^{-1} T$. While the full explanation of this property is beyond the scope of the present manuscript, roughly speaking this is due to the fact that the Frobenius map has a natural action on $\IQ(1)_\ell$ acting by multiplication by $p$. This corresponds to multiplying the matrix $\mtU_p(X)$ by $p^{-1}$, or equivalently rescaling $T \mapsto p^{-1} T$. This also explains the shift of the argument of the $L$-function, $L(M\otimes\IQ(1)_\ell,s) \mapsto L(M,s+1)$.
\subsection{Critical motives}
\vskip-10pt
Given the pure motive $M$ discussed in section \ref{sect:Delignes_periods}, one can consider its étale realisation, which is the étale cohomology group $H^4_\et(X,\IQ_\ell)$. This furnishes a representation of the absolute Galois group $\text{Gal}(\overline{\IQ},\IQ)$, and in particular has a well defined action of the Frobenius map $\Fr_p^{-1}$.\footnote{Here we are ignoring the issue of ramification, which would introduce a slight subtlety in the definition of the motivic $L$-function, see for example refs.~\cite{Yang:2020sfu,Frenkel:2005pa}.} We can then define the characteristic polynomials
\begin{align}
P_4^{(p)}(X,T) \= \det \left( \II - T \Fr_p^{-1} | H^4_\et(X,\IQ_\ell) \right)~,
\end{align}
which are essentially just the polynomials $R_4^{(p)}(X,T)$ appearing in eq. \eqref{eq:P_i_definition}, apart from the cases where $p$ is a bad prime. The $L$-function associated to the motive $M$ is then given by
\begin{align}
L(M,s) \= \prod_{p} \frac{1}{P_4^{(p)}(X,p^{-s})}~.
\end{align}
It is also natural to define a factor $L_\infty(M,s)$ corresponding to the `infinite' (or Archimedean) prime. To form this factor, let us first define the gamma factor 
\begin{align}
\Gamma_\IC(s) \defineas 2(2\pi)^{-s} \Gamma(s)~,
\end{align}
where $\Gamma(s)$ is the gamma function. We only consider the cases where the subspace $H^{w/2}$ of the Hodge decomposition of $H_\BB^4(X,\IQ) \otimes_\IQ \IC$ is trivial, where $w$ is the weight of the Hodge structure on $H_\BB^4(X,\IQ)\otimes_\IQ \IC$. This is in particular the case for the (conjectural) motive $M^\AK{}$ and the motive associated to the transcendental lattice $T(S)$ of an attractive K3~surface, which are the only two cases we study in the main text. For such a motive, the L-factor $L_\infty(M,s)$ is defined by
\begin{align}
L_\infty(M,s) \= \prod_{p < q} \Gamma_\IC(s - p)^{h^{p,q}(H_\BB^4(X,\IQ))}~.
\end{align}
Criticality of the motive $M$ is defined as a condition on the function $L_\infty(M,s)$. To state this condition, we still need to define the \textit{dual motive} of $M^\vee$, which can be defined via a Tate twist
\begin{align}
M^\vee \defineas M \otimes (2\pi \ii)^w \IQ~.
\end{align}
The motive $M$ is said to be critical if neither $L_\infty(M,s)$ nor $L_\infty(M^\vee,1-s)$ has a pole at $s=0$. In general, an integer $n$ such that neither of these function has a pole at $s=n$, is called \textit{critical}. 

Recalling that the Tate twist by $(2\pi\ii)^w \IQ$ has the effect of changing the Hodge type of the motive $M$ by $(-w,-w)$, a simple computation shows that 
\begin{align}
L_\infty(M^\vee,1-s) \= L_\infty(M,1-s+w)~.
\end{align}
As an example, let us consider a motive $M^\AK{}$ of Hodge type $(3,1)+(1,3)$: a straightforward computation gives us
\begin{align}
\begin{split}
L_\infty(M^\AK{},s)\phantom{(1)} &\propto \Gamma(s-1)~, \\
L_\infty(M^\AK{}(1),s) &\propto \Gamma(s)~, \\
L_\infty(M^\AK{}(2),s) &\propto \Gamma(s+1)~,\\
&\hskip-20pt\dots
\end{split}
\qquad
\begin{split}
L_\infty((M^\AK{})^\vee,1-s) \phantom{(1)}&\propto \Gamma(4-s)~,\\
L_\infty((M^\AK{}(1))^\vee,1-,s) &\propto \Gamma(3-s)~,\\
L_\infty((M^\AK{}(2))^\vee,1-,s) &\propto \Gamma(2-s)~,\\
&\hskip-20pt\dots
\end{split}
\end{align}
This shows that, as claimed in the main text, the motives $M^{\AK{}}(2)$ and $M^{\AK{}}(3)$ of Hodge types $(2,0)+(0,2)$ and $(1,-1)+(-1,1)$ are critical, whereas $M^{\AK{}}$ itself, nor its other Tate twists are.
\newpage
\bibliographystyle{JHEP}
\bibliography{Fourfold_Zeta}

\providecommand{\href}[2]{#2}\begingroup\raggedright\begin{thebibliography}{100}

\bibitem{Moore:1998pn}
G.~W. Moore, \emph{{Arithmetic and attractors}},
  \href{https://arxiv.org/abs/hep-th/9807087}{{\ttfamily hep-th/9807087}}.

\bibitem{Moore:1998zu}
G.~W. Moore, \emph{{Attractors and arithmetic}},
  \href{https://arxiv.org/abs/hep-th/9807056}{{\ttfamily hep-th/9807056}}.

\bibitem{Candelas:2007mb}
P.~Candelas and X.~de~la Ossa, \emph{{The Zeta-Function of a p-Adic Manifold,
  Dwork Theory for Physicists}},
  \href{https://doi.org/10.4310/CNTP.2007.v1.n3.a2}{\emph{Commun. Num. Theor.
  Phys.} {\bfseries 1} (2007) 479}
  [\href{https://arxiv.org/abs/0705.2056}{{\ttfamily 0705.2056}}].

\bibitem{Candelas:2019llw}
P.~Candelas, X.~de~la Ossa, M.~Elmi and D.~Van~Straten, \emph{{A One Parameter
  Family of Calabi-Yau Manifolds with Attractor Points of Rank Two}},
  \href{https://doi.org/10.1007/JHEP10(2020)202}{\emph{JHEP} {\bfseries 10}
  (2020) 202} [\href{https://arxiv.org/abs/1912.06146}{{\ttfamily
  1912.06146}}].

\bibitem{Kachru:2020abh}
S.~Kachru, R.~Nally and W.~Yang, \emph{{Flux Modularity, F-Theory, and Rational
  Models}},  \href{https://arxiv.org/abs/2010.07285}{{\ttfamily 2010.07285}}.

\bibitem{Candelas:2021tqt}
P.~Candelas, X.~De~La~Ossa and D.~Van~Straten, \emph{{Local Zeta Functions From
  Calabi-Yau Differential Equations}},
  \href{https://arxiv.org/abs/2104.07816}{{\ttfamily 2104.07816}}.

\bibitem{Bonisch:2022slo}
K.~B\"onisch, M.~Elmi, A.-K. Kashani-Poor and A.~Klemm, \emph{{Time reversal
  and CP invariance in Calabi-Yau compactifications}},
  \href{https://doi.org/10.1007/JHEP09(2022)019}{\emph{JHEP} {\bfseries 09}
  (2022) 019} [\href{https://arxiv.org/abs/2204.06506}{{\ttfamily
  2204.06506}}].

\bibitem{Yui2003a}
N.~Yui, \emph{Update on the modularity of {C}alabi-{Y}au varieties},  in
  \emph{Calabi-{Y}au varieties and mirror symmetry ({T}oronto, {ON}, 2001)},
  vol.~38 of \emph{Fields Inst. Commun.}, pp.~307--362.
\newblock Amer. Math. Soc., Providence, RI, 2003.
\newblock \href{https://doi.org/10.1090/fic/038/17}{DOI}.

\bibitem{Yui2012a}
N.~Yui, \emph{Modularity of calabi--yau varieties: 2011 and beyond},
  \href{https://arxiv.org/abs/1212.4308}{{\ttfamily 1212.4308}}.

\bibitem{Candelas:2023yrg}
P.~Candelas, X.~de~la Ossa, P.~Kuusela and J.~McGovern, \emph{{Flux vacua and
  modularity for $\mathbb{Z}_2$ symmetric Calabi-Yau manifolds}},
  \href{https://doi.org/10.21468/SciPostPhys.15.4.146}{\emph{SciPost Phys.}
  {\bfseries 15} (2023) 146}
  [\href{https://arxiv.org/abs/2302.03047}{{\ttfamily 2302.03047}}].

\bibitem{Schimmrigk:2020dfl}
R.~Schimmrigk, \emph{{On flux vacua and modularity}},
  \href{https://doi.org/10.1007/JHEP09(2020)061}{\emph{JHEP} {\bfseries 09}
  (2020) 061} [\href{https://arxiv.org/abs/2003.01056}{{\ttfamily
  2003.01056}}].

\bibitem{Kachru:2020sio}
S.~Kachru, R.~Nally and W.~Yang, \emph{{Supersymmetric Flux Compactifications
  and Calabi-Yau Modularity}},
  \href{https://arxiv.org/abs/2001.06022}{{\ttfamily 2001.06022}}.

\bibitem{Gukov:1999ya}
S.~Gukov, C.~Vafa and E.~Witten, \emph{{CFT's from Calabi-Yau four folds}},
  \href{https://doi.org/10.1016/S0550-3213(00)00373-4}{\emph{Nucl. Phys. B}
  {\bfseries 584} (2000) 69}
  [\href{https://arxiv.org/abs/hep-th/9906070}{{\ttfamily hep-th/9906070}}].

\bibitem{Taylor:1999ii}
T.~R. Taylor and C.~Vafa, \emph{{RR flux on Calabi-Yau and partial
  supersymmetry breaking}},
  \href{https://doi.org/10.1016/S0370-2693(00)00005-8}{\emph{Phys. Lett. B}
  {\bfseries 474} (2000) 130}
  [\href{https://arxiv.org/abs/hep-th/9912152}{{\ttfamily hep-th/9912152}}].

\bibitem{Haack:2001jz}
M.~Haack and J.~Louis, \emph{{M theory compactified on Calabi-Yau fourfolds
  with background flux}},
  \href{https://doi.org/10.1016/S0370-2693(01)00464-6}{\emph{Phys. Lett. B}
  {\bfseries 507} (2001) 296}
  [\href{https://arxiv.org/abs/hep-th/0103068}{{\ttfamily hep-th/0103068}}].

\bibitem{Berg:2002es}
M.~Berg, M.~Haack and H.~Samtleben, \emph{{Calabi-Yau fourfolds with flux and
  supersymmetry breaking}},
  \href{https://doi.org/10.1088/1126-6708/2003/04/046}{\emph{JHEP} {\bfseries
  04} (2003) 046} [\href{https://arxiv.org/abs/hep-th/0212255}{{\ttfamily
  hep-th/0212255}}].

\bibitem{Witten:1996md}
E.~Witten, \emph{{On flux quantization in M theory and the effective action}},
  \href{https://doi.org/10.1016/S0393-0440(96)00042-3}{\emph{J. Geom. Phys.}
  {\bfseries 22} (1997) 1}
  [\href{https://arxiv.org/abs/hep-th/9609122}{{\ttfamily hep-th/9609122}}].

\bibitem{deWit:2003ja}
B.~de~Wit, I.~Herger and H.~Samtleben, \emph{{Gauged locally supersymmetric D =
  3 nonlinear sigma models}},
  \href{https://doi.org/10.1016/j.nuclphysb.2003.08.022}{\emph{Nucl. Phys. B}
  {\bfseries 671} (2003) 175}
  [\href{https://arxiv.org/abs/hep-th/0307006}{{\ttfamily hep-th/0307006}}].

\bibitem{Becker:1996gj}
K.~Becker and M.~Becker, \emph{{M theory on eight manifolds}},
  \href{https://doi.org/10.1016/0550-3213(96)00367-7}{\emph{Nucl. Phys. B}
  {\bfseries 477} (1996) 155}
  [\href{https://arxiv.org/abs/hep-th/9605053}{{\ttfamily hep-th/9605053}}].

\bibitem{Weil1949a}
A.~Weil, \emph{Numbers of solutions of equations in finite fields},
  \href{https://doi.org/10.1090/S0002-9904-1949-09219-4}{\emph{Bull. Amer.
  Math. Soc.} {\bfseries 55} (1949) 497}.

\bibitem{Dwork1960a}
B.~Dwork, \emph{On the rationality of the zeta function of an algebraic
  variety}, \href{https://doi.org/10.2307/2372974}{\emph{Amer. J. Math.}
  {\bfseries 82} (1960) 631}.

\bibitem{Grothendieck1995a}
A.~Grothendieck, \emph{Formule de {L}efschetz et rationalit\'{e} des fonctions
  {$L$}},  in \emph{S\'{e}minaire {B}ourbaki, {V}ol. 9}, pp.~Exp. No. 279,
  41--55.
\newblock Soc. Math. France, Paris, 1995.

\bibitem{Deligne1974a}
P.~Deligne, \emph{La conjecture de {W}eil. {I}}, {\emph{Inst. Hautes \'{E}tudes
  Sci. Publ. Math.} (1974) 273}.

\bibitem{Deligne1980a}
P.~Deligne, \emph{La conjecture de {W}eil. {II}}, {\emph{Inst. Hautes
  \'{E}tudes Sci. Publ. Math.} (1980) 137}.

\bibitem{multiparameter_zeta}
P.~Candelas, X.~de~la Ossa and P.~Kuusela, \emph{{The Local Zeta Functions of
  Multiparameter Calabi-Yau Threefolds from Picard-Fuchs Equations}},
  {\emph{{\rm To Appear}} (2023) }.

\bibitem{Honma:2013hma}
Y.~Honma and M.~Manabe, \emph{{Exact Kahler Potential for Calabi-Yau
  Fourfolds}}, \href{https://doi.org/10.1007/JHEP05(2013)102}{\emph{JHEP}
  {\bfseries 05} (2013) 102} [\href{https://arxiv.org/abs/1302.3760}{{\ttfamily
  1302.3760}}].

\bibitem{Gerhardus:2015sla}
A.~Gerhardus and H.~Jockers, \emph{{Dual pairs of gauged linear sigma models
  and derived equivalences of Calabi\textendash{}Yau threefolds}},
  \href{https://doi.org/10.1016/j.geomphys.2016.12.005}{\emph{J. Geom. Phys.}
  {\bfseries 114} (2017) 223}
  [\href{https://arxiv.org/abs/1505.00099}{{\ttfamily 1505.00099}}].

\bibitem{Gerhardus:2016iot}
A.~Gerhardus and H.~Jockers, \emph{{Quantum periods of Calabi\textendash{}Yau
  fourfolds}},
  \href{https://doi.org/10.1016/j.nuclphysb.2016.09.021}{\emph{Nucl. Phys. B}
  {\bfseries 913} (2016) 425}
  [\href{https://arxiv.org/abs/1604.05325}{{\ttfamily 1604.05325}}].

\bibitem{Hulek2005a}
K.~Hulek and H.~Verrill, \emph{On modularity of rigid and nonrigid
  {C}alabi-{Y}au varieties associated to the root lattice {$A_4$}},
  \href{https://doi.org/10.1017/S0027763000025617}{\emph{Nagoya Math. J.}
  {\bfseries 179} (2005) 103}
  [\href{https://arxiv.org/abs/math/0304169}{{\ttfamily math/0304169}}].

\bibitem{Candelas:2021lkc}
P.~Candelas, X.~de~la Ossa, P.~Kuusela and J.~McGovern, \emph{{Mirror symmetry
  for five-parameter Hulek-Verrill manifolds}},
  \href{https://doi.org/10.21468/SciPostPhys.15.4.144}{\emph{SciPost Phys.}
  {\bfseries 15} (2023) 144}
  [\href{https://arxiv.org/abs/2111.02440}{{\ttfamily 2111.02440}}].

\bibitem{Serre1975a}
J.-P. Serre, \emph{Valeurs propres des op\'{e}rateurs de {H}ecke modulo {$l$}},
   in \emph{Journ\'{e}es {A}rithm\'{e}tiques de {B}ordeaux ({C}onf., {U}niv.
  {B}ordeaux, {B}ordeaux, 1974)}, Ast\'{e}risque, Nos. 24-25, pp.~109--117.
\newblock Soc. Math. France, Paris, 1975.

\bibitem{Serre1987a}
J.-P. Serre, \emph{Sur les repr\'{e}sentations modulaires de degr\'{e} {$2$} de
  {${\rm Gal}(\overline{\mathbf{Q}}/\mathbf{Q})$}},
  \href{https://doi.org/10.1215/S0012-7094-87-05413-5}{\emph{Duke Math. J.}
  {\bfseries 54} (1987) 179}.

\bibitem{Deligne1979a}
P.~Deligne, \emph{Valeurs de fonctions {$L$} et p\'{e}riodes d'int\'{e}grales},
   in \emph{Automorphic forms, representations and {$L$}-functions ({P}roc.
  {S}ympos. {P}ure {M}ath., {O}regon {S}tate {U}niv., {C}orvallis, {O}re.,
  1977), {P}art 2}, vol.~XXXIII of \emph{Proc. Sympos. Pure Math.},
  pp.~313--346.
\newblock Amer. Math. Soc., Providence, RI, 1979.

\bibitem{Bonisch:2022mgw}
K.~B\"onisch, A.~Klemm, E.~Scheidegger and D.~Zagier, \emph{{D-brane masses at
  special fibres of hypergeometric families of Calabi-Yau threefolds, modular
  forms, and periods}},  \href{https://arxiv.org/abs/2203.09426}{{\ttfamily
  2203.09426}}.

\bibitem{Yang:2019kib}
W.~Yang, \emph{{Periods of CY $n$-folds and mixed Tate motives, a numerical
  study}},  \href{https://arxiv.org/abs/1908.09965}{{\ttfamily 1908.09965}}.

\bibitem{Yang:2020lhd}
W.~Yang, \emph{{Rank-2 attractors and Deligne\textquoteright{}s conjecture}},
  \href{https://doi.org/10.1007/JHEP03(2021)150}{\emph{JHEP} {\bfseries 03}
  (2021) 150} [\href{https://arxiv.org/abs/2001.07211}{{\ttfamily
  2001.07211}}].

\bibitem{Yang:2020sfu}
W.~Yang, \emph{{Deligne's conjecture and mirror symmetry}},
  \href{https://doi.org/10.1016/j.nuclphysb.2020.115245}{\emph{Nucl. Phys. B}
  {\bfseries 962} (2021) 115245}
  [\href{https://arxiv.org/abs/2001.03283}{{\ttfamily 2001.03283}}].

\bibitem{Bonisch:2020qmm}
K.~B\"onisch, F.~Fischbach, A.~Klemm, C.~Nega and R.~Safari, \emph{{Analytic
  structure of all loop banana integrals}},
  \href{https://doi.org/10.1007/JHEP05(2021)066}{\emph{JHEP} {\bfseries 05}
  (2021) 066} [\href{https://arxiv.org/abs/2008.10574}{{\ttfamily
  2008.10574}}].

\bibitem{Kuusela:2022hga}
P.~Kuusela, \emph{{Modular Calabi-Yau manifolds, attractor points, and flux
  vacua}}, Ph.D. thesis, Oxford University, 2022.

\bibitem{vanderPut1986a}
M.~van~der Put, \emph{The cohomology of {M}onsky and {W}ashnitzer},  No.~23,
  pp.~4, 33--59.
\newblock 1986.

\bibitem{Intriligator:2012ue}
K.~Intriligator, H.~Jockers, P.~Mayr, D.~R. Morrison and M.~R. Plesser,
  \emph{{Conifold Transitions in M-theory on Calabi-Yau Fourfolds with
  Background Fluxes}},
  \href{https://doi.org/10.4310/ATMP.2013.v17.n3.a2}{\emph{Adv. Theor. Math.
  Phys.} {\bfseries 17} (2013) 601}
  [\href{https://arxiv.org/abs/1203.6662}{{\ttfamily 1203.6662}}].

\bibitem{Goresky2004a}
M.~Goresky, \emph{{Langlands' Conjectures for Physicists}}, {\emph{Lectures at
  IAS} (2004) }.

\bibitem{Dwork1962a}
B.~Dwork, \emph{On the zeta function of a hypersurface}, {\emph{Inst. Hautes
  \'{E}tudes Sci. Publ. Math.} (1962) 5}.

\bibitem{Dwork1964a}
B.~Dwork, \emph{On the zeta function of a hypersurface. {II}},
  \href{https://doi.org/10.2307/1970392}{\emph{Ann. of Math. (2)} {\bfseries
  80} (1964) 227}.

\bibitem{Lauder2004a}
A.~G.~B. Lauder, \emph{Deformation theory and the computation of zeta
  functions}, \href{https://doi.org/10.1112/S0024611503014461}{\emph{Proc.
  London Math. Soc. (3)} {\bfseries 88} (2004) 565}.

\bibitem{Lauder2004b}
A.~G.~B. Lauder, \emph{Counting solutions to equations in many variables over
  finite fields}, \href{https://doi.org/10.1007/s10208-003-0093-y}{\emph{Found.
  Comput. Math.} {\bfseries 4} (2004) 221}.

\bibitem{Candelas:2024vzf}
P.~Candelas, X.~de~la Ossa and P.~Kuusela, \emph{{Local Zeta Functions of
  Multiparameter Calabi-Yau Threefolds from the Picard-Fuchs Equations}},
  \href{https://arxiv.org/abs/2405.08067}{{\ttfamily 2405.08067}}.

\bibitem{Candelas:2000fq}
P.~Candelas, X.~de~la Ossa and F.~Rodriguez-Villegas, \emph{{Calabi-Yau
  manifolds over finite fields. 1.}},
  \href{https://arxiv.org/abs/hep-th/0012233}{{\ttfamily hep-th/0012233}}.

\bibitem{Koblitz1984a}
N.~Koblitz, \emph{{$p$}-adic numbers, {$p$}-adic analysis, and zeta-functions},
  vol.~58 of \emph{Graduate Texts in Mathematics}. Springer-Verlag, New York,
  second~ed., 1984,
  \href{https://doi.org/10.1007/978-1-4612-1112-9}{10.1007/978-1-4612-1112-9}.

\bibitem{Zhou:2017vhw}
Y.~Zhou, \emph{{Wick rotations, Eichler integrals, and multi-loop Feynman
  diagrams}}, \href{https://doi.org/10.4310/CNTP.2018.v12.n1.a5}{\emph{Commun.
  Num. Theor. Phys.} {\bfseries 12} (2018) 127}
  [\href{https://arxiv.org/abs/1706.08308}{{\ttfamily 1706.08308}}].

\bibitem{Borisov1993a}
L.~Borisov, \emph{Towards the mirror symmetry for calabi-yau complete
  intersections in gorenstein toric fano varieties},
  \href{https://arxiv.org/abs/alg-geom/9310001}{{\ttfamily alg-geom/9310001}}.

\bibitem{Batyrev:1994pg}
V.~V. Batyrev and L.~A. Borisov, \emph{{On Calabi-Yau complete intersections in
  toric varieties}},  \href{https://arxiv.org/abs/alg-geom/9412017}{{\ttfamily
  alg-geom/9412017}}.

\bibitem{Aspinwall:1994ev}
P.~S. Aspinwall, \emph{{Resolution of orbifold singularities in string
  theory}}, {\emph{AMS/IP Stud. Adv. Math.} {\bfseries 1} (1996) 355}
  [\href{https://arxiv.org/abs/hep-th/9403123}{{\ttfamily hep-th/9403123}}].

\bibitem{Strominger:1996it}
A.~Strominger, S.-T. Yau and E.~Zaslow, \emph{{Mirror symmetry is T duality}},
  \href{https://doi.org/10.1016/0550-3213(96)00434-8}{\emph{Nucl. Phys. B}
  {\bfseries 479} (1996) 243}
  [\href{https://arxiv.org/abs/hep-th/9606040}{{\ttfamily hep-th/9606040}}].

\bibitem{Libgober:1999aaa}
A.~Libgober, \emph{Chern classes and the periods of mirrors}, {\emph{Math. Res.
  Lett.} {\bfseries 6} (1999) 141}
  [\href{https://arxiv.org/abs/math.AG/9803119}{{\ttfamily math.AG/9803119}}].

\bibitem{Iritani:2007aaa}
H.~Iritani, \emph{Real and integral structures in quantum cohomology {I}:
  {T}oric orbifolds},  \href{https://arxiv.org/abs/0712.2204}{{\ttfamily
  0712.2204}}.

\bibitem{MR2553377}
H.~Iritani, \emph{An integral structure in quantum cohomology and mirror
  symmetry for toric orbifolds},
  \href{https://doi.org/10.1016/j.aim.2009.05.016}{\emph{Adv. Math.} {\bfseries
  222} (2009) 1016} [\href{https://arxiv.org/abs/0903.1463}{{\ttfamily
  0903.1463}}].

\bibitem{MR2483750}
L.~Katzarkov, M.~Kontsevich and T.~Pantev, \emph{Hodge theoretic aspects of
  mirror symmetry},  in \emph{From {H}odge theory to integrability and {TQFT}
  tt*-geometry}, vol.~78 of \emph{Proc. Sympos. Pure Math.}, pp.~87--174.
\newblock Amer. Math. Soc., Providence, RI, 2008.
\newblock \href{https://arxiv.org/abs/0806.0107}{{\ttfamily 0806.0107}}.

\bibitem{Halverson:2013qca}
J.~Halverson, H.~Jockers, J.~M. Lapan and D.~R. Morrison, \emph{{Perturbative
  Corrections to K\"ahler Moduli Spaces}},
  \href{https://doi.org/10.1007/s00220-014-2157-z}{\emph{Commun.Math.Phys.}
  {\bfseries 333} (2015) 1563}
  [\href{https://arxiv.org/abs/1308.2157}{{\ttfamily 1308.2157}}].

\bibitem{Hori:2013ika}
K.~Hori and M.~Romo, \emph{{Exact Results In Two-Dimensional (2,2)
  Supersymmetric Gauge Theories With Boundary}},
  \href{https://arxiv.org/abs/1308.2438}{{\ttfamily 1308.2438}}.

\bibitem{CaboBizet:2014ovf}
N.~Cabo~Bizet, A.~Klemm and D.~Vieira~Lopes, \emph{{Landscaping with fluxes and
  the E8 Yukawa Point in F-theory}},
  \href{https://arxiv.org/abs/1404.7645}{{\ttfamily 1404.7645}}.

\bibitem{Witten:1998cd}
E.~Witten, \emph{{D-branes and K-theory}},
  \href{https://doi.org/10.1088/1126-6708/1998/12/019}{\emph{JHEP} {\bfseries
  12} (1998) 019} [\href{https://arxiv.org/abs/hep-th/9810188}{{\ttfamily
  hep-th/9810188}}].

\bibitem{Diaconescu:1997br}
D.-E. Diaconescu, M.~R. Douglas and J.~Gomis, \emph{{Fractional branes and
  wrapped branes}},
  \href{https://doi.org/10.1088/1126-6708/1998/02/013}{\emph{JHEP} {\bfseries
  02} (1998) 013} [\href{https://arxiv.org/abs/hep-th/9712230}{{\ttfamily
  hep-th/9712230}}].

\bibitem{Diaconescu:1999dt}
D.-E. Diaconescu and J.~Gomis, \emph{{Fractional branes and boundary states in
  orbifold theories}},
  \href{https://doi.org/10.1088/1126-6708/2000/10/001}{\emph{JHEP} {\bfseries
  10} (2000) 001} [\href{https://arxiv.org/abs/hep-th/9906242}{{\ttfamily
  hep-th/9906242}}].

\bibitem{Brunner:2001eg}
I.~Brunner and J.~Distler, \emph{{Torsion D-branes in nongeometrical phases}},
  \href{https://doi.org/10.4310/ATMP.2001.v5.n2.a3}{\emph{Adv. Theor. Math.
  Phys.} {\bfseries 5} (2002) 265}
  [\href{https://arxiv.org/abs/hep-th/0102018}{{\ttfamily hep-th/0102018}}].

\bibitem{Candelas:2021mwz}
P.~Candelas, P.~Kuusela and J.~McGovern, \emph{{Attractors with large complex
  structure for one-parameter families of Calabi-Yau manifolds}},
  \href{https://doi.org/10.1007/JHEP11(2021)032}{\emph{JHEP} {\bfseries 11}
  (2021) 032} [\href{https://arxiv.org/abs/2104.02718}{{\ttfamily
  2104.02718}}].

\bibitem{Ferrara:1995ih}
S.~Ferrara, R.~Kallosh and A.~Strominger, \emph{{N=2 extremal black holes}},
  \href{https://doi.org/10.1103/PhysRevD.52.R5412}{\emph{Phys. Rev. D}
  {\bfseries 52} (1995) R5412}
  [\href{https://arxiv.org/abs/hep-th/9508072}{{\ttfamily hep-th/9508072}}].

\bibitem{Pioline:2006ni}
B.~Pioline, \emph{{Lectures on black holes, topological strings and quantum
  attractors}}, \href{https://doi.org/10.1088/0264-9381/23/21/S05}{\emph{Class.
  Quant. Grav.} {\bfseries 23} (2006) S981}
  [\href{https://arxiv.org/abs/hep-th/0607227}{{\ttfamily hep-th/0607227}}].

\bibitem{Khare2009a}
C.~Khare and J.-P. Wintenberger, \emph{Serre's modularity conjecture. {I}},
  \href{https://doi.org/10.1007/s00222-009-0205-7}{\emph{Invent. Math.}
  {\bfseries 178} (2009) 485}.

\bibitem{Khare2009b}
C.~Khare and J.-P. Wintenberger, \emph{Serre's modularity conjecture. {II}},
  \href{https://doi.org/10.1007/s00222-009-0206-6}{\emph{Invent. Math.}
  {\bfseries 178} (2009) 505}.

\bibitem{Kisin2009a}
M.~Kisin, \emph{Modularity of 2-adic {B}arsotti-{T}ate representations},
  \href{https://doi.org/10.1007/s00222-009-0207-5}{\emph{Invent. Math.}
  {\bfseries 178} (2009) 587}.

\bibitem{Livne1995a}
R.~Livn\'{e}, \emph{Motivic orthogonal two-dimensional representations of
  {${\rm Gal}(\overline {\bf Q}/ {\bf Q})$}},
  \href{https://doi.org/10.1007/BF02762074}{\emph{Israel J. Math.} {\bfseries
  92} (1995) 149}.

\bibitem{Milne2013a}
J.~S. Milne, \emph{Motives---{G}rothendieck's dream},  in \emph{Open problems
  and surveys of contemporary mathematics}, vol.~6 of \emph{Surv. Mod. Math.},
  pp.~325--342.
\newblock Int. Press, Somerville, MA, 2013.

\bibitem{Kadir:2010dh}
S.~Kadir, M.~Lynker and R.~Schimmrigk, \emph{{String Modular Phases in
  Calabi-Yau Families}},
  \href{https://doi.org/10.1016/j.geomphys.2011.04.010}{\emph{J. Geom. Phys.}
  {\bfseries 61} (2011) 2453}
  [\href{https://arxiv.org/abs/1012.5807}{{\ttfamily 1012.5807}}].

\bibitem{Schimmrigk:2006dy}
R.~Schimmrigk, \emph{{The Langlands program and string modular K3 surfaces}},
  \href{https://doi.org/10.1016/j.nuclphysb.2007.01.027}{\emph{Nucl. Phys. B}
  {\bfseries 771} (2007) 143}
  [\href{https://arxiv.org/abs/hep-th/0603234}{{\ttfamily hep-th/0603234}}].

\bibitem{Voisin2007a}
C.~Voisin, \emph{Hodge theory and complex algebraic geometry. {I}}, vol.~76 of
  \emph{Cambridge Studies in Advanced Mathematics}. Cambridge University Press,
  Cambridge, english~ed., 2007.

\bibitem{Griffiths1978a}
P.~Griffiths and J.~Harris, \emph{Principles of algebraic geometry}, Pure and
  Applied Mathematics. Wiley-Interscience [John Wiley \& Sons], New York, 1978.

\bibitem{Kontsevich2001a}
M.~Kontsevich and D.~Zagier, \emph{Periods},  in \emph{Mathematics
  unlimited---2001 and beyond}, pp.~771--808.
\newblock Springer, Berlin, 2001.

\bibitem{Frenkel:2005pa}
E.~Frenkel, \emph{{Lectures on the Langlands program and conformal field
  theory}},  in \emph{{Les Houches School of Physics: Frontiers in Number
  Theory, Physics and Geometry}}, pp.~387--533, 2007,
  \href{https://arxiv.org/abs/hep-th/0512172}{{\ttfamily hep-th/0512172}},
  \href{https://doi.org/10.1007/978-3-540-30308-4_11}{DOI}.

\bibitem{Dokchitser2004a}
T.~Dokchitser, \emph{Computing special values of motivic {$L$}-functions},
  {\emph{Experiment. Math.} {\bfseries 13} (2004) 137}
  [\href{https://arxiv.org/abs/math/0207280}{{\ttfamily math/0207280}}].

\bibitem{sagemath}
{The Sage Developers}, \emph{{S}ageMath, the {S}age {M}athematics {S}oftware
  {S}ystem ({V}ersion 9.7)}, 2023.

\bibitem{Verrill1996a}
H.~A. Verrill, \emph{Root lattices and pencils of varieties},
  \href{https://doi.org/10.1215/kjm/1250518557}{\emph{J. Math. Kyoto Univ.}
  {\bfseries 36} (1996) 423}.

\bibitem{Hulek2006a}
K.~Hulek and H.~Verrill, \emph{On the modularity of {C}alabi-{Y}au threefolds
  containing elliptic ruled surfaces},  in \emph{Mirror symmetry. {V}}, vol.~38
  of \emph{AMS/IP Stud. Adv. Math.}, pp.~19--34.
\newblock Amer. Math. Soc., Providence, RI, 2006.
\newblock \href{https://arxiv.org/abs/math/0502158}{{\ttfamily math/0502158}}.
\newblock \href{https://doi.org/10.1090/amsip/038/02}{DOI}.

\bibitem{Weigand:2010wm}
T.~Weigand, \emph{{Lectures on F-theory compactifications and model building}},
  \href{https://doi.org/10.1088/0264-9381/27/21/214004}{\emph{Class. Quant.
  Grav.} {\bfseries 27} (2010) 214004}
  [\href{https://arxiv.org/abs/1009.3497}{{\ttfamily 1009.3497}}].

\bibitem{Donagi:2009ra}
R.~Donagi and M.~Wijnholt, \emph{{Higgs Bundles and UV Completion in
  F-Theory}}, \href{https://doi.org/10.1007/s00220-013-1878-8}{\emph{Commun.
  Math. Phys.} {\bfseries 326} (2014) 287}
  [\href{https://arxiv.org/abs/0904.1218}{{\ttfamily 0904.1218}}].

\bibitem{Collinucci:2010gz}
A.~Collinucci and R.~Savelli, \emph{{On Flux Quantization in F-Theory}},
  \href{https://doi.org/10.1007/JHEP02(2012)015}{\emph{JHEP} {\bfseries 02}
  (2012) 015} [\href{https://arxiv.org/abs/1011.6388}{{\ttfamily 1011.6388}}].

\bibitem{Braun:2011zm}
A.~P. Braun, A.~Collinucci and R.~Valandro, \emph{{G-flux in F-theory and
  algebraic cycles}},
  \href{https://doi.org/10.1016/j.nuclphysb.2011.10.034}{\emph{Nucl. Phys. B}
  {\bfseries 856} (2012) 129}
  [\href{https://arxiv.org/abs/1107.5337}{{\ttfamily 1107.5337}}].

\bibitem{Marsano:2011hv}
J.~Marsano and S.~Sch{\"a}fer-Nameki, \emph{{Yukawas, G-flux, and Spectral
  Covers from Resolved Calabi-Yau's}},
  \href{https://doi.org/10.1007/JHEP11(2011)098}{\emph{JHEP} {\bfseries 11}
  (2011) 098} [\href{https://arxiv.org/abs/1108.1794}{{\ttfamily 1108.1794}}].

\bibitem{Krause:2011xj}
S.~Krause, C.~Mayrhofer and T.~Weigand, \emph{{$G_4$ flux, chiral matter and
  singularity resolution in F-theory compactifications}},
  \href{https://doi.org/10.1016/j.nuclphysb.2011.12.013}{\emph{Nucl. Phys. B}
  {\bfseries 858} (2012) 1} [\href{https://arxiv.org/abs/1109.3454}{{\ttfamily
  1109.3454}}].

\bibitem{Jockers:2016bwi}
H.~Jockers, S.~Katz, D.~R. Morrison and M.~R. Plesser, \emph{{SU(N) Transitions
  in M-Theory on Calabi\textendash{}Yau Fourfolds and Background Fluxes}},
  \href{https://doi.org/10.1007/s00220-016-2741-5}{\emph{Commun. Math. Phys.}
  {\bfseries 351} (2017) 837}
  [\href{https://arxiv.org/abs/1602.07693}{{\ttfamily 1602.07693}}].

\bibitem{Kim:2022jvv}
M.~Kim, \emph{{D-instanton superpotential in string theory}},
  \href{https://doi.org/10.1007/JHEP03(2022)054}{\emph{JHEP} {\bfseries 03}
  (2022) 054} [\href{https://arxiv.org/abs/2201.04634}{{\ttfamily
  2201.04634}}].

\bibitem{Schafer-Nameki:2016cfr}
S.~Sch\"afer-Nameki and T.~Weigand, \emph{{F-theory and 2d $(0, 2)$ theories}},
  \href{https://doi.org/10.1007/JHEP05(2016)059}{\emph{JHEP} {\bfseries 05}
  (2016) 059} [\href{https://arxiv.org/abs/1601.02015}{{\ttfamily
  1601.02015}}].

\bibitem{Yang:2020gwr}
W.~Yang, \emph{{K3 mirror symmetry, Legendre family and Deligne's conjecture
  for the Fermat quartic}},
  \href{https://doi.org/10.1016/j.nuclphysb.2020.115303}{\emph{Nucl. Phys. B}
  {\bfseries 963} (2021) 115303}
  [\href{https://arxiv.org/abs/2004.00820}{{\ttfamily 2004.00820}}].

\bibitem{Sethi:1996es}
S.~Sethi, C.~Vafa and E.~Witten, \emph{{Constraints on low dimensional string
  compactifications}},
  \href{https://doi.org/10.1016/S0550-3213(96)00483-X}{\emph{Nucl. Phys. B}
  {\bfseries 480} (1996) 213}
  [\href{https://arxiv.org/abs/hep-th/9606122}{{\ttfamily hep-th/9606122}}].

\bibitem{Greene:1993vm}
B.~R. Greene, D.~R. Morrison and M.~R. Plesser, \emph{{Mirror manifolds in
  higher dimension}}, \href{https://doi.org/10.1007/BF02101657}{\emph{Commun.
  Math. Phys.} {\bfseries 173} (1995) 559}
  [\href{https://arxiv.org/abs/hep-th/9402119}{{\ttfamily hep-th/9402119}}].

\bibitem{Becker:1996ay}
K.~Becker, M.~Becker, D.~R. Morrison, H.~Ooguri, Y.~Oz and Z.~Yin,
  \emph{{Supersymmetric cycles in exceptional holonomy manifolds and Calabi-Yau
  4 folds}}, \href{https://doi.org/10.1016/S0550-3213(96)00491-9}{\emph{Nucl.
  Phys. B} {\bfseries 480} (1996) 225}
  [\href{https://arxiv.org/abs/hep-th/9608116}{{\ttfamily hep-th/9608116}}].

\bibitem{Silverman2009a}
J.~H. Silverman, \emph{The arithmetic of elliptic curves}, vol.~106 of
  \emph{Graduate Texts in Mathematics}. Springer, Dordrecht, second~ed., 2009,
  \href{https://doi.org/10.1007/978-0-387-09494-6}{10.1007/978-0-387-09494-6}.

\bibitem{Zagier2008a}
D.~Zagier, \emph{Elliptic modular forms and their applications},  in \emph{The
  1-2-3 of modular forms}, Universitext, pp.~1--103.
\newblock Springer, Berlin, 2008.
\newblock \href{https://doi.org/10.1007/978-3-540-74119-0\_1}{DOI}.

\bibitem{Wiles:1995ig}
A.~J. Wiles, \emph{{Modular elliptic curves and Fermat's last theorem}},
  \href{https://doi.org/10.2307/2118559}{\emph{Annals Math.} {\bfseries 141}
  (1995) 443}.

\bibitem{Breuil2001a}
C.~Breuil, B.~Conrad, F.~Diamond and R.~Taylor, \emph{On the modularity of
  elliptic curves over {$\mathbf Q$}: wild 3-adic exercises},
  \href{https://doi.org/10.1090/S0894-0347-01-00370-8}{\emph{J. Amer. Math.
  Soc.} {\bfseries 14} (2001) 843}.

\bibitem{Diamond1996a}
F.~Diamond, \emph{On deformation rings and {H}ecke rings},
  \href{https://doi.org/10.2307/2118586}{\emph{Ann. of Math. (2)} {\bfseries
  144} (1996) 137}.

\bibitem{Taylor1995a}
R.~Taylor and A.~Wiles, \emph{Ring-theoretic properties of certain {H}ecke
  algebras}, \href{https://doi.org/10.2307/2118560}{\emph{Ann. of Math. (2)}
  {\bfseries 141} (1995) 553}.

\bibitem{Huybrechts2016a}
D.~Huybrechts, \emph{Lectures on {K}3 surfaces}, vol.~158 of \emph{Cambridge
  Studies in Advanced Mathematics}. Cambridge University Press, Cambridge,
  2016,
  \href{https://doi.org/10.1017/CBO9781316594193}{10.1017/CBO9781316594193}.

\bibitem{Aspinwall:1996mn}
P.~S. Aspinwall, \emph{{K3 surfaces and string duality}},  in
  \emph{{Theoretical Advanced Study Institute in Elementary Particle Physics
  (TASI 96): Fields, Strings, and Duality}}, pp.~421--540, 11, 1996,
  \href{https://arxiv.org/abs/hep-th/9611137}{{\ttfamily hep-th/9611137}}.

\bibitem{Bonisch:2023}
K.~B\"onisch, \emph{{Modularity of special motives of rank four associated with
  Calabi-Yau threefolds}}, Ph.D. thesis, University of Bonn, 2022.

\bibitem{MR33411}
G.~Bol, \emph{{I}nvarianten {L}inearer {D}ifferentialgleichungen},
  \href{https://doi.org/10.1007/BF03343515}{\emph{Abh. Math. Sem. Univ.
  Hamburg} {\bfseries 16} (1949) 1}.

\bibitem{MR2385372}
J.~H. Bruinier, G.~van~der Geer, G.~Harder and D.~Zagier, \emph{The 1-2-3 of
  modular forms}, Universitext. Springer-Verlag, Berlin, 2008,
  \href{https://doi.org/10.1007/978-3-540-74119-0}{10.1007/978-3-540-74119-0}.

\bibitem{Eichler1957a}
M.~Eichler, \emph{Eine {V}erallgemeinerung der {A}belschen {I}ntegrale},
  \href{https://doi.org/10.1007/BF01258863}{\emph{Math. Z.} {\bfseries 67}
  (1957) 267}.

\bibitem{Shimura1971a}
G.~Shimura, \emph{Introduction to the arithmetic theory of automorphic
  functions}, vol.~No. 1 of \emph{Kan\^o{} Memorial Lectures}. Iwanami Shoten
  Publishers, Tokyo; Princeton University Press, Princeton, NJ, 1971.

\bibitem{Pasol2013a}
V.~Pa\c{s}ol and A.~A. Popa, \emph{Modular forms and period polynomials},
  \href{https://doi.org/10.1112/plms/pdt003}{\emph{Proc. Lond. Math. Soc. (3)}
  {\bfseries 107} (2013) 713}.

\bibitem{Manin1973a}
J.~I. Manin, \emph{Periods of cusp forms, and {$p$}-adic {H}ecke series},
  {\emph{Mat. Sb. (N.S.)} {\bfseries 92(134)} (1973) 378}.

\bibitem{Schimmrigk:2008mp}
R.~Schimmrigk, \emph{{Emergent spacetime from modular motives}},
  \href{https://doi.org/10.1007/s00220-010-1179-4}{\emph{Commun. Math. Phys.}
  {\bfseries 303} (2011) 1} [\href{https://arxiv.org/abs/0812.4450}{{\ttfamily
  0812.4450}}].

\bibitem{Huber2017a}
A.~Huber and S.~M\"{u}ller-Stach, \emph{Periods and {N}ori motives}, vol.~65 of
  \emph{Ergebnisse der Mathematik und ihrer Grenzgebiete. 3. Folge. A Series of
  Modern Surveys in Mathematics [Results in Mathematics and Related Areas. 3rd
  Series. A Series of Modern Surveys in Mathematics]}. Springer, Cham, 2017.

\bibitem{Brown2016a}
F.~Brown, \emph{Periods, galois theory and particle physics}, {\emph{Gergen
  Lecture} (2016) }.

\bibitem{Filippini2015a}
S.~A. Filippini, H.~Ruddat and A.~Thompson, \emph{An introduction to {H}odge
  structures},  in \emph{Calabi-{Y}au varieties: arithmetic, geometry and
  physics}, vol.~34 of \emph{Fields Inst. Monogr.}, pp.~83--130.
\newblock Fields Inst. Res. Math. Sci., Toronto, ON, 2015.
\newblock \href{https://doi.org/10.1007/978-1-4939-2830-9\_4}{DOI}.

\end{thebibliography}\endgroup

\end{document}